\documentclass[11pt,twoside]{report}
\usepackage{a4,epsf,amssymb,amsmath,latexsym,theorem}
{\theorembodyfont{\rm} }
\def \slas{\kern -6.2pt /}
\def \sla{\kern -5.4pt /}
\def \sl{\kern -4.0pt /}
\def \Cslas{\kern -6.8pt /}
\def \Dslas{\kern -7.4pt /}
\def \slass{\kern -7.4pt /}
\def \OO{{\cal O}}
\def \QQ{{\cal Q}}
\def \ii{{\mathrm{i}}}
\def \d{{\mathrm{d}}}
\def \dd#1{\frac{\mathrm{d}}{\mathrm{d}#1}}
\def \lcd{\tilde{\partial}}
\def \pd{\partial}
\def \e{{\mathrm{e}}}
\def \lcxg{x\slas}

\def \lcx{\tilde{x}}
\newcommand\ka{\kappa_1}
\newcommand\kb{\kappa_2}

\newcommand\xx{\tilde{x}}
\def \tl#1{\overset{\kern 2pt\circ}{#1}}
\def \tll#1{\overset{\kern -1pt\circ}{#1}}
\def \TL#1{\overset{\kern -28pt \circ}{#1}}
\def \TLL#1{\overset{\kern -7pt \circ}{#1}}
\def \relstack#1#2{\mathrel{\mathop{#2}\limits_{#1}}}
\newcommand{\SC}{\scriptstyle}
\def \Tensor#1{\overset{\leftrightarrow}{#1}}
\def \LTensor#1{\overset{\!\!\!\leftrightarrow}{#1}}
\def \LD#1{\overset{\,\leftarrow}{#1}}
\def \RD#1{\overset{\,\,\rightarrow}{#1}}
\def \LDL#1{\overset{\!\!\!\leftarrow}{#1}}
\def \RDL#1{\overset{\!\rightarrow}{#1}}

\newcommand{\gvh}{\hat g_\perp^{(v)}}
\newcommand{\gah}{\hat g_\perp^{(a)}}
\newcommand{\gvn}{g_{\perp n}^{(v)}}

\newcommand{\hsh}{\hat h_\parallel^{(s)}}

\newcommand{\htth}{\hat h_\parallel^{(t)}}

\newcommand{\httn}{h_{\parallel n}^{(t)}}

\renewcommand{\theequation}{\arabic{chapter}.\arabic{section}.\arabic{equation}}

\begin{document}

{\thispagestyle{empty}
\title{ 
\Huge{\bf{Group Theoretical Analysis 
of Light-Cone Dominated Hadronic Processes and Twist Decomposition 
of Nonlocal Operators
in Quantum~Chromodynamics}}\\ 
\vspace*{1.0cm}
}
\author{
\vspace*{1.5cm}
{{\LARGE{\bf  Markus Lazar}}}\\
\hspace*{1cm}
Fakult{\"a}t f{\"u}r Physik und Geowissenschaften \\
\vspace*{1.5cm}
der Universit{\"a}t Leipzig\\
\vspace*{1.0cm}
\vspace*{1.0cm}
{{\Large{\sf Doctoral Thesis}}}\\
June 17, 2002
\date{}}

\maketitle


\begin{abstract}


In the framework of nonlocal light-cone expansion of two 
current operators we construct bilocal as well as trilocal 
QCD light-cone operators with definite geometric twist. 
We are able to decompose uniquely the appearing QCD light-cone operators 
into all their twist parts on the light-cone.

The quark distribution functions and the vector meson distribution
amplitudes which are defined as forward and skewed nucleonic matrix 
elements of nonlocal light-cone operators are classified with respect
to geometric twist. Hereby we found well-kown and new Wandzura-Wilczek 
type relations between the quark distribution functions with dynamical twist and, additionally, 
between the vector meson distribution amplitudes with dynamical twist. 

Using group theoretical methods we are able to construct all bilocal
off-cone QCD operators with twist 2 and 3. Additionally, we calculate their
nucleonic matrix elements (double distribution functions, scalar and 
vector meson distribution amplitudes). In some special cases we give 
also the twist decomposition of off-cone QCD operators
from minimal up to infinite twist. 
\end{abstract}

{\pagenumbering{roman}
\tableofcontents
\cleardoublepage
}
\pagenumbering{arabic}
\setcounter{page}{1}

\newpage
\chapter{Introduction}
\setcounter{equation}{0} 

Quantum Chromodynamics (QCD) which is a 
relativistic local quantum field theory based on the 
non-Abelian colour gauge group $SU_{\rm c}(3)$
is well established nowadays as a microscopic theory of strong interaction.
Because the gauge group is non-Abelian
a gauge theory of this kind is called  Yang-Mills theory.
In a gauge theory exist fields which transform 
according to the adjoint representation (gauge fields) of the gauge group
and spinor fields which transform as a 
fundamental representation (matter fields).
The matter fields in QCD are ``coloured'' quarks. The interaction of quarks 
is mediated by the gauge bosons (gluons). 
Quarks and gluons form composite particles called
hadrons (mesons and baryons).
Due to presently unsolved bound state problem it is not possible to 
predict measurable observables from the quark and gluon fields.
Although 
the high energy behaviour of QCD is believed to be described by the
perturbation theory due to the asymptotically free nature of QCD, 
the cross sections
are complicated combinations of short- and long-distance interactions
and is not directly calculable.
The standard approach to overcome these difficulties is based on the 
factorization theorem which allows to separate (factorize) the soft 
and hard part of the cross section in a systematic fashion.
The soft part (long-distance part) can be described by expection values of 
QCD operators sandwiched between hadronic states.
Due to the confinement this part can be determined only from experimental data
or a suitable lattice calculation. 
On the other hand, the hard subprocess (short-distance part) and 
the evolution of the functions 
which are occurring in the soft part can be systematically calculated 
with the help of perturbation theory. 
The success of perturbative QCD in predicting the momentum dependence of 
hadronic observables serves an important argument for the correctness 
of the theory.
So far all experimental data are consistent with the predictions of QCD. 

The universal, non-perturbative parton distribution amplitudes 
parametrizing, modulo kinematical factors, the matrix elements of 
appropriate non-local quark-antiquark (as well as gluon) operators 
play a central role in phenomenological considerations. These 
operators occur in the quantum field theoretic description 
of light-cone dominated hadronic processes via the non-local light-cone 
expansion~\cite{AZ78,ZAV,mul94}.
For deep inelastic lepton-hadron scattering and Drell-Yan processes 
the parton distributions are given as forward matrix elements of bilocal 
light-ray operators resulting from the time-ordered product of appropriate 
hadronic currents. For (deeply) virtual Compton scattering 
and hadron wave functions the so-called double distributions and hadron
distribution amplitudes, respectively, are given
by corresponding non-forward matrix elements.  Thereby, various processes 
are governed by one and the same (set of) light-ray operators.
The computation of the anomalous dimensions of these 
operators is one of the most important problems of the theory, 
since these quantities determine the 
deviation from the
scaling behaviour of experimental observables.

The success of QCD as fundamental theory of strong
interactions is intimately tied to its ability to describe hard
exclusive and 
inclusive reactions, which has been tested in numerous experiments. In
the corresponding kinematic regime, i.e.,\ at large space-like virtualities,
the relevant amplitudes are dominated by singularities on the
light-cone, which, in the framework of a light-cone expansion, can be
described in terms of contributions of definite twist. The notion of
(geometric) twist, for local operators, 
has been introduced originally by Gross and Treiman~\cite{Gross} as
twist = (canonical) dimension$\,-\,$ (Lorentz) spin;
it uses the irreducible representations of the orthochronous
Lorentz group and, as such, is a Lorentz-invariant concept.
Using the notion of twist, one is able to expand systematically 
the cross section in inverse powers of a characteristic large momentum 
scale $Q^2$, and extract the leading contributions. 
The higher twist effects describe the quark-gluon behaviour inside the 
nucleon, and thus contain information about the quark-gluon correlations.
Possibly, in the near future the experimental precision data  
will allow for the determination of
non-leading contributions and, therefore, require for a careful
analysis of the various sub-dominant effects contributing to the
physical processes. When considered beyond leading order, i.e.,~beyond
lowest twist operators in tree approximation, one is confronted 
not only with radiative corrections but also with power corrections
resulting from higher twist as well as target mass effects. 
Higher twist contributions are obtained by the decomposition of the
operators with respect to (irreducible tensor) representations 
of the Lorentz group having definite twist $\tau$. 
These representations are characterized
by their symmetry type (under index permutations of the traceless 
tensors) which is determined by corresponding Young tableaux.
With the growing accuracy of the experimental data a
perturbative expansion with respect to the 
experimentally relevant variable, $M^2/Q^2$, $M$ being a 
typical mass of the process, will be very useful. 

An alternative approach to twist-counting is based on the 
light-cone quantization in the infinite momentum frame~\cite{KS70} 
with the decomposition of the quark fields into
"good" and "bad" components, $\psi = \psi_+ + \psi_-$, with
$\psi_\pm = \frac{1}{2}\gamma^\mp\gamma^\pm \psi$.
As has been
pointed out in~\cite{JJ91,JJ92}, a "bad" component introduces one
unit of $M/Q$ because these components are not
dynamically independent and may be expressed through the 
equations of motion by the "good" ones times the above
 kinematical factor. Despite being 
conceptionally different this definition looks similar to the
phenomenologically more convenient one which counts only 
powers of $1/Q$ in the infinite momentum frame~\cite{JJ91,JJ92}.
However, these power--counting concepts 
of "dynamical twist" are not Lorentz invariant and
do not agree with the original
definition. Furthermore, the same power of $1/Q$ may occur for
different values of $\tau$. 
The mismatch between dynamical and
geometric twist becomes relevant once power-suppressed higher-twist
contributions are included and
leads to so-called Wandzura-Wilczek (WW) relations
between matrix-elements of operators of 
different dynamical, but identical geometric
twist, the prototype of which has been derived by Wandzura and 
Wilczek for the nucleon distribution  functions (DFs) 
$g_1$ and $g_2$~\cite{Wandzura}. 

In addition, the ``dynamical twist'' is
{\em applicable only for matrix elements} of the operators
 and, therefore, is not directly related to the operators
itself. Even more, if the matrix elements contain more than
one momentum, as will be the case for the deeply virtual Compton scattering, 
the definition
of $Q$ is quite ambiguous and different definitions are related
according to $Q \rightarrow Q(1 + f(p_\pm, q) M^2/Q^2+ \ldots)$
with some Lorentz invariant factor $f$.
Therefore, from the point of view of renormalized quantum
field theory only the original geometric definition of twist is
a well-defined concept. The problem which remains is
how this concept has to be generalized to the nonlocal operators
appearing in the nonlocal light--cone expansion.

Because the (geometrical) twist is a straightforward scheme to decompose
local operators and to classify generalized parton as well as 
skewed distribution amplitudes we generalize this notation for nonlocal
operators, in principle, for any twist. The considered operators are 
necessary for the investigation of different phenomenological 
QCD processes.
We perform this program for light-cone as well as off-cone QCD operators. 

This thesis is organized as follows. 
The Section~\ref{ope} is mainly introductory.
It collects the necessary definitions, physical motivations and explains the 
basic ideas. 
In Section~\ref{proc} 
we give a motivation why we use the concept of geometric twist for 
nonlocal operators.
We collect and explain the steps 
for the procedure of twist decomposition of nonlocal operators.
In Section~\ref{forward} and \ref{meson} we give the straightforward 
classification with respect to (geometric) twist 
of quark distribution amplitudes and 
vector meson distribution amplitudes on the light-cone.
These both classifications are quite similar because
the  same light-cone operators are related to 
different hadronic processes.
A further problem, which we will solve, 
is how the (geometrical) twist decomposition of parton distributions
which has been given in this work relates to the (dynamical) twist
decomposition of parton distributions given, e.g.,~by
Jaffe and Ji \cite{JJ91,JJ92} and to the decomposition of 
vector meson distribution amplitudes given by 
Ball and Braun~\cite{ball98,ball99}. 
The interrelations between the dynamical and geometrical distribution
amplitudes are calculated
and used to derive Wandzura-Wilczek type
relations between the dynamical twist distribution functions. 
Most of them are new and might be of phenomenological interest.
The material presented in these two Sections was first published 
in Ref.~\cite{GL01} and  in Refs.~\cite{Lazar01a,Lazar01b}.
In Section~\ref{harmonic} we calculated the power corrections of
double distribution amplitudes, and
meson as well as pion distribution amplitudes with definite geometric twist. 
With the help of the harmonic extension we are able to resum the 
target mass or
power corrections in general non-forward matrix elements.
The material presented in this Section is based on Ref.~\cite{GLR01}.

The Section~\ref{tensor} contains all formulas of the so-called 
tensor polynomials 
which are relevant for the twist or spin decomposition in $2h$-dimensional 
spacetime in this work. The constructed tensor polynomials are irreducible
with respect to $SO(1,2h-1)$ and $SO(2h)$, respectively, on-cone as well as 
off-cone. 
They are given on-cone as well as off-cone. 

In Section~\ref{gluon} we 
give systematically the complete twist decomposition of the various gluonic 
bilocal light--ray tensor operators of second rank being specified 
by different symmetry classes; their twist content ranges from 2 
up to 6. 

In Section~\ref{trilocal} we extend these results with minor 
modifications to the trilocal light--ray operators: 
trilocal quark--gluon correlation operators 
(related to so-called Shuryak-Vainshtein operators) and 
four--fermion operator, as well as multilocal quark and gluon operators.

In Section~\ref{off-cone} we present the general twist decomposition
of off-cone operators. 
The procedure is very close to
the so-called Nachtmann method for the mass corrections in
inclusive processes.
This approach is based on (irreducible) harmonics of $SO(1,3)$ and $SO(4)$, respectively.
From the group theoretical point of view it also works for
the case of non-forward processes.
The power of that approach consists in the fact that one determines
the twist decomposition at first for the local and nonlocal operators
and only afterwards takes the matrix elements.
For totally symmetric (local) operators we are able to study the decomposition
up to infinite twist.
Additionally we give the twist decomposition of operators 
with other symmetry types  up to twist 2 and 3. 

The technical details which are needed for the twist decomposition
are collected in the Appendixes.
In Appendix~\ref{young} a short exposition about the characterization of 
irreducible tensor representations through the Young tableaux is given and
the Appendix~\ref{lorentz} contains  
some basic notation with respect to the Lorentz group. 
Appendix~\ref{inner} is devoted to the concept of an interior differential operator
on the light-cone which was originally introduced by Bargmann and Todorov~\cite{BT77}.
In Appendix~\ref{mass} we investigate the target mass corrections for
a scalar theory in $2h$-dimensional spacetime. In this framework we calculate the
corresponding scattering amplitude.


\chapter{Quantum field theoretical treatment of light-cone dominated 
hadronic processes}
\section{Light-cone dominated scattering processes and  
light-cone expansion in QCD}
\label{ope}
In this Section we discuss the quantum field theoretical description 
of light-cone dominated scattering processes. 
In fact, we consider some scattering amplitudes, QCD-operators and 
corresponding nucleonic matrix elements. 
Unfortunately, so far it has not been possible to make an ab initio calculation 
of a QCD scattering amplitude on hadrons. The main reason is due to the fact
that hadrons are bound states of both quarks and gluons (confinement). 
Anyway, until the confinement is not better understood,
a way out to calculate QCD processes is the light-cone expansion together
with the renormalization group equation.

\setcounter{equation}{0}

\subsection{Phenomenology of hard scattering processes in QCD}
The scattering amplitudes of light-cone dominated hadronic processes, 
e.g. DIS, 
according to the factorization hypothesis -- are usually represented
by the convolution of the hard (process-dependent) scattering amplitude 
of the partons with appropriate soft (process-independent) parton distribution
amplitudes. 
For many hard scattering processes in QCD which are dominated 
on the light-cone, 
the nonlocal light--cone (LC) expansion, together with the 
renormalization group equation, is a powerful tool to 
determine the dependence of the nonperturbative 
distribution amplitudes on the experimentally relevant
momentum transfer $Q^2$. This is the case for,
e.g.,~the parton distributions in deep inelastic
scattering (DIS), the pion as well as vector meson distribution
amplitudes and, as growing up more recently, the
non--forward distribution functions in deeply virtual
Compton scattering (DVCS). 

First, we discuss the relation between the forward virtual Compton scattering 
and the DIS.
The virtual Compton amplitude for the forward case is defined by
\begin{equation}
\label{COMP}
T_{\mu\nu}(P,Q) 
=\ii \int \d^4x \,\e^{\ii qx}\,
\langle P, S\,|T (J_{\mu}(x/2) J_{\nu}(-x/2))|\,P,S\rangle,
\end{equation}
as shown graphically in Fig.~\ref{blob}.
\begin{figure}[h]
\centerline{\epsfxsize=5cm\epsffile{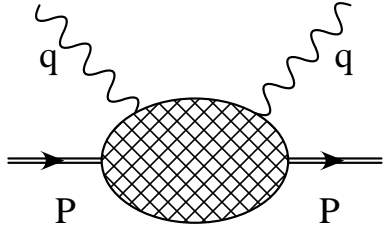}}
\caption{\label{blob} Kinematic notation for the forward Compton amplitude.}
\end{figure}
Here, $|P,S\rangle$ is the hadron state with momentum $P$ and spin $S$.
The imaginary part of the virtual Compton scattering amplitude, 
$T^{\mu\nu}(P,Q^2)$, is related to the hadronic tensor $W^{\mu\nu}$ 
with the help of the optical theorem according to
\begin{align}
W^{\mu\nu}=\frac{1}{2\pi}\, {\rm Im}\, T^{\mu\nu}.
\end{align}
The hadronic tensor is given by
\begin{align}\label{hadten}
W^{\mu\nu}=\frac{1}{4\pi}\int\d^4 x\,\e^{\ii qx}
\langle P,S|\left[J^\mu(x/2),J^\nu(-x/2)\right]|P,S\rangle .
\end{align}
It describes the internal structure of the nucleon in deep inelastic scattering (DIS).
DIS is a scattering process between leptons and a fixed hadronic target. The incoming
lepton ($l$) with momentum $k$ interacts with the hadron ($N$) target with momentum $P$ 
through the exchange of a virtual photon of momentum $q=k-k^\prime$
if we consider only the QED interaction between the lepton and the nucleon and keep
only the lowest order (see Fig.~\ref{dep}).
\begin{figure}[h]
\centerline{\epsfxsize=10cm\epsffile{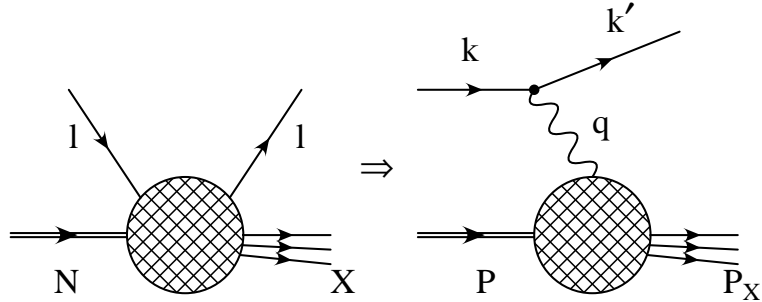}}
\caption{\label{dep} The DIS process in Born approximation.}
\end{figure}
Thus, it is possible to calculate the hadronic tensor by means of the virtual
Compton amplitude in the framework of OPE. 
Let $S_\alpha$ be the spin vector of the nucleon so
that
\begin{align}
\label{kin1}
P^2=M^2,\qquad S^2=-M^2,\qquad (P S)=0, 
\end{align}
where $M$ denotes the mass of the nucleon target.

On the other hand, using Lorentz covariance, gauge invariance, parity conservation and discrete
symmetries of the strong interaction
the hadronic tensor for the case of a nucleon target 
can be expressed in 
the Bjorken limit:
\begin{eqnarray}
\label{}
\nu =  qP \longrightarrow \infty, \qquad
Q^2 = - q^2 \longrightarrow \infty
\end{eqnarray}
with the scaling variable
\begin{align}
\xi  =  \frac{Q^2 }{2\nu}\qquad\text{fixed}
\end{align}
by the four
scalar structure functions $F_1$, $F_2$, $g_1$ and $g_2$ according to
\begin{align}
\label{HT}
W^{\mu\nu}&=\left(\frac{q^\mu q^\nu}{q^2}-g^{\mu\nu}\right)F_1(\xi,Q^2)+
\left(P^\mu-\frac{(Pq)}{q^2}q^\mu\right)\left(P^\nu-\frac{(Pq)}{q^2}q^\nu\right)
F_2(\xi,Q^2)\nonumber\\
&\qquad -\ii\epsilon^{\mu\nu\alpha\beta} q_\alpha
\left(\frac{S_\beta}{(Pq)}G_1(\xi,Q^2)+\frac{S_\beta(Pq)-P_\beta(Sq)}{(Pq)^2}
G_2(\xi,Q^2)\right)\ .
\end{align}
The symmetric (antisymmetric) part of $W^{\mu\nu}$ is related to the 
unpolarized (polarized) scattering.
The structure functions $F_1$ and $F_2$ can be measured by using an unpolarized 
beam and target,
the structure functions $g_2$ and $g_2$ require both a polarized beam and 
a polarized target.
The differential cross-section for inclusive scattering 
$l(k)+N(P)\rightarrow l(k')+X(P_X)$ (see Fig.~\ref{dep}) 
can be separated into the leptonic ($L^{\mu\nu}$)
and the hadronic tensor. The leptonic tensor can be calculated completely
in the framework of QED. Unlike the leptonic tensor, the
hadronic tensor cannot be computed directly from QCD because of non-perturbative 
effects in the strong interaction dynamics.

In the Bjorken limit the structure functions are functions of $\xi$ and independent
of $Q^2$ to leading order, a property known as scaling. 
Radiative corrections in QCD produce a small logarithmic $Q^2$ dependence 
of the structure functions which is calculable for large $Q^2$ since QCD is 
asymptotically free. Accordingly, the scale invariance (dilation symmetry) 
in QCD is broken
due to quantum corrections and introduces scaling violations (anomalous dimensions).
The $Q^2$ evaluation of the structure functions or parton distribution functions
is calculable in perturbative QCD.  It is governed by the anomalous dimensions of the corresponding
operators and the renormalization group equation.

Second, we discuss the off-forward virtual Compton scattering.
The Compton amplitude for the scattering of a virtual photon off
a hadron provides another possibility in QCD 
to understand the short-distance behaviour of the theory.
The Compton amplitude for this general case of non-forward scattering is given
by
\begin{equation}
\label{COMP-nf}
T_{\mu\nu}(P_+,P_-,Q) 
=\ii \int \d^4x \,\e^{\ii qx}\,
\langle P_2, S_2\,|T (J_{\mu}(x/2) J_{\nu}(-x/2))|\,P_1,S_1\rangle.
\end{equation}
Here,
\begin{align}
&P_+ = P_2 + P_1,\qquad P_- = P_2 - P_1 = q_1 - q_2,\\
&q = (q_1 + q_2)/2,\qquad P_1+q_1=P_2+q_2,
\end{align}
where $q_1(q_2)$ and $P_1(P_2)$ denote the four-momenta of the incoming(outgoing)
photon and hadron, respectively, and $S_1$, $S_2$ are the spins of the 
initial- and final-state hadron.
The kinematics of this process is depicted schematically in Fig.~\ref{blob-nf}.
\begin{figure}[h]
\centerline{\epsfxsize=5cm\epsffile{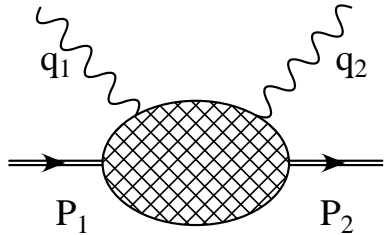}}
\caption{\label{blob-nf} Kinematic notation for the off-forward Compton amplitude.}
\end{figure}
To specify the asymptotics of 
virtual Compton scattering the generalized Bjorken 
region will be defined as follows
\begin{eqnarray}
\label{gBr1}
\nu =  \frac{qP_+}{2} \longrightarrow \infty, \qquad
Q^2 = - q^2 \longrightarrow \infty
\end{eqnarray}
with the two independent scaling variables
\begin{align}
\xi  =  \frac{Q^2 }{qP_+},\qquad
\eta = \frac{qP_-}{qP_+} = \frac{q_1^2 - q_2^2}{2\nu},
\end{align}
to be held fixed by the experimental setup. These variables
which are not restricted to the interval $[-1, +1]$ are
obvious generalizations of the usual Bjorken variable
and the ``skewedness'' parameter. 
Of special experimental importance are the cases of deep inelastic
scattering (DIS) described by the absorptive part of the forward Compton
amplitude (see above), $\eta=0$, and the deeply virtual Compton 
scattering (DVCS)\cite{Ji97}
with one real outgoing photon $q^2_2=0$ corresponding to $\xi=-\eta$.
In the generalized Bjorken region the amplitude~(\ref{COMP-nf}) is dominated 
by the light-cone singularities which allows to apply the OPE. 
For DIS as well as DVCS
the relevant information about the non-perturbative structure of the nucleon
is contained in the expectation values of QCD operators in the nucleon, which can 
be interpreted as the (generalized) parton distributions and ``skewed'' 
parton distributions, respectively.

Similar considerations also hold for hard (light-cone dominated) exclusive processes
in QCD. The amplitudes of those processes are expressed by
the factorization formulas which separate the short-distance part from the
long-distance one. 
Vector mesons ($V=\rho$, $\omega$, $K^*$, $\phi$) are produced in processes 
like exclusive semileptonic (or radiative) $B$ decays 
($B\rightarrow v e\nu$, $B\rightarrow V+\gamma$)
and hard electroproduction of vector mesons 
($\gamma^*+N\rightarrow V+ N'$). 
The physical interest in this case arises from the fact that the following 
vertex function is directly observable in QCD:
\begin{align}
\label{TA}
T^V_{\mu\nu}(q,P;\lambda) = \ii \int \d^4 x\, \e^{\ii qx}
\langle 0| RT\left(J_\mu(x/2)J_\nu(-x/2){\cal S}\right)
|V(P,\lambda) \rangle,
\end{align}
where $J_\mu$ can be axial or vector currents and $|V(P,\lambda)\rangle$
is the vector meson state of momentum $P$ and helicity $\lambda$.
Here, the universal nonperturbative quantities, the meson distribution 
amplitudes (DAs), sometimes also called meson wave functions,
are given by means of the vacuum-to-meson matrix elements of bi-local
operators on the light-cone,
$\langle 0|O_\Gamma(\lcx,-\lcx) |V(P,\lambda)\rangle$,
with $\Gamma = \{1, \gamma_5; \gamma_\alpha, \gamma_\alpha \gamma_5;
\sigma_{\alpha\beta}, \sigma_{\alpha\beta} \gamma_5 \}$ 
and describe the long-distance part.

\subsection{Light-cone expansion}
The light-cone expansion (LCE) is a straightforward scheme to treat different
light-cone dominated scattering processes. 
It is possible to obtain asymptotically valid informations about
these processes in this framework.
There exist different versions of the LCE, e.g., the local and the nonlocal
LCE.

The local LCE is a generalization of Wilson's short distance expansion~\cite{Wilson69}.
For scalar currents $j(x)=\phi(x)\phi(x)$  the local OPE is given by~\cite{BP71}:
\begin{equation}\label{l_ope}
T(j(x)j(0))\relstack{x^2\rightarrow 0}{\approx}\sum_{n=0}^\infty C_{n}(x^2)
x^{\mu_1}\ldots x^{\mu_n}\,O_{\mu_1\ldots\mu_n}(0)+\text{higher order terms} .
\end{equation}
Here the considered operator product is represented as an infinite sum 
(tower) of local operators, 
\begin{align}
O_{\mu_1\ldots\mu_n}(0)=\phi(0)\pd_{\mu_1}\ldots\pd_{\mu_n}\phi(0),
\end{align}
with canonical dimension $d_n$ which is equal to the scale 
dimension\footnote{The scale dimension 
of the local operator $O_{\mu_1\ldots\mu_n}(y)$
is given by the transformation behaviour with respect to dilations 
at the point $y=0$.} 
in free-field theory.
Because this operator is completely symmetric
and traceless ($x^2=0$) on the light-cone it has the maximal Lorentz-spin $j$.
Usually local operators are classified with respect to their
(geometrical) twist $\tau=d_n-j$~\cite{Gross}. 
Here $C_{n}(x^2)$ are local singular c-number coefficient functions 
which can be classified according to their degree of singularity
at $x^2=0$. 
Operators of the same singularity at $x^2=0$ are of the same
importance. 
Coefficient functions with maximal singularity appear 
in leading order.
A disadvantage of the local LCE is that Eq.~(\ref{l_ope})
is not a true operator identity. 
It exists on a dense subspace of the Fock space only~\cite{ZAV,BR80}.

The standard form of nonlocal light-cone expansion in scalar field theory
was introduced by Anikin and Zavialov~\cite{AZ78}.
It is given by the following expression:
\begin{equation}\label{nl_ope}
R\, T(j(x)j(0) S)\relstack{x^2\rightarrow 0}{\approx} 
\int_0^1\d\kappa_1\int_0^1\d\kappa_2\,
C(x^2,\kappa_1,\kappa_2)\,O(\kappa_1\lcx,\kappa_2\lcx)+
\text{higher order terms},
\end{equation}
which represents a true operator identity.
A rigorous proof of the nonlocal LCE has been given 
in \cite{AZ78} for the scalar field theory.
This nonlocal OPE holds on the whole Hilbert space in contrast 
to local OPE which is only justified in a restricted 
region of the Hilbert space~\cite{ZAV,BR80}.
Here $O(\kappa_1\lcx,\kappa_2 \lcx)$ are bilocal light-ray operators 
on a straight light-like path
between $\kappa_1\lcx$ and $\kappa_2\lcx$. 
The light-like vector
\begin{align}
\label{x-lc}
\lcx=x+\rho\,\frac{(x\rho)}{\rho^2}
\left(\sqrt{1-\frac{x^2\rho^2}{(x\rho)^2}}-1\right),
\quad\text{with}\quad \lcx^2=0
\end{align}
is related to $x$ and a fixed non-null four-vector $\rho$ whose dependence
drops out in leading order expressions.
The nonlocal LCE is in some sense an integral or summed up representation 
of the local LCE. 
The nonlocal and local operators are related through the following formulas:
\begin{eqnarray}\label{summ}
O(\kappa_1 \lcx,\kappa_2 \lcx)&=&\sum_{n_1,n_2}^\infty\frac{\kappa_1^{n_1}}{n_1!}\frac{\kappa_2^{n_2}}{n_2!}\,
O_{n_1,n_2}(\lcx)\, ,\\
O_{n_1,n_2}(\lcx)&=&\left(\frac{\partial}{\partial\kappa_1}\right)^{n_1}
\left(\frac{\partial}{\partial\kappa_2}\right)^{n_2}\,
O(\kappa_1 \lcx,\kappa_2 \lcx)|_{\kappa_1=\kappa_2=0}\, .\nonumber
\end{eqnarray}

In Refs.~\cite{Karch83,geyer85,mul94,balitsky83,BB88} 
some aspects of the nonlocal LCE for the case of QCD has been studied.
Following the Refs.~\cite{geyer85,mul94}
we use now the nonlocal OPE for the case of QCD.
In this way we consider the renormalized ($R$) time-ordered ($T$) product of
two electromagnetic currents
\begin{align}
R\,T\big(J^\mu(x/2)J^\nu(-x/2){\cal S}\big),
\end{align}
where the renormalized $S$-matrix is given by
\begin{align}
{\cal S}=R\,T\,\exp\left(\ii\int\d^4 x\, {\cal L}_{\rm eff}\right).
\end{align}
Here ${\cal L}_{\rm eff}$ denotes the effective Lagrangian of QCD~\cite{IZ}
\begin{align}
{\cal L}_{\rm eff}=
-\frac{1}{4}\, F^a_{\mu\nu} F^{a\mu\nu}
+\bar\psi\left(\ii\gamma^\mu D_\mu(A)-m\right)\psi
-\frac{1}{2\alpha}\,\big(\pd^\mu A_\mu^a\big)^2
+\bar c ^a\pd^\mu D^{ab}_\mu(A) c^b,
\end{align}
which contains the gluon field $A^a_\mu(x)$, the ghost fields $c^a(x)$
and $\bar c^a(x)$, 
the quark field $\psi(x)$, and $\alpha$ is the gauge fixing parameter. 
The indices $a$, $b$ are used for the field and the covariant 
derivative $D^{ab}_\mu(A)$ in the adjoint representation 
of the gauge group $SU(3)_{\rm colour}$. 
On the other hand,
the quark fields and the covariant derivative 
\begin{align}
D_\mu(A)=\pd_\mu +\ii g A^a_\mu(x) t^a
\end{align}
are given in the fundamental representation 
of this gauge group.

The electromagnetic current of the hadrons reads
\begin{align}
J^\mu(x)=\bar{\psi}(x)\gamma^\mu\,
\frac{1}{2}\left(\lambda^3_f+\frac{1}{\sqrt{3}}\lambda^8_f\right)\psi(x) ,
\qquad\text{for}\qquad SU(3)_{\rm flavour}.
\end{align}
In lowest order in the coupling constant (${\cal S}=1$) we obtain for the 
$T$-product of the two vector currents
\begin{align}\label{T-Prod}
T\big(J^\mu(x/2)J^\nu(-x/2)\big)&= -\bar{\psi}(x/2)\gamma^\mu\ii S_F(x)\gamma^\nu c^a\lambda^a_f\psi(-x/2)
                          -\bar{\psi}(-x/2)\gamma^\nu\ii S_F(x)\gamma^\mu c^a\lambda^a_f\psi(x/2)\nonumber\\
                          &\quad+\text{higher order terms},
\end{align}
where
\begin{equation}\label{Prop}
 S_F(x)\approx-\frac{\lcxg}{2\pi^2(x^2-\ii 0)^2}
-\frac{\ii m_q }{4\pi^2(x^2-\ii 0)}+{\cal O}(m_q^2)
\end{equation}
is the free quark-propagator in the $x$-space near the light-cone
with leading quark mass correction. Obviously, the first part in 
Eq.~(\ref{Prop}) is the massless quark-propagator.
The $SU(3)_{\rm flavour}$ vector
\begin{align}
c^a=\left(\frac{2}{9}\delta^{a0}+\frac{1}{6}\delta^{a3}+\frac{1}{6\sqrt{3}}\delta^{a8}\right) e^2
\end{align}
determines the flavour content and $e$ denotes the electric charge.
Here $\lambda^a_f$ is a generator of the flavour group
corresponding to the considered hadron.
If we use the Chisholm-identity,
\begin{align}
\gamma^\mu\gamma^\alpha\gamma^\nu=\left(g^{\mu\alpha}g^{\nu\beta}+g^{\mu\beta}g^{\nu\alpha}
-g^{\mu\nu}g^{\alpha\beta}\right)\gamma_\beta
-\ii\epsilon^{\mu\nu\alpha\beta}\gamma^5\gamma_\beta,
\end{align}
where $\epsilon^{\mu\nu\alpha\beta}$ denotes the Levi-Civita symbol,
we obtain for the massless part
\begin{align}
\label{T-mless1}
T\big(J^\mu(x/2)J^\nu(-x/2)\big)&=
\frac{c^a x_\alpha}{2\pi^2(x^2-\ii 0)^2}
\Big\{
\epsilon^{\mu\nu\alpha\beta}
\left(\bar{\psi}(x/2)\gamma^5\gamma_\beta\lambda^a_f \psi(-x/2)+
\bar{\psi}(-x/2)\gamma^5\gamma_\beta \lambda^a_f\psi(x/2)\right)\nonumber\\
&\qquad
+\ii S^{\mu\nu\alpha\beta}
\left(\bar{\psi}(x/2)\gamma_\beta\lambda^a_f \psi(-x/2)-
\bar{\psi}(-x/2)\gamma_\beta \lambda^a_f\psi(x/2)\right)
\Big\}+\ldots, 
\end{align}

The nonlocal LCE of QCD-operators is given by means of the 
T-product of two vector currents as an integral representation of two 
auxiliary variables $\kappa_i$:
\begin{align}
\label{T-mless2}
&T\big(J^\mu(x/2)J^\nu(-x/2))\relstack{x^2\rightarrow 0}{\approx} 
\int_{-1}^1\!\d\kappa_1\int_{-1}^1\!\d\kappa_2
\bigg(C^a(x^2,\kappa_1,\kappa_2,\mu^2)
S^{\mu\nu\alpha\beta} \lcx_\alpha (O^a_\beta(\kappa_1,\kappa_2)
-O^{a}_\beta(\kappa_2,\kappa_1))
\nonumber\\
&\qquad\qquad -C^{a,5}(x^2,\kappa_1,\kappa_2,\mu^2)\ii\epsilon^{\mu\nu\alpha\beta} \lcx_\alpha
(O^{a,5}_\beta(\kappa_1,\kappa_2)
+O^{a,5}_\beta(\kappa_2,\kappa_1))\bigg)
+\ldots\ 
\end{align}
with the coefficient functions which in Born approximation read:
\begin{align}
C^a(x^2,\kappa_1,\kappa_2,\mu^2)&=\frac{\ii c^a}{(2\pi)^2(x^2-\ii 0)^2}
\left(\delta(\kappa_1-1/2)\delta(\kappa_2+1/2)-\delta(\kappa_2-1/2)\delta(\kappa_1+1/2)\right)
\nonumber\\
C^{a,5}(x^2,\kappa_1,\kappa_2,\mu^2)&=\frac{\ii c^a}{(2\pi)^2(x^2-\ii 0)^2}
\left(\delta(\kappa_1-1/2)\delta(\kappa_2+1/2)+\delta(\kappa_2-1/2)\delta(\kappa_1+1/2)\right) 
\nonumber.
\end{align}

In the flavour non-singlet case $\lambda^a_f\neq 1$, the bilocal operators 
appearing in the operator-product expansion
of two electromagnetic currents are 
the gauge-invariant (vector) quark operators:
\begin{align}
\label{O-q}
O^a_\alpha(\kappa_1,\kappa_2)&=\bar{\psi}(\kappa_1 \lcx)\gamma_\alpha\lambda^a U(\kappa_1 \lcx,
\kappa_2 \lcx)\psi(\kappa_2 \lcx)\\
\label{O-q5}
O^{a,5}_\alpha(\kappa_1,\kappa_2)&=\bar{\psi}(\kappa_1\lcx)\gamma^5\gamma_\alpha\lambda^a
U(\kappa_1\lcx,\kappa_2\lcx)\psi(\kappa_2\lcx),
\end{align}
where $\lambda^a$ are the generators of the flavour group $SU_f(N)$. 
To guarantee the gauge invariance both operators contain the 
path-ordered phase factor along the straight line connecting the 
points $\kappa_1\lcx$ and $\kappa_2\lcx$
\begin{align}
\label{phase}
U(\ka\lcx, \kb\lcx) = P \exp\left\{\ii g
\int_{\ka}^{\kb} \d\tau \,\lcx^\mu A_\mu (\tau \lcx)\right\},
\end{align}
where $P$ denotes the path ordering, 
$g$ is the strong coupling parameter 
and $ A_\mu = A_\mu^a t^a$ 
with $t^a$ being the generators of $SU(3)_{\rm color}$
in the fundamental representation spanned 
by the quark fields $\psi$.

In the flavour singlet case ($\lambda^0_f\equiv 1$), the situation is much more complicated.
The quark operators~(\ref{O-q}) 
mix with operators that are made of two
gluon fields (bilocal gluon operators):
\begin{align}
\label{Gtensor}
G_{\alpha\beta}(\kappa_1,\kappa_2)&=F^{a\rho}_\alpha (\kappa_1\lcx) U^{ab}(\kappa_1\lcx,
\kappa_2\lcx)F^b_{\rho\beta}(\kappa_2\lcx)\\
\label{Gtensor5}
G^5_{\alpha\beta}(\kappa_1,\kappa_2)&=F^{a\rho}_\alpha (\kappa_1\lcx) U^{ab}(\kappa_1\lcx,
\kappa_2\lcx)\widetilde F^b_{\rho\beta}(\kappa_2\lcx).
\end{align}
Here
\begin{align}
F_{\alpha\beta}\equiv F^a_{\alpha\beta} t^a
=\pd_{\alpha}A_{\beta}-\pd_{\beta} A_\alpha-\ii g[A_\alpha, A_\beta]
,\qquad
\widetilde F_{\alpha\beta}=\frac{1}{2}\,\epsilon_{\alpha\beta\mu\nu} F^{\mu\nu}
\end{align}
is the field strength and the dual field strength of gluons. 
Additionally, $U^{ab}(\kappa_1\lcx,\kappa_2\lcx)$ is the phase factor
(\ref{phase}) in the adjoint representation.
It should be noted that these operators are a complete set of operators
in the massless quark theory $m_q=0$. 

Now we calculate the part which is given by the second part of 
Eq.~(\ref{Prop}).
It is known that
for $m_q\neq 0$ there appears
a further set of gauge invariant operators~\cite{Kodaira79,Lazar98}.
If we use the relation
\begin{align}
\gamma^\mu\gamma^\nu &= g^{\mu\nu}-\ii\sigma^{\mu\nu}\nonumber\\
                     &= g^{\mu\nu}+\frac{1}{2}\epsilon^{\mu\nu\alpha\beta}\gamma_5
\sigma_{\alpha\beta},
\end{align}
we obtain 
\begin{align}
&T\big(J^\mu(x/2)J^\nu(-x/2)\big)=(\ref{T-mless1})\nonumber\\
&\hspace{2cm} -\frac{m_q c^a }{8\pi^2(x^2-\ii 0)}
\Big\{
g^{\mu\nu}
\left(\bar{\psi}(x/2)\lambda^a_f \psi(-x/2)+
\bar{\psi}(-x/2)\lambda^a_f\psi(x/2)\right)\\
&\hspace{2cm}-\epsilon^{\mu\nu\alpha\beta}
\left(\bar{\psi}(x/2)\gamma^5\sigma_{\alpha\beta}\lambda^a_f \psi(-x/2)-
\bar{\psi}(-x/2)\gamma^5\sigma_{\alpha\beta}\lambda^a_f\psi(x/2)\right)
\Big\}+\ldots,\nonumber
\end{align}
and in terms of the nonlocal LCE:
\begin{align}
&T\big(J^\mu(x/2)J^\nu(-x/2))\relstack{x^2\rightarrow 0}{\approx} 
(\ref{T-mless2})\nonumber\\
&\hspace{1cm}-m_q\int_{-1}^1\!\d\kappa_1\int_{-1}^1\!\d\kappa_2
\bigg(C^a_m(x^2,\kappa_1,\kappa_2,\mu^2)
g^{\mu\nu} (N^a(\kappa_1,\kappa_2)
+N^{a}(\kappa_2,\kappa_1))
\nonumber\\
&\hspace{2cm} +C^{a,5}_m(x^2,\kappa_1,\kappa_2,\mu^2)\ii\epsilon^{\mu\nu\alpha\beta} 
(M^{a,5}_{[\alpha\beta]}(\kappa_1,\kappa_2)
-M^{a,5}_{[\alpha\beta]}(\kappa_2,\kappa_1))\bigg)
+\ldots\ 
\end{align}
with the coefficient functions:
\begin{align}
C^a_m(x^2,\kappa_1,\kappa_2,\mu^2)&=\frac{ c^a}{(4\pi)^2(x^2-\ii 0)}
\left(\delta(\kappa_1-1/2)\delta(\kappa_2+1/2)+\delta(\kappa_2-1/2)\delta(\kappa_1+1/2)\right)
\nonumber\\
C^{a,5}_m(x^2,\kappa_1,\kappa_2,\mu^2)&=\frac{ c^a}{(4\pi)^2(x^2-\ii 0)}
\left(\delta(\kappa_1-1/2)\delta(\kappa_2+1/2)-\delta(\kappa_2-1/2)\delta(\kappa_1+1/2)\right) 
\nonumber.
\end{align}
The additional bilocal operators 
appearing in the operator-product expansion
of two electromagnetic currents are 
the gauge-invariant quark operators:
\begin{align}
\label{N-q}
N^a(\kappa_1,\kappa_2)&=\bar{\psi}(\kappa_1\lcx)\lambda^a U(\kappa_1\lcx,
\kappa_2 \lcx)\psi(\kappa_2 \lcx)\\
\label{M-q5}
M^{a,5}_{[\alpha\beta]}(\kappa_1,\kappa_2)&=\bar{\psi}(\kappa_1\lcx)\ii\gamma^5
\sigma_{\alpha\beta}\lambda^aU(\kappa_1\lcx,\kappa_2\lcx)\psi(\kappa_2\lcx).
\end{align}
In the flavour singlet case,
the quark operators~(\ref{N-q}) and (\ref{M-q5}) which are multiplied by $m_q$
may be mix with operators that are made of two
gluon fields:
\begin{align}
\label{Gtensor-2}
G^\mu_{\ \mu}(\kappa_1,\kappa_2)&=F^{a\mu\nu}(\kappa_1\lcx) U^{ab}(\kappa_1\lcx,
\kappa_2\lcx)F^b_{\mu\nu}(\kappa_2\lcx)\\
\label{Gtensor5-2}
G^5_{[\alpha\beta]}(\kappa_1,\kappa_2)&=F^{a\rho}_{\ \ [\alpha} (\kappa_1\lcx) U^{ab}(\kappa_1\lcx,
\kappa_2\lcx)\widetilde F^b_{\beta]\rho}(\kappa_2\lcx).
\end{align}

In general, the coefficient functions $C^a(x^2,\kappa_1,\kappa_2,\mu^2)$ and
$C^{a,5}(x^2,\kappa_1,\kappa_2,\mu^2)$ 
are generalized functions. They are determined perturbatively 
and can be ordered according to their degree of singularity at $x^2=0$.
The Fourier transform $F_\Gamma(x^2, x q_i; \mu^2)$ 
of $C_\Gamma(x^2, \kappa_i; \mu^2)$ with respect to 
$\kappa_i$ are an entire functions of the new arguments 
$\xx q_i$ \cite{AZ78} according to
\begin{align}
C^a(x^2,\kappa_1,\kappa_2,\mu^2)&=\int\d (\lcx q_1)\d (\lcx q_2)\,\e^{-\ii\kappa_1(\lcx q_1)
-\ii\kappa_2(\lcx q_2)}F^a(x^2,(\lcx q_1),(\lcx q_2),\mu^2)\nonumber\\
C^{a,5}(x^2,\kappa_1,\kappa_2,\mu^2)&=\int\d (\lcx q_1)\d (\lcx q_2)\,\e^{-\ii\kappa_1(\lcx q_1)
-\ii\kappa_2(\lcx q_2)}F^{a,5}(x^2,(\lcx q_1),(\lcx q_2),\mu^2).\ 
\end{align}
In Born approximation we obtain
(choice of the renormalization point of $\mu^2=Q^2$): 
\begin{eqnarray}
F^a(x^2,(\lcx q_1),(\lcx q_2),\mu^2)&=&\frac{\ii c^a}{(2\pi)^2(x^2-\ii 0)^2}
\left(\e^{\ii\kappa_1(\lcx q_1)}-\e^{\ii\kappa_2(\lcx q_2)}\right)
\nonumber\\
F^{a,5}(x^2,(\lcx q_1),(\lcx q_2),\mu^2)&=&\frac{\ii c^a}{(2\pi)^2(x^2-\ii 0)^2}
\left(\e^{\ii\kappa_1(\lcx q_1)}+\e^{\ii\kappa_2(\lcx q_2)}\right).
\end{eqnarray}
The range of 
$\kappa_i$ is restricted to $-1 \leq \kappa_i \leq +1$.
Similar relations exist between $F^\Gamma_m(x^2, x q_i; \mu^2)$ 
of $C^\Gamma_m(x^2, \kappa_i; \mu^2)$.

Finally the Compton amplitude for the general case of non-forward scattering
is given by means of the LCE as follows
\begin{align}
T^{\mu\nu}(P_+,P_-,Q)\relstack{x^2\rightarrow 0}{\approx}\ 
&\frac{1}{(2\pi)^2}\int\d^4x \frac{c^a\e^{\ii xq}}{(x^2-\ii 0)^2}
\int_{-1}^1\d\kappa_1 \int_{-1}^1\d\kappa_2\\\
&\times\Big\{\ii\Delta_+(\kappa_1,\kappa_2)
\epsilon^{\mu\nu\alpha\beta}\lcx_\alpha
\big\langle P_2,S_2|O^{a,5}_\beta (\kappa_1,\kappa_2)
+O^{a,5}_\beta(\kappa_2,\kappa_1)|P_1,S_1\big\rangle\nonumber\\
&\quad-\Delta_-(\kappa_1,\kappa_2)
S^{\mu\nu\alpha\beta} \lcx_\alpha
\left\langle P_2,S_2|O^a_\beta (\kappa_1,\kappa_2)-O^a_\beta(\kappa_2,\kappa_1)|P_1,S_1\right\rangle
\Big\}+\ldots,\nonumber
\end{align}
where
\begin{align}
\Delta_\pm(\kappa_1,\kappa_2)=
\big[\delta(\kappa_1-1/2)\delta(\kappa_2+1/2)
\pm
\delta(\kappa_2-1/2)\delta(\kappa_2+1/2)\big].
\end{align}

\subsection{Distribution amplitudes}
Distribution amplitudes of hadrons are the key ingredients of the QCD approach
of exclusive, diffractive, and deep inelastic scattering processes. 
They provide the universal nonperturbative input in physical amplitudes.
Accordingly, they have to be addressed using nonperturbative methods.

The parton distribution functions (PDFs) 
are the DAs in forward scattering processes (DIS)
and are usually defined by the Fourier transformation of nucleon matrix 
elements of light-cone operators.
We begin with the simple case -- the chiral-even scalar operators on the 
light-cone\footnote{In Section~\ref{forward} we will prove that these 
scalar LC-operators are twist-2 operators.}
\begin{align}
O^a(\kappa\lcx,-\kappa\lcx)&\equiv
\bar\psi(\kappa\lcx)(\lcx\gamma)\lambda^a U(\kappa\lcx,-\kappa\lcx)\psi(-\kappa\lcx),\\
O^{a,5}(\kappa\lcx,-\kappa\lcx)&\equiv
\bar\psi(\kappa\lcx)\gamma^5(\lcx\gamma)\lambda^a U(\kappa\lcx,-\kappa\lcx)\psi(-\kappa\lcx).
\end{align}
In the unpolarized case, the forward matrix element of this operator is given 
by~\cite{JJ91} (see also Section~\ref{forward})
\begin{align}
\label{pd1}
\langle P|O^a(\kappa\lcx,-\kappa\lcx) |P\rangle
&=2\, (\lcx P)\int\!\d z\, \e^{2\ii\kappa z (\lcx P)} \, q^a(z,\mu^2)\nonumber\\
&=2\, (\lcx P)\sum_{n=0}^\infty\frac{(2\ii\kappa (\lcx P))^n}{n!} \, q^a_{n}(\mu^2).
\end{align}
Here $|P\rangle$ denotes the nucleon state with momentum $P$ and 
$q^a(z,\mu^2)$ is a parton distribution function\footnote{The PDF $q(z,\mu^2)$
is equal to the PDFs of twist-2 $f_1(z,\mu^2)$ and $F^{(2)}(z,\mu^2)$, 
see Eqs.~(\ref{q_eq:favda}) and (\ref{q_matrix_O_tw2_sca}).}
which is related to a reduced matrix element $q^a_{ n}(\mu^2)$ by means of
a Mellin transformation,
\begin{align}
q^a_{n}(\mu^2)=\int\d z\, z^n q^a(z,\mu^2),
\end{align}
with respect to the distribution parameter $z$ with $-1\le z\le 1$.
This parton distribution function is defined through the inverse Fourier transformation
of Eq.~(\ref{pd1}) and reads~\cite{mul94}
\begin{align}
q^a(z,\mu^2)=
\frac{1}{2} \int\frac{\d \kappa}{2\pi} \, \e^{-2\ii\kappa z (\lcx P)} \,
\frac{|\lcx P|}{(\lcx P)}\,
\langle P|O^a(\kappa\lcx,-\kappa\lcx) |P\rangle|_{\mu^2=Q^2}.
\end{align}
The factor $1/(\lcx P)$ is due to the structure $\lcx\gamma$ of the
operator $O^a(\kappa\lcx,-\kappa\lcx)$ in the matrix element (\ref{pd1}).
Physically, this distribution function describes the probability to find 
quarks with the momentum fraction $z P$ in an unpolarized nucleon. 
For $z>0$ ($z<0$) it can be
interpreted as a quark (antiquark) distribution function.
The (unpolarized) structure function $F_1$ is given by the linear combination of quark and antiquark 
distribution functions in the following way~\cite{Jaffe96}
\begin{align}
F_1(z,Q^2)=\frac{1}{2}\sum_{a=1}^{N_f} e^2_a\big(q^a(z,Q^2)-q^a(-z,Q^2)\big).
\end{align}
In the polarized case, the forward matrix element of the axial chiral-even 
bilocal scalar operator is given by
by~\cite{JJ91}
\begin{align}
\label{pd2}
\langle PS|O^{a,5}(\kappa\lcx,-\kappa\lcx) |PS\rangle
&=2\, (\lcx S)\int\!\d z\, \e^{2\ii\kappa z (\lcx P)} \, \Delta q^{a}(z,\mu^2)\nonumber\\
&=2\, (\lcx S)\sum_{n=0}^\infty\frac{(2\ii\kappa (\lcx P))^n}{n!} \, \Delta q^{a}_{n}(\mu^2),
\end{align}
where $S$ is the spin vector of the nucleon.
This parton distribution function\footnote{Here $\Delta q(z,\mu^2)$
is equivalent to the PDFs of twist-2 $g_1(z,\mu^2)$ and $G^{(2)}(z,\mu^2)$, 
see Eqs.~(\ref{q_eq:fvda}) and (\ref{q_matrix_O_t2_sca}).}
is defined through the inverse Fourier transformation
according to
\begin{align}
\Delta q^{a}(z,\mu^2)=
\frac{1}{2} \int\frac{\d \kappa}{2\pi} \, \e^{-2\ii\kappa z (\lcx P)} \,
\frac{|\lcx P|}{(\lcx S)}\, \langle P|O^a(\kappa\lcx,-\kappa\lcx) |P\rangle|_{\mu^2=Q^2}.
\end{align}
The factor $1/(\lcx S)$ is due to the structure $\gamma^5\lcx\gamma$ of the
operator $O^{a,5}(\kappa\lcx,-\kappa\lcx)$ in the matrix element (\ref{pd2}).
Physically, this distribution function describes the probability to find 
quarks with the momentum fraction $z P$ in a polarized nucleon. 
The (polarized) structure function $G_1$ is given in terms of quark
and antiquark distribution functions as follows~\cite{Jaffe96}
\begin{align}
G_1(z,Q^2)=\frac{1}{2}\sum_{a=1}^{N_f} e^2_a\big(\Delta q^{a}(z,Q^2)+\Delta q^{a}(-z,Q^2)\big).
\end{align}

Analogously, 
we may define the scalar meson distribution amplitude $\hat\phi^a(\xi,\mu^2)$
by means of the vacuum to meson matrix element of the bilocal
operator~\cite{mul94}:
\begin{align}
\hat\phi^a(\xi,\mu^2)=
\int\frac{\d \kappa}{2\pi} \, \e^{-\ii\kappa \xi (\lcx P)} \,
\frac{|\lcx P|}{(\lcx P)}\,
\langle 0|O^a(\kappa\lcx,-\kappa\lcx) |P\rangle|_{\mu^2=Q^2}.
\end{align}
Here $|P\rangle$ is the one-particle state of a scalar meson and $P$
denotes the momentum of the scalar meson.
The meson distribution amplitude is the probability amplitude to find 
a quark-antiquark pair in the meson in dependence of the momentum fraction 
parameter $\xi$ and the momentum transfer $Q^2$. 

Consequently,
in the unpolarized case, the nonforward matrix element of the scalar operator 
on the light-cone is given 
by~\cite{mul94} 
\begin{align}
\label{pd3}
\langle P_2|O^a(\kappa\lcx,-\kappa\lcx) |P_1\rangle
&= (\lcx P_+)\int\d z_+\int\d z_-\, \e^{\ii\kappa \lcx( P_+ z_+ + P_- z_-)} \, f^a(z_+,z_-,\mu^2)\nonumber\\
&= (\lcx P_+)\sum_{n=0}^\infty\frac{(\ii\kappa)^n}{n!} 
   \sum_{m=0}^n \binom{n}{m} (\lcx P_+)^{n-m} (\lcx P_-)^{m}\,
   f^a_{nm}(\mu^2),\\
&= \sum_{n=0}^\infty\frac{(\ii\kappa)^n}{n!} \,(\lcx P_+)^{n+1}
   \sum_{m=0}^n \binom{n}{m} \tilde\tau^{m}\, f^a_{nm}(\mu^2),\nonumber
\end{align}
where 
\begin{align}
\tilde\tau=\frac{\lcx P_-}{\lcx P_+},\quad
P_+ z_+ +P_-  z_-=P_1 z_1+P_2 z_2
\quad\text{with}\quad
z_{\pm}=\frac{1}{2}(z_2\pm z_1),
\end{align}
and
$\binom{n}{m}=n!/(n-m)!m!$ is the binomial coefficient.
Here $f^a(z_+,z_-,\mu^2)$ is a double distribution function which is related to a 
reduced matrix element $f^a_{n m}(\mu^2)$ by the help of
a twofold Mellin transformation,
\begin{align}
f^a_{nm }(\mu^2)=\int\d z_+\int\d z_- \, z_{+}^{n-m}\, z_-^{m} f^a(z_+,z_-,\mu^2),
\end{align}
with respect to the distribution parameters $z_+$ and $z_-$ 
and the integration domain is: $-1\le z_+\le 1$, $-1+|z_+|\le z_-\le 1-|z_+|$.
Hence, $f^a_{nm}(\mu^2)$ are double moments of the double distribution function
$f^a(z_+,z_-,\mu^2)$ which only depend on the distribution parameters and not
on kinematical factors like $P_+$, $P_-$, etc..
The double distributions are related to the ordinary 
parton distributions.
By taking the matrix element (\ref{pd3}) with $P_1=P_2=P$,
we obtain~\cite{R97,GLR01}
\begin{align}
q^a(z,\mu^2)=\int\d z_- f^a(z_+=z, z_-,\mu^2).
\end{align}

But on the other hand we may parametrize the matrix element in terms of a
skewed parton distribution function~\cite{Ji97,R97,RW00,BM00,BM01}: 
\begin{align}
\label{pd4}
\langle P_2|O^a(\kappa\lcx,-\kappa\lcx) |P_1\rangle
&= (\lcx P_+)\int_{-1}^{1}\d z\, \e^{\ii\kappa \lcx P_+ z} \, H^a(z,\tilde\tau,\mu^2)\nonumber\\
&= \sum_{n=0}^\infty\frac{(\ii\kappa)^n}{n!} \,(\lcx P_+)^{n+1}
   H^a_{n}(\tilde\tau,\mu^2),
\end{align}
where $H^a(z,\tilde\tau,\mu^2)$ is the skewed parton distribution 
function\footnote{Ji~\cite{Ji97} originally introduced a 
skewed parton distribution $H(z,\tilde\tau,P_-^2)$ which depends on the 
kinematical factor $P_-^2$. 
Forming the first moment of this skewed distribution, one gets the sum
rule~\cite{Ji97}: $\int_{-1}^1\d z H(z,\tilde\tau,P_-^2)=F_1 (P_-^2)$, where
$F_1(P_-^2)$ is the Dirac form factor.
In Section~\ref{harmonic} we will see
that the kinematical factors like $P_-^2$ are proportional to $x^2$ and can be
generated through the twist projectors of geometric twist.}
and $H^a_{n}(\tilde\tau,\mu^2)$ are the corresponding moments which depend on the 
variable $\tilde\tau$.
By comparing Eqs.~(\ref{pd3}) and (\ref{pd4}) we observe the following 
relation~\cite{R97}
between the skewed parton distribution $H^a(z,\tilde\tau,\mu^2)$ and the double
distribution $f^a(z_+,z_-,\mu^2)$:
\begin{align}
\label{pdf-dd}
H^a_{n}(\tilde\tau,\mu^2)=
\int_{-1}^1 \d z\, z^n H^a(z,\tilde\tau,\mu^2)=\sum_{m=0}^n \binom{n}{m} \tilde\tau^m 
\int\d z_+\int\d z_- \, z_{+}^{n-m}\, z_-^{m} f^a(z_+,z_-,\mu^2).
\end{align}
Thus, skewed parton distributions are obtained from double distributions 
after an appropriate integration (see Eq.~(\ref{pdf-dd})).  
Additionally, the skewed parton distributions are related to the ordinary 
parton distributions.
In fact, we have
\begin{align}
H^a(z,0,\mu^2)=q^a(z,\mu^2).
\end{align}

Similar relations can be derived for the double distribution function
$\Delta f^a(z_+,z_-,\mu^2)$, the skewed parton distribution 
$\widetilde{H}^a (z,\tilde\tau,\mu^2)$, and the parton distribution
$\Delta q^a(z,\mu^2)$ which are corresponding to the axial operators
$O^{a,5}(\kappa\lcx,-\kappa\lcx)$.

Let me note that we will discuss double distribution amplitudes
as off-cone nonforward matrix elements of more complicated operators with 
definite geometric twist in Section~\ref{harmonic}.

\subsection{Anomalous dimensions of QCD operators}
We discuss now some facts about the anomalous dimension of 
(twist-2) QCD-operators. Physically, the $Q^2$-evolution of DAs is
governed by the anomalous dimensions, which are determined by the 
renormalization of the corresponding operators. The anomalous dimensions
are necessary to connect and compare experimental data 
which are obtained at different values of $Q^2$.

For simplicity we symbolically use the notation ${\cal O}_n(\lcx)$ for
any local operator on the light-cone (see Section~\ref{proc}). 
Here ${\cal O}^{\rm bare}_n(\lcx)$ 
denotes the
unrenormalized operator and the renormalized operator ${\cal O}_n(\lcx;\mu^2)$ is 
given by
\begin{align}
{\cal O}_n(\lcx;\mu^2)=\sum_{n'=0}^n Z^{-1}_{n n'}(\mu^2) {\cal O}^{\rm bare}_{n'}(\lcx),
\end{align}
where $Z_{n n'}(\mu^2)$ is the renormalization matrix and $\mu^2$ is the 
renormalization point.
The invariance of the physics under 
the choice of the point of renormalization can be
characterized by the so-called renormalization group equation
\begin{align}
\mu \dd{\mu}\, {\cal O}_n(\lcx;\mu^2) =
\sum_{n'=0}^n\gamma_{n n'}(g(\mu^2)){\cal O}_{n'}(\lcx;\mu^2),
\end{align}
where $\gamma_{n n'}(g(\mu^2))$ is the (local) anomalous dimension of the 
operator ${\cal O}_n(\lcx;\mu^2)$. It is defined by
\begin{align}
\gamma_{n n'}(g(\mu^2))=Z_{nm}^{-1}\, \mu\frac{\pd}{\pd\mu}\,  Z_{m n'}(\mu^2).
\end{align}
In this connection, the $\beta$-function is introduced:
\begin{align}
\beta(g(\mu^2))=\mu\frac{\pd}{\pd\mu}\, g(\mu^2).
\end{align}
The scale dependence of the  operator is obtained by the solution
of the renormalization group equation as
\begin{align}
{\cal O}_n(\lcx;Q^2)=\sum_{n'=0}^n{\cal O}_{n'}(\lcx;\mu^2)\, 
\exp\left[-\int^{g(Q^2)}_{g(\mu^2)}\d g\, \frac{\gamma_{n n'}(g)}{\beta(g)}\right].
\end{align}
The anomalous dimension and the $\beta$-function are calculable using 
perturbation theory.
They can be expanded as power series in $g^2$ as
\begin{align}
\gamma_{n n'}(g)&=\frac{g^2}{16 \pi^2}\, \gamma^0_{n n'}+
\frac{g^4}{(16 \pi^2)^2}\, \gamma^1_{n n'}+ O(g^6),\\
\beta(g)&=-\frac{g^3}{16 \pi^2}\, \beta^0-
\frac{g^5}{(16 \pi^2)^2}\, \beta^1+ O(g^7),
\end{align}
where the coefficients of the $\beta$ functions are:
\begin{align}
\beta_0=11-\frac{2}{3}\, N_f,\qquad 
\beta_1=102-\frac{38}{3}\, N_f,
\end{align}
with the number of quark flavours $N_f$.

If we build the forward matrix element of the local operator, 
the renormalization group equation of the corresponding quark distribution
is obtained as ($\gamma_n\equiv\gamma_{n n}$)
\begin{align}
\mu \dd{\mu}\, q^a_n(\mu^2) =
\gamma^{qq}_{n}(g(\mu^2))q^a_n(\mu^2).
\end{align}

In the flavour singlet case, the quark and gluon operators mix with each 
other under renormalization.
The renormalization group equation for the local 
operators\footnote{The local operators are obtained after local expansion
of the nonlocal ones (see Section~\ref{proc}).}
\begin{align}
O_n(\lcx)&=\bar\psi(0)(\lcx\gamma)(\lcx\Tensor D)^{n-1} \psi(0),\\
G_n(\lcx)&= \lcx^\mu \lcx^\nu F_{\mu}^{\ \rho}(0) (\lcx\Tensor D)^{n-2} F_{\rho\nu}(0)
\end{align}
is given by
\begin{align}
\mu \dd{\mu}
\begin{pmatrix}
O_n(\lcx;\mu^2)\\
G_n(\lcx;\mu^2)
\end{pmatrix}
=
\sum_{n'=0}^n
\begin{pmatrix}
\gamma^{qq}_{nn'}&\gamma^{qG}_{nn'}\\
\gamma^{Gq}_{nn'}&\gamma^{GG}_{nn'}
\end{pmatrix}
\begin{pmatrix}
O_{n'}(\lcx;\mu^2)\\
G_{n'}(\lcx;\mu^2)
\end{pmatrix}
\end{align}
where the local (one-loop) unpolarized anomalous dimensions are~\cite{BGR87}:
\begin{align}
\label{uad}
\gamma^{qq}_{nn'}
&=\frac{g^2}{4\pi^2}\,C_F\bigg\{
\bigg[ \frac{1}{2} - \frac{1}{(n+1)n}
       + 2 \sum^{n}_{j=2} \frac{1}{j} \bigg] \delta_{nn'}
     - \bigg[\frac{1}{(n+1)n}
       + \frac{2}{n-n'}\frac{n'}{n}\bigg]
\theta_{nn'}\bigg\}
\\
\gamma^{qG}_{nn'}
&= - \frac{g^2}{4\pi^2}\, N_f T_R \,
\frac{2}{(n+2)(n+1)n}\Big[(n^2+n+2)-(n-n')n\Big],
\\
\gamma^{Gq}_{nn'}
&=- \frac{g^2}{4\pi^2}\, C_F\,
\frac{1}{n(n^2-1)}
\Big[(n^2+n+2)\delta_{nn'}+2\theta_{nn'}\Big],
\\
\label{x112}
\gamma^{GG}_{nn'}
&=\frac{g^2}{4\pi^2}\,C_A
\bigg\{
\bigg[\frac{1}{6} - \frac{2}{n(n-1)} - \frac{2}{(n+2)(n+1)}
     + 2 \sum^{n}_{j=2}\frac{1}{j} + \frac{2N_f T_R}{3 C_A}
\bigg]\delta_{nn'}
\\
&\quad +\bigg[ 2\bigg(\frac{2n-1}{n(n-1)}
     - \frac{1}{n-n'}\bigg)
     - (n-n'+2)\bigg(\frac{1}{n(n-1)}
     + \frac{1}{(n+2)(n+1)}\bigg)\bigg] \theta_{nn'}\bigg\},
\nonumber
\end{align}
where $C_F=(N_c^2-1)/2N_c\equiv 4/3$, $T_R=1/2$, $C_A=N_c\equiv 3$.
We used the following notation
\begin{align}
\theta_{n\,n'}
=
\left\{
\begin{array}{ll}
     1    &    {\rm for}\;\; n'< n,     \\
     0    &    {\rm otherwise}~.
\end{array}\right.
\nonumber
\end{align}

The local (axial) operators
\begin{align}
O^5_n(\lcx)&=\bar\psi(0)\gamma^5(\lcx\gamma)(\lcx\Tensor D)^{n-1} \psi(0),\\
G^5_n(\lcx)&= \lcx^\mu \lcx^\nu F_{\mu}^{\ \rho}(0) (\lcx\Tensor D)^{n-2} \widetilde F_{\rho\nu}(0)
\end{align}
satisfy the renormalization group equation
\begin{align}
\mu \dd{\mu}
\begin{pmatrix}
O^5_n(\lcx;\mu^2)\\
G^5_n(\lcx;\mu^2)
\end{pmatrix}
=
\sum_{n'=0}^n
\begin{pmatrix}
\Delta\gamma^{qq}_{nn'}&\Delta\gamma^{qG}_{nn'}\\
\Delta\gamma^{Gq}_{nn'}&\Delta\gamma^{GG}_{nn'}
\end{pmatrix}
\begin{pmatrix}
O^5_{n'}(\lcx;\mu^2)\\
G^5_{n'}(\lcx;\mu^2)
\end{pmatrix}
\end{align}
with the local (one-loop)
polarized anomalous dimensions~\cite{BGR99}:
\begin{align}
\label{pad}
\Delta\gamma^{qq}_{nn'}
&=\gamma^{qq}_{nn'},
\\
\Delta\gamma^{qG}_{nn'}
&=- \frac{g^2}{4\pi^2}\, N_f T_R \,
\frac{2(n'-1)}{(n+1)n},
\\
\Delta\gamma^{Gq}_{nn'}
&=\frac{g^2}{4\pi^2}\, C_F\,
\frac{1}{(n+1)n}
\Big[(n+2)\delta_{nn'}- \frac{2}{n-1}\, \theta_{nn'}\Big],
\\
\label{x131}
\Delta\gamma^{GG}_{nn'}
&=\frac{g^2}{4\pi^2}\,C_A\bigg\{
\bigg[\frac{1}{6} - \frac{4}{(n+1)n}
     + 2 \sum^{n}_{j=2}\frac{1}{j} + \frac{2N_f T_R}{3 C_A}
\bigg]\delta_{nn'}
\\
& +\bigg[ 2\bigg(\frac{2n-1}{n(n-1)}
     - \frac{1}{n-n'}\bigg)
     - (n-n'+2)\,\frac{2}{(n+1)n}
     \bigg] \theta_{nn'}\bigg\}.
\nonumber
\end{align}
The polarized and unpolarized forward
anomalous dimensions~\cite{GW74,GP74,AR} are obtained as the diagonal 
elements of 
Eqs.~(\ref{uad})\,--\,(\ref{x112}),\,(\ref{pad})\,--\,(\ref{x131}).\footnote{This
(diagonal) form of the anomalous dimensions is also valid for 
the so-called conformal operators 
(see Refs.~\cite{rady80,shifman,makeenko81,Ohrndorf82,craigie,Craigie85}).}

The nonlocal anomalous dimensions are obtained from the local ones 
by means of an inverse Mellin transformation.
The corresponding nonlocal twist-2 anomalous dimensions calculated in the framework
of nonlocal OPE are given in Refs.~\cite{BGR87,BGR97,BGR99,BB88,BR97,R97}.

\newpage
\section{The concept of geometric twist of nonlocal operators}
\label{proc}
In this Section we present a straightforward scheme to discuss 
nonlocal operators with respect to definite geometric twist.
This method is based by applying the representation theory of the 
Lorentz group.
The advantages of this method are: 
First, the nonlocal operators of definite geometric twist are Lorentz covariant 
tensors (contrary to the phenomenological concept of twist).
Second, this twist decomposition is unique, process-- and model--independent.
Furthermore, the twist decomposition is independent from the dimension 
$2h$ of spacetime.
Up to now, no clear scheme for higher geometric twist was worked out 
in the literature.

On the other hand, Jaffe and Ji~\cite{JJ91,JJ92} proposed the notion of 
{\it dynamical} twist ($t$) by counting powers $Q^{2-t}$ which is directly related to 
the power by which the corresponding distributions contribute to the 
scattering amplitudes.
Additionally, the dynamical twist is only defined for 
the {\it matrix elements} of operators and is not Lorentz invariant.

If one uses QCD operators with definite geometric twist, 
the different phenomenological distribution functions of definite geometric twist 
are given as matrix elements of these operators.
It should be
emphasized that the LC operators of definite geometric twist are to be
preferred also from the point of view of renormalization theory.
Therefore, we are using the concept of geometric twist instead of 
dynamical twist.

\subsection{The procedure of twist decomposition of nonlocal operators}
\setcounter{equation}{0}
This Section is devoted to explain how the QCD operators may
by decomposed according to their (geometric) twist which is defined~\cite{Gross}:
\begin{align}
\text{twist}\, (\tau)\ =\ \text{canonical dimension}\, (d) - \text{Lorentz spin}\, (j).
\nonumber
\end{align}
We use the notion of twist because it is a straightforward scheme to classify
DAs and, additionally, operators with equal twist mix with each other under
renormalization.

The nonlocal operators which are relevant for the 
QCD processes (see previous Section)
are obtained by the nonlocal LC expansion~\cite{AZ78} of the (renormalized)
time-ordered product of two electromagnetic hadronic currents:
\begin{align}
\label{LCE1}
RT\,\big(J^\mu(\hbox{$y+\large\frac{x}{2}$})
J^\nu(\hbox{$y-\large \frac{x}{2}$})S\big) 
\relstack{x^2\rightarrow 0}{\approx}
\int
\d^2 \underline\kappa\,
C^{\mu\nu}_\Gamma(x^2, \underline\kappa) 
RT\big(\OO^\Gamma(y+\ka\lcx, y+\kb\lcx)S\big) +\cdots,
\end{align}
where 
$\Gamma=\{1,\gamma_\mu,\sigma_{\mu\nu};
\gamma_5,\gamma_\mu\gamma_5,\sigma_{\mu\nu}\gamma_5\}$
indicates the tensor structure of the nonlocal QCD operators. 
Strictly speaking, the expression in Eq.~(\ref{LCE1}) is an operator valued 
distribution.

In general, the twist decomposition of an arbitrary nonlocal 
operator 
may be formulated for any $2h$-dimensional spacetime. Let us denote 
such operators for arbitrary values of $x$ as follows:
\begin{equation}
\label{O_Gamma}
\OO^\Gamma(\kappa_1  x,\kappa_2 x)
=
{\Phi}'(\kappa_1  x)\Gamma 
U(\kappa_1  x,\kappa_2 x)
\Phi(\kappa_2 x),
\end{equation}
where, suppressing any indices indicating the group representations, $\Phi$ 
generically denotes the various local fields 
with a scale dimension\footnote{The scale dimension, $d_\Phi$, of a field $\Phi$
is determined from the infinitesimal transformation law under dilations
at the point $y=0$: $[\hat D,\Phi(y)]|_{y=0}=\ii d_\Phi\, \Phi(0)$, where $\hat D$ is the 
dilation generator. For the unrenormalized operators
the scale dimension agrees with the canonical dimension. 
For the renormalized operators
the difference from the canonical one
is called anomalous dimension.}, e.g.,~scalar ($d_\varphi=h-1$), 
Dirac spinor ($d_\psi=h-\frac{1}{2}$) as well as gauge field strength 
($d_F=h$). The gauge potential $A_\mu$ has dimension $d_A=h-1$. Furthermore, 
$\Gamma$ labels the tensor structure as well as additional quantum numbers, if 
necessary.\\
Now, the twist decomposition of operators (\ref{O_Gamma}) consists of the 
following steps:\\
\noindent
(1)\quad{\em Taylor expansion} of the nonlocal operators for 
{\em arbitrary} values of $x$ at the point $y$ into an infinite 
tower of local tensor operators having definite rank $n$ and 
canonical dimension $d$.
If we choose the expansion point $y=0$, we obtain the expansion:
\begin{align}
\label{O_ent1}
\OO^\Gamma (\kappa_1  x,\kappa_2 x)
&=\sum_{n=0}^{\infty} \frac{1}{n!}\,
x^{\mu_1}\ldots x^{\mu_n}
\Big[\Phi'(y)\Gamma\Tensor D_{\mu_1}\!(\kappa_1, \kappa_2)
\ldots \Tensor D_{\mu_n}\!(\kappa_1, \kappa_2)
\Phi(y)\Big]_{y=0}\nonumber\\
&\equiv\sum_{n=0}^{\infty}\frac{1}{n!}\,
\Big[\Phi'(y)\Gamma
\big( x \Tensor D(\kappa_1, \kappa_2) \big)^n
\Phi(y)\Big]_{y=0}
\end{align}
with the generalized covariant derivatives
\begin{eqnarray}
\label{D_kappa}
\Tensor D_\mu(\kappa_1, \kappa_2)
&\equiv& 
\kappa_1 \LD D_\mu+
\kappa_2 \RD D_{\mu}\,,
\\
\RD D_\mu=\RDL{\pd^y_\mu} +\ii g A_\mu(y),
&&
\LD D_\mu=\LDL{\pd^y_\mu} -\ii g A_\mu(y)\,. 
\nonumber
\end{eqnarray}
We introduce, for later convenience, the variables
\begin{align}
\kappa_{\pm}=\frac{1}{2}(\kappa_2\pm\kappa_1)
\quad\text{with}\quad
\kappa_{1,2}=\kappa_+\mp\kappa_-\,,
\end{align}
and the following derivatives
\begin{align}
D^\mu_{\pm}= \RDL{D^\mu} \pm \LDL{D^\mu},
\end{align}
with\footnote{The derivative $\pd_+$ is sometimes called 
``total'' derivative~\cite{BB88}.}
\begin{align}
D^\mu_+\equiv\pd^\mu_+=\LDL{\pd^\mu_y}+\RDL{\pd^\mu_y},\quad
D^\mu_-\equiv\LTensor{D^\mu}=-\LDL{\pd^\mu_y}+\RDL{\pd^\mu_y} +2\ii g A^\mu(y).
\end{align}
Now we are able to rewrite Eq.~(\ref{O_ent1}) in terms of these
variables and derivatives according to
\begin{align}
\label{O_ent2}
\OO^\Gamma (\kappa_1  x,\kappa_2 x)
&=\sum_{n=0}^{\infty} \frac{1}{n!}\,
\Big[\Phi'(y)\Gamma
(\kappa_- x D_-+\kappa_+ x D_+)^n
\Phi(y)\Big]_{y=0}\nonumber\\
&=\sum_{n=0}^{\infty}\sum_{k=0}^n
\frac{\kappa_+^k\, \kappa_-^{n-k}}{(n-k)! k!}\,
\OO^\Gamma_{nk}(y,x)|_{y=0},
\end{align}
where we introduced the following local operator
\begin{align}
\label{O_nk}
\OO^\Gamma_{nk}(y,x)=
\Phi'(y)\Gamma\big( x D_+\big)^k \big(x D_-)^{n-k}\Phi(y).
\end{align}
The local operators~(\ref{O_nk}) mix under renormalization for the 
same value of $k$. Therefore, their anomalous are independent
of $k$~\cite{BGR87}.
Another possibility of local expansion is to choose the expansion point
$y=\kappa_+ x$. We get
\begin{align}
\label{O_ent4}
\OO^\Gamma (\kappa_1  x,\kappa_2 x)
&\equiv
\OO^\Gamma (\kappa_+x -\kappa_- x  ,\kappa_+ x+\kappa_- x)\nonumber\\
&=\sum_{n=0}^{\infty}
\frac{\kappa_-^{n}}{n!}\,
\OO^\Gamma_{n}(y,x)|_{y=\kappa_+ x},
\end{align}
with the local operator:
\begin{align}
\label{O_n2}
\OO^\Gamma_{n}(y,x)=
\Phi'(y)\Gamma \big(x D_-)^{n}\Phi(y).
\end{align}
In this way the variables $\kappa_+ x$ and
$2\kappa_- x$, respectively, are centre and  the relative coordinates of the
two points $\kappa_1 x$ and $\kappa_2 x$.
Now we discuss the bilocal operator $\OO^\Gamma (-\kappa  x,\kappa x)$
which is centred around $y=0$.
For the local expansion of the ``centred'' operator we obtain
the simpler formula 
\begin{align}
\label{O_ent3}
\OO^\Gamma (-\kappa  x,\kappa x)
&=\sum_{n=0}^{\infty}
\frac{\kappa^{n}}{n!}\,
\OO^\Gamma_{n}(y,x)|_{y=0},
\end{align}
with the local operator:
\begin{align}
\label{O_n}
\OO^\Gamma_{n}(y,x)=
\Phi'(y)\Gamma \big(x D_-)^{n}\Phi(y).
\end{align}
Obviously, (\ref{O_ent3}) is obtained from (\ref{O_ent4}) 
with $\kappa_+=0$ and $\kappa_-=\kappa$. 
In the following we often use the abbreviation 
$\OO^\Gamma_{n}(0,x)\equiv \OO^\Gamma_{n}(x)$.
Some remarks are necessary with respect to these relations.
The local operators in Eqs.~(\ref{O_ent1}), (\ref{O_ent2}), and
(\ref{O_ent3}) have to be infinitely differentiable with respect
to the expansion point $y\in {\Bbb R}^{2h}$ and are homogeneous
polynomials of degree $n$ in $x\in {\Bbb R}^{2h}$.
Additionally, the operators (\ref{O_nk}) and (\ref{O_n}) have the same
anomalous dimensions.

The Fourier transforms of the `centred' operator
${\cal O}^{\Gamma}(-\kappa x, \kappa x)$ and its $n$th moments
${\cal O}^\Gamma_n(x)$, i.e., its Taylor coefficients
w.r.t.~$\kappa$, are related as follows:
\begin{align}
\label{O^Gint}
&{\cal O}^{\Gamma}(-\kappa x,\kappa x)
=\int \d^{2h} p \,{\cal O}^{\Gamma}(p)\, \e^{-\ii\kappa xp}
=\sum_{n=0}^\infty \frac{(-\ii \kappa)^n}{n!}\,{\cal O}^\Gamma_ n(x),\\
\label{O^Gloc}
&{\cal O}^\Gamma_n(x)
= \int \d^{2h} p \,{\cal O}^{\Gamma}(p)\, (xp)^n
= (\ii)^n \frac{\pd^n}{\pd\kappa^n}
{\cal O}^{\Gamma}(-\kappa x, \kappa x)\Big|_{\kappa=0}.
\end{align}
For notational simplicity we wrote the Fourier
measure without the usual factor $1/(2\pi)^{2h}$.

\noindent
(2)\quad{\em Decomposition} of local operators with respect to 
{\em irreducible tensor representations} of the Lorentz group $SO(1,2h-1)$ 
or, equivalently, the orthogonal group 
$SO(2h)$.\footnote{
The corresponding representations are related through 
analytic continuation, cf. \cite{BR,VK,GLR99}}
These representations are built up by traceless tensors of rank $m$
($m=n+\text{number of indices of } \Gamma$)
whose symmetry class is determined
by some (normalized) Young operators ${\cal Y}_{[m]}=(f_{[m]}/m!){\cal QP}$,
where $[m] = (m_1, m_2, \ldots m_r)$
with $m_1 \geq m_2\geq \ldots \geq m_r$
and $\sum^r_{i=1}\, m_i = m$ denotes the corresponding Young pattern. 
$\cal P$ and $\cal Q$, as usual, denote symmetrization and antisymmetrization 
with respect to that pattern. The allowed Young patterns for $SO(2h)$, which 
because of the tracelessness are restricted by 
$\ell_1+\ell_2\le 2h$ ($\ell$: length of {\em columns} of $[m]$), 
are for a fixed value of $m$:\\
\begin{enumerate}
\item[(i)] \unitlength0.4cm
\begin{picture}(30,1)
\linethickness{0.15mm}
\multiput(1,0)(1,0){13}{\line(0,1){1}}
\put(1,1){\line(1,0){12}}
\put(1,0){\line(1,0){12}}
\put(15,0){$j=m,m-2,m-4,\ldots$}
\end{picture}

\item[(ii)]\unitlength0.4cm
\begin{picture}(5,1)
\linethickness{0.15mm}
\multiput(3,0)(1,0){10}{\line(0,1){1}}
\multiput(1,-1)(1,0){2}{\line(0,1){2}}
\put(1,1){\line(1,0){11}}
\put(1,0){\line(1,0){11}}
\put(1,-1){\line(1,0){1}}
\put(15,0){$j=m-1,m-2,m-3,\ldots$}
\end{picture}

\item[(iii)] \unitlength0.4cm
\begin{picture}(5,2)
\linethickness{0.15mm}
\multiput(3,0)(1,0){9}{\line(0,1){1}}
\multiput(1,-2)(1,0){2}{\line(0,1){3}}
\put(1,1){\line(1,0){10}}
\put(1,0){\line(1,0){10}}
\put(1,-1){\line(1,0){1}}
\put(1,-2){\line(1,0){1}}
\put(15,0){$j=m-2,m-3,m-4,\ldots$}
\end{picture}

\newpage

\item[(iv)] \unitlength0.4cm
\begin{picture}(5,2)
\linethickness{0.15mm}
\multiput(4,0)(1,0){8}{\line(0,1){1}}
\multiput(1,-1)(1,0){3}{\line(0,1){2}}
\put(1,1){\line(1,0){10}}
\put(1,0){\line(1,0){10}}
\put(1,-1){\line(1,0){2}}
\put(15,0){$j=m-2,m-3,m-4,\ldots$}
\end{picture}
\end{enumerate}
\vspace*{4mm}
$\qquad\qquad\vdots$
\\
\\
(Here, we depicted only those patterns which appear for $h=2$ where 
$\ell_1+\ell_2\le 4$.)
Loosely speaking, the lower spins in the above series correspond to the
various trace terms which define, in general, reducible higher twist
operators. These higher twist operators again have to be 
decomposed according to the considered symmetry classes.
For more details see Appendices~\ref{young} and \ref{lorentz}.

Now, two remarks are in order. At first, this decomposition of the 
local operators 
appearing in (\ref{O_ent1}) into irreducible components is independent 
of the special values of $\ka, \kb$. 
Furthermore, the 
totally symmetric indices of
irreducible local tensor operators are multiplied by
${x}^{\mu_1}\ldots {x}^{\mu_n}$ 
in order to obtain harmonic (tensorial) polynomials of order $n$.
Obviously, in this turn we use the vector $x$ as a device
for writing tensors with special Young symmetry in analytic form (see 
Section~\ref{tensor}).

A local operator without definite twist and spin
is given as a finite sum of
local operator with definite  geometric twist,
\begin{equation}
\label{O_local}
\OO_{\Gamma\, n}(y,x)=
\bigoplus_{\tau=\tau_{\text{min}}}^{\tau_{\text{max}}} 
c^{(\tau)\Gamma' n'}_{\Gamma n}\!(x)\,\OO^{(\tau)}_{\Gamma'\,n'}(y,x),
\end{equation}
where $c^{(\tau)\Gamma' n'}_{\Gamma n}(x)$ are coefficient functions 
depending on $x^2$ and $x_\alpha$ etc..
The local operators of geometric twist can be written 
\begin{align}
\label{twist-l}
{\cal O}^{(\tau)}_{\Gamma\,n}(y, x)
&= {\cal P}^{(\tau)\Gamma' n'}_{~~\Gamma n}
{\cal O}_{\Gamma'\,n}(y,x),
\\
\label{proj-l}
\big({\cal P}^{(\tau)} \times
{\cal P}^{(\tau')}\big)^{~~\Gamma' n'}_{\Gamma n}
&= \delta^{\tau \tau'}{\cal P}^{(\tau)\Gamma' n'}_{~\Gamma n},
\end{align}
where ${\cal P}^{(\tau)}_{\Gamma n\Gamma' n'}$ are orthogonal projectors
of definite geometric twist and definite spin, respectively. 
These projectors are given in terms of $x^2$, $\square$, $x_\alpha$, 
$\pd_\alpha$ etc..
It is obvious that the twist decomposition corresponds to a Lorentz
spin decomposition of the local operator.

\noindent
(3)\quad{\em Resummation} of the local operators 
$\OO^{(\tau)}_{\Gamma\,n}(y,x)$
belonging to the same symmetry class (for any $n$) and having equal 
twist $\tau$. That infinite tower of local 
operators~(\ref{twist-l}) creates
a {\em nonlocal} operator of definite twist,
\begin{equation}
\label{O_nonlocal}
\OO^{(\tau)}_\Gamma(\kappa_1 x,\kappa_2 x)=\bigoplus_{n=0}^\infty 
\frac{\kappa_-^n}{n!}\,
\OO^{(\tau)}_{\Gamma\,n}(y,x)|_{y=\kappa_+x}.
\end{equation}
Obviously, from a group theoretical point of view, this nonlocal 
operator is built up as direct sum of irreducible local tensor operators
which have the same twist but different Lorentz spin. 
In fact, the nonlocal operator~(\ref{O_nonlocal}) has no definite Lorentz
spin because it is an infinite tower of local operators with definite 
Lorentz spin. 
Therfore, the notion of geometric twist is a well-defined concept to 
classify nonlocal operators.  
The nonlocal operators $\OO^{(\tau)}_\Gamma(\kappa_1 x,\kappa_2 x)$ are 
harmonic tensor functions and they are given by
\begin{align}
\label{twist-nl}
{\cal O}^{(\tau)}_{\Gamma}(\kappa_1 x,\kappa_2 x)
&= {\cal P}^{(\tau)\Gamma'}_{~~\Gamma}
{\cal O}_{\Gamma'}(\kappa_1 x,\kappa_2 x),
\\
\label{proj-nl}
\big({\cal P}^{(\tau)} \times
{\cal P}^{(\tau')}\big)^{~~\Gamma'}_{\Gamma}
&= \delta^{\tau \tau'}{\cal P}^{(\tau)\Gamma'}_{~\Gamma},
\end{align}
where ${\cal P}^{(\tau)}_{\Gamma\Gamma'}$ are orthogonal projection
operators of definite geometric twist for the nonlocal operators.

Now, putting together all contributions, i.e.,~including the higher twist
contributions resulting from the trace terms, we get the (infinite) 
twist decomposition of the nonlocal operator we started with:
\begin{equation}
\label{O-nl-t}
\OO_\Gamma(\kappa_1 x,\kappa_2 x)
=\bigoplus_{\tau=\tau_{\rm min}}^\infty 
c^{(\tau)\Gamma'}_{\Gamma }\!(x)\,
\OO^{(\tau)}_{\Gamma'}(\kappa_1 x,\kappa_2 x)\,,
\end{equation}
where $c^{(\tau)\Gamma'}_{\Gamma }\!(x)$ are the coefficient functions
of the nonlocal operators with geometric twist which are 
obtained from the coefficients $c^{(\tau)\Gamma' n'}_{\Gamma n}(x)$.


\noindent
(4)\quad Finally, the {\em projection onto the light--cone},
$x\rightarrow\tilde{x}$, leads to the required light--cone operator with 
well defined geometric twist. However, since the harmonic tensor
polynomials essentially depend on (infinite) sums of powers of $x^2$ and 
$\square$ as well as some specific differential operators in front
of it, in that limit only a finite number of terms survive:
\begin{equation}
\OO_\Gamma(\ka\xx,\kb\xx)=
\bigoplus_{\tau=\tau_{\rm min}}^{\tau_{\rm max}} 
c^{(\tau)\Gamma'}_{\Gamma }\!(\lcx)\,
\OO^{(\tau)}_{\Gamma'}(\ka\xx,\kb\xx)\,.
\end{equation}
The resulting light--ray operators of definite twist
are tensor functions on the light--cone 
which fulfill another kind of tracelessness conditions to be formulated 
with the help of the interior (on the cone) derivative 
(see Appendix~\ref{inner}). 
Let us remark 
that step~(3) and (4) can be interchanged without changing the result.


\subsection{Transformation properties of (non)local operators with definite geometric twist}

Because the notion of geometric twist is related to $(j,d)$ 
of the group $SO(1,2h-1)\times {\Bbb R}_+$, where ${\Bbb R}_+$ is the group 
of dilations, 
it is useful to give the transformation rules of the (unrenormalized) 
nonlocal operators
with respect to this group. We induce the transformation rules of the 
nonlocal operators from the transformation properties of local
fields which are given in~\cite{Dobrev76b,Dobrev77}. 
Let $\OO^{(\tau)}_{\Gamma n}(y, x)$ be the local operator
of geometric twist $\tau$.

The transformation properties of these local operators 
are~~\cite{Dobrev76b,Dobrev77}\\
a) Poincar{\'e} transformations:
\begin{align}
\hat U(a,\Lambda) \OO^{(\tau)}_{\Gamma n}(y, x)\hat U^{-1}(a,\Lambda)
= V_{\Gamma\Gamma'}(\Lambda)\OO^{(\tau)}_{\Gamma' n}(\Lambda^{-1}(y-a),\Lambda^{-1} x),
\quad a\in{\Bbb R}^{2h}, \Lambda\in SO(1,2h-1),
\end{align}
where $V(\Lambda)$ is a finite dimensional 
representation of the Lorentz group.\\
b) Dilations:
\begin{align}
\hat U(\rho) \OO^{(\tau)}_{\Gamma n}(y, x)\hat U^{-1}(\rho)
= \rho^d \OO^{(\tau)}_{\Gamma n}(\rho y, x),
\quad \rho > 0,
\end{align}
where $d$ is the (scale) dimension of the local operator (in mass units).
From group theoretical point of view, the anomalous dimensions do not
destroy the invariance under dilations, they just change the representation 
with respect to the dilation group~\cite{Craigie85}.

The transformation properties for the nonlocal operators are\\
a) Poincar{\'e} transformations:
\begin{align}
\hat U(a,\Lambda) \OO^{(\tau)}_{\Gamma}(y- x,y+ x)\hat U^{-1}(a,\Lambda)
= V_{\Gamma\Gamma'}(\Lambda)\OO^{(\tau)}_{\Gamma'}(\Lambda^{-1}(y-a- x),\Lambda^{-1}(y-a+ x)),
\end{align}
b) Dilations:
\begin{align}
\hat U(\rho) \OO^{(\tau)}_{\Gamma}(y- x,y+ x)\hat U^{-1}(\rho)
= \rho^d \OO^{(\tau)}_{\Gamma }(\rho (y- x),\rho (y+ x)).
\end{align}
Let us point out that in the case of nonlocal operators, i.e., for an
infinite series of local ones, we deal 
with an infinite-dimensional representation of the dilation and 
the Lorentz group.         
Obviously, the translation $\hat U(a,0)=\e^{-\ii a\cdot\Hat P}$ with
$\Hat P_\mu=\ii\frac{\pd}{\pd y^\mu}$ just shifts the expansion point
$y$ by the vector $a$. 
If we had chosen $a= x$, the twist of these
operators would be not invariant under this translation, 
because the differential operators for definite twist do not commute
with $ x$.~\footnote{The total derivative $\pd_+$ is related to translations 
in the following way: $[\hat P,\OO^{(\tau)}_{\Gamma n}(y,x)]=\ii \pd_+ \OO^{(\tau)}_{\Gamma n}(y,x)$.}
For the non-forward matrix element of an uncentred operator 
with definite twist we obtain the relation to the corresponding
centred operator as follows
\begin{align}
\label{transl-matrix}
\langle P_2|\OO^{(\tau)}_\Gamma (\kappa_1  x,\kappa_2 x)|P_1\rangle
&\equiv
\langle P_2|\OO^{(\tau)}_\Gamma (\kappa_+x -\kappa_- x  ,\kappa_+ x+\kappa_- x)|P_1\rangle\nonumber\\
&=
{\cal P}_\Gamma^{(\tau) \Gamma'}\left\{ \e^{\ii\kappa_+ (xP_-)}
\langle P_2| 
\OO^{}_{\Gamma'} ( -\kappa_- x  , \kappa_- x)|P_1\rangle\right\}.
\end{align}
We see that the twist projector acts on the factor $\exp(\ii \kappa_+ (xP_-))$ 
in Eq.~(\ref{transl-matrix}). In the forward case ($P_1=P_2$) this factor gives
no contribution due to $P_-=0$.

Let me 
mention that it is possible to construct conformal 
operators with geometric twist which carry elementary representations
of the conformal group $SO(2,2h)$. The condition of conformal invariance
is given by the infinitesimal transformation law under special conformal transformations at 
$y=0$: 
\begin{align}
[\hat K_\mu,\OO^{(\tau)}_{\Gamma n}(y,x)]|_{y=0}=0,
\end{align}
where $\hat K_\mu$ is the generator for special
conformal transformations~\cite{Craigie85,Karchev83,FGG71}. The conformal
LC-operators with minimal twist $\tau=2h-2$ are~\cite{Craigie85}:
\begin{align}
\OO^\phi_n(y,\lcx)&=(\lcx\pd_+)^n\phi(y)C^{h-3/2}_n(\lcx D_-/\lcx\pd_+)\phi(y),\\
\OO^\psi_n(y,\lcx)&=(\lcx\pd_+)^{n-1}\bar\psi(y) (\lcx\gamma)C^{h-1/2}_{n-1}(\lcx D_-/\lcx\pd_+)\psi(y),\\
\OO^F_n (y,\lcx)&=(\lcx\pd_+)^{n-2}\lcx^\mu F_{\mu}^{a \rho}(y) C^{h+1/2}_{n-2}(\lcx D_-/\lcx\pd_+)
F^a_{\rho\nu}(y)\lcx^\nu.
\end{align}

\newpage
\setcounter{equation}{0}
\chapter{Application of the twist decomposition to different
aspects of the hadron phenomenology}
\section{Quark distribution functions of definite geometric twist}
\label{forward}



The aim of this Section is to present new {\em forward quark distribution 
functions} which are related to the nonlocal LC-operators of different
geometric twist. In that framework, it is possible to investigate 
in a unique manner the contributions resulting from the traces of the 
operators having well-defined twist. In addition, the classification of the 
various distribution functions appears to be quite straightforward.

The definitions of geometric and dynamical twist do not coincide at higher orders.
Thus, the definition of dynamical twist has a mismatch with the conventional 
definition of geometrical twist. This mismatch gives rise to relations of  
Wandzura-Wilczek type which show that dynamical twist distributions contain 
various parts of
different geometric twist. 


\subsection{Quark distribution functions of definite dynamical twist}
Quark distribution functions are usually defined as the forward matrix
elements of bilocal quark-antiquark operators.
Jaffe and Ji~\cite{JJ91,JJ92} observed that one can generate all dynamical
quark distribution functions up to (dynamical) twist-4 by decomposing
the bilocal operators into independent tensor structures of their matrix
elements. 
To implement the light-cone expansion in a systematic way, it is convenient 
to use light-like vectors.
Using the relations (\ref{kin1}),
one may introduce light-like vectors $p_\alpha$ and $\lcx_\alpha$ with
\begin{align}
p^2=0,\qquad \lcx^2=0,
\end{align}
according to (see Eq.~(\ref{x-lc}) with $\rho_\alpha=P_\alpha/M$ and $\rho^2=1$)
\begin{align}
\lcx_\alpha = x_\alpha -\frac{P_\alpha}{M^2} \Big(
(xP) - \sqrt{(xP)^2 - x^2 M^2}
\Big),\qquad
p_\alpha=P_\alpha-\frac{1}{2} \lcx_\alpha\frac{M^2}{\lcx P}.
\end{align}
The spin vector of the nucleon can be decomposed as
\begin{align}
S_\alpha=S^\bot_\alpha+p_\alpha \frac{\lcx S}{\lcx P}-
\frac{1}{2} \lcx_\alpha \frac{\lcx S}{(\lcx P)^2} M^2,
\end{align}
where $S^\bot_\alpha$ is the transversal spin-polarization ($\lcx S^\bot=0$). 
Some useful scalar products are
\begin{align}
\lcx P=\lcx p=\sqrt{(xP)^2 - x^2 M^2},\qquad
p S=-\frac{M^2}{2(\lcx p)}\, \lcx S\, .
\end{align}

Accordingly, one finds nine independent quark distribution 
functions associating with each tensor structure. 
Using the Lorentz-tensor decomposition of the matrix elements of 
the gauge-invariant bilocal quark-antiquark operators
(\ref{O-q5}), (\ref{O-q}), (\ref{M-q5}), and (\ref{N-q}),
the definitions of the chiral-even quark distributions 
are:\footnote{Here we suppress the flavour index.}
\begin{align}
&\langle PS|O_{5\alpha}(\kappa_1\lcx,\kappa_2\lcx)|PS\rangle  
= 2\bigg[ p_{\alpha}
\frac{\lcx S}{\lcx P}
\int_{0}^{1} \d z\, \e^{\ii\kappa z(\lcx P)} g_{1}(z,\mu^2) 
+ S^{\perp}_\alpha\int_{0}^{1} \d z\, \e^{\ii\kappa z(\lcx P)} g_{T}(z,\mu^2)
\nonumber\\ 
&\hspace{8cm}
+ \lcx_{\alpha}\frac{\lcx S}{(\lcx P)^{2}} M^{2}
\int_{0}^{1} \d z\,\e^{\ii\kappa z(\lcx P)} g_{3}(z,\mu^2) \bigg]
\label{q_eq:fvda}
\end{align}
and
\begin{align}
\langle P|O_{\alpha}(\kappa_1\lcx,\kappa_2\lcx)|P\rangle
 =  2 \left[ p_\alpha \int_{0}^{1} \d z\, \e^{\ii\kappa z(\lcx P)}
f_1(z,\mu^2) + \lcx_\alpha \frac{M^2}{\lcx P}  \int_{0}^{1} \d z\, 
\e^{\ii\kappa z (\lcx P)} f_4(z,\mu^2)\right],
\label{q_eq:favda}
\end{align}
while the chiral-odd quark distributions are defined as
\begin{align}
\langle P|N(\kappa_1\lcx,\kappa_2\lcx)|P\rangle
= 2M \int_{0}^{1} \d z\, \e^{\ii\kappa z (\lcx P)} e(z,\mu^2),
\label{q_eq:fsda}
\end{align}
and
\begin{align}
\label{q_eq:ftda}
&\langle PS|M_{5[\alpha\beta]}(\kappa_1\lcx,\kappa_2\lcx)|PS\rangle 
=\frac{4}{M}\bigg[S^{\perp}_{[\alpha} p_{\beta]}
\int_{0}^{1} \d z\, \e^{\ii \kappa z(\lcx P)} h_{1}(z,\mu^2)\\ 
&\hspace{2cm}
-\lcx_{[\alpha} p_{\beta]} 
\frac{\lcx S}{(\lcx P)^{2}}\, M^2 
\int_{0}^{1} \d z\, \e^{\ii\kappa z(\lcx P)} h_{L}(z,\mu^2) 
-\lcx_{[\alpha}S^{\perp}_{\beta]} 
\frac{M^{2}}{\lcx P}  
\int_{0}^{1} \d z\, \e^{\ii\kappa z(\lcx P)} h_{3}(z,\mu^2) \bigg].\nonumber
\end{align}
These distribution functions are dimensionsless functions of the Bjorken 
variable which we here denote by $z$. Additionally, they depend on the renormalization point $\mu^2$ 
as well because the bilocal light-cone operators on the l.h.s. are 
renormalized at $\mu^2$. We defined $\kappa=\kappa_1-\kappa_2$.

In Tab.~\ref{q_tab:1} the classification of these
dynamical quark distribution functions according to  
their spin, dynamical twist and chirality and the
relation to the geometric twist (see Section~\ref{qf-gd})
are given.
\begin{table}
\begin{center}
\renewcommand{\arraystretch}{1.3}
\begin{tabular}{|c|ccc|}
\hline
Twist $t$   & 2 & 3 & 4 \\
    & $O(1)$  & $O(1/Q)$& $O(1/Q^{2})$ \\ \hline
spin ave.& $f_{1}=F^{(2)}$ & $\underline{e}=E^{(3)}$ & $f_{4}=F^{(4)}/2+F_f(F^{(2)},F^{(4)})/2$ \\
$S_{\parallel}$ & $g_{1}=G^{(2)}$ & $\underline{h_{L}}=H^{(3)}+F_L(H^{(2)},H^{(3)})$ & $g_{3}=-G^{(4)}/2+F_g(G^{(2)},G^{(3)},G^{(4)})/2$ \\
$S_{\perp}$ & $\underline{h_{1}}=H^{(2)}$ & $g_{T}=G^{(3)}+F_T(G^{(2)},G^{(3)})$ & $\underline{h_{3}}=H^{(4)}/2+F_h(H^{(2)},H^{(3)},H^{(4)})/2$
\\[2pt]\hline
\end{tabular}
\renewcommand{\arraystretch}{1}
\end{center}
\caption{Spin, dynamical twist and chiral classification of the 
nucleon structure functions.}
\label{q_tab:1}
\end{table}%
The parton distributions in 
the first row are spin-independent, those in the
second and third row describe
longitudinally ($S_{\parallel}$) and transversely ($S_{\perp}$)
polarized nucleons, respectively.
The underlined distributions are referred to as 
chiral-odd, 
because they correspond to 
chirality-violating Dirac matrix structures 
$\Gamma = \{\ii\sigma_{\alpha\beta}  \gamma_{5},\, 1\}$. 
The other distributions are termed chiral-even, because of the 
chirality-conserving structures 
$\Gamma = \{\gamma_{\alpha},\, \gamma_\alpha\gamma_5\}$. 

The nucleon (spin-$1/2$ hadron) has thus three independent dynamical twist-2
quark distributions $f_1$, $g_1$, and $h_1$. These distributions
have a simple parton model interpretation and appear as a leading 
contribution to various hard inclusive processes. 
For example, $g_1$ measures the quark helicity distribution in a 
longitudinally polarized nucleon and is called the
{\it helicity} distribution. 
The both distributions 
$f_1$ and $g_1$ can be measured through deep inelastic lepton-nucleon scattering (DIS).
The function $h_1$ appears as a leading contribution to the transversely 
polarised nucleon-nucleon Drell-Yan process. This distribution measures the 
probability to find a quark with spin polarized along the transverse spin of a nucleon 
minus the probability to find it polarized oppositely and was called
the {\it transversity} distribution in Refs.~\cite{JJ91,JJ92}. 
Additionally, the nucleon has three dynamical twist-3 distributions
$g_T$, $h_L$, and $e$. Physically higher-twist distributions represent 
complicated quark-gluon correlations in the hadrons and cannot be interpreted
in a simple parton model. Therefore, higher-twist or multiparton distributions
are useful to probe quark-gluon dynamics as well as nonperturbative aspect of 
QCD. The distributions $g_T$ and $h_L$ are contributions to proper 
asymmetries in polarized DIS and Drell-Yan processes, respectively.
The direct measurement of $e$ seems very difficult~\cite{JJ92}. The
dynamical twist-4 distributions represent rather complicated two-quark-one-gluon,
two-quark-two-gluon, and four-quark correlations so that they are very 
difficult to handle with. However, in principle, $f_4$ and $h_3$ are 
observable in unpolarized DIS and polarized Drell-Yan processes, 
respectively~\cite{JJ92}. Up to now, no measurement of $g_3$ is 
known in the literature.

\subsection{Twist decomposition of bilinear quark operators}
In order to be able to classify the quark distributions with respect
to geometrical twist, we have to use the corresponding operators
with geometric twist.
Now we apply the bilocal quark operators on the light-cone 
and
the derived scalar and vector operators, 
$O_{(5)}(\kappa_1\lcx,\kappa_2\lcx) = \lcx^\alpha
O_{(5)\alpha}(\kappa_1\lcx,\kappa_2\lcx)$ and 
$M_{(5)\alpha}(\kappa_1\lcx,\kappa_2\lcx)= \lcx^\beta
M_{(5)[\alpha\beta]}(\kappa_1\lcx,\kappa_2\lcx)$, respectively.

The resulting decomposition for the vector and skew tensor operators
-- which is independent of the presence or absence of $\gamma_5$ --
reads:\footnote{The explicite twist decomposition of these light-cone 
operators will be computed in Section~\ref{gluon} 
(see also Ref.~\cite{GLR99}). The twist decomposition of off-cone
operators will be discussed in Sections~\ref{tensor} and \ref{off-cone}.}
\begin{align}
\label{q_O2sca}
O(\kappa_1\lcx,\kappa_2\lcx)
&=
\lcx^\alpha O^{\rm tw2}_\alpha(\kappa_1\lcx,\kappa_2\lcx),\\
\label{q_O_tw_nl}
O_{\alpha}(\kappa_1\lcx,\kappa_2\lcx)&=
 O^{\mathrm{tw2}}_{\alpha}(\kappa_1\lcx,\kappa_2\lcx)
+O^{\mathrm{tw3}}_{\alpha}(\kappa_1\lcx,\kappa_2\lcx)
+O^{\mathrm{tw4}}_{\alpha}(\kappa_1\lcx,\kappa_2\lcx)
,\\
\label{q_M_tw_nl}
M_{[\alpha\beta]}(\kappa_1\lcx,\kappa_2\lcx)&=
M^{\mathrm{tw2}}_{[\alpha\beta]}(\kappa_1\lcx,\kappa_2\lcx)
+M^{\mathrm{tw3}}_{[\alpha\beta]}(\kappa_1\lcx,\kappa_2\lcx)
+M^{\mathrm{tw4}}_{[\alpha\beta]}(\kappa_1\lcx,\kappa_2\lcx)
,\\
\label{q_M_vec}
M_{\alpha}(\kappa_1\lcx,\kappa_2\lcx)&
=M^{\mathrm{tw2}}_{\alpha}(\kappa_1\lcx,\kappa_2\lcx)
+M^{\mathrm{tw3}}_{\alpha}(\kappa_1\lcx,\kappa_2\lcx),\\
N(\kappa_1\lcx,\kappa_2\lcx)&
=N^{\mathrm{tw3}}(\kappa_1\lcx,\kappa_2\lcx)
\end{align}
with 
\begin{align}
\label{q_O2vec}
O^{\mathrm{tw2}}_{\alpha}(\kappa_1\lcx,\kappa_2\lcx)
&=
\int_{0}^{1} \d\lambda
\Big(\pd_\alpha +
\hbox{\Large$\frac{1}{2}$}(\ln\lambda)\,x_\alpha\square\Big)
x^\mu
O_\mu(\kappa_1\lambda x, \kappa_2\lambda x)
\big|_{x=\tilde{x}}
\\
\label{q_O3vec}
O^{\mathrm{tw3}}_{\alpha}
(\kappa_1\lcx,\kappa_2\lcx)
&=
\int_{0}^{1}\d\lambda
\Big(\delta_\alpha^\mu(x\pd)-
x^\mu\pd_\alpha-(1+2\ln\lambda)x_\alpha\pd^\mu
-(\ln\lambda)\, x_\alpha x^\mu\square
\Big) O_\mu(\kappa_1\lambda x, \kappa_2\lambda x)
\big|_{x=\tilde{x}}
\\
\label{q_O4vec}
O^{\mathrm{tw4}}_{\alpha}
(\kappa_1\lcx,\kappa_2\lcx)
&=\lcx_\alpha
\int_{0}^{1}\d\lambda\Big(
(1+\ln\lambda)\pd^\mu+
\hbox{\Large$\frac{1}{2}$}(\ln\lambda)\,x^\mu \square
\Big)
O_\mu(\kappa_1\lambda x, \kappa_2\lambda x)\big|_{x=\tilde{x}}
\\
\label{q_M_tw2_ten}
M^{\mathrm{tw2}}_{[\alpha\beta]}(\kappa_1\lcx,\kappa_2\lcx)
&=
 \int_{0}^{1}\d\lambda\Big\{2\lambda\,
\pd_{[\beta}\delta_{\alpha]}^\mu
-(1-\lambda)\big(2x_{[\alpha}\pd_{\beta]}\pd^\mu
-x_{[\alpha}\delta_{\beta]}^\mu\square\big)\Big\}
x^\nu M_{[\mu\nu]}(\kappa_1\lambda x, \kappa_2\lambda x)
\big|_{x=\tilde{x}}
\\
\label{q_M_tw3_ten}
M^{\rm tw3}_{[\alpha\beta]}(\kappa_1\lcx,\kappa_2\lcx)
&=
\int_{0}^{1}\d\lambda
\Big\{\lambda\big((x\pd)\delta_{[\beta}^\nu
- 2x^\nu\pd_{[\beta}\big)\delta_{\alpha]}^\mu
+\hbox{\Large$\frac{1-\lambda^2}{\lambda}$}
\Big(x_{[\alpha}\big(
\delta_{\beta]}^{[\mu} (x\pd)
-
x^{[\mu}\pd_{\beta]}
\big)\pd^{\nu]}
\nonumber\\
&\qquad-x_{[\alpha}\delta_{\beta]}^{[\mu}x^{\nu]}\square
-x_{[\alpha}\pd_{\beta]}x^{[\mu}\pd^{\nu]}
\Big)\Big\}
M_{[\mu\nu]}(\kappa_1\lambda x, \kappa_2\lambda x)\big|_{x=\lcx}
\\
\label{q_M_tw4_ten}
M^{\rm tw4}_{[\alpha\beta]}(\kappa_1\lcx,\kappa_2\lcx)
&=
\int_{0}^{1}{\d\lambda}
\hbox{\Large$\frac{1-\lambda}{\lambda}$}
\Big\{x_{[\alpha}\delta_{\beta]}^{[\mu}
x^{\nu]}\square
-2x_{[\alpha}\big(
\delta_{\beta]}^{[\mu} (x\pd)
-
x^{[\mu}\pd_{\beta]}
\big)\pd^{\nu]}
\Big\}
M_{[\mu\nu]}(\kappa_1\lambda x, \kappa_2\lambda x)\big|_{x=\lcx}
\\
\label{q_M_tw2_vec}
M^{\mathrm{tw2}}_{\alpha}(\kappa_1\lcx,\kappa_2\lcx)
&=
M_\alpha(\kappa_1\lcx, \kappa_2\lcx)
-\lcx_\alpha \pd^\mu
\int_0^1\d\lambda\,\lambda\,
M_{\mu}(\kappa_1\lambda x, \kappa_2\lambda x)
\big|_{x=\tilde{x}}
\\
\label{q_M_tw3_vec}
M^{\mathrm{tw3}}_{\alpha}(\kappa_1\lcx,\kappa_2\lcx)
&=
\lcx_\alpha\pd^\mu
\int_0^1\d\lambda\,\lambda
M_\mu(\kappa_1\lambda x, \kappa_2\lambda x)\big|_{x=\tilde{x}}.
\end{align}
Let us remark that the (axial) vector and skew tensor operators of  
twist $\tau$, $O^{(\tau)}_{(5)\alpha}(\kappa_1\lcx, \kappa_2 \lcx)$ and
$M^{(\tau)}_{(5)[\alpha\beta]}(\kappa_1\lcx, \kappa_2 \lcx)$,
are obtained from the (undecomposed)
operators $O_{(5)\alpha}(\kappa_1\lcx, \kappa_2 \lcx)$ and 
 $M_{(5)[\alpha\beta]}(\kappa_1\lcx, \kappa_2 \lcx)$
by the application of the 
twist projectors on the light-cone 
${\cal P}^{(\tau)\mu}_\alpha$ and
${\cal P}^{(\tau)[\mu\nu]}_{[\alpha\beta]}$ 
(see also Eqs.~(\ref{twist-nl}) and (\ref{proj-nl}))
defined by Eqs.~(\ref{q_O2vec}) -- (\ref{q_O4vec})
and (\ref{q_M_tw2_ten}) -- (\ref{q_M_tw4_ten}), respectively:
\begin{align}
\label{q_OPROJ}
O^{(\tau)}_{(5)\alpha}(\kappa_1\lcx, \kappa_2 \lcx) &= 
({\cal P}^{(\tau)\mu}_\alpha O_{(5)\mu})(\kappa_1\lcx, \kappa_2 \lcx)
\\
\label{q_MPROJ}
M^{(\tau)}_{(5)[\alpha\beta]}(\kappa_1\lcx, \kappa_2 \lcx) &= 
({\cal P}^{(\tau)[\mu\nu]}_{[\alpha\beta]} 
M_{(5)[\mu\nu]})(\kappa_1\lcx, \kappa_2 \lcx)
\\
\label{q_MVPROJ}
M^{(\tau)}_{(5)\alpha}(\kappa_1\lcx, \kappa_2 \lcx) &= 
({\cal P}^{(\tau)\mu}_{(v)\alpha} 
M_{(5)\mu})(\kappa_1\lcx, \kappa_2 \lcx).
\end{align}
The twist--2 vector and skew tensor 
operators, Eqs.~(\ref{q_O2vec}) and (\ref{q_M_tw2_ten}), are related to the
corresponding scalar and vector operators, (\ref{q_O2sca}) and (\ref{q_M_tw2_vec}),
respectively, in group theoretical way.


Here, we give the corresponding local expressions of the 
nonlocal operators, Eqs. (\ref{q_O2sca}) -- (\ref{q_M_tw3_vec}). 
The relation between the local and nonlocal operators are 
\begin{align}
O_\alpha(\kappa_1\lcx, \kappa_2\lcx)
=\sum_{n=0}^\infty \frac{\kappa_-^n}{n!}\,O_{\alpha n}(y,\lcx)|_{y=\kappa_+\lcx},
\qquad\qquad
M_{[\alpha\beta]}(\kappa_1\lcx, \kappa_2\lcx)
=\sum_{n=0}^\infty \frac{\kappa_-^n}{n!}\,M_{[\alpha\beta] n}(y,\lcx)|_{y=\kappa_+\lcx}.
\nonumber
\end{align}
The expressions for the local operators are: 
\begin{align}
\label{q_O2sca_l}
O^{\rm tw2}_{n+1}(y,\lcx)
&\equiv
\lcx^\mu O_{\mu n}(y,\lcx)
=
\bar{\psi}(y)\gamma^5(\gamma\lcx)
(\lcx \Tensor D)^n \psi(y)
\\
\label{q_O2vec_l}
O^{\mathrm{tw2}}_{\alpha n}(y,\lcx)
&=\hbox{\Large$\frac{1}{n+1} $}
\Big(\pd_\alpha -\hbox{\Large$\frac{1}{2(n+1)}$}
\,x_\alpha\square\Big)
O_{n+1}(y,x)
\big|_{x=\tilde{x}}
\\
\label{q_O3vec_l}
O^{\mathrm{tw3}}_{\alpha n}(y,\lcx)
&=\hbox{\Large$\frac{1}{n+1}$}
\Big(n\delta_\alpha^\mu-x^\mu\pd_\alpha
-\hbox{\Large$\frac{1}{n+1}$}
x_\alpha\big((n-1)\pd^\mu-x^\mu\square\big)
\Big) O_{\mu n}(y,x)\big|_{x=\tilde{x}}
\\
\label{q_O4vec_l}
O^{\mathrm{tw4}}_{\alpha n}(y,\lcx)
&=\hbox{\Large$\frac{1}{(n+1)^2}$}
\,\lcx_\alpha\Big(n\pd^\mu-\hbox{\Large$\frac{1}{2}$}
x^\mu \square\Big)O_{\mu n}(y,x)\big|_{x=\tilde{x}}\\
\label{q_M_tw2_ten_l}
M^{\mathrm{tw2}}_{[\alpha\beta]n}(y,\lcx)
&=\hbox{\Large$\frac{1}{n+2}$}
 \Big\{2\pd_{[\beta}\delta_{\alpha]}^\mu
-\hbox{\Large$\frac{1}{n+1}$}
\big(2x_{[\alpha}\pd_{\beta]}\pd^\mu
-x_{[\alpha}\delta_{\beta]}^\mu\square\big)\Big\}M_{\mu n+1}(y,x)\big|_{x=\tilde{x}}
\\
\label{q_M_tw3_ten_l}
M^{\rm tw3}_{[\alpha\beta]n}(y,\lcx)
&=\hbox{\Large$\frac{1}{n+2}$}
\Big\{\big(n\delta_{[\beta}^\nu
- 2x^\nu\pd_{[\beta}\big)\delta_{\alpha]}^\mu
+ \hbox{\Large$\frac{2}{n}$}
\Big(x_{[\alpha}\big((n-1)\delta_{\beta]}^{[\mu} 
-x^{[\mu}\pd_{\beta]}\big)\pd^{\nu]}
-x_{[\alpha}\delta_{\beta]}^{[\mu}x^{\nu]}\square\nonumber\\
&\qquad
-x_{[\alpha}\pd_{\beta]}x^{[\mu}\pd^{\nu]}\Big)\Big\}M_{[\mu\nu] n}(y,x)\big|_{x=\lcx}
\\
\label{q_M_tw4_ten_l}
M^{\rm tw4}_{[\alpha\beta]n}(y,\lcx)
&=\hbox{\Large$\frac{1}{(n+1)n}$}
\Big\{x_{[\alpha}\delta_{\beta]}^{[\mu}
x^{\nu]}\square
-2x_{[\alpha}\big((n-1)\delta_{\beta]}^{[\mu} 
-x^{[\mu}\pd_{\beta]}\big)\pd^{\nu]}\Big\}
M_{[\mu\nu] n}(y,x)\big|_{x=\lcx}
\\
\label{q_M_tw2_vec_l}
M^{\mathrm{tw2}}_{\alpha n+1}(y,\lcx)
&=
M_{\alpha n+1}(\lcx)-\hbox{\Large$\frac{1}{n+2}$}
\,\lcx_\alpha \pd^\mu M_{\mu n+1}(y,x)\big|_{x=\tilde{x}}
\\
\label{q_M_tw3_vec_l}
M^{\mathrm{tw3}}_{\alpha n+1}(y,\lcx)
&=\hbox{\Large$\frac{1}{n+2}$}
\,\lcx_\alpha\pd^\mu M_{\mu n+1}(y,x)\big|_{x=\tilde{x}}\\
\label{N_sca_l}
N^{\rm tw3}_{n}(y,\lcx)
&=\bar{\psi}(y)(\lcx \Tensor D)^n \psi(y).
\end{align}
The twist projectors of the  nonlocal 
as well as local LC operators can be formulated 
in terms of {\em inner derivatives} on the light--cone 
(see Section~\ref{tensor} and Ref.~\cite{GL01}).

\subsection{Forward matrix elements of LC--operators with definite geometric twist}
Now, we define the (polarized) quark distribution functions 
for the bilinear quark operators with definite twist. As usual,
the matrix elements of the nucleon targets, 
$\bar{U}(P,S)\gamma^\mu U(P,S)=2P^\mu$ and 
$\bar{U}(P,S)\gamma^5\gamma^\mu U(P,S)=2S^\mu$, are related to the 
nucleon momentum $P_\mu$ and nucleon spin vector $S_\mu$, respectively.
Here $U(P,S)$ denotes the free hadronic spinor.

Let us first consider the chiral-even {\em pseudo scalar operator} in the 
polarized case. The (forward) matrix element of this twist-2 operator, 
Eq.~(\ref{q_O2sca}), taken between hadron states $|PS\rangle$ is 
trivially represented as 
\begin{align}
\label{q_matrix_O_t2_sca}
\langle PS|O_5^{\text{tw2}}(\kappa_1\lcx,\kappa_2\lcx)|PS\rangle
=
2(\lcx S)\int_0^1\d z\, G^{(2)}(z,\mu^2)\,{\e^{\ii\kappa z(\lcx P)}}
=2(\lcx S)\sum_{n=0}^\infty \frac{(\ii\kappa(\lcx P))^n}{n!}\,G^{(2)}_n(\mu^2).
\end{align}
Here, $G^{(2)}(z,\mu^2)$ is the twist-2 parton 
distribution function which by a Mellin 
transformation is obtained from the corresponding moments 
\begin{align}
G^{(2)}_n(\mu^2)=\int_0^1\d z\, z^n G^{(2)}(z,\mu^2).
\nonumber
\end{align} 

Now we consider the chiral-even {\em  axial vector operator}.
Using the projection properties (\ref{q_OPROJ}), 
we introduce the  parton distribution functions 
$G^{(\tau)}(z, \mu^2)$ of twist $\tau$ by
\begin{align}
\label{q_Gfct}
\langle PS|O^{(\tau)}_{5\alpha}(\kappa_1\lcx, \kappa_2 \lcx)|PS\rangle 
&\equiv 
\langle PS|({\cal P}^{(\tau)\beta}_\alpha 
O^{(\tau)}_{5\beta})(\kappa_1\lcx, \kappa_2 \lcx)|PS\rangle\nonumber\\
&= {\cal P}^{(\tau)\beta}_\alpha
\Big(2 S_\beta\int_0^1\d z\, G^{(\tau)}(z,\mu^2)\,
{\e^{\ii\kappa z(\lcx P)}}\Big).
\end{align} 
For $\tau =2$ this is  consistent with (\ref{q_matrix_O_t2_sca}), it reads
\begin{align}
\langle PS|O^{\text{tw2}}_{5\alpha}(\kappa_1\lcx,\kappa_2\lcx)|PS\rangle
&=2\int_0^1\d\lambda\Big[\pd_\alpha+\frac{1}{2} \ln(\lambda) 
x_\alpha\square\Big]
(xS)\int_0^1\d z\, G^{(2)}(z,\mu^2)\,{\e^{\ii\kappa \lambda z(xP)}}
\big|_{x=\tilde{x}}\ .
\nonumber
\end{align}
Using the projection operators as they are determined by Eqs.~(\ref{q_O2vec})
-- (\ref{q_O4vec}) we obtain (from now on we suppress $\mu^2$)
\begin{align}
\label{q_matrix_O_t2_nl}
&\langle PS|O^{\text{tw2}}_{5\alpha}(\kappa_1\lcx,\kappa_2\lcx)|PS\rangle
=2\int_0^1\d\lambda\int_0^1\d z\,G^{(2)}(z)
\Big[S_\alpha+\ii\kappa\lambda z 
P_\alpha(\tilde{x}S),\nonumber\\
&\hspace{7cm}+\frac{\tilde{x}_\alpha}{2}(\tilde{x}S)M^2
(\ii\kappa\lambda z)^2\big(\ln\lambda\big)\Big]
\e^{\ii\kappa \lambda z(\tilde{x}P)}\nonumber\\
&\qquad=2\sum_{n=0}^\infty \frac{(\ii\kappa(\lcx P))^n}{n!}G^{(2)}_n\Big\{
\frac{1}{n+1}\Big( S_\alpha+n P_\alpha \frac{\lcx S}{\lcx P}\Big)
-\frac{n(n-1)}{2(n+1)^2}\lcx_\alpha\frac{\lcx S}{(\lcx P)^2}M^2\Big\},
\\
\label{q_matrix_O_t3_nl}
&\langle PS|O^{\mathrm{tw3}}_{5\alpha}(\kappa_1\lcx,\kappa_2\lcx)|PS\rangle
=2\int_0^1\!\d\lambda\int_0^1\!\d z\, G^{(3)}(z)
\Big[\Big(S_\alpha (\tilde{x}P)
-P_\alpha (\tilde{x}S)\Big)\ii\kappa\lambda z\nonumber\\
&\hspace{7cm}-\tilde{x}_\alpha M^2(\tilde{x}S)
(\ii\kappa\lambda z)^2\ln\lambda\Big]
\e^{\ii\kappa\lambda z(\tilde{x}P)},\nonumber\\
&\qquad=2\sum_{n=1}^\infty \frac{(\ii\kappa(\lcx P))^n}{n!}G^{(3)}_n\Big\{
\frac{n}{n+1}\Big( S_\alpha-P_\alpha \frac{\lcx S}{\lcx P}\Big)
+\frac{n(n-1)}{(n+1)^2}\lcx_\alpha\frac{\lcx S}{(\lcx P)^2}M^2\Big\},
\\
\label{q_matrix_O_t4_nl}
&\langle PS|O^{\mathrm{tw4}}_{5\alpha}(\kappa_1\lcx,\kappa_2\lcx)|PS\rangle
=
\int_0^1\d\lambda\int_0^1\d z\, G^{(4)}(z)
\Big[\lcx_\alpha(\tilde{x}S)M^2
(\ii\kappa\lambda z)^2\ln\lambda\Big]
\e^{\ii\kappa\lambda z(\tilde{x}P)},\nonumber\\
&\qquad=-\sum_{n=2}^\infty \frac{(\ii\kappa(\lcx P))^n}{n!}G^{(4)}_n
\frac{n(n-1)}{(n+1)^2}\lcx_\alpha\frac{\lcx S}{(\lcx P)^2}M^2.
\end{align}

In the first expression of any equation we have written the nonlocal LC-operators
of definite twist, 
and in the second line, after expanding the exponential, we introduced the 
moments of the structure functions. The {\em local} 
twist-2 and twist-3 matrix elements of traceless operators are given 
off--cone in Refs.~\cite{ehrnsperger94,maul97} for $n\leq 3$ and in
Refs.~\cite{PR98,BT99} for any $n$. Obviously,
the trace terms which have been explicitly subtracted  
are proportional to $M^2$. According to the terminology of
Jaffe and Ji, they contribute to {\it dynamical} twist-4. 
For the twist--2 operator we observe that the terms proportional to 
$S_\alpha$, $P_\alpha$ and $\lcx_\alpha$ have contributions starting 
with the zeroth, first and second moment, respectively.
The twist--3 operator starts with the first moment, and the
twist--4 operator starts with the second moment. Analogous statements
also hold for the twist--$\tau$ operators below.

Putting together the different twist contributions we obtain,
after replacing $\ii\kappa\lambda z(\lcx P)$ by $\lambda\pd/\pd\lambda$
and performing partial integrations, the following forward
matrix element of the original operator ($\zeta=\kappa(\lcx P)$)
\begin{align}
\label{q_O_full}
\langle PS|O_{5\alpha}(\kappa_1\lcx,\kappa_2\lcx)|PS\rangle
&=2P_\alpha\frac{\lcx S}{\lcx P}\int_0^1\d z
\Big(G^{(2)}(z)-G^{(3)}(z)\Big)[e_0(\ii\zeta z)-e_1(\ii\zeta z)]\\
&+2S_\alpha\int_0^1\d z
\Big(G^{(2)}(z)e_1(\ii\zeta z)+G^{(3)}(z)
[e_0(\ii\zeta z)-e_1(\ii\zeta z)]\Big)\nonumber\\
&-\lcx_\alpha
\frac{\lcx S}{(\lcx P)^2} M^2\int_0^1\!\d z
\Big(G^{(2)}(z)-2G^{(3)}(z)+G^{(4)}(z)\Big)\nonumber\\
&\qquad\times
\Big[e_0(\ii\zeta z)-3e_1(\ii\zeta z)
+2\int_0^1\!\d\lambda e_1(\ii\zeta\lambda z)\Big], \nonumber
\end{align}
where we introduced the following ``truncated exponentials''
\begin{align}
e_0(\ii\zeta z)=\e^{\ii\zeta z},\quad 
e_1(\ii\zeta z)=\int_0^1\!\d\lambda\,\e^{\ii\zeta z\lambda}
=\frac{\e^{\ii\zeta z}-1}{\ii\zeta z},
\quad\cdots\quad,
e_{n+1}(\ii\zeta z)
=\frac{(-1)^{n}}{n!}\int_0^1\!\d\lambda\,\lambda^n\,\e^{\ii\zeta z\lambda}.
\end{align}
As it should be the application of the projection operators
${\cal P}^{(\tau)\beta}_\alpha$ onto (\ref{q_O_full}) reproduces
the matrix elements (\ref{q_matrix_O_t2_nl}) --
(\ref{q_matrix_O_t4_nl}). 
In comparison with the dynamical distribution functions
we observe that the geometric distribution functions are accompanied
not simply by the exponentials, $e_0(\ii\zeta z)$, but by more 
involved combinations whose series expansion directly leads to the
representations with the help of moments.

Now we consider the chiral-even {\em vector operator}
$O_{\alpha}(\kappa_1\lcx,\kappa_2\lcx)$,
which
obeys relations Eqs.~(\ref{q_O2sca}) -- (\ref{q_O4vec}), as well as 
(\ref{q_OPROJ}) with the {\em same} projection operator
as the axial vector operator. Let us introduce the corresponding
parton distribution functions $F^{(\tau)}(z)$ of twist $\tau$ by
\begin{align}
\label{q_Ffct}
\langle PS| O^{(\tau)}_\alpha(\kappa_1\lcx, 
\kappa_2 \lcx)|PS\rangle 
&= {\cal P}^{(\tau)\mu}_\alpha\Big(
2 P_\mu\int_0^1\d z\, F^{(\tau)}(z)\,{\e^{\ii\kappa z(\lcx P)}}
\Big).
\end{align}
Again, this is consistent with the definition of the unpolarized 
distribution function $F^{(2)}(z)$ by the matrix element of the 
corresponding twist-2 {\em scalar operator} 
\begin{align}
\label{q_matrix_O_tw2_sca}
\langle P| O^{\text{tw2}}(\kappa_1\lcx,\kappa_2\lcx)|P\rangle
=
2(\lcx P)\int_0^1\d z\, F^{(2)}(z)\,\e^{\ii\kappa z(\lcx P)}
=2(\lcx P)\sum_{n=0}^\infty \frac{(\ii\kappa(\lcx P))^n}{n!}F^{(2)}_n .
\end{align}
The forward matrix elements of the {\em vector operators} of twist $\tau$
are obtained as follows:
\begin{align}
\label{q_matrix_O_tw2_nl}
\hspace{-.3cm}
\langle P|
O^{\text{tw2}}_{\alpha}(\kappa_1\lcx,\kappa_2\lcx)|P\rangle
&=2\int_0^1\!\d\lambda\!\int_0^1\!\d z\,F^{(2)}(z)
\Big[P_\alpha\big(1+\ii\kappa\lambda z(\tilde{x}P)\big)\nonumber\\
&\qquad\qquad+\lcx_\alpha M^2\ii\kappa\lambda z\ln\lambda
\Big(1+\frac{1}{2}\ii\kappa\lambda z(\tilde{x}P)\Big)\Big]
\e^{\ii\kappa \lambda z(\tilde{x}P)}\nonumber\\
\hspace{-.3cm}
&=2\sum_{n=0}^\infty \frac{(\ii\kappa(\lcx P))^n}{n!}F^{(2)}_n\Big\{
P_\alpha -\frac{n}{2(n+1)}\lcx_\alpha\frac{M^2}{\lcx P}\Big\},
\\
\hspace{-.3cm}
\label{q_matrix_O_tw3_nl}
\langle P|
 O^{\mathrm{tw3}}_{\alpha}(\kappa_1\lcx,\kappa_2\lcx)|P\rangle
&=-2\lcx_\alpha M^2\int_0^1\!\d\lambda
\int_0^1\!\d z\, F^{(3)}(z)
\ii\kappa\lambda z\,\Big[
\Big(1+2\ln\lambda\big)\nonumber\\
&\qquad\qquad+\ii\kappa\lambda z\big(\ln\lambda\big)
\left(\tilde{x}P\right)\Big]
\e^{\ii\kappa\lambda z(\tilde{x}P)} 
\equiv 0\,,
\\
\hspace{-.3cm}
\langle P|
O^{\mathrm{tw4}}_{\alpha}(\kappa_1\lcx,\kappa_2\lcx)|P\rangle
&=-2\lcx_\alpha M^2
\int_0^1\!\d\lambda\int_0^1\!\d z\,F^{(4)}(z)
\ii\kappa\lambda z\, \big(\ln\lambda\big)
\Big(1+\frac{1}{2}\ii\kappa\lambda z(\tilde{x}P)\Big)
\e^{\ii\kappa \lambda z(\tilde{x}P)}\nonumber
\\
\hspace{-.3cm}
&=\sum_{n=1}^\infty \frac{(\ii\kappa(\lcx P))^n}{n!}F^{(4)}_n
\frac{n}{(n+1)}\lcx_\alpha\frac{M^2}{\lcx P}.
\end{align}

The off--cone traceless local twist-2 matrix element has already been 
obtained in~\cite{guth,GP}.
The vanishing of the twist-3 distribution function $F^{(3)}(z)$ which
follows here by partial integration proves in the {\em nonlocal} case
have already proven for the {\em local} twist-3 operator
by Jaffe and Soldate~\cite{JS82}. 
The trace terms of the twist-2 operator and the twist-4 
operator itself both contribute to dynamical twist-4 and, consequently, 
to the same $1/Q^2$-behaviour in the cross section. 

Putting together all the contributions of different twist and 
performing partial $\lambda$--integrations, we obtain the following 
representation for the forward matrix element of the original operator
\begin{align}
\label{q_matrix_O_nl}
&\langle P| O^{}_{\alpha}(\kappa_1\lcx,\kappa_2\lcx)|P\rangle
=2P_\alpha\int_0^1\d z\, F^{(2)}(z) e_0(\ii\zeta z)
-
\lcx_\alpha\frac{M^2}{\lcx P}\int_0^1\d z
\Big(F^{(2)}(z)-F^{(4)}(z)\Big)\nonumber\\
&\hspace{8cm}
\times
\big[e_0(\ii\zeta z)- e_1(\ii\zeta z)\big].
\end{align}
Let us now consider the 
simplest bilocal {\em scalar operator},
$N(\kappa_1\lcx,\kappa_2\lcx)\equiv N^{\rm tw3}(\kappa_1\lcx,\kappa_2\lcx)$,
which is trivially a twist-3 operator on the light-cone.
Its matrix element arises as
\begin{align}
\langle P|N(\kappa_1\lcx,\kappa_2\lcx)|P\rangle
=2M\int_0^1\d z\, E^{(3)}(z)\,{\e^{\ii\kappa z(\lcx P)}}
=2M\sum_{n=0}^\infty \frac{(\ii\kappa(\lcx P))^n}{n!}E^{(3)}_n ,
\end{align}
where $E^{(3)}(z)$ is another spin-independent twist-3 structure 
function.

Now, we calculate the matrix elements of the chiral-odd {\em  vector 
operators}. In fact they may be obtained from the chiral-odd skew-tensor 
operator by contraction with $\lcx_\beta$ (see Eq.~(\ref{q_M_vec})). 
This also determines the 
corresponding structure functions which are introduced by
\begin{align}
\label{q_Hfct}
\langle PS|M^{(\tau)}_{5[\alpha\beta]}(\kappa_1\lcx, \kappa_2 \lcx)|PS\rangle 
&=\frac{2}{M}{\cal P}^{(\tau)[\mu\nu]}_{[\alpha\beta]}
\Big( (S_\mu P_\nu-S_\nu P_\mu) 
\int_0^1\d z\, H^{(\tau)}(z)\,{\e^{\ii\kappa  z(\lcx P)}}
\Big).
\end{align}
Applying the same procedure as above we obtain
\begin{align}
\label{q_matrix_Mv_tw2}
\langle PS|M^{\mathrm{tw2}}_{5\alpha}(\kappa_1\lcx,\kappa_2\lcx)|PS\rangle
&=
\frac{2}{M}\int_0^1\!\d z\, H^{(2)}(z)
\bigg[\Big(S_{\alpha}(\tilde{x}P)-P_\alpha(\tilde{x}S)\Big)
\e^{\ii\kappa  z(\tilde{x}P)}\nonumber\\
&\qquad\qquad\qquad\qquad
+\tilde{x}_\alpha(\tilde{x}S)M^2
\int_0^1\ \d \lambda\,\ii\kappa z\lambda^2\,
\e^{\ii\kappa \lambda z(\tilde{x}P)}\bigg],\\
&=\frac{2}{M}\sum_{n=0}^\infty \frac{(\ii\kappa(\lcx P))^n}{n!}H^{(2)}_n\Big\{
 S_\alpha(\lcx P)- P_\alpha (\lcx S)
+\frac{n}{n+2}\lcx_\alpha\frac{\lcx S}{\lcx P} M^2\Big\},
\nonumber\\
\label{q_matrix_Mv_tw3}
\langle PS|M^{\mathrm{tw3}}_{5\alpha}(\kappa_1\lcx,\kappa_2\lcx)|PS\rangle
&=-2\int_0^1\!\d\lambda
\int_0^1 \d z\,H^{(3)}(z)
\Big[\tilde{x}_\alpha (\tilde{x}S)M\ii\kappa z \lambda^2\Big]
\e^{\ii\kappa \lambda z(\tilde{x}P)},\nonumber\\
&=-2\sum_{n=1}^\infty \frac{(\ii\kappa(\lcx P))^n}{n!}H^{(3)}_n
\frac{n}{n+2}\lcx_\alpha\frac{\lcx S}{\lcx P} M.
\end{align}
The twist--2 part is in agreement with Jaffe and Ji's local 
expression~(see Eq.~(45) in \cite{JJ92}) and also with~\cite{Koike95}. 
Adding up both local terms the matrix element of 
$M_{\alpha}(\kappa_1\lcx,\kappa_2\lcx)$ 
coincides with the corresponding one of \cite{JJ92}; its nonlocal 
version reads:
\begin{align}\label{q_M_vector_matrix}
\langle PS|M_{5\alpha}(\kappa_1\lcx,\kappa_2\lcx)|PS\rangle
=&\,
\frac{2}{M}\Big(S_{\alpha}(\tilde{x}P)-P_\alpha(\tilde{x}S)\Big)
\int_0^1\!\d z H^{(2)}(z)\, e_0(\ii\zeta z)
\\
&+\tilde{x}_\alpha \frac{\tilde{x}S}{\lcx P} M
\int_0^1\!\d z \Big(H^{(2)}(z)-H^{(3)}(z)\Big)
\Big( e_0(\ii\zeta z) +2 e_2(\ii\zeta z)\Big)\nonumber.
\end{align}
Obviously, the twist-2 distribution function $H^{(2)}(z)$ and 
the twist-3 distribution function $H^{(3)}(z)$ contribute 
to the same $1/Q^2$-behaviour of the cross section.
A similar {\em nonlocal} twist--2 matrix element (modulo correction
of the trace term) has been given in~\cite{Kodaira99}.

Now, we consider the more complicated chiral-odd {\em  skew tensor operator}. 
Let us note that Jaffe and Ji~\cite{JJ92} have not considered the matrix
elements of the traceless skew tensor operator. They have just used the 
corresponding traceless vector operator in their investigations.
We obtain the matrix elements of the skew tensor operators of twist $\tau$ 
by using the projectors determined by Eqs.~(\ref{q_M_tw2_ten})
-- (\ref{q_M_tw4_ten}) and Eq.~(\ref{q_Hfct}):
\begin{align}
\label{q_matrix_M_tw2_nl}
&\langle PS|
M^{\mathrm{tw2}}_{5[\alpha\beta]}(\kappa_1\lcx,\kappa_2\lcx)|PS\rangle
=\frac{2}{M}\int_0^1\d\lambda\int_0^1\d z\, H^{(2)}(z)
\Big[2\lambda\, S_{[\alpha} P_{\beta]}\big(2+\ii\kappa\lambda z(\lcx P)\big)
\nonumber\\
&\qquad\qquad\qquad
+\left(1-\lambda\right)M^2
\lcx_{[\alpha}\Big\{4(\ii\kappa\lambda z)S_{\beta]}
+(\ii\kappa\lambda z)^2\big(S_{\beta]}(\lcx P)
+P_{\beta]}(\lcx S)\big)\Big\}
\Big]\e^{\ii\kappa\lambda z(\lcx P)},\nonumber\\
&\quad
=\frac{2}{M}\sum_{n=0}^\infty \frac{(\ii\kappa(\lcx P))^n}{n!}H^{(2)}_n\Big\{
 2S_{[\alpha}P_{\beta]}
+\frac{4n}{(n+2)(n+1)}\frac{M^2}{\lcx P}\lcx_{[\alpha}S_{\beta]}
\nonumber\\
&\hspace{6cm}
+\frac{n(n-1)}{(n+2)(n+1)}\frac{M^2}{\lcx P}\lcx_{[\alpha}\Big(S_{\beta]}
+P_{\beta]}\frac{\lcx S}{\lcx P}\Big)\Big\},
\\
&\langle PS|
M^{\mathrm{tw3}}_{5[\alpha\beta]}(\kappa_1\lcx,\kappa_2\lcx)|PS\rangle
=-2\int_0^1\d\lambda\frac{1-\lambda^2}{\lambda}
\int_0^1\d z\,\ii\kappa\lambda z H^{(3)}(z) M
\lcx_{[\alpha}\Big\{ S_{\beta]}
+\ii\kappa\lambda  z P_{\beta]}(\lcx S)\Big\}\nonumber\\
&\hspace{10cm}\times\e^{\ii\kappa\lambda z(\lcx P)},\\
&\phantom{\langle PS|
M^{\mathrm{tw2}}_{\alpha\beta}(\kappa_1\lcx,\kappa_2\lcx)|PS\rangle}
=-2\sum_{n=1}^\infty \frac{(\ii\kappa(\lcx P))^n}{n!}H^{(3)}_n
\frac{2}{n+2}\frac{M}{\lcx P}\lcx_{[\alpha}\Big(S_{\beta]}
+(n-1)P_{\beta]}\frac{\lcx S}{\lcx P}\Big),\nonumber\\
\label{q_matrix_M_tw4_nl}
&\langle PS|
M^{\mathrm{tw4}}_{5[\alpha\beta]}(\kappa_1\lcx,\kappa_2\lcx)|PS\rangle
=-2\int_0^1\d\lambda\,\frac{1-\lambda}{\lambda}
\int_0^1\d z\, (\ii \lambda\kappa z)^2 H^{(4)}(z)M
\lcx_{[\alpha}\Big\{S_{\beta]}(\lcx P)-P_{\beta]}(\lcx S)\Big\}\nonumber\\
&\hspace{12cm}\times\e^{\ii\kappa\lambda z(\lcx P)},\nonumber\\
&\phantom{\langle PS|
M^{\mathrm{tw2}}_{[\alpha\beta]}(\kappa_1\lcx,\kappa_2\lcx)|PS\rangle}
=-2\sum_{n=2}^\infty \frac{(\ii\kappa(\lcx P))^n}{n!}H^{(4)}_n
\frac{n-1}{n+1}\frac{M}{\lcx P}\lcx_{[\alpha}\Big(S_{\beta]}
-P_{\beta]}\frac{\lcx S}{\lcx P}\Big).
\end{align}

Again, the moments of distributions functions of twist $\tau = 2,3$
and $4$ begin with $n=0,1$ and $2$, respectively. In addition,
we remark that only those terms of the operator (\ref{q_M_tw3_ten}) 
contribute to the twist--3 structure function which result from
the trace terms of (\ref{q_M_tw2_ten}). Analogous to the vector case
the forward matrix element of the `true' twist--3 part of (\ref{q_M_tw3_ten})
vanishes. 
In Eq.~(\ref{q_matrix_M_tw4_nl}) 
only the twist--4 operator contributes which result from
the trace terms of (\ref{q_M_tw2_ten}).
Let us also mention that after multiplication of (\ref{q_matrix_M_tw4_nl}) with
$\lcx_\alpha$ (or $\lcx_\beta$) the matrix element vanishes because the 
corresponding vector operator does not contain any twist-4 contribution.
This is a simple but important property which may be traced back to the fact
that the corresponding Young pattern $[n+2] = (n,1,1)$ does allow only 
$n$ symmetrizations; it is therefore characteristic for any twist--4 
skew tensor operator.  

The matrix element of the original skew tensor operator is obtained as
\begin{align}
\label{q_M_full_JJ}
\langle PS|M_{5[\alpha\beta]}(\kappa_1\lcx,\kappa_2\lcx)&|PS\rangle
=
\frac{4}{M}S_{[\alpha}P_{\beta]}
\int_0^1\d z H^{(2)}(z)e_0(\ii\zeta z)\\
&+2\lcx_{[\alpha}P_{\beta]}\frac{\lcx S}{(\lcx P)^2} M
\int_0^1\d z \Big\{
  H^{(2)}(z)\big[e_0(\ii\zeta z) + 2e_1(\ii\zeta z) + 6e_2(\ii\zeta z)\big]
\nonumber\\
&
- H^{(3)}(z)\big[1 +2e_0(\ii\zeta z) + 6e_2(\ii\zeta z)\big]
+ H^{(4)}(z)\big[1 + e_0(\ii\zeta z) - 2e_1(\ii\zeta z)\big]
\Big\}\nonumber\\
&
+2\lcx_{[\alpha}S_{\beta]}\frac{M}{\lcx P}
\int_0^1\d z \Big\{
  H^{(2)}(z)\big[e_0(\ii\zeta z)-2e_1(\ii\zeta z)-2e_2(\ii\zeta z)\big]
\nonumber\\
&
+ H^{(3)}(z)\big[1+2e_2(\ii\zeta z)\big]
- H^{(4)}(z)\big[1+e_0(\ii\zeta z)-2e_1(\ii\zeta z)\big]
\Big\}.
\nonumber
\end{align}

\subsection{Relations between quark distribution functions of dynamical and 
geometric twist}
\label{qf-gd}
Obviously, since the geometric twist distribution functions are related to 
traceless operators they differ from the dynamical distribution 
functions.
As far as the scalar LC--operators are concerned which definitely are
of twist--2 the new and the old distributions functions coincide.

In order to be able to compare dynamical and geometrical distribution 
functions, we rewrite the matrix elements (\ref{q_O_full}), (\ref{q_matrix_O_nl}) and
(\ref{q_M_full_JJ}) in terms of $p_\alpha$ and $S^\bot_\alpha$.
Using this parametrization we obtain for the chiral-even quark distributions
\begin{align}
\label{q_JJ_full}
&\langle PS|O_{5\alpha}(\kappa_1\lcx,\kappa_2\lcx)|PS\rangle
=2p_\alpha\frac{\lcx S}{\lcx P}\int_0^1\d z\, G^{(2)}(z)e_0(\ii\zeta z)
\\
&\qquad\qquad+2S^\bot_\alpha\int_0^1\d z
\Big\{G^{(2)}(z)e_1(\ii\zeta z)+G^{(3)}(z)
[e_0(\ii\zeta z)-e_1(\ii\zeta z)]\Big\}
\nonumber\\
&\qquad\qquad
-\lcx_\alpha\frac{\lcx S}{(\lcx P)^2}M^2\int_0^1\d z
\Big\{G^{(4)}(z)\Big[e_0(\ii\zeta z)-3e_1(\ii\zeta z)
+2\int_0^1\d\lambda\, e_1(\ii\zeta\lambda z)\Big]\nonumber\\
&\qquad\qquad
-G^{(2)}(z)\Big[e_1(\ii\zeta z)
-2\int_0^1\d\lambda\, e_1(\ii\zeta\lambda z)\Big]
+4G^{(3)}(z)\Big[e_1(\ii\zeta z)
-\int_0^1\d\lambda\, e_1(\ii\zeta\lambda z )\Big]\Big\},
\nonumber
\end{align}
and
\begin{align}
\label{q_JJ_O_nl}
&\langle P| O^{}_{\alpha}(\kappa_1\lcx,\kappa_2\lcx)|P\rangle
=2p_\alpha\int_0^1\d z\, F^{(2)}(z) e_0(\ii\zeta z)\nonumber\\
&\qquad\qquad
+\lcx_\alpha\frac{M^2}{\lcx P}\int_0^1\d z
\Big\{F^{(4)}(z)e_0(\ii\zeta z)+
\big[F^{(2)}(z)-F^{(4)}(z)\big]e_1(\ii\zeta z)\Big\} ,
\end{align}
while the chiral-odd distributions are given by
\begin{align}
&\langle P|N(\kappa_1\lcx,\kappa_2\lcx)|P\rangle
=2M\int_0^1\d z\, E^{(3)}(z){e_0{(\ii\zeta z)}},
\end{align}
and
\begin{align}
\label{q_JJM_full}
&\langle PS|M_{5[\alpha\beta]}(\kappa_1\lcx,\kappa_2\lcx)|PS\rangle
=
\frac{4}{M}S^\bot_{[\alpha}p_{\beta]}
\int_0^1\d z\, H^{(2)}(z)e_0(\ii\zeta z)
\\
&\qquad\qquad
+2\lcx_{[\alpha}p_{\beta]}\frac{\lcx S}{(\lcx P)^2}M
\int_0^1\d z \Big\{
 4H^{(2)}(z)e_2(\ii\zeta z)
-2H^{(3)}(z)\big[e_0(\ii\zeta z)+2e_2(\ii\zeta z)\big]
\Big\}
\nonumber\\
&\qquad\qquad
-2\lcx_{[\alpha}S^\bot_{\beta]}\frac{M}{\lcx P}
\int_0^1\d z \Big\{
2 H^{(2)}(z)\big[e_1(\ii\zeta z)+e_2(\ii\zeta z)\big]
- H^{(3)}(z)\big[1+2e_2(\ii\zeta z)\big]
\nonumber\\
&\hspace{5cm}
+ H^{(4)}(z)\big[1+e_0(\ii\zeta z)-2e_1(\ii\zeta z)\big]
\Big\}.\nonumber
\end{align}
Let us note that after multiplication of (\ref{q_JJM_full}) by $\lcx_\alpha$
(or $\lcx_\beta$) the last part of the matrix element vanishes. This means that the dynamical twist-4 function
$h_3(z)$ can be ignored in the case of the vector operator 
$M_{5\alpha}(\kappa_1\lcx,\kappa_2\lcx)$ but not in the case of the 
more general tensor operator $M_{5[\alpha\beta]}(\kappa_1\lcx,\kappa_2\lcx)$.

From the expressions (\ref{q_JJ_full}) -- (\ref{q_JJM_full}) and 
(\ref{q_eq:fvda}) -- (\ref{q_eq:ftda})
it is obvious that the conventional structure            
functions of twist $t \geq 3$ contain contributions also of lower
geometric twist. 
By re-expressing the truncated exponentials
and performing appropriate variable transformations or using 
the local expansion of the bilocal operators,
we obtain the following relations, which give the relations between the 
distributions of geometric and dynamical twist. The dynamical twist distributions 
are given in terms of geometrical twist distributions as follows:
\begin{align}  
\label{q_rel-g1}
g_1(z)&=G^{(2)}(z),\\
\label{q_rel-gT}
g_T(z)&=G^{(3)}(z) + \int_z^1 \frac{\d y}{y}
\Big(G^{(2)}-G^{(3)}\Big)\left(y\right),\\
\label{q_rel-g3}
2g_3(z)&=-G^{(4)}(z) + \int_z^1 \frac{\d  y}{y}\Big\{
\Big(G^{(2)}-4G^{(3)}+3G^{(4)}\Big)\left(y\right)
\nonumber\\
&\hspace{6cm}
+ 2 \ln \Big(\frac{z}{y}\Big)
\Big(G^{(2)}-2G^{(3)}+G^{(4)}\Big)\left(y\right)\Big\},\\
\label{q_rel-f1}
f_1(z)&=F^{(2)}(z),\\
\label{q_rel-f4}
2f_4(z)&=F^{(4)}(z) + \int_z^1 \frac{\d y}{y}
\Big(F^{(2)}-F^{(4)}\Big)\left(y\right),\\
e(z)&=E^{(3)}(z),\\
\label{q_rel-h1}
h_1(z)&=H^{(2)}(z),\\
\label{q_rel-hL}
h_L(z)&=H^{(3)}(z) + 2z \int_z^1 \frac{\d y}{y^2}
\Big(H^{(2)}-H^{(3)}\Big)\left(y\right),\\
\label{q_rel-h3}
2h_3(z)&=H^{(4)}(z)+ \int_z^1 \frac{\d y}{y}\Big\{
2\Big(H^{(2)}-H^{(4)}\Big)\left(y\right)
-2\frac{z}{y}\Big(H^{(2)}-H^{(3)}\Big)\left(y\right)\nonumber\\
&\hspace{7cm}-\delta\Big(\frac{z}{y}\Big)\Big(H^{(3)}-H^{(4)}\Big)\left(y\right)
\Big\}.
\end{align}
The relations between the dynamical and geometric quark distribution
functions hold for $1 \geq z \geq 0$; the corresponding antiquark
distribution functions are obtained for $z \rightarrow -z$.
We observe that both decompositions coincide in the leading twist, but 
differ at higher order. 
For instance, $g_1(z)$ and $h_1(z)$ are proper geometric 
twist-2 structure functions and $e(z)$ is a proper geometric twist-3 
distribution.
The distribution functions 
$g_T(z)$, $f_1(z)$ and $h_L(z)$ with dynamical twist $t=3$ also contain 
contributions of geometrical twist $\tau = 2$ and $3$. Additionally, 
the dynamical twist
$t=4$ functions $g_3(z)$ and $h_3(z)$ contain geometrical twist-2, twist-3, 
as well as twist-4 parts and the dynamical twist
$t=4$ function $f_4(z)$ contains geometrical twist-2 and twist-4.

These relations may be inverted in order to express
the geometric twist distributions 
in terms of dynamical twist distributions.
The nontrivial inverse relations are:
\begin{align}
\label{q_rel-G3}
G^{(3)}(z)&=g_T(z) +\frac{1}{z} \int_z^1 \d y
\big(g_T-g_1\big)\left(y\right),\\
G^{(4)}(z)&=-\Big\{2g_3(z) 
+\frac{1}{z^2} \int_z^1 \d y\, y
\big(6g_3+4g_T-g_1\big)\left(y\right)\nonumber\\
&\hspace{5cm}+\frac{1}{z^2} \int_z^1 \d y\, y \Big(1-\frac{z}{y}\Big)
\big(2g_3+4g_T-3g_1\big)\left(y\right)\Big\},\\
F^{(4)}(z)&=2f_4(z) + \frac{1}{z}\int_z^1 \d y
\big(2f_4-f_1\big)\left(y\right),\\
\label{q_rel-H3}
H^{(3)}(z)&=h_L(z) + \frac{2}{z}\int_z^1 \d y
\big(h_L-h_1\big)\left(y\right)\,\\
\label{q_rel-H4}
H^{(4)}(z)&=2\Big\{h_3(z) 
+\frac{1}{z^2} \int_z^1 \d y\, y
\big(2h_3-h_L\big)\left(y\right)
-\frac{1}{z^2} \int_z^1 \d y\, y \Big(1-\frac{z}{y}\Big)
\big(h_L-h_1\big)\left(y\right)\Big\}.
\end{align}

The relations between the moments may be read off from Eqs.~(\ref{q_JJ_full})
-- (\ref{q_JJM_full}) as follows:
\begin{align}  
g_{1n}&=G^{(2)}_n,\\
g_{Tn}&=G^{(3)}_n +  \frac{1}{n+1}
\Big(G^{(2)}_n-G^{(3)}_n\Big),\\
2g_{3n}&=-G^{(4)}_n +  \frac{1}{n+1}
\Big(G^{(2)}_n-4G^{(3)}_n+3G^{(4)}_n\Big)
-\frac{2}{(n+1)^2}
\Big(G^{(2)}_n-2G^{(3)}_n+G^{(4)}_n\Big),\\
f_{1n}&=F^{(2)}_n,\\
2f_{4n}&=F^{(4)}_n +  \frac{1}{n+1}
\Big(F^{(2)}_n-F^{(4)}_n\Big),\\
e_n&=E^{(3)}_n,\\
h_{1n}&=H^{(2)}_n,\\
h_{Ln}&=H^{(3)}_n +  \frac{2}{n+2}
\Big(H^{(2)}_n-H^{(3)}_n\Big),\\
2h_{3n}&=H^{(4)}_n+  \frac{2}{n+1}
\Big(H^{(2)}_n-H^{(4)}_n\Big)
-\frac{2}{n+2}\Big(H^{(2)}_n-H^{(3)}_n\Big)
-\delta_{n0}\Big(H^{(3)}_n-H^{(4)}_n\Big).
\end{align}
In terms of the moments the relations between conventional and new distribution
functions may be easily inverted. The nontrivial inverse relations are:
\begin{align}  
G^{(3)}_{n}&=g_{Tn} +  \frac{1}{n}
\Big(g_{Tn}-g_{1n}\Big),\qquad n>0\\
G^{(4)}_{n}&=-
\Big\{ 2g_{3n}+  \frac{1}{n-1}
\Big(6g_{3n}+4g_{Tn}-g_{1n}\Big)
+\frac{1}{n(n-1)}\Big(2g_{3n}+4g_{Tn}-3g_{1n}\Big)\Big\},\qquad n>1\\
F^{(4)}_n&=2f_{4n} +  \frac{1}{n}
\Big(2f_{4n}-f_{1n}\Big),\qquad n>0\\
H^{(3)}_{n}&=h_{Ln}+  \frac{2}{n}
\Big(h_{Ln}-h_{1n}\Big),\qquad n>0\\
H^{(4)}_n&=
2\Big\{ h_{3n}+  \frac{1}{n-1}
\Big(2h_{3n}-h_{Ln}\Big)
-\frac{1}{n(n-1)}\Big(h_{Ln}-h_{1n}\Big)\Big\},\qquad n>1.
\end{align}

The nontrivial relationships between the conventional and the new
distribution functions are much simpler than to be assumed by a first
glance at expressions such as (\ref{q_O_full}), (\ref{q_matrix_O_nl}), and
(\ref{q_M_full_JJ}). They show that the conventional distribution functions 
are determined by the new ones of the same as well as lower geometrical 
twist, and vice versa (with respect to dynamical twist).
Obviously, the same holds for their moments. In principle, this
allows to determine, e.g., the new distribution functions from
the experimental data if these are known for the conventional
ones. At least, this should be possible for the lowest moments.

Another very important consequence is, if the dynamical and the geometrical
distribution functions coincide then also their anomalous dimensions coincide.
Accordingly, we are able to predict the anomalous dimensions of the 
geometric distribution functions $F^{(2)}(x,Q^2)$, $G^{(2)}(x,Q^2)$, 
$H^{(2)}(x,Q^2)$ from
the anomalous dimensions of the dynamical ones. 
The leading order (LO) $Q^2$ evolution of $f_1(x,Q^2)$ 
and $g_1(x,Q^2)$ has been known for a long time~\cite{GP,GW74,AR} and
the LO $Q^2$ evolution for $h_1(x,Q^2)$ was also studied
in Refs.~\cite{shifman,artru,Koike95,blum01}
(in 1-loop calculation with $\mu^2=Q^2$).
For simplicity, we use the local anomalous dimensions. The local geometric
twist-2 anomalous dimensions are given by\footnote{The corresponding local
operators are $O^{\rm tw2}_{(5) n}(\lcx)$ and $M^{\rm tw2}_{(5)\alpha n}(\lcx)$.}
 ($C_F=4/3$)
\begin{align}
&\gamma^{F^{(2)}}_n\equiv\gamma^{f_1}_n =
\frac{g^2}{8\pi^2}\, C_F\left(1-\frac{2}{(n+1)n}+4\sum_{j=2}^{n}\frac{1}{j}\right),\\
&\gamma^{G^{(2)}}_n\equiv\gamma^{g_1}_n=
\frac{g^2}{8\pi^2}\, C_F\left(1-\frac{2}{(n+1)n}+4\sum_{j=2}^{n}\frac{1}{j}\right),\\
&\gamma^{H^{(2)}}_n\equiv\gamma^{h_1}_n=
\frac{g^2}{8\pi^2}\, C_F\left(1+4\sum_{j=2}^{n}\frac{1}{j}\right),
\end{align}
and hence $\gamma^{H^{(2)}}_n>\gamma^{F^{(2)}}_n$ for all $n$.  
The expressions for the anomalous dimensions of higher geometrical twist
will be more complicated.

\subsection{Wandzura-Wilczek type relations}
As an immediate consequence of the interrelations (\ref{q_rel-g1}) --
(\ref{q_rel-h3}) and (\ref{q_rel-G3}) -- (\ref{q_rel-H4}) between the conventional 
structure functions and the new ones of proper geometric twist
we are able to derive the decomposition of the conventional structure
functions into its parts of genuine twist. Thereby, we also obtain new
Wandzura-Wilczek--like relations for the conventional structure functions
and also new sum rules of the type of
the Burkhardt-Cottingham sum rule.

Let us use the 
decomposition of $g_T(z)=g^{(2)}_T(z)+g^{(3)}_T(z)$ into its parts
$g^{(2)}_T(z)$ and $g^{(3)}_T(z)$ of genuine twist-2 and twist-3,
respectively, parts of $g_2(z)$. Then,
substituting (\ref{q_rel-g1}) into (\ref{q_rel-gT}), we get
\begin{align}
\label{q_WW-tw2}
g^{(2)}_T(z)&=\int_z^1 \frac{\d y}{y}\,g_1(y)\\
\label{q_WW-tw3}
g^{(3)}_T(z)&=g_T(z) - \int_z^1 \frac{\d y}{y}\,g_1(y),
\end{align}
where (\ref{q_WW-tw2}) is the Wandzura-Wilczek relation\footnote{With 
$g_T=g_2+g_1$ this relation can be rewritten in its original form.} 
for the twist-2 part~\cite{Wandzura}. On the other hand, the 
analogous relation (\ref{q_WW-tw3}) for the twist-3 part, which is an
immediate consequence of the above definition, has already been obtained 
in the framework of the local OPE~\cite{blum97}
and the nonlocal OPE~\cite{Lazar98}.
Wandzura and Wilczek noted in their original derivation that the twist-3
is small~\cite{Wandzura}. They obtained in the MIT bag model the approximation
\begin{align}
\label{WW-approx}
\big|G^{(3)}_n/G^{(2)}_n\big|\simeq 0.10\,.
\end{align}
So far the SLAC data~\cite{SLAC} agree very well with the Wandzura-Wilczek
prediction. However the accuracy of the data is such that a relevant
$g_T^{(3)}$-term would not yet be detectable~\cite{Jaffe96}. Therefore,
if more precise data of $g_T$ become available, one can investigate
the geometric twist-3 part of $g_T$ in more detail.

Using QCD equations of motion operator relations (see~\cite{BB88,ball98,Kodaira99}), 
one can decompose the 
geometric twist-3 part of $g_T(z)$, Eq.~(\ref{q_WW-tw3}), into a quark mass dependent term
and a quark-gluon correlation part~\cite{TM94}
\begin{align}
\label{q_WW-tw3m}
g^{(3)}_T(z)=\frac{m}{M}\left[\frac{h_1(z)}{z}-\int_z^1 \frac{\d y}{y^2}\,h_1(y)\right]
              +\tilde{g}^{(3)}_T(z),
\end{align}
where $\tilde{g}^{(3)}_T(z)$ denotes the twist-3 quark-gluon correlation 
function of the so-called Shuryak-Vainshtein operator. Therefore, in the 
massless case $m=0$ the function $g^{(3)}_T(z)$ measures the quark-gluon
correlation in the nucleon.

For the moments we obtain
\begin{align}
\label{q_WW-tw2-n}
g^{(2)}_{Tn}&=\frac{1}{n+1}\,g_{1 n},\\
\label{q_WW-tw3-n}
g^{(3)}_{T n}&=g_{T n} - \frac{1}{n+1}\, g_{1n },\qquad n>0.
\end{align}
Obviously, if we 
introduce the usual notation $g_T(z)\equiv g_1(z)+g_2(z)$,
we get from the relation (\ref{q_WW-tw2-n}) for $n=0$ the 
Burkhardt-Cottingham sum rule~\cite{BC70}:
\begin{align}
\int_0^1\d z\, g_2(z)=0,
\end{align}
which is consistent with the SLAC data. 

Of course, these relationships are well known. However, with the help of
the results of the preceeding Section we are able to generalize these
relationships also to the other nontrivial structure functions.
Using the formulas (\ref{q_rel-g1}), (\ref{q_rel-gT}), (\ref{q_rel-g3}), and
(\ref{q_rel-G3}), we obtain Wandzura-Wilczek--like integral relations for 
geometric twist parts of
the function $g_3(z)=g^{(2)}_3(z)+g^{(3)}_3(z)+g^{(4)}_3(z)$ as follows:
\begin{align}
\label{q_WW-g3-tw2}
g^{(2)}_3(z)&=\int_z^1\frac{\d y}{y}\Big\{\frac{g_1(y)}{2}
+\ln\Big(\frac{z}{y}\Big)g_1(y)\Big\},\\
g^{(3)}_3(z)&=-2\int_z^1\frac{\d y}{y}\Big\{g_T(y)
	      +\ln\Big(\frac{z}{y}\Big)g_1(y)\Big\},\\
g^{(4)}_3(z)&=g_3(z)+\frac{1}{2}\int_z^1\frac{\d y}{y}\Big\{\big(4g_T-g_1\big)(y)
	      +2\ln\Big(\frac{z}{y}\Big)g_1(y)\Big\}.
\end{align}
Due to the fact that $g_3(z)$ contains twist-2, twist-3 as well as twist-4
we have obtained three integral relations. For example, Eq.~(\ref{q_WW-g3-tw2})
demonstrates that the twist-2 part $g^{(2)}_3(z)$ can be expressed in terms
of the twist-2 function $g_1(z)$.
The relations for the moments are:
\begin{align}
\label{q_WW-g3-tw2-n}
g^{(2)}_{3 n}&=\frac{1}{2(n+1)}\, g_{1 n}-\frac{1}{(n+1)^2}\,g_{1 n},\\
\label{q_WW-g3-tw3-n}
g^{(3)}_{3 n}&=-\frac{2}{n+1}\,g_{T n}
	       +\frac{2}{(n+1)^2}\, g_{1n} ,\quad n>0\\
g^{(4)}_{3 n}&=g_{3 n}
		+\frac{1}{2(n+1)}\Big(4g_{T n}-g_{1 n}\Big)
	        -\frac{1}{(n+1)^2}\, g_{1n} ,\quad n>1.
\end{align}
Using again the approximation (\ref{WW-approx}), one should expect
$g^{(3)}_{3n}\ll g^{(2)}_{3n}$. Additionally, we expect that $g^{(4)}_{3n}$ 
is small.
Additionally, from (\ref{q_WW-g3-tw3-n}) and 
(\ref{q_WW-tw3-n}) we observe the twist-3 relation
\begin{align}
g^{(3)}_{3 n}=-\frac{2}{n+1}\,g^{(3)}_{T n},\quad
g^{(3)}_3(z)=-2\int_z^1\frac{\d y}{y}\,g^{(3)}_T(y).
\end{align}

For $n=0$ in (\ref{q_WW-g3-tw2-n}) we find the sum rule
\begin{align}
\label{g3-sr1}
\int_0^1\d z\, g_3(z) = - \frac{1}{2} \int_0^1 \d z\, g_1(z).
\end{align}
Additionally, we obtain for $n=1$ in (\ref{q_WW-g3-tw2-n}) and 
(\ref{q_WW-g3-tw3-n}) the sum rules
\begin{align}
\int_0^1\d z\, z\, g^{(2)}_3(z)=0,\quad
\int_0^1 \d z\, z\, g_3^{(3)}(z) =
-\frac{1}{2} \int_0^1 \d z\, z (2g_T - g_1)(z),
\end{align}
which give the following sum rule for $g_3(z)$
\begin{align}
\label{g3-sr2}
\int_0^1 \d z\, z\, g_3(z) =
-\frac{1}{2} \int_0^1 \d z\, z (2g_T -  g_1)(z).
\end{align}
If we use the Efremov-Leader-Teryaev sum rule
\begin{align}
\int_0^1 \d z\, z (2g_T - g_1)(z)=0,
\end{align}
which was discussed in~\cite{ELT96} in the context of a QCD field theoretical
framework,
we discover the interesting sum rule from Eq.~(\ref{g3-sr2})
\begin{align}
\int_0^1 \d z\, z\, g_3(z)=0.
\end{align}
Let me note that the sum rules (\ref{g3-sr1}) and (\ref{g3-sr2}) should
be experimentally verified
because they give information about the dynamical twist-4 function $g_3(z)$.
Although the function $g_3(z)$ does not contribute to either measurable 
distribution function including electromagnetic interaction~\cite{JJ91b}, it 
gives contributions to the three polarized distribution functions in
weak interaction processes which are introduced by Ji~\cite{Ji93}.
Therefore, the {\it new} sum rules and WW-type relations of $g_3(z)$ will 
be relevant in these processes. 

Substituting (\ref{q_rel-f1}) into (\ref{q_rel-f4}), we get
the integral relations for the twist-2 and twist-4 part of 
$f_4(z)=f^{(2)}_4(z)+f^{(4)}_4(z)$
\begin{align}
f^{(2)}_4(z)&=\frac{1}{2} \int_z^1 \frac{\d y}{y}f_1(y),\\
f^{(4)}_4(z)&=f_4(z) -\frac{1}{2}\int_z^1 \frac{\d y}{y}f_1(y).
\end{align}
The corresponding relations for the moments read
\begin{align}
\label{q_WW-f4-tw2-n}
f^{(2)}_{4 n}&=\frac{1}{2(n+1)}\, f_{1 n},\\
f^{(4)}_{4 n}&=f_{4 n} -\frac{1}{2(n+1)}\,f_{1 n},\qquad n>0.
\end{align}
For $n=0,1$ in (\ref{q_WW-f4-tw2-n}) we observe the following sum rules
\begin{align}
\int_0^1\d z\, f_4(z) = \frac{1}{2} \int_0^1\d z\, f_1(z),
\qquad
\int_0^1\d z\, z\, f^{(2)}_4(z) = \frac{1}{4} \int_0^1\d z\, z\, f_1(z).
\end{align}

The distributions $h_1(z)$, $h_L(z)$, and $h_3(z)$ can be treated in 
a similar fashion. 
Unfortunately, Jaffe and Ji used two different relations between 
$h_L(z)$ and $h_2(z)$. In their earlier paper~\cite{JJ91} they used
the notation $h_L(z)\equiv h_2(z)+h_1(z)/2$. But later~\cite{JJ92} they 
introduced the relation $h_L(z)\equiv h_1(z)+h_2(z)/2$. 
In order to avoid any misunderstanding we use $h_L(z)$ instead of $h_2(z)$.

If we now substitute (\ref{q_rel-h1}) into (\ref{q_rel-hL}) and use
$h_L(z)=h^{(2)}_L(z)+h^{(3)}_L(z)$,
we derive
\begin{align}
\label{q_WW-JJ-tw2}
h^{(2)}_L(z)&= 2z\int_z^1 \frac{\d y}{y^2}\,h_1(y),\\
\label{q_WW-JJ-tw3}
h^{(3)}_L(z)&=h_L(z) - 2z\int_z^1 \frac{\d y}{y^2}\,h_1(y),
\end{align}
where (\ref{q_WW-JJ-tw2}) is a twist-2 Wandzura-Wilczek--like relation which 
was obtained earlier by Jaffe and Ji~\cite{JJ92} 
(using the notation $h_L(z)\equiv h_2(z)+h_1(z)/2$ we recover the 
Wandzura-Wilczek relation of the function $h_2(z)$ which was obtained in Ref.~\cite{JJ91}). 
Obviously, (\ref{q_WW-JJ-tw3}) is the corresponding twist-3 relation which 
can be decomposed by means of QCD equations of motion operator relations 
as follows~(see \cite{JJ92,TM94,Kodaira99})
\begin{align}
\label{q_WW-JJ-tw3m}
h^{(3)}_L(z)=\frac{m}{M}\left[\frac{g_1(z)}{z}-2z\int_z^1 \frac{\d y}{y^3}\,g_1(y)\right]
              +\tilde{h}^{(3)}_L(z),
\end{align}
where $\tilde{h}^{(3)}_L(z)$ denotes the quark-gluon correlation
function of the twist-3 part of the trilocal quark-gluon operator
$V_{\alpha\beta}^{21}(x,t x, -x)$ (see Section~\ref{trilocal}).

For the moments we obtain
\begin{align}
\label{q_WW-JJ-tw2-n}
h^{(2)}_{L n}&= \frac{2}{n+2}\,h_{1 n},\\
\label{q_WW-JJ-tw3-n}
h^{(3)}_{L n}&=h_{L n} -\frac{2}{n+2}\, h_{1 n},\qquad n>0.
\end{align}
For $n=0,1$ in (\ref{q_WW-JJ-tw2-n}) we observe the following sum rule
\begin{align}
\label{hL-sr1}
\int_0^1\d z\, h_L(z) &=  \int_0^1\d z\ h_1(z),\\
\int_0^1\d z\, z\, h^{(2)}_L(z) &= \frac{2}{3} \int_0^1\d z\, z\, h_1(z).
\end{align}
The sum rule~(\ref{hL-sr1}) was earlier derived in Ref.~\cite{Bur95}.
If we use the notation $h_L(z)\equiv h_1(z)+h_2(z)/2$, we recover in
(\ref{hL-sr1}) a sum rule
very similar to the Burkhardt-Cottingham sum rule, namely
$\int_0^1\d z\, h_2(z)=0$, which was originally derived by Tangerman and 
Mulders~\cite{TM94}.

Furthermore, 
using the formulas (\ref{q_rel-h1}), (\ref{q_rel-hL}) (\ref{q_rel-h3}) and
(\ref{q_rel-H3}), we obtain the integral relations for the structure function
$h_3(z)=h^{(2)}_3(z)+h^{(3)}_3(z)+h^{(4)}_3(z)$ as follows:
\begin{align}
\label{q_WW-h3-tw2}
h^{(2)}_3(z)&=\frac{1}{z}\int_z^1\d y\Big(\frac{z}{y}-\frac{z^2}{y^2}\Big)h_1(y),\\
h^{(3)}_3(z)&=\frac{1}{z}\int_z^1 \d y\Big\{ h_L(y)
	      -\Big(1-\frac{z^2}{y^2}\Big) h_1(y)\Big\},\\
h^{(4)}_3(z)&= h_3(z)-\frac{1}{z}\int_z^1 \d y\Big\{ h_L(y)
	      -\Big(1-\frac{z}{y}\Big) h_1(y)\Big\},
\end{align}
and the relations for the corresponding moments
\begin{align}
\label{q_WW-h3-tw2-n}
h^{(2)}_{3 n}&=\frac{1}{n+1}\,h_{1 n}-\frac{1}{n+2}\,h_{1 n},\\
\label{q_WW-h3-tw3-n}
h^{(3)}_{3 n}&=\frac{1}{n}\,h_{L n}
	      -\frac{2}{(n+2)n}\,h_{1 n},\qquad n>0\\
h^{(4)}_{3 n}&=h_{3 n}-\frac{1}{n}\,h_{L n}
		+\frac{1}{(n+1)n}\,h_{1 n},\qquad n>1.
\end{align}
From (\ref{q_WW-h3-tw3-n}) and (\ref{q_WW-JJ-tw3-n}) we observe the 
following twist-3 relation
\begin{align}
h^{(3)}_{3 n}=\frac{1}{n}\,h^{(3)}_{L n},\quad
h^{(3)}_3(z)=\frac{1}{z}\int_z^1 \d y\, h^{(3)}_L(y).
\end{align}
Therefore we conclude that $h^{(2)}_3$ has a simple parton interpretation
like $h_1$, $h^{(3)}_3$ represent the quark-gluon correlation, like $g_T$, 
in the nucleon and last but not least the distribution $h^{(4)}_3$ is determined
by the rather complicated multi quark-gluon and quark-quark correlation in the
nucleon. Similar arguments are valid for the different parts of $g_3$.

Here we also obtain a interesting sum rule from (\ref{q_WW-h3-tw2-n}) in the
case $n=0$, namely,
\begin{align}
\label{h3-sr1}
\int_0^1\d z\, h_3(z) = \frac{1}{2}\int_0^1 \d z\, h_1(z),
\end{align}
which was also obtained in Ref.~\cite{Bur95}.
For $n=1$ we get from Eqs.~(\ref{q_WW-h3-tw2-n}) and (\ref{q_WW-h3-tw3-n})
the following sum rules
\begin{align}
\int_0^1\d z\,z\, h^{(2)}_3(z)& = \frac{1}{6}\int_0^1 \d z\,z\, h_1(z),\\
\int_0^1\d z\, z\,h^{(3)}_3(z)& = \int_0^1 \d z\,z \Big( h_L(z)-\frac{2}{3}\,h_1(z)\Big),
\end{align}
which give the phenomenological interesting sum rule
for $h_3(z)$:
\begin{align}
\label{h3-sr2}
\int_0^1\d z\, z\, h_3(z) = \int_0^1 \d z\, z\Big(h_L(z)-\frac{1}{2}\,h_1(z)\Big).
\end{align}
The sum rules (\ref{h3-sr1}) and (\ref{h3-sr2}) give information about the 
dynamical twist-4 function $h_3(z)$ from the functions $h_1(z)$ and $h_2(z)$
and should be phenomenologically tested.

Let us note that Jaffe and Ji ignored in their investigations~\cite{JJ92} the function $h_3(z)$ 
because they claimed that it is twist-4 and can be neglected.
However, this assumption is only true in the case of the vector operator 
$M_{5\alpha}(\kappa_1\lcx,\kappa_2\lcx)$ but not in the
case of the tensor operator $M_{5[\alpha\beta]}(\kappa_1\lcx,\kappa_2\lcx)$ 
which appear in polarized deep inelastic scattering by means of the massive
quark propagator.
As we have seen in Eq. (\ref{q_rel-h3}),
then the function $h_3(z)$ contains twist-2, twist-3 as well as twist-4 parts.
Moreover, the twist-2 part $h_3^{(2)}(z)$ is given in terms of the twist-2
function $h_1(z)$ in (\ref{q_WW-h3-tw2}). Therefore, the structure function 
$h_3(z)$ is experimentally relevant also at the twist-2 and twist-3 level and 
cannot be ignored in these cases. 
The dynamical twist-4 function $h_3(z)$ is, in principle, observable in the
polarised Drell-Yan process~\cite{HJ94}.

\subsection{Conclusions}
Using the notion of geometric twist, we discussed the calculation 
of the forward matrix elements for those nonlocal LC-operators which 
correspond to the independent tensor structures in polarized 
nucleon matrix elements
of the type $\langle PS| \bar\psi(\kappa_1 \lcx) \Gamma 
U(\kappa_1 \lcx, \kappa_2 \lcx) \psi(\kappa_2\lcx)|PS \rangle$.
We have found nine independent forward distribution functions with well
defined twist $\tau$ which, for twist $\tau \geq 3$, differ from the
conventional ones. From the field theoretical point of view this Lorentz 
invariant classification is the most appropriate frame of introducing
distribution functions since the separation of different (geometric) twist 
is unique and independent from the special kinematics of the process.
Only the operators of definite geometric twist will have the correct mixing 
behaviour under renormalization.
A very useful property of the nonlocal (and local) LC operators of definite
twist $\tau$ is that they are obtained from the original, i.e.,~undecomposed
LC--operators by the application of corresponding projection operators
of geometric twist.
An essential result of our calculations is the relation between the new
parton distribution functions and those given by Jaffe and Ji~\cite{JJ91,JJ92}.
In addition we have also given the moments of all the distribution
functions. 
These relations demonstrate
the interrelations between the different twist definitions.
One principal result are Wandzura-Wilczek--like relations
between the dynamical twist distributions which we have obtained by means of 
interrelation rules in a natural manner. 
An advantage in our approach was that we used operators with geometric twist 
which allowed us to reveal the
Wandzura-Wilczek--like relations between the dynamical twist distribution
functions.

\newpage
\setcounter{equation}{0}

\section{Vector meson distribution amplitudes of definite geometric twist}
\label{meson}
In this Section,
we centre our discussion around the specific case of light
vector-meson distribution amplitudes (DAs), which are relevant for describing
light-cone dominated processes, e.g., the vector meson production
off nucleons,
$\gamma^* + N\to V+N$
and can be expressed in terms of matrix-elements of gauge-invariant
non-local operators sandwiched between the vacuum and the meson
state,
$\langle 0 | \bar u(x) \Gamma U(x,-x) d(-x) | \rho(P,\lambda)\rangle$.
In particular, we investigate the (two-particle) 
$\rho$-{\em meson light-cone distribution amplitudes} of
geometric twist. 
We will find eight meson distribution amplitudes of geometric twist.

On the other hand, Ball {\it et al.} used the notion of {\it dynamical twist} $t$
for the classification of (two-particle) vector meson distribution amplitudes and 
vector meson 
distribution amplitudes, respectively~\cite{ball96,ball98,ball99}. 
They classified the meson distribution amplitudes 
which correspond to the independent tensor structure of the matrix 
elements of bilocal quark-antiquark operators.
One key ingredient in their approach was the use of QCD equations of motion
in order to obtain dynamical Wandzura-Wilczek-type relations for 
DAs that are not dynamically independent.
The pion distribution functions were investigated in similar way~\cite{braun89,braun90,ball99b}.

Meson distribution amplitudes and form factors have been discussed in the 
framework of local operator product expansion and in the infinite 
momentum frame by~\cite{brodsky79,brodsky80,rady80,chern84,shifman,craigie}
and in the framework of nonlocal operator product expansion~\cite{geyer85,geyer87,mul94}.

The aim of this Section is to present the (two-particle) 
{\em meson distribution amplitudes} which are related to the nonlocal LC-operators of different
geometric twist. 
In the framework of geometric twist, it is possible to investigate 
the interrelations and the mismatch between the different twist 
definitions of meson distribution amplitudes as a kind of group theoretical relations. 
By means of this relations, we are able to obtain geometric
Wandzura-Wilczek-type relations between the the dynamical twist function
which differ from the dynamical Wandzura-Wilczek-type relations. 

\subsection{The $\rho$-meson distribution amplitudes of definite dynamical twist}
Usually, the meson light-cone distributions are defined as the 
vacuum-to-meson matrix elements of quark-antiquark nonlocal gauge invariant 
operators on the light-cone.
In this way,
Ball~{\it et al.}~\cite{ball96,ball98,ball99} found eight independent two-particle distributions.
The classification of these dynamical twist $\rho$-meson distribution amplitudes with respect to spin, dynamical twist, chirality and the relations to geometrical twist (see Section~\ref{sec-geoDA})
are summarized in Tab.~\ref{tab:2}. 
\begin{table}[b]
\begin{center}
\renewcommand{\arraystretch}{1.3}
\begin{tabular}{|c|ccc|}
\hline 
Twist $t$   & 2 & 3 & 4 \\
    & $O(1)$  & $O(1/Q)$& $O(1/Q^{2})$ \\ \hline
$e_{\parallel}$ & $\hat\phi_{\parallel}\equiv\hat\Phi^{(2)}$ & 
$\underline{\htth}=\hat\Phi^{(3)}+F_t(\hat\Phi^{(2)},\hat\Phi^{(3)})$& 
$\hat g_{3}=\hat\Phi^{(4)}+F_g(\hat\Phi^{(2)},\hat\Phi^{(3)},\hat\Phi^{(4)})$ \\
& & $\underline{\hsh}\equiv\hat\Upsilon^{(3)}$ &\\
$e_{\perp}$ & $\underline{\hat\phi_{\perp}}\equiv\hat\Psi^{(2)}$ & 
$\hat g_{\perp}^{(v)}=\hat\Psi^{(3)}+F_v(\hat\Psi^{(2)},\hat\Psi^{(3)})$& 
$\underline{\hat h_{3}}=\hat\Psi^{(4)}+F_h(\hat\Psi^{(2)},\hat\Psi^{(3)},\hat\Psi^{(4)})$\\
& & $\hat g_{\perp}^{(a)}\equiv\hat\Xi^{(3)}$ & 
\\[2pt] \hline
\end{tabular}
\renewcommand{\arraystretch}{1}
\end{center}
\caption{Spin, dynamical twist and chiral classification of the $\rho$-meson
distribution 
amplitudes.}
\label{tab:2}
\end{table}%
One distribution amplitude was obtained 
for longitudinally ($e_{\parallel}$) and transversely ($e_{\perp}$)
polarized $\rho$-mesons of twist-2 and twist-4, respectively.
Whereas, the number of  twist-3
distribution amplitudes is doubled for each polarization.
The higher twist distribution amplitudes
contribute to a hard exclusive amplitude with
additional powers of $1/Q$ compared to the leading twist-2 ones.
The underlined distribution amplitudes 
are chiral-odd, the others chiral-even.

Using the Lorentz-tensor decomposition of the matrix elements of relevant 
bilocal operators, 
\begin{align}
\label{O_ent}
O_{\alpha}(\lcx,-\lcx)&=\bar{u}(\lcx)\gamma_{\alpha}U(\lcx,-\lcx)d(-\lcx),\\
\label{O5_ent}
O_{5\alpha}(\lcx,-\lcx)&=\bar{u}(\lcx)\gamma_{\alpha}\gamma_5 U(\lcx,-\lcx)d(-\lcx),\\
\label{M_ent}
M_{[\alpha\beta]}(\lcx,-\lcx)
&=\bar{u}(\lcx)\sigma_{\alpha\beta}U(\lcx,-\lcx)d(-\lcx),\\
\label{N_ent}
N(\lcx,-\lcx)&=\bar{u}(\lcx)U(\lcx,-\lcx)d(-\lcx),
\end{align}
in analogy to nucleon structure functions, the explicit definitions of the chiral-even $\rho$-distributions 
are:
\begin{align}
\langle 0|O_{\alpha}(\lcx,-\lcx)|\rho(P,\lambda)\rangle 
&= f_{\rho} m_{\rho} \left[ p_{\alpha}
\frac{e^{(\lambda)}\lcx}{\lcx P}
\int_{-1}^{1} \!\d \xi\, \e^{\ii \xi(\lcx P)} \hat\phi_{\parallel}(\xi, \mu^{2})  
+ e^{(\lambda)}_{\perp \alpha}
\int_{-1}^{1} \!\d \xi\, \e^{\ii \xi(\lcx P)} \hat g_{\perp}^{(v)}(\xi, \mu^{2}) \right.
\nonumber \\
&\qquad\left.-\frac{1}{2}\lcx_{\alpha}
\frac{e^{(\lambda)}\lcx }{(\lcx P)^{2}} m_{\rho}^{2}
\int_{-1}^{1} \!\d \xi\, \e^{\ii \xi (\lcx P)} \hat g_{3}(\xi, \mu^{2}) \right]
\label{eq:vda}
\end{align}
and 
\begin{eqnarray}
\langle 0|O_{5\alpha}(\lcx,-\lcx)|\rho(P,\lambda)\rangle = 
\frac{1}{2}\left(f_{\rho} - f_{\rho}^{\rm T}
\frac{m_{u} + m_{d}}{m_{\rho}}\right)
m_{\rho} \epsilon_{\alpha}^{\phantom{\alpha}\beta \mu \nu}
e^{(\lambda)}_{\perp \beta} p_{\mu} \lcx_{\nu}
\int_{-1}^{1} \!\d \xi\, \e^{\ii \xi(\lcx P)}\hat g^{(a)}_{\perp}(\xi, \mu^{2}),
\label{eq:avda}
\end{eqnarray}
while the chiral-odd distributions are defined as 
\begin{align}
\langle 0|M_{[\alpha \beta]}(\lcx,-\lcx)|\rho(P,\lambda)\rangle 
&= \ii f_{\rho}^{\rm T} \left[ ( e^{(\lambda)}_{\perp \alpha}p_\beta -
e^{(\lambda)}_{\perp \beta}p_\alpha )
\int_{-1}^{1} \!\d \xi\, \e^{\ii \xi(\lcx P)} \hat\phi_{\perp}(\xi,\mu^2) \right.
\nonumber \\
&\qquad + (p_\alpha \lcx_\beta - p_\beta \lcx_\alpha )
\frac{e^{(\lambda)}\lcx}{(\lcx P)^{2}}
m_{\rho}^{2} 
\int_{-1}^{1} \!\d \xi\, \e^{\ii \xi(\lcx P)} \htth (\xi, \mu^{2}) 
\nonumber \\
& \qquad \left.
+ \frac{1}{2}
(e^{(\lambda)}_{\perp \alpha} \lcx_\beta -e^{(\lambda)}_{\perp \beta} \lcx_\alpha) 
\frac{m_{\rho}^{2}}{\lcx P} 
\int_{-1}^{1} \!\d \xi\, \e^{\ii \xi(\lcx P)}\hat h_{3}(\xi, \mu^{2}) \right]
\label{eq:tda}
\end{align}
and
\begin{equation}
\langle 0|N(\lcx,-\lcx)|\rho(P,\lambda)\rangle
= -\ii \left(f_{\rho}^{\rm T} - f_{\rho}\frac{m_{u} + m_{d}}{m_{\rho}}
\right)(e^{(\lambda)}\lcx) m_{\rho}^{2}
\int_{-1}^{1} \!\d \xi\, \e^{\ii \xi(\lcx P)}\hsh (\xi, \mu^{2}).
\label{eq:sda}
\end{equation}
In the above definitions we used the two light-cone vectors 
(see Eq.~(\ref{x-lc}) with $\rho_\alpha=P_\alpha/m_\rho$)
\begin{align}
\label{Osc}
\lcx_\alpha = x_\alpha -\frac{P_\alpha}{m_\rho^2} \Big(
(xP) - \sqrt{(xP)^2 - x^2 m_\rho^2}
\Big), \qquad
p_\alpha=P_\alpha-\frac{1}{2} \lcx_\alpha\frac{m_\rho^2}{\lcx P},
\end{align}
with $p^2=0$, $\lcx^2=0$ and $p\lcx=P\lcx$.
Here $P_\alpha$ is the $\rho$-meson momentum vector, so that
$P^2=m_\rho^2$, $e^{(\lambda)} e^{(\lambda)}=-1$, $P e^{(\lambda)}=0$, 
$m_\rho$ denotes the $\rho$-meson mass.
The polarization vector of the $\rho$-meson $e^{(\lambda)}_\alpha$ 
can be decomposed into projections onto the two light-like vectors and the 
orthogonal plane:
\begin{align}
e^{(\lambda)}_\alpha=p_\alpha \frac{e^{(\lambda)}\lcx}{\lcx P}-
\frac{1}{2} \lcx_\alpha \frac{e^{(\lambda)}\lcx}{(\lcx P)^2} m_\rho^2
+e^{(\lambda)}_{\bot\alpha}.
\end{align}

The distribution amplitudes are dimensionless functions of $\xi$ and 
describe
the probability amplitudes to find the $\rho$-meson in a state with minimal
number of constituents (quark and antiquark) which carry
momentum fractions $\xi$.
For example, the functions $\hat\phi_\parallel$ and $\hat \phi_\perp$ give 
the leading twist distributions in the fraction of total momentum carried 
by the quark in longitudinally and transversely polarized mesons, 
respectively~\cite{ABS94}.
The nonlocal operators are renormalized at scale $\mu$, so that
the distribution amplitudes depend on $\mu$ as well. This
dependence can be calculated in perturbative QCD (from now on we suppress
$\mu^2$).

The vector and tensor  decay constants $f_\rho$ and $f_\rho^{\rm T}$ are defined 
as usually as
\begin{eqnarray}
\langle 0|\bar u(0) \gamma_{\alpha}
d(0)|\rho(P,\lambda)\rangle & = & f_{\rho}m_{\rho}
e^{(\lambda)}_{\alpha},
\label{eq:fr}\\
\langle 0|\bar u(0) \sigma_{\alpha \beta} 
d(0)|\rho(P,\lambda)\rangle &=& \ii f_{\rho}^{\rm T}
(e_{\alpha}^{(\lambda)}P_{\beta} - e_{\beta}^{(\lambda)}P_{\alpha}).
\label{eq:frp}
\end{eqnarray}
All eight distributions $\hat\phi=\{\hat\phi_\parallel,\hat \phi_\perp,
\hat g_\perp^{(v)},\hat g_\perp^{(a)},\htth,\hsh,\hat h_3,\hat g_3\}$ 
are normalized as
\begin{equation}
\int_{-1}^1\!\d \xi\, \hat\phi(\xi) =1.
\label{eq:norm}
\end{equation}
The distribution amplitude moments related to the reduced matric elements
are given as
\begin{align}
\label{moment}
\phi_n=\int_{-1}^1\!\d \xi\, \xi^n \hat\phi(\xi)
\equiv\int_{0}^1\!\d u\, \xi^n \phi(u),\qquad \xi=2u-1
\end{align}
with both types of meson distribution amplitudes  
\begin{align}
\hat\phi(\xi)=\frac{1}{2}\,\phi\left(\frac{\xi+1}{2}\right),\qquad
\phi(u)=2\hat\phi(2u-1).
\end{align}
Eq.~(\ref{moment}) means, having information about the distribution amplitude moments, one can 
reconstruct the distribution amplitude itself.

\subsection{The $\rho$-meson matrix elements of LC--operators with definite 
geometric twist}
In this Section we define the meson distribution amplitudes 
for the bilinear LC-quark operators with definite twist sandwiched between the 
vacuum and the meson state. 
The twist projectors for the bilocal operators (\ref{O_ent})--(\ref{N_ent})
are given in 
Eqs.~(\ref{q_O2vec}) -- (\ref{q_M_tw3_vec}).
As usual,
the matrix elements of the meson are related to the 
meson momentum $P_\mu$ and meson polarization vector $e^{(\lambda)}_\mu$, 
respectively and $\lcx_\alpha$ is generated by the twist projectors.
According to the above projection properties we are able 
to introduce one (and only one) distribution amplitude for any operator of 
definite twist. 

We start with the chiral-even {\em scalar operator} 
$O^{\rm tw2}(\lcx,-\lcx)=\bar{u}(\lcx)(\gamma\lcx)U(\lcx, -\lcx) d(-\lcx)$.
The  matrix element of this nonlocal twist-2 operator taken between the vacuum $\langle 0|$ and the meson 
state $|\rho(P,\lambda)\rangle$ reads
\begin{align}
\label{matrix_O_t2_sca}
\langle 0|O^{\text{tw2}}(\lcx,-\lcx)|\rho(P,\lambda)\rangle
=
f_\rho m_\rho(e^{(\lambda)}\lcx)\int_{-1}^1\d\xi\,\hat\Phi^{(2)}(\xi)\,{\e^{\ii \xi(\lcx P)}}
=f_\rho m_\rho(e^{(\lambda)}\lcx)\sum_{n=0}^\infty \frac{(\ii(\lcx P))^n}{n!}\,\Phi^{(2)}_n.
\end{align}
Here, $\hat\Phi^{(2)}(\xi)$ is the twist-2 meson 
distribution amplitude which is related to the corresponding moments 
\begin{align}
\Phi^{(2)}_n=\int_{-1}^1\d\xi\, \xi^n \hat\Phi^{(2)}(\xi).
\nonumber
\end{align} 

Now we consider the chiral-even {\em  vector operator}.
Using the projection property (\ref{q_OPROJ}), 
we introduce the  meson distribution amplitudes 
$\hat\Phi^{(\tau)}(\xi)$ of geometric twist $\tau$by
\begin{align}
\label{Gfct}
\langle 0|O^{(\tau)}_\alpha(\lcx, -\lcx)|\rho(P,\lambda)\rangle 
&\equiv 
\langle 0|({\cal P}^{(\tau)\beta}_\alpha 
O^{(\tau)}_\beta)(\lcx,-\lcx)|\rho(P,\lambda)\rangle 
= f_\rho m_\rho{\cal P}^{(\tau)\beta}_\alpha
\Big(e^{(\lambda)}_\beta\int_{-1}^1\d\xi\, \hat\Phi^{(\tau)}(\xi)\,
{\e^{\ii\xi(\lcx P)}}\Big)\ ,
\end{align} 
which, for $\tau =2$,  is  consistent with (\ref{matrix_O_t2_sca}).
Using the twist projection operators as they are determined by Eqs.~(\ref{q_O2vec})
-- (\ref{q_O4vec}) we obtain for the twist-2 operator 
\begin{align}
\label{matrix_O_t2_nl}
&\langle 0|O^{\text{tw2}}_{\alpha}(\lcx,-\lcx)|\rho(P,\lambda)\rangle
=f_\rho m_\rho\int_0^1\d t \Big[\pd_\alpha+\frac{1}{2} \big(\ln t\big) 
x_\alpha\square\Big]
(e^{(\lambda)}x)\int_{-1}^1\d\xi\, \hat\Phi^{(2)}(\xi)\,{\e^{\ii\xi t(xP)}}
\big|_{x=\tilde{x}},\nonumber\\
&\qquad=f_\rho m_\rho\int_0^1\d t \int_{-1}^1\d\xi\,\Phi^{(2)}(\xi)
\Big[e^{(\lambda)}_\alpha+\ii\xi t
P_\alpha(e^{(\lambda)}\lcx)
+\frac{\tilde{x}_\alpha}{2}(e^{(\lambda)}\lcx)m_\rho^2
(\ii\xi t)^2\big(\ln t \big)\Big]
\e^{\ii\xi t(\tilde{x}P)},\nonumber\\
&\qquad=f_\rho m_\rho
\sum_{n=0}^\infty \frac{(\ii(\lcx P))^n}{n!}\Phi^{(2)}_n\Big\{
\frac{1}{n+1}\Big( e^{(\lambda)}_\alpha+n P_\alpha \frac{e^{(\lambda)}\lcx}
{\lcx P}\Big)
-\frac{n(n-1)}{2(n+1)^2}\lcx_\alpha\frac{m_\rho^2 (e^{(\lambda)}\lcx)}{(\lcx P)^2}\Big\},
\end{align}
and for the higher twist operators
\begin{align}
\label{matrix_O_t3_nl}
&\langle 0|O^{\mathrm{tw3}}_{\alpha}(\lcx,-\lcx)|\rho(P,\lambda)\rangle
=f_\rho m_\rho \int_0^1\!\!\d t \int_{-1}^1\!\!\d\xi\, \hat\Phi^{(3)}(\xi)
\Big[\Big(e^{(\lambda)}_\alpha (\tilde{x}P)
-P_\alpha (e^{(\lambda)}\lcx)\Big)\ii\xi t
\nonumber\\
&\hspace{7cm}
-\tilde{x}_\alpha m_\rho^2(e^{(\lambda)}\lcx)
(\ii\xi t)^2\ln t \Big]
\e^{\ii\xi t(\tilde{x}P)},\nonumber\\
&\qquad=f_\rho m_\rho\sum_{n=1}^\infty \frac{(\ii(\lcx P))^n}{n!}\Phi^{(3)}_n\Big\{
\frac{n}{n+1}\Big( e^{(\lambda)}_\alpha-P_\alpha \frac{e^{(\lambda)}\lcx}{\lcx P}\Big)
+\frac{n(n-1)}{(n+1)^2}\lcx_\alpha\frac{m_\rho^2 (e^{(\lambda)}\lcx)}{(\lcx P)^2}\Big\},
\\
\label{matrix_O_t4_nl}
&\langle 0|O^{\mathrm{tw4}}_{\alpha}(\lcx,-\lcx)|\rho(P,\lambda)\rangle
=
\frac{1}{2}f_\rho m_\rho
\lcx_\alpha(e^{(\lambda)}\lcx)m_\rho^2
\int_0^1\d t \int_{-1}^1\d\xi\, \hat\Phi^{(4)}(\xi)
(\ii\xi t)^2(\ln t) \,\e^{\ii\xi t(\tilde{x}P)},\nonumber\\
&\qquad=-f_\rho m_\rho \sum_{n=2}^\infty \frac{(\ii(\lcx P))^n}{n!}\Phi^{(4)}_n
\frac{n(n-1)}{2(n+1)^2}\lcx_\alpha\frac{m_\rho^2 (e^{(\lambda)}\lcx)}{(\lcx P)^2}.
\end{align}

Obviously, the trace terms which have been explicitly subtracted  
are proportional to $m_\rho^2$. 
They give rise to light-cone mass correction terms of mesons and
contribute to {\it dynamical} twist-4. 
For the twist--2 operator we observe that the terms proportional to 
$e^{(\lambda)}_\alpha$, $P_\alpha$ and $\lcx_\alpha$ have contributions starting 
with the zeroth, first and second moment, respectively.
The twist--3 operator starts with the first moment, and the
twist--4 operator starts with the second moment.

If we put together the different twist contributions we obtain,
after replacing $\ii \xi t(\lcx P)$ by $ t \pd/\pd t $
and performing partial integrations, the following 
matrix element of the original operator 
\begin{align}
\label{O_full}
&\langle 0|O_{\alpha}(\lcx,-\lcx)|\rho(P,\lambda)\rangle
=f_\rho m_\rho\Bigg[
P_\alpha\frac{e^{(\lambda)}\lcx }{\lcx P}\int_{-1}^1\d\xi
\Big(\hat\Phi^{(2)}(\xi)-\hat\Phi^{(3)}(\xi)\Big)[e_0(\ii\zeta\xi)-e_1(\ii\zeta\xi)]\\
&\quad
+e^{(\lambda)}_\alpha\int_{-1}^1\d\xi
\Big(\hat\Phi^{(2)}(\xi)e_1(\ii\zeta\xi)+\hat\Phi^{(3)}(\xi)
[e_0(\ii\zeta\xi)-e_1(\ii\zeta\xi)]\Big)\nonumber\\
&\quad
-\frac{1}{2}\lcx_\alpha
\frac{e^{(\lambda)}\lcx }{(\lcx P)^2}m_\rho^2\int_{-1}^1\!\!\d\xi
\Big(\hat\Phi^{(2)}(\xi)-2\hat\Phi^{(3)}(\xi)+\hat\Phi^{(4)}(\xi)\Big)
\Big[e_0(\ii\zeta\xi)-3e_1(\ii\zeta\xi)
+2\int_0^1\!\!\d t\,  e_1(\ii\zeta \xi t)\Big]
\Bigg], \nonumber
\end{align}
where we used again the  ``truncated exponentials'' and $\zeta=(\lcx P)$.
Let us mention that the twist-2 part of Eq.~(\ref{O_full}) and the local 
expression (\ref{matrix_O_t2_nl}) are in agreement with 
Stoll's result~\cite{Stoll} for the light-cone twist-2 operator.

Now we consider the chiral-even {\em axial vector operator}
$O_{5\alpha}(\lcx,-\lcx)$, which
obeys relations Eqs.~(\ref{q_O2vec}) -- (\ref{q_O4vec}), as well as 
(\ref{q_OPROJ}) with the {\em same} projection operator
as the vector operator. Let us introduce the corresponding
meson distribution amplitudes $\Xi^{(\tau)}(z)$ of twist $\tau$ by
\begin{align}
\label{Ffct}
\langle 0|O^{(\tau)}_{5\alpha}(\lcx, -\lcx)|\rho(P,\lambda)\rangle 
= \frac{1}{2}\Big(f_\rho -f_\rho^{\rm T}\frac{m_u+m_d}{m_\rho}\Big)m_\rho
{\cal P}^{(\tau)\beta}_\alpha
\Big(\epsilon_\beta^{\ \,\gamma\mu\nu}
e^{(\lambda)}_\gamma P_\mu\lcx_\nu\int_{-1}^1\d\xi\, \hat\Xi^{(\tau)}(\xi)\,
{\e^{\ii\xi(\lcx P)}}\Big),
\end{align} 
where $f^{\rm T}_\rho$ denotes the tensor decay constant.
The vacuum-to-meson matrix elements of these {\em vector operators} of 
geometric twist are obtained as follows:
\begin{align}
\label{matrix_O_tw3_nl}
\langle 0|
O^{\text{tw3}}_{5\alpha}(\lcx,-\lcx)|\rho(P,\lambda)\rangle
&=
\frac{1}{2}\Big(f_\rho -f_\rho^{\rm T}\frac{m_u+m_d}{m_\rho}\Big)m_\rho
\epsilon_\alpha^{\ \,\beta\mu\nu}
e^{(\lambda)}_\beta P_\mu \lcx_\nu
\int_0^1\!\!\d t \int_{-1}^1\!\!\d\xi\, \hat\Xi^{(3)}(\xi)\nonumber\\
&\hspace{5cm}
\times\big(1+\ii\xi t(\lcx P)\big)
\e^{\ii\xi t(\tilde{x}P)}\nonumber\\
&=\frac{1}{2}\Big(f_\rho -f_\rho^{\rm T}\frac{m_u+m_d}{m_\rho}\Big)m_\rho
\epsilon_\alpha^{\ \,\beta\mu\nu}
e^{(\lambda)}_\beta P_\mu\lcx_\nu
\sum_{n=0}^\infty \frac{(\ii(\lcx P))^n}{n!}\,\Xi^{(3)}_n .
\end{align}
We see that only the twist-3 operator gives a contribution and the twist-2 
and twist-4 operator as well as 
all trace terms vanish. 
Thus, the matrix element of the original operator reads
\begin{align}
\label{matrix_O_nl}
&\langle 0|O^{}_{5\alpha}(\lcx,-\lcx)|\rho(P,\lambda)\rangle
=\frac{1}{2}\Big(f_\rho -f_\rho^{\rm T}\frac{m_u+m_d}{m_\rho}\Big)m_\rho
\epsilon_\alpha^{\ \,\beta\mu\nu}
e^{(\lambda)}_\beta P_\mu\lcx_\nu
\int_{-1}^1\d\xi\, \hat\Xi^{(3)}(\xi) e_0(\ii\zeta \xi).
\end{align}

The matrix element of the simplest bilocal {\em scalar operator}
arises as
\begin{align}\label{}
\langle 0|N(\lcx,-\lcx)|\rho(P,\lambda)\rangle
&=-\ii\Big( f^{\rm T}_\rho-f_\rho \frac{m_u+m_d}{m_\rho}\Big)
\big(e^{(\lambda)}\lcx\big)m_\rho^2
\int_{-1}^1\d\xi\, \hat\Upsilon^{(3)}(\xi){\e^{\ii\xi(\lcx P)}},
\nonumber\\
&=-\ii\Big( f^{\rm T}_\rho-f_\rho \frac{m_u+m_d}{m_\rho}\Big)
\big(e^{(\lambda)}\lcx\big)m_\rho^2
\sum_{n=0}^\infty \frac{(\ii(\lcx P))^n}{n!}\,\Upsilon^{(3)}_n ,
\end{align}
where $\hat\Upsilon^{(3)}(\xi)$ is another spin-independent twist-3 meson 
DA.

Now, we consider the matrix elements of the chiral-odd {\em  skew tensor 
operators}.
The corresponding distribution amplitudes are introduced by
\begin{align}
\label{Hfct}
\langle 0|M^{(\tau)}_{[\alpha\beta]}(\lcx, -\lcx)|\rho(P,\lambda)\rangle 
&=\ii f^{\rm T}_\rho {\cal P}^{(\tau)[\mu\nu]}_{[\alpha\beta]}
\Big((e^{(\lambda)}_\mu P_\nu-e^{(\lambda)}_\nu P_\mu) 
\int_{-1}^1\d\xi\, \hat\Psi^{(\tau)}(\xi)\,{\e^{\ii\xi(\lcx P)}}\Big).
\end{align}
The matrix elements of the skew tensor operators of twist $\tau$ 
are obtained using the projectors determined by Eqs.~(\ref{q_M_tw2_ten})
-- (\ref{q_M_tw4_ten}):
\begin{align}
\label{matrix_M_tw2_nl}
&\langle 0|
M^{\mathrm{tw2}}_{[\alpha\beta]}(\lcx,-\lcx)|\rho(P,\lambda)\rangle
=\ii f_\rho^{\rm T}\int_0^1\d t \int_{-1}^1\d\xi\,\hat\Psi^{(2)}(\xi)
\Big[2 t \, e^{(\lambda)}_{[\alpha} P_{\beta]}\big(2+\ii\xi t(\lcx P)\big)
\\
&\qquad\qquad\qquad\qquad
+\left(1- t \right)m_\rho^2
\lcx_{[\alpha}\Big\{4(\ii\xi t)e^{(\lambda)}_{\beta]}
+(\ii\xi t)^2\big(e^{(\lambda)}_{\beta]}(\lcx P)
+P_{\beta]}(e^{(\lambda)}\lcx)\big)\Big\}
\Big]\e^{\ii\xi t(\lcx P)},\nonumber\\
&\quad
= \ii f_\rho^{\rm T}\sum_{n=0}^\infty \frac{(\ii(\lcx P))^n}{n!}\Psi^{(2)}_n\Big\{
 2e^{(\lambda)}_{[\alpha}P_{\beta]}
+\frac{4n}{(n+2)(n+1)}\frac{m_\rho^2}{\lcx P}\lcx_{[\alpha}e^{(\lambda)}_{\beta]}
\nonumber\\
&\qquad\qquad\qquad\qquad
+\frac{n(n-1)}{(n+2)(n+1)}\frac{m_\rho^2}{\lcx P}\lcx_{[\alpha}\Big(e^{(\lambda)}_{\beta]}
+P_{\beta]}\frac{e^{(\lambda)}\lcx}{\lcx P}\Big)\Big\},
\nonumber\\
&\langle 0|M^{\mathrm{tw3}}_{[\alpha\beta]}(\lcx,-\lcx)|\rho(P,\lambda)\rangle
=-\ii f_\rho^{\rm T}\int_0^1\d t \, \frac{1- t ^2}{ t }
\int_{-1}^1\d\xi\,\ii\xi t\, \hat\Psi^{(3)}(\xi) m_\rho^2
\lcx_{[\alpha}\Big\{ e^{(\lambda)}_{\beta]}
+\ii\xi t P_{\beta]}(e^{(\lambda)}\lcx)\Big\}\nonumber\\
&\hspace{12cm}\times
\e^{\ii\xi t(\lcx P)},\\
&\phantom{\langle 0|M^{\mathrm{tw2}}_{\alpha\beta}(\lcx,-\lcx)|\rho(P,\lambda)\rangle}
=-\ii f_\rho^{\rm T}\sum_{n=1}^\infty \frac{(\ii\kappa(\lcx P))^n}{n!}\Psi^{(3)}_n
\frac{2}{n+2}\frac{m_\rho^2}{\lcx P}\lcx_{[\alpha}\Big(e^{(\lambda)}_{\beta]}
+(n-1)P_{\beta]}\frac{e^{(\lambda)}\lcx}{\lcx P}\Big),
\nonumber\\
\label{matrix_M_tw4_nl}
&\langle 0|
M^{\mathrm{tw4}}_{[\alpha\beta]}(\lcx,-\lcx)|\rho(P,\lambda)\rangle
=-\ii f_\rho^{\rm T}\int_0^1\d t \,\frac{1- t }{ t }
\int_{-1}^1\d\xi\, (\ii  t \xi )^2 \hat\Psi^{(4)}(\xi)\,m_\rho^2
\lcx_{[\alpha}\Big\{e^{(\lambda)}_{\beta]}(\lcx P)-P_{\beta]}(e^{(\lambda)}\lcx)\Big\}
\nonumber\\
&\hspace{12cm}\times
\e^{\ii \xi t(\lcx P)},\nonumber\\
&\phantom{\langle 0|M^{\mathrm{tw2}}_{\alpha\beta}(\lcx,-\lcx)|\rho(P,\lambda)\rangle}
=-\ii f_\rho^{\rm T}\sum_{n=2}^\infty \frac{(\ii(\lcx P))^n}{n!}\Psi^{(4)}_n
\frac{n-1}{n+1}\frac{m_\rho^2}{\lcx P}\lcx_{[\alpha}\Big(e^{(\lambda)}_{\beta]}
-P_{\beta]}\frac{e^{(\lambda)}\lcx}{\lcx P}\Big).
\end{align}
Again, the moments of distribution amplitudes of twist $\tau = 2,3$
and $4$ start with $n=0,1$ and $2$, respectively. In addition,
we remark that only those terms of the operator (\ref{q_M_tw3_ten}) 
contribute to the twist--3 distribution amplitude which result from
the trace terms of (\ref{q_M_tw2_ten}). 
 
The matrix element of the original skew tensor operator is obtained as
\begin{align}
\label{M_full_JJ}
\langle 0|M_{[\alpha\beta]}(\lcx,-\lcx)&|\rho(P,\lambda)\rangle
=
\ii f_\rho^{\rm T}\Big[
2e^{(\lambda)}_{[\alpha}P_{\beta]}
\int_{-1}^1\d\xi\, \hat\Psi^{(2)}(\xi)e_0(\ii\zeta \xi)\\
&
+\lcx_{[\alpha}P_{\beta]}\frac{m_\rho^2(e^{(\lambda)}\lcx)}{(\lcx P)^2}
\int_{-1}^1\d\xi \Big\{
 \hat\Psi^{(2)}(\xi)\big[e_0(\ii\zeta \xi) + 2e_1(\ii\zeta \xi) + 6e_2(\ii\zeta \xi)\big]
\nonumber\\
&\qquad
- \hat\Psi^{(3)}(\xi)\big[1 +2e_0(\ii\zeta \xi) + 6e_2(\ii\zeta \xi)\big]
+ \hat\Psi^{(4)}(\xi)\big[1 + e_0(\ii\zeta \xi) - 2e_1(\ii\zeta \xi)\big]
\Big\}\nonumber\\
&
+\lcx_{[\alpha}e^{(\lambda)}_{\beta]}\frac{m_\rho^2}{\lcx P}
\int_{-1}^1\d\xi \Big\{
  \hat\Psi^{(2)}(\xi)\big[e_0(\ii\zeta \xi)-2e_1(\ii\zeta \xi)-2e_2(\ii\zeta \xi)\big]
\nonumber\\
&\qquad
+ \hat\Psi^{(3)}(\xi)\big[1+2e_2(\ii\zeta \xi)\big]
- \hat\Psi^{(4)}(\xi)\big[1+e_0(\ii\zeta \xi)-2e_1(\ii\zeta \xi)\big]
\Big\}\Big] .
\nonumber
\end{align}
Now we finish the determination of the eight meson wave
functions which result from the nonlocal
light--cone quark operators of definite twist.

\subsection{The relations between 
$\rho$-meson distribution amplitudes of 
 dynamical and geometric twist}
\label{sec-geoDA}
By means of operators of geometric twist and using the vectors $p_\alpha$ and 
$e^{(\lambda)}_{\perp\alpha}$, 
the chiral-even $\rho$-distributions 
with definite (geometric) twist are:
\begin{align}
\label{O_v}
&\langle 0|O_{\alpha}(\lcx,-\lcx)|\rho(P,\lambda)\rangle
=f_\rho m_\rho\bigg[ p_\alpha\frac{e^{(\lambda)}\lcx}{\lcx P}\int_{-1}^1\d \xi\, \hat\Phi^{(2)}(\xi) e_0(\ii\zeta\xi)\\
&\qquad+e^{(\lambda)}_{\perp\alpha}\int_{-1}^1\!\!\d \xi
\Big\{\hat\Phi^{(2)}(\xi)e_1(\ii\zeta \xi)+\hat\Phi^{(3)}(\xi)
[e_0(\ii\zeta \xi)-e_1(\ii\zeta \xi)]\Big\}
\nonumber\\
&\qquad
-\frac{1}{2}\lcx_\alpha\frac{e^{(\lambda)}\lcx}{(\lcx P)^2} m_\rho^2
\int_{-1}^1\!\!\d \xi
\Big\{\hat\Phi^{(4)}(\xi)\Big[e_0(\ii\zeta \xi)-3e_1(\ii\zeta \xi)
+2\int_0^1\!\!\d t\,  e_1(\ii\zeta \xi t)\Big]\nonumber\\
&\qquad\qquad\qquad-\hat\Phi^{(2)}(\xi)\Big[e_1(\ii\zeta \xi)
-2\int_0^1\!\!\d t\, e_1(\ii\zeta \xi t)\Big]
+4\hat\Phi^{(3)}\Big[e_1(\ii\zeta \xi)
-\int_{0}^1\!\!\d t\, e_1(\ii\zeta \xi t)\Big]\Big\}\bigg]\nonumber
\end{align}
and
\begin{align}
\label{O_5v}
\langle 0|O_{5\alpha}(\lcx,-\lcx)|\rho(P,\lambda)\rangle
&=
\frac{1}{2}\bigg(f_\rho -f_\rho^{\rm T}\frac{m_u+m_d}{m_\rho}\bigg)m_\rho
\epsilon_\alpha^{\ \,\beta\mu\nu}
e^{(\lambda)}_{\perp\beta} p_\mu\lcx_\nu
\int_{-1}^1\d \xi\, \hat\Xi^{(3)}(\xi) e_0(\ii\zeta \xi),
\end{align}
while the chiral-odd distributions are defined as
\begin{align}
\label{M_ten}
&\langle 0|M_{[\alpha \beta]}(\lcx,-\lcx)|\rho(P,\lambda)\rangle
=
\ii f_\rho^{\rm T}\bigg[
2e^{(\lambda)}_{\bot[\alpha}p_{\beta]}
\int_{-1}^1\!\!\d \xi \hat\Psi^{(2)}(\xi)e_0(\ii\zeta \xi)
\\
&\qquad
+2\lcx_{[\alpha}p_{\beta]}\frac{e^{(\lambda)}\lcx}{(\lcx P)^2}m_\rho^2
\int_{-1}^1\!\!\d \xi \Big\{
 2\hat\Psi^{(2)}(\xi)e_2(\ii\zeta \xi)
-\hat\Psi^{(3)}(\xi)\big[e_0(\ii\zeta \xi)+2e_2(\ii\zeta \xi)\big]
\Big\}
\nonumber\\
&\qquad
-\lcx_{[\alpha}e^{(\lambda)}_{\beta]\perp}\frac{m_\rho^2}{\lcx P}
\int_{-1}^1\!\!\d \xi \Big\{
2 \hat\Psi^{(2)}(\xi)\big[e_1(\ii\zeta \xi)+e_2(\ii\zeta \xi)\big]
- \hat\Psi^{(3)}(\xi)\big[1+2e_2(\ii\zeta \xi)\big]\nonumber\\
&\hspace{5cm}
+ \hat\Psi^{(4)}(\xi)\big[1+e_0(\ii\zeta \xi)-2e_1(\ii\zeta \xi)\big]
\Big\}\bigg] \nonumber
\end{align}
and
\begin{align}
\label{O_s}
\langle 0|N(\lcx,-\lcx)|\rho(P,\lambda)\rangle
&=-\ii\bigg( f^{\rm T}_\rho-f_\rho \frac{m_u+m_d}{m_\rho}\bigg)
\big(e^{(\lambda)}\lcx\big)m_\rho^2
\int_{-1}^1\d \xi\, \hat\Upsilon^{(3)}(\xi)e_0(\ii\zeta \xi).
\end{align}

Comparing these expressions~(\ref{O_v})--(\ref{O_s})
with the meson distribution amplitudes of dynamical twist
(\ref{eq:vda}) -- (\ref{eq:sda}),
we observe that it is necessary to re-express the truncated exponentials
and perform appropriate variable transformations. After such manipulations
we obtain the following relations, which allow to reveal the interrelations
between the different twist definitions of meson distribution amplitudes:
\begin{align}  
\label{rel-phi1}
\hat\phi_\|(\xi)&\equiv\hat\Phi^{(2)}(\xi),\\
\label{rel-gv}
\gvh(\xi)&=\hat\Phi^{(3)}(\xi) + \int_\xi^{{\rm sign}(\xi)} \frac{\d \tau}{\tau}
\Big(\hat\Phi^{(2)}-\hat\Phi^{(3)}\Big)\left(\tau\right),\\
\label{rel-g3}
\hat g_3(\xi)&=\hat\Phi^{(4)}(\xi) - \int_\xi^{{\rm sign}(\xi)} \frac{\d\tau}{\tau}\Big\{
\Big(\hat\Phi^{(2)}-4\hat\Phi^{(3)}+3\hat\Phi^{(4)}\Big)\left(\tau\right)\nonumber\\
&\hspace{6cm}+2 \ln \Big(\frac{\xi}{\tau}\Big)
\Big(\hat\Phi^{(2)}-2\hat\Phi^{(3)}+\hat\Phi^{(4)}\Big)\left(\tau\right)\Big\},\\
\label{rel-ga}
\hat g_\perp^{(a)}(\xi)&\equiv\hat\Xi^{(3)}(\xi),\\
\label{rel-phi2}
\hat\phi_\perp(\xi)&\equiv\hat\Psi^{(2)}(\xi),\\
\label{rel-ht}
\hat h^{(t)}_\|(\xi)&=\hat\Psi^{(3)}(\xi) + 2\xi \int_\xi^{{\rm sign}(\xi)}\frac{\d \tau}{\tau^2}
\Big(\hat\Psi^{(2)}-\hat\Psi^{(3)}\Big)\left(\tau\right),\\
\label{rel-h3}
\hat h_3(\xi)&=\hat\Psi^{(4)}(\xi)+ \int_\xi^{{\rm sign}(\xi)} \frac{\d \tau}{\tau}\Big\{
2\Big(\hat\Psi^{(2)}-\hat\Psi^{(4)}\Big)\left(\tau\right)
-2\frac{\xi}{\tau}\Big(\hat\Psi^{(2)}-\hat\Psi^{(3)}\Big)\left(\tau\right)
\nonumber\\
&\hspace{8cm}
-\delta\Big(\frac{\xi}{\tau}\Big)\Big(\hat\Psi^{(3)}-\hat\Psi^{(4)}\Big)\left(\tau\right)
\Big\},\\
\hat h_\|^{(s)}(\xi)&\equiv\hat\Upsilon^{(3)}(\xi).
\end{align}
Obviously, both decompositions coincide in the leading terms, e.g., 
the meson distribution amplitudes $\hat\phi_\|(\xi)$ and $\hat\phi_\perp(\xi)$ are genuine
geometric twist-2 and $\gah(\xi)$ and $\hsh(\xi)$ are genuine 
geometric twist-3 functions.
But the different twist definitions of meson distribution amplitudes differ at 
higher twist. For instance, the meson distribution amplitudes 
$\gvh(\xi)$ and $\hat h^{(t)}_\|(\xi)$ with dynamical twist $t=3$ contain 
contributions with geometrical twist $\tau = 2$ and $3$. Additionally,
dynamical twist $t=4$ meson distribution amplitudes  
$\hat g_3(\xi)$ and $\hat h_3(\xi)$ contain 
contributions with geometrical twist $\tau = 2$, $3$ as well as $4$.

The conventional distribution amplitudes can be written in
terms of the new $\rho$-meson distribution amplitudes.
The nontrivial relations are:
\begin{align}
\label{rel-Phi3}
\hat\Phi^{(3)}(\xi)&=\gvh(\xi) +\frac{1}{\xi} \int_\xi^{{\rm sign}(\xi)} \d \tau
\big(\gvh-\hat\phi_\|\big)\left(\tau\right),\\
\hat\Phi^{(4)}(\xi)&=\hat g_3(\xi) 
+\frac{1}{\xi^2} \int_\xi^{{\rm sign}(\xi)} \d \tau\, \tau
\big(3\hat g_3-4\gvh+\hat\phi_\|\big)\left(\tau\right)\\
&\hspace{5cm}
+\frac{1}{\xi^2} \int_\xi^{{\rm sign}(\xi)} \d \tau\, \tau \Big(1-\frac{\xi}{\tau}\Big)
\big(\hat g_3-4\gvh+3\hat\phi_\|\big)\left(\tau\right),\nonumber\\
\label{rel-Psi3}
\hat\Psi^{(3)}(\xi)&=\hat h_\|^{(t)}(\xi) + \frac{2}{\xi}\int_\xi^{{\rm sign}(\xi)} \d \tau
\big(\hat h_\|^{(t)}-\hat\phi_\perp\big)\left(\tau\right), \\
\hat\Psi^{(4)}(\xi)&=\hat h_3(\xi) 
+\frac{2}{\xi^2} \int_\xi^{{\rm sign}(\xi)} \d \tau\, \tau
\big(\hat h_3-\hat h_\|^{(t)}\big)\left(\tau\right)
-\frac{2}{\xi^2} \int_\xi^{{\rm sign}(\xi)} \d \tau\, \tau \Big(1-\frac{\xi}{\tau}\Big)
\big(\hat h^{(t)}_\|-\hat\phi_\perp\big)\left(\tau\right).
\end{align}

The relation between the moments may be read off from Eqs.~(\ref{O_v})
-- (\ref{O_s}) as follows:
\begin{align}  
\phi_{\|n}&\equiv\Phi^{(2)}_n,\\
g^{(v)}_{\perp n}&=\Phi^{(3)}_n +  \frac{1}{n+1}
\Big(\Phi^{(2)}_n-\Phi^{(3)}_n\Big),\\
g_{3n}&=\Phi^{(4)}_n -  \frac{1}{n+1}
\Big(\Phi^{(2)}_n-4\Phi^{(3)}_n+3\Phi^{(4)}_n\Big)
+\frac{2}{(n+1)^2}
\Big(\Phi^{(2)}_n-2\Phi^{(3)}_n+\Phi^{(4)}_n\Big),\\
\label{ga_n}
g_{\perp n}^{(a)}&\equiv\Xi^{(3)}_n,\\
\phi_{\perp n}&\equiv\Psi^{(2)}_n,\\
h_{\|n}^{(t)}&=\Psi^{(3)}_n +  \frac{2}{n+2}
\Big(\Psi^{(2)}_n-\Psi^{(3)}_n\Big),\\
h_{3n}&=\Psi^{(4)}_n+  \frac{2}{n+1}
\Big(\Psi^{(2)}_n-\Psi^{(4)}_n\Big)
-\frac{2}{n+2}\Big(\Psi^{(2)}_n-\Psi^{(3)}_n\Big)
-\delta_{n0}\Big(\Psi^{(3)}_n-\Psi^{(4)}_n\Big),\\
h_{\|n}^{(s)}&\equiv\Upsilon^{(3)}_n.
\end{align}
In terms of the moments the relations between old and new wave
functions may be easily inverted; for the distribution amplitudes itself
the expression of the new distribution amplitudes through the old ones is more
involved. The inverse relations are:
\begin{align}  
\Phi^{(3)}_{n}&=g_{\perp n}^{(v)} +  \frac{1}{n}
\Big(g_{\perp n}^{(v)}-\phi_{\|n}\Big),\qquad n>0\\
\Phi^{(4)}_{n}&=
g_{3n}+  \frac{1}{n-1}
\Big(3g_{3n}-4g_{\perp n}^{(v)}+\phi_{\| n}\Big)
+\frac{1}{n(n-1)}\Big(g_{3n}-4g_{\perp n}^{(v)}+3\phi_{\|n}\Big),\qquad  n>1\\
\Psi^{(3)}_{n}&=h_{\|n}^{(t)}+  \frac{2}{n}
\Big(h_{\|n}^{(t)}-\phi_{\perp n}\Big),\qquad  n>0\\
\Psi^{(4)}_n&=
h_{3n}+  \frac{2}{n-1}
\Big(h_{3n}-h_{\|n}^{(t)}\Big)
-\frac{2}{n(n-1)}\Big(h_{\| n}^{(t)}-\phi_{\perp n}\Big),\qquad  n>1.
\end{align}

The relations between the dynamical and geometrical twist meson 
DAs show that the dynamical distribution amplitudes can be used to determine 
the new geometrical ones and vice versa.
Obviously, the same holds for their moments. In principle, this
allows to determine, e.g., the new distribution amplitudes from
the experimental data if these are known for the conventional 
ones. Thus, all of the physical and experimental correlations obtained 
by Ball~{\it et al.}~\cite{ball96,ball98,ball99}, e.g., the asymptotic wave
functions, can be used for the 
geometrical twist distribution amplitudes.

\subsection{Wandzura-Wilczek type relations}
This Section is devoted to the general discussion of geometric Wandzura-Wilczek-type
relations between the conventional distribution amplitudes which we have 
obtained by means of the distribution amplitudes with a definite geometric twist.
These geometric Wandzura-Wilczek-type relations are independent from the QCD field equations
and the corresponding operator relations. 
The mismatch between the definition of dynamical and geometric twist gives
rise to relations of geometric Wandzura-Wilczek-type which show that the dynamical 
twist functions contain various parts of different geometric twist. 
Consequently, a genuine geometric twist distribution amplitude do not give rise
to any geometric Wandzura-Wilczek-type relation. Thereby, we obtain new 
(geometric) Wandzura-Wilczek-type relations and sum rules for the meson distribution amplitudes 
of dynamical twist.

We use the notations $\gvh(\xi)=\hat g^{(v),\rm{tw2}}_\perp(\xi)+\hat g^{(v),\rm{tw3}}_\perp(\xi)$, 
where $\hat g^{(v),\rm{tw2}}_\perp(\xi)$  is the genuine twist-2 and 
$\hat g^{(v),\rm{tw3}}_\perp(\xi)$ the genuine
twist-3 part of $\hat g^{(v)}_\perp(\xi)$. 
Substituting (\ref{rel-phi1}) into (\ref{rel-gv}),
we get
\begin{align}
\label{WW-tw2}
\hat g^{(v),\rm{tw2}}_\perp(\xi)&= \int_\xi^{{\rm sign}(\xi)} \frac{\d \tau}{\tau}\,
\hat\phi_\|(\tau),\\
\label{WW-tw3}
\hat g^{(v),\rm{tw3}}_\perp(\xi)&=\gvh(\xi)-\int_\xi^{{\rm sign}(\xi)} \frac{\d \tau}{\tau}\,\hat\phi_\|(\tau),
\end{align}
where (\ref{WW-tw2}) is really the analogue of the Wandzura-Wilczek relation 
for the twist-2 part~\cite{Wandzura} because they argued that the geometric 
twist-3 contribution should be small.
On the other hand, Eq.~(\ref{WW-tw3})
is the Wandzura-Wilczek-type relation of geometric twist-3 and gives the 
information how much is the physical contribution of the geometric twist-3
operator in this non-forward process. 
These relations show that the transverse distribution $\gvh(\xi)$ is related 
to the longitudinal distribution $\hat\phi_\|(\xi)$.
Obviously, this analogy between the Wandzura-Wilczek relations of parton
distribution amplitudes and of meson distribution amplitudes
is not surprising because the operator structures are the same 
and the $\rho$-meson polarization vector formally substitutes 
the nucleon spin vector in the Lorentz structures.
The generalization of the geometric Wandzura-Wilczek relation~(\ref{WW-tw2}) for
non-forward distribution amplitudes was discussed in Ref.~\cite{BR00}.
Let me note that the Wandzura-Wilczek relations which are obtained 
in Refs.~\cite{ABS94,ball96,ball98}
for $g^{(v),\rm{tw2}}_\perp(u)$ are significantly different (see also~\cite{BL01}). 
The reason is that they used equations of motion operator relations
in order to isolate the dynamical twist-3 part. Therefore, their 
Wandzura-Wilczek relations are dynamical relations 
(dynamical Wandzura-Wilczek-type relations). 
The corresponding geometric Wandzura-Wilczek-type relations for the moments 
are: 
\begin{align}
\label{WW-tw2-n}
g^{(v),\rm{tw2}}_{\perp n}&= \frac{1}{n+1}\,\phi_{\| n},\\
\label{WW-tw3-n}
g^{(v),\rm{tw3}}_{\perp n}&=g^{(v)}_{\perp n}-\frac{1}{n+1}\,\phi_{\| n}, \qquad n>0.
\end{align}
Note that Eq.~(\ref{WW-tw2-n}) was earlier obtained in Ref.~\cite{ball96}. 
The reason for this agreement between the twist-2 part of  
the dynamical and geometric Wandzura-Wilczek relation is that the dynamical 
and geometric twist-2 meson distribution amplitude coincides. 
Obviously, from the relation (\ref{WW-tw2-n}) for $n=0$ 
the condition of the normalization
and for $n=1$ the 
following sum rule follow:
\begin{align}
\int_{-1}^1\d \xi\,\hat g^{(v)}_{\perp}(\xi) =\int_{-1}^1\d \xi\, \hat\phi_{\|}(\xi)=1,
\qquad
\int_{-1}^1\d \xi\,\xi \hat g^{(v),\rm{tw2}}_{\perp}(\xi) =\frac{1}{2}\int_{-1}^1\d \xi\,\xi \hat\phi_{\|}(\xi).
\end{align}
Using the formulas (\ref{rel-phi1}), (\ref{rel-gv}), (\ref{rel-g3}) and
(\ref{rel-Phi3}), we obtain the integral relations for the function
$\hat g_3(\xi)=\hat g^{\rm tw2}_3(\xi)+\hat g^{\rm tw3}_3(\xi)+\hat g^{\rm tw4}_3(\xi)$ as
\begin{align}
\label{WW-g3-tw2}
\hat g^{\rm tw2}_3(\xi)&=-\int_\xi^{{\rm sign}(\xi)}\frac{\d \tau}{\tau}            
\Big\{\hat\phi_\|(\tau)+2\ln\Big(\frac{\xi}{\tau}\Big)\hat\phi_\|(\tau)\Big\},\\
\label{WW-g3-tw3}
\hat g^{\rm tw3}_3(\xi)&=4\int_\xi^{{\rm sign}(\xi)}\frac{\d \tau}{\tau}            
\Big\{\gvh(\tau)+\ln\Big(\frac{\xi}{\tau}\Big)\hat\phi_\|(\tau)\Big\},\\
\label{WW-g3-tw4}
\hat g^{\rm tw4}_3(\xi)&=\hat g_3(\xi)-\int_\xi^{{\rm sign}(\xi)} \frac{\d \tau}{\tau}
    	    \Big\{\big(4\gvh-\hat\phi_\|\big)(\tau)
	   +2\ln\Big(\frac{\xi}{\tau}\Big)\hat\phi_\|(\tau)\Big\}.
\end{align}
Due to the fact that $\hat g_3(\xi)$ contains twist-2, twist-3 as well as twist-4,
we have obtained three integral relations. For example, Eq.~(\ref{WW-g3-tw2})
demonstrates that the twist-2 part $\hat g^{\rm tw2}_3(\xi)$ can be expressed in terms
of the twist-2 function $\hat\phi_\|(\xi)$. Additionally, the twist-3 part 
$\hat g^{\rm tw3}_3(\xi)$ 
is given in Eq.~(\ref{WW-g3-tw3}) in terms of the functions 
$\hat\phi_\|(\xi)$ and $\gvh(\xi)$.
Eq.~(\ref{WW-g3-tw4}) expresses how can
be obtained the twist-4 part $\hat g^{\rm tw4}_3(\xi)$ from $\hat g_3(\xi)$.
For the distribution amplitude moments we obtain:
\begin{align}
\label{WW-g3-tw2-n}
g^{\rm tw2}_{3 n}&=-\frac{1}{n+1}\,\phi_{\| n}+\frac{2}{(n+1)^2}\,\phi_{\| n},\\
\label{WW-g3-tw3-n}
g^{\rm tw3}_{3 n}&=\frac{4}{n+1}\, \gvn
              -\frac{4}{(n+1)^2}\,\phi_{\| n},\qquad n>0\\
g^{\rm tw4}_{3 n}&=g_{3 n}
              -\frac{1}{n+1}\Big(4\gvn-\phi_{\| n}\Big)
              +\frac{2}{(n+1)^2}\,\phi_{\| n},\qquad n>1.
\end{align}
For 
$n=1$ in (\ref{WW-g3-tw2-n}) and (\ref{WW-g3-tw3-n}) we find the sum rules
\begin{align}
\int_{-1}^1\d \xi\, \xi \hat g^{\rm tw2}_3(\xi)&=0,\\
\int_{-1}^1\d \xi\, \xi \hat g^{\rm tw3}_3(\xi)& = \int_{-1}^1 \d \xi\,\xi\big(
                                         2\hat g_\perp^{(v)}-\hat\phi_{\|}\big)(\xi),
\end{align}
which give the interesting sum rule for the dynamical twist-4 function 
$\hat g_3(\xi)$:
\begin{align}
\int_{-1}^1\d \xi\, \xi \hat g_3(\xi)& = \int_{-1}^1 \d \xi\,\xi\big(
                                         2\hat g_\perp^{(v)}-\hat\phi_{\|}\big)(\xi).
\end{align}

Let us now discuss the geometric Wandzura-Wilczek-type relations for the
chiral-odd meson distribution amplitudes.
If we substitute (\ref{rel-phi2}) into (\ref{rel-ht}) and using
$\hat h^{(t)}_\| (\xi)=\hat h^{(t),\rm{tw2}}_\|(\xi)+\hat h^{(t),\rm{tw3}}_\|(\xi)$
we derive
\begin{align}
\label{WW-JJ-tw2}
\hat h^{(t),\rm{tw2}}_\|(\xi)&= 2\xi\int_\xi^{{\rm sign}(\xi)} \frac{\d \tau}{\tau^2}\,\hat \phi_\perp(\tau),\\
\label{WW-JJ-tw3}
\hat h^{(t),\rm{tw3}}_\|(\xi)&=\hat h^{(t)}_\|(\xi)- 2\xi\int_\xi^{{\rm sign}(\xi)} 
\frac{\d\tau}{\tau^2}\,\hat \phi_\perp(\tau),
\end{align}
where (\ref{WW-JJ-tw2}) is the analogue of the twist-2 Wandzura-Wilczek-type relation of 
the nucleon structure function $h_L(z)$ which was obtained by Jaffe and Ji ~\cite{JJ92}. 
These Wandzura Wilczek-type relations  point out that the longitudinal distribution 
$\hat h^{(t)}_\| (\xi)$ is related 
to the transverse distribution $\hat\phi_\perp(\xi)$.
Dynamical Wandzura-Wilczek relations were obtained in Ref.~\cite{ball98}. 
Obviously, (\ref{WW-JJ-tw3}) is the corresponding twist-3 relation. 
The relations for the moments read:
\begin{align}
\label{WW-JJ-tw2-n}
h^{(t),\rm{tw2}}_{\| n}&=  \frac{2}{n+2}\,\phi_{\perp n},\\
\label{WW-JJ-tw3-n}
h^{(t),\rm{tw3}}_{\| n}&=h^{(t)}_{\| n}-\frac{2}{n+2}\,\phi_{\perp n},\qquad n>0.
\end{align}
For $n=1$ in (\ref{WW-JJ-tw2-n}) we observe the following sum rule
\begin{align}
\int_{-1}^1\d \xi\, \xi \hat h^{(t),\rm tw2}_{\|}(\xi) = \frac{2}{3} \int_{-1}^1\d \xi\, \xi \hat\phi_\perp(\xi).
\end{align}

Now, using the formulas (\ref{rel-phi2}), (\ref{rel-ht}), (\ref{rel-h3}) and
(\ref{rel-Psi3}), we obtain the integral relations for the function
with $\hat h_3(\xi)=\hat h^{\rm tw2}_3(\xi)+\hat h^{\rm tw3}_3(\xi)+\hat h^{\rm tw4}_3(\xi)$ as
\begin{align}
\label{WW-h3-tw2}
\hat h^{\rm{tw2}}_3(\xi)&=\frac{2}{\xi}\int_\xi^{{\rm sign}(\xi)}\d\tau
	\Big(\frac{\xi}{\tau}-\frac{\xi^2}{\tau^2}\Big)\hat\phi_\perp(\tau),\\
\label{WW-h3-tw3}
\hat h^{\rm tw3}_3(\xi)&=\frac{2}{\xi}\int_\xi^{{\rm sign}(\xi)} \d\tau
	\Big\{\htth (\tau)
	-\Big(1-\frac{\xi^2}{\tau^2}\Big)\hat\phi_\perp(\tau)\Big\},\\
\label{WW-h3-tw4}
\hat h^{\rm tw4}_3(\xi)&=\hat h_3(\xi)-
\frac{2}{\xi}\int_\xi^{{\rm sign}(\xi)} \d\tau
	\Big\{\htth (\tau)
	-\Big(1-\frac{\xi}{\tau}\Big)\hat\phi_\perp(\tau)\Big\}.
\end{align}
and for the moments
\begin{align}
\label{WW-h3-tw2-n}
h^{\rm{tw2}}_{3 n}&=\frac{2}{n+1}\,\phi_{\perp n}
                     -\frac{2}{n+2}\,\phi_{\perp n}, \\
\label{WW-h3-tw3-n}
h^{\rm tw3}_{3 n}&=\frac{2}{n}\,\httn
              -\frac{4}{(n+2)n}\,\phi_{\perp n},\qquad n>0\\
h^{\rm tw4}_{3n}&=h_{3n}-\frac{2}{n}\,\httn	
               +\frac{2}{(n+1)n}\,\phi_{\perp n},\qquad n>1.
\end{align}
Thus, Eq.~(\ref{WW-h3-tw2})
means that the twist-2 part $\hat h^{\rm tw2}_3(\xi)$ can be expressed in terms
of the twist-2 function $\hat\phi_\perp(\xi)$.
Additionally, the twist-3 part $\hat h^{\rm tw3}_3(\xi)$ 
is given in Eq.~(\ref{WW-h3-tw3}) in terms of the functions $\hat\phi_\perp(\xi)$ 
and $\htth(\xi)$.
Eq.~(\ref{WW-h3-tw4}) gives the subtraction rule in order to obtain the 
twist-4 part $\hat h^{\rm tw4}_3(\xi)$ from the original function $\hat h_3(\xi)$. 
Obviously, the same is valid for the corresponding distribution amplitude moments.
For $n=1$ in (\ref{WW-h3-tw2-n}) and (\ref{WW-h3-tw3-n}) we observe the following sum rules
\begin{align}
\int_{-1}^1\d \xi\, \xi \hat h^{\rm tw2}_{3}(\xi)& = \frac{1}{3} \int_{-1}^1\d \xi\, \xi \hat\phi_\perp(\xi),\\
\int_{-1}^1\d \xi\, \xi\hat h^{\rm tw3}_{3}(\xi) &= \frac{1}{3} \int_{-1}^1\d \xi\, \xi \big( 6\htth(\xi)-4\hat\phi_\perp(\xi)\big),
\end{align}
which give the sum rule for the dynamical twist-4 function $\hat h_3(\xi)$:
\begin{align}
\int_{-1}^1\d \xi\, \xi\hat h_{3}(\xi) =  \int_{-1}^1\d \xi\, \xi \big( 2\htth(\xi)-\hat\phi_\perp(\xi)\big).
\end{align}

Because the functions $\gah(\xi)$ and $\hsh(\xi)$ are pure geometric as well as dynamical 
twist-3 functions, there is no mismatch between the dynamical and geometric twist
and no geometric Wandzura-Wilczek-type relation occurs. 
However, dynamical Wandzura-Wilczek relations for $\gah(\xi)$ and $\hsh(\xi)$
were obtained by Ball {\it et al.}~\cite{ball96,ball98}
by using the QCD equations of motion. Here I am discussing the relation for
$g^{(a)}_{\perp n}$. Neglecting quark masses and three-particle quark-antiquark-gluon 
operators of twist-3 one recovers the operator relation~(see~\cite{BB88})
\begin{align}
\label{total1}
\big[\bar{u}(\lcx)\gamma_\alpha\gamma_5 d(-\lcx)\big]^{\rm tw3}=
\ii\epsilon_\alpha^{\ \,\beta\mu\nu}\lcx_\beta \int_{0}^1\d u\, u\, \hat\pd_\mu
\big[\bar{u}(u\lcx)\gamma_\nu d(-u\lcx)\big]+\ldots\, ,
\end{align}
where $\hat\pd_\mu$ is the so-called total derivative which translates the expansion point 
$y$:
\begin{align}
\label{tot-der}
\hat\pd_\mu
\big[\bar{u}(u\lcx)\gamma_\nu d(-u\lcx)\big]\equiv
\frac{\pd}{\pd y^\mu}
\big[\bar{u}(y+u\lcx)\gamma_\nu d(y-u\lcx)\big]\big|_{y\rightarrow 0}.
\end{align}
The total derivative on the r.h.s. of Eq.~(\ref{total1}) is calculated as
\begin{align}
\label{total2}
\hat\pd_\mu
\big[\bar{u}(u\lcx)\gamma_\nu d(-u\lcx)\big]=
\bar{u}(u\lcx)\big(\LD D\!_\mu+\RD D\!_\mu\big)\gamma_\nu d(-u\lcx)
-\ii\int_{-u}^{u}\d v\, \bar{u}(u\lcx)\lcx^\rho F_{\rho\mu}(v\lcx)\gamma_\nu d(-u\lcx),
\end{align}
where $F_{\mu\nu}$ is the gluon field strength.
Obviously, the second operator on the r.h.s. of Eq.~(\ref{total2}) is a 
Shuryak-Vainshtein type operator having minimal twist-3. The other operator
on the r.h.s. of Eq.~(\ref{total2}) has, in general, twist-2, 3 and 4. 
But, after multiplying by the Levi-Civita tensor $\epsilon_{\alpha\beta\mu\nu}$
only the twist-3 part survives and gives contributions which are not small. 
Therefore, the operator identity (\ref{total1}) is a genuine twist-3 relation. 
The twist-3 matrix element of the first operator on the r.h.s. of Eq.~(\ref{total2}) 
is given by
\begin{align}
\label{O_total}
\ii \epsilon_\alpha^{\ \,\beta\mu\nu}\lcx_\beta
&\langle 0|
\bar{u}(\lcx)\big(\LD D\!_\mu+\RD D\!_\mu\big)\gamma_\nu d(-\lcx)
|\rho(P,\lambda)\rangle
= f_\rho m_\rho
\epsilon_\alpha^{\ \,\beta\mu\nu}
e^{(\lambda)}_{\perp\beta} p_\mu\lcx_\nu
\int_{-1}^1\d \xi\, \hat\Omega^{(3)}(\xi) e_0(\ii\zeta \xi),
\end{align}
where $\hat\Omega^{(3)}(\xi)$  is the corresponding meson distribution amplitude 
of twist-3.

Taking the matrix element of the operators on both sides of Eq.~(\ref{total1}), 
using Eqs.~(\ref{O_5v}), (\ref{rel-ga}), and (\ref{O_total}),
and neglecting quark masses and trilocal operators, one obtains a geometric 
relation between the meson distribution amplitudes $\hat g^{(a)}_\perp(\xi)$ and $\hat\Omega^{(3)}(\xi)$
imposed by the QCD equations of motion as 
\begin{align}
\label{dyn-rel1}
\hat g^{(a)}_\perp(\xi)=2\xi \int_\xi^{\text{sign}(\xi)}\frac{\d\tau}{\tau^2}\, \hat\Omega^{(3)}(\tau),
\end{align}
and for the meson distribution amplitude moments 
\begin{align}
\label{dyn-rel2}
g^{(a)}_{\perp n}=\frac{2}{n+2}\, \Omega^{(3)}_{ n},
\end{align}
where $\Omega^{(3)}_{ n}$ are the geometric twist-3 distribution amplitude moments.
This relation (\ref{dyn-rel2}) reproduces the fact that the moments 
$g^{(a)}_{\perp n}$ have genuine geometric twist-3 (see Eq.~(\ref{ga_n}))
and no geometric twist-2 contributions from the total derivative come into the game. 

But this is not what Ball {\it et al.}~\cite{ball96,ball98} did.
They have calculated the r.h.s. of Eq.~(\ref{total1}) by means of
\begin{align}
\hat\pd_\mu
\langle 0|\bar{u}(\lcx)\gamma_\nu d(-\lcx)
|\rho(P,\lambda)\rangle
\end{align}
and Eq.~(\ref{eq:vda}).
Eventually, they obtained the dynamical Wandzura-Wilczek-type relation~\cite{ball96}
\begin{align}
\label{dyn-rel3}
\hat g^{(a)}_\perp(\xi)=2\xi \int_\xi^{\text{sign}(\xi)}\frac{\d\tau}{\tau^2}\, \hat g^{(v)}_\perp(\tau),
\end{align}
and for the  moments 
\begin{align}
\label{dyn-rel4}
g^{(a)}_{\perp n}=\frac{2}{n+2}\, g^{(v)}_{\perp n}.
\end{align}
Obviously, Eqs.~(\ref{dyn-rel3}) and (\ref{dyn-rel4}) are relations between
equal dynamical twist. But, the l.h.s. of these relations has geometric twist-3
and the r.h.s. has geometric twist-2 and twist-3. Therefore, a dynamical 
Wandzura-Wilczek relation is a relation which is derived by means of the QCD
equations of motion in contrast to geometric one.
Note that the Wandzura-Wilczek-like relations used in 
Refs.~\cite{APT00,BM00,KPST01,RW00,AT01} are dynamical ones.

Similar arguments are valid for the other QCD equations of motion operator relations
used in Ref.~\cite{ball98}.
Thus, from the group theoretical point of view, it is a bit misleading to 
claim that an additional geometric twist-2 contribution is produced by 
geometric twist-3 operators with a total derivative
and that the QCD equations of motion give a relation 
between distribution amplitudes of different geometric twist
in exclusive processes and off-forward scattering. 
From this point of view, the geometric twist is a proper concept for 
classifying non-forward inclusive or general exclusive matrix-elements.


\subsection{Conclusions}
We discussed the model-independent 
classification of meson distribution amplitudes with respect to geometric twist. 
We gave the relations between 
the geometric twist and Ball and Braun's dynamical twist distribution amplitudes. 
These relations demonstrate the interrelations between the different twist definitions.
Consequently, these relations are ``transformation rules'' between the dynamical 
and geometric twist distribution amplitudes.

Another results of this Section are geometric Wandzura-Wilczek-type relations
between the dynamical twist distributions which we have obtained by means of 
our ``transformation rules''. 
The reason of these geometric Wandzura-Wilczek-type relations is the mismatch 
of the dynamical twist with respect to geometric twist. 
Because we have not used the QCD equations of motion, 
the geometric Wandzura-Wilczek-type relations are based on a no-dynamical level 
and are model-independent. 
Additionally, I have discussed the difference between dynamical and geometric
Wandzura-Wilczek-type relations.
 

\section{Power corrections of non-forward quark distributions and
harmonic operators with definite geometric twist}
\setcounter{equation}{0}
\label{harmonic}
In this Section we develop a formalism for the resummation of the target
mass corrections for off-forward processes by means of harmonic 
off-cone operators of geometric twist. The harmonic operators of a given spin
or twist have a fixed tensor structure and their matrix elements have
an explicit target mass dependence, determined by the condition that the 
(local) operators have to be traceless.

We determine the twist decomposition
of the following chiral-even (axial) vector operators,
$O_{\alpha}(\kappa x,-\kappa x)$, and
$O_{5\alpha}(\kappa x,-\kappa x)$,
together with the corresponding (pseudo) scalar operators, 
$O_{(5)}(\kappa x,-\kappa x)$,
and the chiral-odd scalar and skew tensor operator,
$N(\kappa x,-\kappa x)$, and
$M_{[\alpha\beta]}(\kappa x,-\kappa x)$
together with the vector and scalar operators,
$M_{\alpha}(\kappa x,-\kappa x) 
= x^\beta M_{[\alpha\beta]}(\kappa x,-\kappa x)$,
$M(\kappa x,-\kappa x) 
= x^\beta \pd^\alpha M_{[\alpha\beta]}(\kappa x,-\kappa x)$.
The operators 
constitute a basis not only for the (usual) parton
distributions as well as scalar and vector meson distribution amplitudes~\cite{GL01,Lazar01a,Lazar01b}
but also for the consideration of double distribution amplitudes  
being relevant for the various light-cone dominated QCD processes 
under consideration. Here, their twist decomposition
off the light-cone will be given up to twist 3 and, in the case of
scalar operators, also for any twist.
In fact, the `external' operation of contracting with $x^\alpha$ or 
$x^\beta\pd^\alpha$ also influences the possible symmetry type of these
operators and, therefore, of their twist decomposition. 

These off-cone operators of definite twist are given 
for the local as well as the resummed nonlocal operators. The local 
operators given as the $n$-th Taylor coefficients of the non-local ones, 
are represented 
by (a finite series of) Gegenbauer polynomials $C^\nu_n(z),\, \nu \geq 1$. 
The nonlocal operators, being obtained by resummation
with respect to $n$, are represented by (a related series of) Bessel 
functions or, more exactly, either $J_{\nu-\frac{1}{2}}(z)$ or 
$I_{\nu-\frac{1}{2}}(z)$ depending on the values of their arguments. 

The group theoretical method for the determination of 
target mass corrections in unpolarized deep inelastic scattering
using harmonic scalar operators of definite spin and the corresponding 
matrix elements in terms of Gegenbauer polynomials has been used for
the first time by Nachtmann~\cite{Nachtmann73} and later by 
Baluni and Eichten~\cite{BE76}.  
Using the same procedure,
the target mass contributions for polarized deep inelastic scattering 
were studied in Refs.~\cite{Wandzura77,MU80,KU95}. 
A key ingredient in this framework was that they obtained
so-called ``Nachtmann moments'' for the unpolarized as well as polarized
structure functions.
A short review of 
Nachtmann's method in deep inelastic lepton-hadron scattering was 
given in Ref.~\cite{GRW79}. All these investigations 
are based on the notion of geometric twist.
A different  method is due to Georgi and 
Politzer~\cite{GP} for the target mass corrections in unpolarized
deep inelastic scattering. This method 
has been applied also to the study of target mass
corrections of polarized structure functions~\cite{PR98,BT99}.
In non-forward processes, the structure of mass corrections is more
complicated than for deep-inelastic lepton-hadron 
scattering~\cite{ball99b,BM01}.

Here, we generalize Nachtmann's procedure for the case of non-forward
matrix elements applying it to off-cone quark-antiquark operators.
Additionally, I extend the analysis of $\rho$-meson as well as pion DAs to 
include all meson-mass terms of harmonic twist-2 and twist-3 operators. 
Vector meson mass corrections of order $x^2$ 
were discussed by Ball and Braun~\cite{ball99,ball99b}. A resummed meson mass correction
was just given in the case of a scalar theory by Ball~\cite{ball99b} 
(see also Ref.~\cite{BB91}).

We calculate the operators and the matrix elements in the $x$-space.
The corresponding matrix elements in the $q$-space which are needed to
calculate the mass corrections in the Compton amplitude are given
if we carry out the twist decomposition in the $q$-space.


\subsection{Non-forward matrix elements of harmonic operators: The method}
\subsubsection{Parametrization of non-forward matrix elements
by independent double distributions of definite twist}

Now we consider the bilocal off-cone quark-antiquark operators 
i.e.,~operators {\em without} external operations which, generically,
will be denoted by ${\cal O}_{\Gamma}(\kappa x, -\kappa x)$. 

We will define the (off-cone) non-forward quark distributions of 
definite geometric twist 
as straightforward generalizations of the DAs discussed in Section~\ref{forward}
and \ref{meson}
by writing the matrix elements of bilocal 
operators of definite geometric twist.
Before discussing double distributions of geometric twist let us consider 
the most general parametrization of non-forward (off-cone) matrix elements:
\begin{align}
\label{NFME}
\langle P_2,S_2 |{\cal O}_{\Gamma}(\kappa x,-\kappa x)|P_1,S_1 \rangle
= {\cal K}_\Gamma^a({\mathbb P})
\int {\mathrm D}{\mathbb Z}\, \e^{\ii\kappa (x{\mathbb P}){\mathbb Z}}\,
f_a({\mathbb Z}, {\mathbb P}_i {\mathbb P}_j, x^2;\mu^2).
\end{align}
We use the notation
${\mathbb P} = \{P_+, P_-\}$ and ${\mathbb Z} = 
\{z_+,  z_-\}$ with  $P_\pm = P_2\pm P_1$ and
$z_\pm=\hbox{\large$\frac{1}{2}$} (z_2 \pm z_1)$
thereby defining some (2-dimensional) vector space with scalar product
${\mathbb P}{\mathbb Z} \equiv \sum P_i z_i = P_+z_+ + P_-z_-$.
In addition, the integration measure is defined by
${\mathrm D}{\mathbb Z}= dz_1 dz_2 \theta(1-z_1)\theta(z_1+1)
\theta(1-z_2)\theta(z_2+1)$. The support restriction for the double distributions
reads: $-1\leq z_i \leq 1$.
Here, ${\cal K}_\Gamma^a({\mathbb P})$ denote the linear independent spin
structures being defined by the help of the (free) hadron wave functions 
and governed by the $\Gamma-$structure of the corresponding
nonlocal operator ${\cal O}_{\Gamma}$. For example, in the case
of the virtual Compton scattering, there are two independent 
spin structures, the Dirac and the Pauli structure, 
${\cal K}_\mu^1=\bar U(P_2,S_2) \gamma_\mu U(P_1,S_1)$ and 
${\cal K}_\mu^2=\bar U(P_2,S_2) \sigma_{\mu\nu} P^\nu_- U(P_1,S_1)/M$,
respectively.
Here $f_a({\mathbb Z}, {\mathbb P}_i {\mathbb P}_j, x^2;\mu^2)$ are
(renormalized) two-variable distribution amplitudes which have no
definite geometric twist. The distributions are entire analytic functions
with respect to $(x {\mathbb P}_i)$. In addition these distributions 
are, in principle, the (off-cone) generalizations of the distributions
given in Ref.~\cite{mul94,Ji97}.

Finally, let us comment on operators {\em with} external operations, 
i.e.~when the
(axial) vectors or the (skew) tensors are multiplied `externally' by 
$x^\beta$ and/or $\pd^\alpha$. In these cases the `external' vector 
$x^\beta$ or tensor $x^\beta\pd^\alpha$ is assumed {\em not} to be Fourier 
transformed and the external operations have to be applied onto both sides 
of Eq.~(\ref{NFME}). Thereby the tensor structure of these external  
operations matches with the tensor structure of ${\cal K}^a_\Gamma$ which 
by itself are independent of the coordinates. Despite of this the
external operations heavily influence the possible symmetry type of
the local operators and their decomposition into irreducible tensor
representations of the Lorentz group. 

Now, let us take into account that the non-local operators
${\cal O}_{\Gamma}(\kappa x,-\kappa x)$, formally, are given by infinite 
series of operators of growing (geometric) twist $\tau$, Eq.~(\ref{O-nl-t}).
By substituting the decomposition~(\ref{O-nl-t}) into the representation~(\ref{NFME})
we get an analogous decomposition of the 
distributions $f_a({\mathbb Z}, {\mathbb P}_i {\mathbb P}_j, x^2;\mu^2)$:
\begin{align}
{\cal K}_\Gamma^a({\mathbb P})
f_a({\mathbb Z}, {\mathbb P}_i {\mathbb P}_j, x^2;\mu^2)=
\sum_{\tau=\tau_{\text{min}}}^{\infty}
c^{(\tau) \Gamma'}_{\Gamma}\!(x)\,
{\cal K}_{\Gamma'}^a({\mathbb P})
f^{(\tau)}_a({\mathbb Z}, {\mathbb P}_i {\mathbb P}_j, x^2;\mu^2),
\end{align}
where $f^{(\tau)}_a({\mathbb Z}, {\mathbb P}_i {\mathbb P}_j, x^2;\mu^2)$ 
are distributions of geometric twist which still contain kinematical
factors, like $P_-^2$.
On the other side, the non-forward matrix elements  of operators with
geometric twist $\tau$ read:
\begin{align}
\label{equiv1}
\langle P_2,S_2|{\cal O}^{(\tau)}_{\Gamma}(\kappa x,-\kappa x)|P_1,S_1 \rangle
&=
{\cal K}^a_\Gamma({\mathbb P})\int {\mathrm D}{\mathbb Z}\,
\e^{\ii\kappa (x{\mathbb P}){\mathbb Z}} 
f^{(\tau)}_a({\mathbb Z}, {\mathbb P}_i {\mathbb P}_j, x^2;\mu^2)\nonumber\\
&={\cal P}^{(\tau)\Gamma'}_{~~\Gamma}\!(x,\pd_x) \,
{\cal K}_{\Gamma'}^a({\mathbb P})
\int {\mathrm D}{\mathbb Z}\,
\e^{\ii\kappa (x{\mathbb P}){\mathbb Z}} \,
\,f^{(\tau)}_a({\mathbb Z}; \mu^2 ).
\end{align}
In the second line of Eq.~(\ref{equiv1}) we used the explicite twist 
decomposition by the help of orthogonal projectors 
(see Eq.~(\ref{twist-nl}).
In this way one is able to introduce distributions of geometric twist 
$f^{(\tau)}_a({\mathbb Z};\mu^2)$ 
which are independent form kinematical factors.
Thus $f^{(\tau)}_a({\mathbb Z};\mu^2)$ are
already determined by the decomposition of
the non-local matrix elements on the light-cone. Also their renormalization
properties are only determined by the light-cone operators.
These distribution amplitudes $f^{(\tau)}_a({\mathbb Z};\mu^2)$
depend from the distribution parameters $z_1$ and $z_2$, only. 
Therefore, the power corrections of the distribution amplitudes 
$f^{(\tau)}_a({\mathbb Z}, {\mathbb P}_i {\mathbb P}_j, x^2;\mu^2)$ 
are related to the distributions $f^{(\tau)}_{a}({\mathbb Z}; \mu^2 )$
according to:
\begin{align}
f^{(\tau)}_a({\mathbb Z}, {\mathbb P}_i {\mathbb P}_j, x^2;\mu^2)=
{\cal F}^{(\tau) a'}_a\big((x{\mathbb P}{\mathbb Z}), x^2 ({\mathbb P}{\mathbb Z} )^2\big)
f^{(\tau)}_{a'}({\mathbb Z}; \mu^2),
\end{align}
where ${\cal F}^{(\tau) a'}_a\big(((x{\mathbb P}{\mathbb Z}), x^2 ({\mathbb P}{\mathbb Z} )^2\big)$
is uniquely related to the twist decomposition of the off-cone operators and
contains the information about the power corrections of the distributions
amplitudes.

Let us now present the arguments leading to the r.h.s.~of Eq.~(\ref{equiv1}) 
more explicitly. Namely, the non-forward matrix elements after performing the 
twist projection of relations (\ref{O^Gint}) are given as follows:
\begin{align}
\label{equiv}
\langle P_2,S_2|{\cal O}^{(\tau)}_{\Gamma}(\kappa x,-\kappa x)|P_1,S_1 \rangle
&=
\langle P_2,S_2 |{\cal P}^{(\tau)\Gamma'}_{~~\Gamma}(x,\pd_x)\int \d^4p 
\,{\cal O}_{\Gamma'}(p)\,\e^{\ii\kappa xp} | P_1,S_1 \rangle
\nonumber\\
&=
\int \d^4p\,
\Big({\cal P}^{(\tau)\Gamma'}_{~~\Gamma}(x,\pd_x)\,\e^{\ii\kappa xp}\Big)
\langle P_2,S_2 |{\cal O}_{\Gamma'}(p) | P_1,S_1 \rangle
\\
&= 
\int {\mathrm D}{\mathbb Z}
\Big({\cal P}^{(\tau)\Gamma'}_{~~\Gamma}\!(x,\pd_x) \,
\e^{\ii\kappa (x{\mathbb P}){\mathbb Z}}\Big) \,
{\cal K}_{\Gamma'}^a({\mathbb P})\,f^{(\tau)}_a({\mathbb Z}; \mu^2 ).
\nonumber
\end{align}
Here, in the second line we simply reordered the operations of integration,
taking matrix elements and twist projections appropriately. In order to get 
the third line we observe the symmetry in $x$ and $p$ of the local twist 
projections,
${\cal P}^{(\tau)\Gamma'}_{n~\Gamma}(x,\pd_x) (xp)^n \equiv
{\cal P}^{(\tau)\Gamma'}_{n~\Gamma}(p,\pd_p) (xp)^n$, 
from which one derives for the second line:
\begin{align}
\sum_{n=0}^\infty \frac{(\ii\kappa)^n}{n!}
\Big({\cal P}^{(\tau)\Gamma'}_{n~\Gamma}(x,\pd_x)\,
x^{\mu_1} \ldots x^{\mu_n}\Big)
\int \d^4p\,
\Big({\cal P}^{(\tau)\Gamma''}_{n~\Gamma'}(p,\pd_p)\,
p_{\mu_1} \ldots p_{\mu_n}\Big)
\langle P_2,S_2 |{\cal O}_{\Gamma'' n}(p)| P_1,S_1 \rangle.
\nonumber
\end{align}
Here, the integrand contains the matrix elements of the irreducible local 
operators of definite twist ${\cal O}^{(\tau)}_{\Gamma\mu_1\ldots\mu_n}(p)$.
Their reduced matrix elements $f^{(\tau)}_{n_1n_2},~n_1+n_2=n,$ which are 
related to the decomposition of $(P_2+P_1)^n$ into independent monomials 
may be represented by (double) moments of corresponding double distributions 
$f_a^{(\tau)}(z_1,z_2)$ multiplied by the kinematical factors 
${\cal K}_\Gamma^a(P_1,P_2)$. Finally, since $p$ reflects the dependence
of the (local) operators on $\Tensor D$, after performing the integration
and the resummation over $n$, 
$p$ within the integrand simply has to be replaced by ${\mathbb{PZ}}$.

Formally, in the various experimental situations, in any Fourier integrand 
having the structure of Eq.~(\ref{equiv}), we only have to perform the 
following replacements (denoted by $\doteq$):
\smallskip

\noindent
(A) In the case of {\em non-forward scattering}, as just explained, we obtain 
\begin{align}
\label{NLME}
\langle P_2,S_2 |{\cal O}_{\Gamma}(p) | P_1,S_1 \rangle
\doteq
{\cal K}_\Gamma^a({\mathbb P})
\int {\mathrm D}{\mathbb Z}\, \delta^{(4)}(p-{\mathbb P}{\mathbb Z})\,
f^{(\tau)}_a({\mathbb Z};\mu^2). 
\end{align}

\noindent
(B) In the case of {\em forward scattering}, i.e.,~for $P_1 = P_2 = P$, 
the situation changes into  
\begin{align}
\langle P,S |{\cal O}_{\Gamma}(p) | P,S \rangle
&\doteq
{\cal K}_\Gamma^{a}(P)
\int \d z\, \delta^{(4)}(p-2Pz)\,\hat f^{(\tau)}_a(z; \mu^2),
\end{align}
with $P_+ = 2P, P_- = 0; z_+ = z$. These 
distributions $\hat f^{(\tau)}_a(z,\mu^2)$  are obtained 
from the double distributions $\hat f^{(\tau)}_a({\mathbb Z},\mu^2)$
by integrating out the independent variable $z_-$, i.e.,
\begin{align}
\hat f^{(\tau)}_a(z,\mu^2)= \int \d z_- f^{(\tau)}_a(z_+=z, z_-; \mu^2).
\end{align} 

\noindent
(C) In the case of {\em vacuum-to-hadron transition amplitudes}, 
e.g.,~for the meson distribution amplitudes one obtains
\begin{align}
\label{NLM}
\langle 0 |{\cal O}_{\Gamma}(p) | V(P,\lambda) \rangle
&\doteq
\widetilde{\cal K}_\Gamma^{a}(P)
\int \d \xi\, \delta^{(4)}(p-P\xi)\,\tilde f^{(\tau)}_a(\xi;\mu^2),
\end{align}
with $P_2=P; z_2 = \xi$. Obviously, the spin structures are different
from the above cases and, by construction, only $P$ and $\xi$ occur.

From the foregoing discussion it is obvious that the target mass corrections 
to the various physical processes are completely determined by the twist structure 
of the bilocal operators ${\cal O}_{\Gamma}^{(\tau)}(\kappa x, -\kappa x)
 =\int\d^4 p\,{\cal O}_{\Gamma}^{(\tau)}(p)\,\e^{\ii\kappa xp}$. Performing 
matrix elements according to the replacements Eqs.~(\ref{NLME}) -- 
(\ref{NLM}) we obtain the related expressions for 
the $S$-matrix elements of the physical process under consideration.
Then, a further Fourier transformation with respect to the coordinate $x$ 
leads to a representation of the scattering or transition amplitudes in terms 
of the inverse momentum transfer depending, quite generally, on
$P_iP_j/Q^2$.



\subsubsection{The technique of harmonic polynomials}

Now, we use the form of the harmonic extension in terms of 
Gegenbauer polynomials.
This {\em polynomial technique}
uses the vector $x\in {\Bbb R}^4$ as a device for writing tensors with 
special symmetries in analytic form~\cite{BT77,Dobrev77,Dobrev82}. 
It has the advantage to be directly related to the irreducible tensor
representations of the Lorentz group. Its group theoretical background
as far as it is related to totally symmetric tensors has been given 
in Ref.~\cite{BT77}.

The local operator according to (\ref{O^Gloc}) 
is given by 
\begin{align}
\label{o_n+1}
O_{n+1}(x) \equiv \bar{\psi}(0)(x\gamma)(\ii x\Tensor D)^n \psi(0)
=\int \d^4 p \,\big(\bar{\psi}\gamma^\mu\psi\big)(p)\,x_\mu\,(xp)^n,
\end{align}
which shows that, in this connection, $p$ formally replaces the
covariant derivative $\Tensor D$ sandwiched between the quark operators.  
This local operator has a (finite) twist decomposition whose complete
series will be given later on, cf.,~Eq.~(\ref{O_loc_i}).
The twist-2 part is given by the harmonic 
polynomials~\cite{VK,BT77} (see Section~\ref{tensor}),
\begin{align}
\label{O_tw2_sca}
O^{\rm tw2}_{n+1}(x)=
\sum_{k=0}^{[\frac{n+1}{2}]}\frac{(-1)^k (n+1-k)!}{4^k k!(n+1)!}\, x^{2k}
\square^k O_{n+1}(x)
\equiv {\cal P}^{(2)}_{n+1}(x)\,O_{n+1}(x),
\end{align}
being characterized by traceless, totally symmetric tensors of rank $n+1$
whose indices are completely contracted by $x^{\mu_1} \cdots x^{\mu_{n+1}}$.
They obey the condition of harmonicity: $\square O^{\rm tw2}_{n+1}(x) =0$.
 
Performing the derivatives,
\begin{align}
x^{2k}\square^k (\gamma x) (px)^n=
\frac{n!}{(n-2k)!}\, (\gamma x) (p^2 x^2)^k (px)^{n-2k}
+\frac{2k\,n!}{(n+1-2k)!}\,  x^2 (\gamma p) (p^2 x^2)^{k-1} (px)^{n+1-2k},
\end{align}
and using the series expansion of the Gegenbauer polynomials
(see, e.g.,~Ref.~\cite{PBM}, Appendix II.11),
\begin{align}
\label{GB10}
C_n^\nu(z)=\frac{1}{(\nu-1)!}\sum_{k=0}^{[\frac{n}{2}]}
\frac{(-1)^k(n-k+\nu-1)!}{k!(n-2 k)!}\,(2z)^{n-2k},
\end{align}
we obtain 
\begin{align}
\label{O2_n+1}
O^{\text{tw2}}_{n+1}(x)
=\frac{1}{n+1}
\int\!\d^4 p\, \big(\bar{\psi}\gamma^\mu\psi\big)(p)
\bigg\{& x_\mu
\left(\frac{1}{2}\sqrt{p^2 x^2}\right)^{\!n}
\!C_n^2\bigg(\frac{px}{\sqrt{p^2 x^2}}\bigg)\nonumber\\
&-\frac{1}{2}\, p_\mu\,x^2\,  
\left(\frac{1}{2}\sqrt{p^2 x^2}\right)^{\!n-1}
\!C_{n-1}^2\bigg(\frac{px}{\sqrt{p^2 x^2}}\!
\bigg)\bigg\}.
\end{align}

Eq.~(\ref{O2_n+1}) is the analytic continuation from
Euclidean spacetime --- where the Gegenbauer polynomials obey the
well-known orthonormality relations within the region $-1 \leq z \leq 1$
 --- to the Minkowski spacetime where
the arguments of the square roots and of the Gegenbauer polynomials,
depending on whether $p^2$ and/or $x^2$ are space-like or time-like,
may be imaginary and, furthermore, could take values outside 
that region. This, however, has no influence on the validity of
the above result since these polynomials being entire analytic functions
are well defined on the whole complex plane.

Finally, according to the above mentioned interpretation of $p$ as some kind
of `operator symbol' within the Fourier integral, Eq.~(\ref{O2_n+1}) 
operationally has to be understood as if the polynomial inside the 
curly brackets had been written in terms of $\Tensor D$ instead of $p$ and 
inserted into the operator $\bar\psi(0) \gamma^\mu \psi(0)$.

We now use the formula (see, e.g.,~Ref.~\cite{PBM}, Eq.~II.5.13.1.3),
\begin{align}
\sum_{n=0}^\infty\frac{a^n}{(2\nu)_n}\, C^\nu_n(z)=
\Gamma\left(\nu+\frac{1}{2}\right) 
\left(\frac{a}{2}\sqrt{1-z^2}\right)^{1/2-\nu}
J_{\nu-1/2}\left(a\sqrt{1-z^2}\right) \e^{za},
\end{align}
where 
$(2\nu)_n = 2\nu (2\nu +1) \ldots (2\nu+n-1) =
\Gamma(n+2\nu)/\Gamma(2\nu)$ is the Pochhammer symbol. 
Choosing $z=(px)/\sqrt{p^2 x^2}$ and $a=\ii\sqrt{p^2 x^2}/2$,  
Eq.~(\ref{O2_n+1}) can be summed up 
according to Eq.~(\ref{O^Gint}) to the bilocal operator of twist-2:
\begin{align}
\label{XYZ}
O^{\rm tw2}(\kappa x,-\kappa x)&=
\sum_{n=0}^\infty \frac{(\ii\kappa)^n}{n!}\, O^{\rm tw2}_{n+1}(x)\nonumber\\
&=\sqrt{\pi}\int\!\d^4 p\, \big(\bar{\psi}\gamma^\mu\psi\big)(p)
\left\{ x_\mu\left(2+x\pd\right)
-\frac{1}{2}\,\ii\kappa p_\mu x^2 \right\}
\left(3+x\pd\right)
\nonumber\\
&\qquad\qquad\times
\left(\kappa\sqrt{(px)^2-p^2 x^2}\right)^{-3/2}
\!J_{3/2}\left(\frac{\kappa}{2}\sqrt{(px)^2-p^2 x^2}\right)
\e^{\ii\kappa px/2}.
\end{align}
The homogeneous derivations $(c+x\pd)$ are required to compensate 
the factors $(c+n)$ which are necessary in order to by able to
introduce the Pochhammer symbols in the denominator. Obviously, for the 
second term in Eq.~(\ref{O2_n+1}), after shifting $n \rightarrow n+1$ 
in the series over $n$, only one additional factor is required. ---
Also here $p$ has to be considered as a symbol replacing
the covariant derivatives 
sandwiched between the quark operators.

In order to show the main features of this formalism, we consider the 
forward as well as non-forward matrix elements.
Let us now discuss the forward matrix elements of the twist-2 operators
(\ref{O2_n+1}) and (\ref{XYZ}).
The forward matrix elements are given by the
reduced matrix element $f^{(2)}_{n}$ times the irreducible tensor
written in terms of the momentum $P_+=2P$ and
using $2P\equiv x_\mu\bar U(P)\gamma^\mu U(P)$. 
The final result can be written as
\begin{align}
\label{lo_2}
&\langle P|O^{\text{tw2}}_{n+1}(x)|P\rangle
=f^{(2)}_n\, {\cal P}^{(2)}_{n+1}(x)\, 
\Big\{x_\mu\bar U(P)\gamma^\mu U(P) (2xP)^{n}\Big\}\\
&\qquad=
\frac{f^{(2)}_{n}}{n+1}\,
(2Px)\,\bigg\{\!\!
\left(\sqrt{P^2 x^2}\right)^{\!n}
\!C_n^2\bigg(\frac{Px}{\sqrt{P^2 x^2}}\bigg)
-\frac{P^2\,x^2}{(xP)}\,  
\left(\sqrt{P^2 x^2}\right)^{\!n-1}
\!C_{n-1}^2\bigg(\frac{Px}{\sqrt{P^2 x^2}}\!
\bigg)\!
\bigg\}.\nonumber
\end{align}
The reduced matrix elements $f^{(2)}_n$ are given
as the moments of a distribution $f^{(2)}(z)$ by means of a 
Mellin transformation,
\begin{align}
f^{(2)}_n = \int_{-1}^{1} \d z z^n f^{(2)}(z).
\end{align}
We obtain, after summing up over $n$, the
following expression of the forward matrix elements of the non-local
twist-2 operators 
\begin{align}
\label{xyz}
\langle P|O^{\rm tw2}&(\kappa x,-\kappa x)|P\rangle
=2\sqrt{\pi}(Px)
\int^1_{-1}\d z\,f^{(2)}(z)
\left\{\left(2+x\pd\right)
-\ii\kappa z \frac{P^2 x^2}{(xP)} \right\}
\left(3+x\pd\right)
\nonumber\\
&\times
\left(2\kappa z\sqrt{(Px)^2-P^2 x^2}\right)^{-3/2}
\!J_{3/2}\left({\kappa} z\sqrt{(Px)^2-P^2 x^2}\right)
\e^{\ii\kappa z Px}.
\end{align}
If this expression is restricted to the light-cone, $x^2=0$, the
well-known light-cone parton distribution is obtained. 
Namely, using the Poisson integral for the Bessel functions (cf.,~Ref.~\cite{BE}, Eq.~II.7.12.7),
\begin{align}
\label{Poisson}
\Gamma\left(\nu + \frac{1}{2}\right) J_\nu(z) 
= \frac{1}{\sqrt\pi} \Big(\frac{z}{2}\Big)^\nu
\int^1_{-1} \d t\,(1-t^2)^{\nu-1/2}\, \e^{\ii tz}
\qquad {\rm for} \qquad
{\rm Re~} \nu > - \frac{1}{2},
\end{align}
we obtain ($\xi = \kappa z (\lcx P)$)
\begin{align}
\langle P |O^{\rm tw2}(\kappa \lcx, -\kappa \lcx) | P \rangle
&=
2 (\lcx P) \sqrt{\pi} 
\int \d z\, f^{(2)}(z)\,(2+\xi\pd_\xi)(3+\xi\pd_\xi)
\big(2\xi \big)^{-3/2}
J_{3/2}(\xi)\,\e^{\ii\xi}
\nonumber\\
&=
2(\lcx P) \int \d z\, f^{(2)}(z)\,
\int^1_0 \d t\, t (1-t) (3+t\pd_t)(2+t\pd_t)\,\e^{2 \ii t \xi}
\nonumber\\
\label{lcNLO2}
&=
2(\lcx P) \int \d z\, f^{(2)}(z)\,
\int^1_0 \d t\, t (1+t\pd_t)\,\e^{2 \ii t \xi}\nonumber\\
&=
2 (\lcx P) \int \d z\,  f^{(2)}(z)\,\e^{ 2 \ii \kappa z (\lcx P)};
\end{align}
in the second line we introduced the Poisson integral after shifting the 
integration variable $t \rightarrow (t+1)/2$ and followed by changing
$\xi \pd_\xi  \rightarrow t\pd_t$, then we partially integrated two times
retaining finally only the surface term at $t=1$. As a result we arrived 
at the twist-2 parton distribution as introduced in  \cite{GL01},
$f^{(2)}(z)\equiv F^{(2)}(z)$, which coincides with
$f_1(z)$ in the notation adopted by Jaffe and Ji \cite{JJ91}.

More generally, when non-forward matrix elements are taken the argumentation
is almost the same. But now, in the case of the local operators, there occur
$n+1$ different reduced matrix elements, $f^{(2)}_{n m}$, related to the 
monomials $P_1^m \, P_2^{n-m},~0 \leq m \leq n$ which are obtained from 
$(P_2+P_1)^n$, and the corresponding non-forward matrix element reads: 
\begin{align}
\label{nflo_2}
\langle P_2|O^{\text{tw2}}_{n+1}(x)|P_1\rangle
&=
\sum_{m=0}^n \binom{n}{m}
f^{(2)}_{n m}
\big({\bar U}(P_2)\gamma_\mu U(P_1)\big)\,
{\cal P}^{(2)}_{n+1}(x^2,\pd^2) 
\Big\{x^\mu (xP_1)^m (xP_2)^{n-m}\Big\}.
\end{align}
Now let us rewrite the reduced matrix elements $f^{(2)}_{n m}$ as 
the double moments of a double distribution $f^{(2)}(z_1,z_2)$,
\begin{align}
f^{(2)}_{n m} = \int_{-1}^1 \d z_1 \int_{-1}^1 \d z_2\, 
z_1^m z_2^{n-m} f^{(2)}(z_1,z_2).
\end{align}
After resumming w.r.t.~$n$ and rewriting $f^{(2)}(z_1,z_2) =
F^{(2)}_D(z_+,z_-)$ we finally arrive at the expression for the
non-forward matrix element of the nonlocal operator 
$O^{\text {tw2}}(\kappa x, -\kappa x)$ as
\begin{align}
&\langle P_2 |O^{\rm tw2}(\kappa x, -\kappa x) | P_1 \rangle
=
\sqrt{\pi}\,
\bar U(P_2) \gamma^\mu U(P_1)
\int {\mathrm D}{\mathbb Z}\, F_D^{(2)}({\mathbb Z})\,
\Big\{x_\mu\, (2+x\pd)
- \frac{1}{2}
\ii\kappa \,{\mathbb P}_\mu{\mathbb Z}\, x^2\Big\}(3+x\pd)
\nonumber\\
\label{NLO2}
&\qquad\qquad\times
\Big(\kappa 
\sqrt{(x{\mathbb{PZ}})^2- x^2({\mathbb{PZ}})^2}
\Big)^{-3/2}
J_{3/2}\Big(\frac{\kappa}{2}
\sqrt{(x{\mathbb{PZ}})^2- x^2({\mathbb{PZ}})^2}\Big)
\,\e^{\ii\kappa x{\mathbb{PZ}}/2} .
\end{align}
This holds in the case $(x{\mathbb{PZ}})^2-x^2 {\mathbb{PZ}}^2 \geq 0$, 
otherwise we have to change into $I_\nu(z)$.

From this approach it becomes also
obvious that the mass corrections of different physical processes
are related to each other 
if they can be traced back to the same operator content.
In the following we use this polynomial technique to determine the
local as well as nonlocal quark operators of definite twist. Thereby,
we generalize Nachtmann's approach not only to non-forward matrix
elements of (non)local operators.
In addition we consider also the case of more general
tensor operators having nontrivial symmetry types.

\subsection{Power Corrections of non-forward matrix elements}
In this Section I will give the resummed matrix elements of kinematical mass
corrections in the $x$-space 
for the virtual Compton scattering and for exclusive processes
with vector as well as (pseudo) scalar mesons by means of harmonic operators
considered in the preceding Sections. 

\subsubsection{Double distributions from quark--antiquark operators
with definite twist}
In order to obtain the non-forward matrix elements for the 
quark--antiquark operators
with definite twist $\tau$ we have only to take matrix elements of
the expressions given in the last Section and to use 
formula (\ref{NLME}). 
The spin-dependence of the hadronic states $|P_i, S_i\rangle, i = 1,2$,
is completely contained in the independent kinematical structures
${\cal K}^a_\Gamma({\mathbb P})$ being relevant for the processes
under consideration, e.g., the Dirac and the Pauli structure 
$\bar U(P_2, S_2) \gamma_\mu U(P_1,S_1)$ and 
$\bar U(P_2, S_2) \sigma_{\mu\nu} P_-^\nu U(P_1,S_1)/M$, respectively,
in case of the virtual Compton scattering.

Let me now give the final expressions for matrix elements of
nonlocal as well as local operators up to twist-3
which are relevant for the power corrections of the various
processes. 
For the non-forward matrix element of the nonlocal twist-2
vector operator, Eqs.~(\ref{nl_O2}) with (\ref{H-n})
obeying the replacement (\ref{NLME}), one gets
\begin{align}
&\langle P_2, S_2 |O^{\rm tw2}_\alpha (\kappa x,-\kappa x) |P_1,S_1\rangle
= {\cal K}^{a}_{\mu}({\mathbb P})
\int{\mathrm D}{\mathbb Z}\, F^{(2)}_a({\mathbb Z})\,
\pd_\alpha \int_0^1\!\!\d t
\left\{ x^\mu \left(2+x\pd\right)
-\frac{1}{2}\,\ii\, \kappa t\,
({\mathbb P}^\mu{\mathbb Z})\, x^2\right\}\nonumber\\
&\hspace{10cm}\times\left(3+x\pd\right)\, {\cal H}_2({\mathbb{PZ}}|\kappa tx)
\nonumber\\
&\qquad={\cal K}^{a}_\mu({\mathbb P})
\int{\mathrm D}{\mathbb Z}\, F^{(2)}_a({\mathbb Z})
\int_0^1\!\!\d t\left(2+x\pd\right)
\bigg\{
\Big[
\left(3+x\pd\right)\delta^\mu_{\alpha}
-\ii\,\kappa t\,({\mathbb P}^\mu{\mathbb Z})\,  x_\alpha\Big]
{\cal H}_2({\mathbb{PZ}}|\kappa tx)
\nonumber\\
&\qquad\qquad\qquad
+\Big[
\left(3+x\pd\right)
\left(
\left(4+x\pd\right)\ii\,\kappa t\, ({\mathbb P}_\alpha{\mathbb Z}) x^\mu 
-\frac{1}{2}
(\ii\kappa t)^2 \big(({\mathbb{PZ}})^2 x^\mu x_\alpha
+ ({\mathbb P}^\mu{\mathbb Z}) ({\mathbb P}_\alpha{\mathbb Z})\, x^2 \big) 
\right)
\nonumber\\
&\quad\qquad\qquad\qquad\quad\qquad\qquad
+\frac{1}{4}
(\ii \kappa t)^3\, ({\mathbb P}^\mu{\mathbb Z}) 
({\mathbb{PZ}})^2  x_{\alpha}x^2 \Big]
{\cal H}_3({\mathbb{PZ}}|\kappa tx)
\bigg\}.
\end{align}
The corresponding local matrix element of the twist-2 operator, Eq.~(\ref{XO2}), with
(\ref{h-n}) reads
\begin{align}
\label{O2_n(x)}
&\langle P_2, S_2 |O^{\rm tw2}_{\alpha n}(x)|P_1,S_1\rangle
=\frac{1}{(n+1)^2}\,
{\cal K}^{a}_{\mu}({\mathbb P})
\int{\mathrm D}{\mathbb Z}\, F^{(2)}_a({\mathbb Z})\,
\bigg\{ 
\delta_\alpha^\mu \,h_n^2({\mathbb{PZ}}|x)
- ({\mathbb P}^\mu{\mathbb Z}) x_\alpha \,h_{n-1}^2({\mathbb{PZ}}|x)\nonumber \\
&\hspace{2cm}
+2 x^\mu ({\mathbb P}_\alpha{\mathbb Z}) \,h_{n-1}^3({\mathbb{PZ}}|x)
-\big(x^\mu x_\alpha ({\mathbb{PZ}})^2
+({\mathbb P}^\mu{\mathbb Z}) ({\mathbb P}_\alpha{\mathbb Z}) x^2\big)\,h_{n-2}^3({\mathbb{PZ}}|x)
\nonumber\\
&\hspace{6cm}+\frac{1}{2}\, ({\mathbb P}^\mu {\mathbb Z}) x_\alpha\, 
x^2 ({\mathbb{PZ}})^2 \,h_{n-3}^3({\mathbb{PZ}}|x)
\bigg\}.
\end{align}

For the matrix element of the nonlocal operator~(\ref{nl_O3}) and local
one~(\ref{XO3}), respectively, we obtain
\begin{align}
\label{}
&\langle P_2, S_2 |O^{\rm tw3}_\alpha (\kappa x,-\kappa x)|P_1,S_1\rangle
=2{\cal K}^{a}_{\mu}({\mathbb P})
\int{\mathrm D}{\mathbb Z}\, F^{(3)}_a({\mathbb Z})\,
\int_0^1\!\!\d t\,
\,x^\beta\Big\{
\delta^\mu_{[\alpha}\pd_{\beta]}\,(1+x\pd)
-x_{[\alpha}\pd_{\beta]}\pd^\mu\Big\}\nonumber\\
&\hspace{12cm}
\times{\cal H}_1({\mathbb{PZ}}|\kappa tx)
\nonumber\\
&=
{\cal K}^{a}_{\mu}({\mathbb P})
\int{\mathrm D}{\mathbb Z}\, F^{(3)}_a({\mathbb Z})\,
\left(2+x\pd\right) \int_0^1\!\!\d t\, x^\beta
\bigg\{\Big[
2\,\ii\kappa t\,\delta^\mu_{[\alpha}
({\mathbb P}_{\beta]}{\mathbb Z})(2+x\pd) 
- (\ii\kappa t)^2 \delta^\mu_{[\alpha}x_{\beta]} ({\mathbb{PZ}})^2 \Big]\nonumber\\
&\hspace{12cm}
\times{\cal H}_2({\mathbb{PZ}}|\kappa tx)
\nonumber\\
&\qquad\qquad\qquad\qquad\
-\ii\kappa t\, x_{[\alpha}({\mathbb P}_{\beta]}{\mathbb Z}) 
\Big[2 \ii\kappa t\,({\mathbb P}^\mu{\mathbb Z}) \left(5+x\pd\right)
- (\ii\kappa t)^2 x^\mu ({\mathbb{PZ}})^2 
\Big]\,{\cal H}_3({\mathbb{PZ}}|\kappa tx)
\bigg\},
\end{align}
\begin{align}
\label{O3_n(x)}
&\langle P_2, S_2 |O^{\rm tw3}_{\alpha n} (x)|P_1,S_1\rangle
=\frac{2}{(n+1)^2}\,
{\cal K}^{a}_{\mu}({\mathbb P})
\int{\mathrm D}{\mathbb Z}\, F^{(3)}_a({\mathbb Z})\,
x^\beta\bigg\{ 2(n+1)
\delta_{[\alpha}^\mu ({\mathbb P}_{\beta]}{\mathbb Z})  \,h_{n-1}^2({\mathbb{PZ}}|x)
\nonumber\\
&\hspace{3cm}
+(n+2)\, x_{[\alpha}\delta_{\beta]}^\mu ({\mathbb{PZ}})^2 \,h_{n-2}^2({\mathbb{PZ}}|x)   
-4 x_{[\alpha}({\mathbb P}_{\beta]}{\mathbb Z}) ({\mathbb P}^\mu {\mathbb Z})\,h_{n-2}^3({\mathbb{PZ}}|x) 
\nonumber\\
&\hspace{6cm}
+2 x_{[\alpha} ({\mathbb P}_{\beta]}{\mathbb Z}) x^\mu ({\mathbb{PZ}})^2 \,h_{n-3}^3({\mathbb{PZ}}|x)
\bigg\}.
\end{align}

In the case of the axial vector operators 
$O^{\tau}_{5\alpha}(\kappa x,-\kappa x)$ the kinematical structures
${\cal K}_{5\mu}({\mathbb P})$ contain an additional $\gamma_5$ and the 
double distributions should be denoted by $G^{(\tau)}_{a}(\mathbb Z)$
but the formal structure will be the same.

The non-forward matrix elements of the 
nonlocal twist-3 chiral-odd scalar operator, Eq.~(\ref{nl_N3}), reads
\begin{align}
\label{}
\langle P_2 |N^{\rm tw3}(\kappa x,-\kappa x) |P_1\rangle
={\cal K}^{a}({\mathbb P})
\int{\mathrm D}{\mathbb Z}\, E^{(3)}_a({\mathbb Z})\,
\left(1+x\pd\right)\,{\cal H}_1({\mathbb{PZ}}|\kappa x),
\end{align}
and of the local one~(\ref{XN3})
\begin{align}
\label{N3_n(x)}
\langle P_2 |N^{\rm tw3}_n(x) |P_1\rangle
={\cal K}^{a}({\mathbb P})
\int{\mathrm D}{\mathbb Z}\, E^{(3)}_a({\mathbb Z})\,h^1_n({\mathbb{PZ}}| x).
\end{align}

The non-forward matrix element of the
nonlocal twist-2 chiral-odd skew tensor operator, Eq.~(\ref{nl_M2}), 
is given by
\begin{align}
\label{}
\langle P_2, S_2 |
&M^{\rm tw2}_{[\alpha\beta]}(\kappa x,-\kappa x)|P_1,S_1\rangle
=2{\cal K}^{a}_{[\mu\nu]}({\mathbb P})
\int{\mathrm D}{\mathbb Z}\, H^{(2)}_a({\mathbb Z})\,
\int_0^1\!\!\d t\,t 
\nonumber\\
&\qquad\times\Big\{(2+x\pd) \delta^\mu_{[\alpha}\pd_{\beta]}
-x_{[\alpha}\pd_{\beta]}\pd^\mu \Big\}
\Big[  x^\nu \left(3+x\pd\right)
-\frac{1}{2}\,\ii\kappa t\, 
({\mathbb P}^\nu{\mathbb Z}) x^2 \Big]
{\cal H}_2({\mathbb{PZ}}|\kappa tx),
\end{align}
and of the corresponding local operator~(\ref{XM2}) reads
\begin{align}
\label{}
\langle P_2, S_2 |
&M^{\rm tw2}_{[\alpha\beta]n}(x)|P_1,S_1\rangle
=\frac{2}{(n+2)(n+1)}\,
\bigg\{\delta^\mu_{[\alpha}\pd_{\beta]}
-\frac{1}{n+2}\,x_{[\alpha}\pd_{\beta]}\pd^\mu \bigg\}
{\cal K}^{a}_{[\mu\nu]}({\mathbb P})
\int{\mathrm D}{\mathbb Z}\, H^{(2)}_a({\mathbb Z})\,
\nonumber\\
&\hspace{4cm}
\times\Big\{x^\nu h^2_{n}({\mathbb{PZ}}|x)
-\frac{1}{2}\, ({\mathbb P}^\nu{\mathbb Z}) x^2 
h^2_{n-1}({\mathbb{PZ}}|x)\Big\}.
\end{align}
The related nonlocal twist-3 operator, Eq.~(\ref{nl_M3}), 
and local one~(\ref{XM3}), respectively, reads
\begin{align}
\label{}
\langle P_2, S_2 |M^{\rm tw3}_{[\alpha\beta]}(\kappa x,-\kappa x)|P_1,S_1\rangle
=2 {\cal K}^{a}_{[\mu\nu]}({\mathbb P})
&\int{\mathrm D}{\mathbb Z}\, H^{(3)}_a({\mathbb Z})\,
x_{[\alpha}\pd_{\beta]}\,
 \ii\kappa \, ({\mathbb P}^\mu{\mathbb Z}) x^\nu (2+x\pd)\nonumber\\
&\qquad\times\int_0^1\!\!\d t\,{\cal H}_2({\mathbb{PZ}}|\kappa tx),
\end{align}
\begin{align}
\label{}
\langle P_2, S_2 |M^{\rm tw3}_{[\alpha\beta]n}(x)|P_1,S_1\rangle
=\frac{2}{(n+2)n}\,x_{[\alpha}\pd_{\beta]}\,
{\cal K}^{a}_{[\mu\nu]}({\mathbb P})
&\int{\mathrm D}{\mathbb Z}\, H^{(3)}_a({\mathbb Z})\,
({\mathbb P}^\mu{\mathbb Z}) x^\nu 
h^2_{n-1}({\mathbb{PZ}}|x).
\end{align}

Additionally, the non-forward matrix element of the
independent nonlocal twist-3 skew tensor, Eq.~(\ref{nl_tM3}), and
local operator~(\ref{XtM3}), respectively,
reads
\begin{align}
\label{}
\langle P_2, S_2 |
\widetilde M^{\rm tw3}_{[\alpha\beta]}(\kappa x,-\kappa x)
|P_1,S_1\rangle
&=3 \widetilde{\cal K}^{a}_{[\mu\nu]}({\mathbb P})\!
\int\!{\mathrm D}{\mathbb Z}\, \widetilde H^{(3)}_a({\mathbb Z})\,
x^\gamma 
\Big\{\delta^\mu_{[\alpha}\delta^\nu_{\beta}\pd_{\gamma]} x\pd
-  2\, \delta^\mu_{[\alpha}x_{\beta}\pd_{\gamma]}\pd^{\nu}\Big\}
\nonumber\\
&\qquad
\times(1+x\pd)\int_0^1\!\!\d t\, \frac{1\!-\!t^2}{2t}\,
{\cal H}_1({\mathbb{PZ}}|\kappa tx),
\end{align}
\begin{align}
\label{}
\langle P_2, S_2 |
&\widetilde M^{\rm tw3}_{[\alpha\beta]n}(x)|P_1,S_1\rangle=
\frac{3x^\gamma }{n+2}\,
\bigg\{\delta^\mu_{[\alpha}\delta^\nu_{\beta}\pd_{\gamma]} x\pd
- \frac{2}{n}\, \delta^\mu_{[\alpha}x_{\beta}\pd_{\gamma]}\pd^{\nu}\bigg\}
\widetilde{\cal K}^{a}_{[\mu\nu]}({\mathbb P})\!
\int\!{\mathrm D}{\mathbb Z}\, \widetilde H^{(3)}_a({\mathbb Z})\,
h^1_n({\mathbb{PZ}}|x).
\end{align}

Now, some remarks are in order. First, as indicated, different
tensorial structures of the operators lead to different kinematical 
structures. In the
forward case they simplify or eventually disappear, e.g., the Dirac
structures are to be replaced by $P_\mu$ and $S_\mu$ for the vector
and axial vector case, respectively, whereas the Pauli structures
vanish. 
In the forward case we obtain from
Eqs.~(\ref{O2_n(x)}), (\ref{O3_n(x)}), and (\ref{N3_n(x)})
and the replacement $x\rightarrow q$ the matrix elements of the local
operators in the $q$-space
$\langle PS|O^{\rm tw2}_{\alpha n}(q)|PS\rangle$,
$\langle PS|O^{\rm tw3}_{\alpha n}(q)|PS\rangle$, and 
$\langle PS|N^{\rm tw3}_{n}(q)|PS\rangle$
which are in agreement with Wandzura's forward matrix elements in the 
$q$-space (see Eqs.~(C.6), (C.9), and (C.1) in Ref.~\cite{Wandzura77}).

Second, going on-cone the double distributions remain the same.
Therefore, their evolution is determined by the anomalous dimensions
resulting from the renormalization group equation of the corresponding 
light-cone operators of geometric twist. 

Furthermore, we have to mention that -- contrary to the case when 
the non-forward matrix elements are restricted to the light-cone where
the decomposition of the non-local quark-antiquark operators into 
operators of definite twist terminates at finite values $\tau_{\rm max}$ --
the off-cone decomposition results in an infinite series of any twist.


\subsubsection{Mass corrections of vector meson distributions}
In this Subsection I extend the analysis of the $\rho$-meson DAs
to include all meson momentum resp. mass terms of harmonic
twist-2 and twist-3 off-cone operators in order to get the meson mass corrections 
to the results obtained in Section~\ref{meson}. Vector meson mass 
corrections of order $x^2$  were already discussed by Ball and Braun~\cite{ball99} and resummed 
meson mass corrections in the scalar case was given by Ball~\cite{ball99b} 
(see also Ref.~\cite{BB91}).

The structure of the mass corrections of $\rho$-meson DAs 
of geometric twist are, from the group theoretical point of view, 
similar to the target mass corrections in deep inelastic scattering,
which can be resummed using Nachtmann's method~\cite{Nachtmann73}.
The main tool, in order to obtain the mass corrections, 
are the harmonic operators of definite geometric twist which have been
determined in Section~\ref{meson}. Obviously, they are the harmonic extensions 
of the corresponding light-cone operators which I have already used in 
Section~\ref{meson} for the classification of the
corresponding meson light cone DAs with respect to geometric twist.
 
Let me now introduce the distribution functions for the harmonic operators 
of geometric twist sandwiched between the vacuum and the meson state,
$\langle 0|O^{(\tau)}_{\Gamma}(x,-x)|\rho(P,\lambda)\rangle $. Thereby,
I adopt again the definitions of Chernyak and Zhitnitsky \cite{chern84} in the
terminology of Ref.~\cite{ball98}.
As usual, these matrix elements are related to 
the momentum $P_\alpha$ and polarization vector $e^{(\lambda)}_\alpha$ 
of the meson with helicity $\lambda,\,P^2=m_\rho^2$, 
$(e^{(\lambda)} e^{(\lambda)})=-1$, 
$(P e^{(\lambda)})=0$. 

First, we consider the chiral-even {\em  vector operator}. Using the 
twist projections (\ref{OPROJ-x}) we introduce the moments of the meson DAs,
$\Phi^{(\tau)}_n$, of twist $\tau$ as reduced matrix elements according to
\begin{align}
\label{}
\langle 0|O^{(\tau)}_{\alpha n}(x)|\rho(P,\lambda)\rangle 
= f_\rho m_\rho\,{\cal P}^{(\tau)\beta}_{\alpha n}
\left(e^{(\lambda)}_\beta (Px)^n \Phi^{(\tau)}_n\right),
\end{align} 
where $f_\rho$ is again the vector meson decay constant.
The corresponding meson distribution amplitudes $\hat\Phi^{(\tau)}(\xi)$ 
are given by inverting the moment integral,
\begin{align}
\label{WF}
\Phi^{(\tau)}_n=\int_{-1}^1\d \xi\, \xi^n \hat\Phi^{(\tau)}(\xi),\quad
\xi=u-(1-u)=2u-1.
\end{align} 
These distribution amplitudes are dimensionsless functions of $\xi$ and 
describe the probability
amplitudes to find the $\rho$-meson in a state with minimal number of 
constituents (quark and antiquark) which carry the momentum fractions 
$u$ (quark) and $(1-u)$ (antiquark).

Taking into account expression~(\ref{XO2}) we obtain the local  
twist-2 matrix element in the $x$-space as follows:
\begin{align}
\label{lmatrix_O_tw2}
&\langle 0|O^{\rm tw2}_{\alpha n}(x)|\rho(P,\lambda)\rangle=\\
&=\frac{1}{(n+1)^2}\,f_\rho m_\rho\,\Phi_n^{(2)}
\Big\{
e^{(\lambda)}_\alpha h^2_n(P|x)
+2P_\alpha(e^{(\lambda)}x) h^3_{n-1}(P|x)
-m_\rho^2 x_\alpha (e^{(\lambda)}x)h^3_{n-2}(P|x)
\Big\},\nonumber
\end{align}
which is analogous to Wandzura's expression, 
$\langle PS|O^{\rm tw2}_{5\alpha n}(q)|PS\rangle$,
(Eq.~(C.6) in~\cite{Wandzura77}) 
for the target mass corrections in deep inelastic scattering in the 
$q$-space. After resummation  
we get the bilocal matrix element of geometric twist-2:
\begin{align}
\label{nmatrix_O_tw2}
&\langle 0|O^{\rm tw2}_{\alpha}(x,-x)|\rho(P,\lambda)\rangle 
=f_\rho m_\rho\int_0^1\d t\int_{-1}^1\d \xi\, \hat\Phi^{(2)}(\xi)
\left(3+\xi\pd_\xi\right)\left(2+\xi\pd_\xi\right)\\
&\qquad\qquad\times
\bigg\{e^{(\lambda)}_\alpha
{\cal H}_2(P\xi| t x)
+\bigg(\left(4+\xi\pd_\xi\right)(\ii\xi t) P_\alpha
-\frac{1}{2}(\ii\xi t)^2 m_\rho^2 x_\alpha\bigg) 
(e^{(\lambda)}x){\cal H}_3(P\xi| t x)\bigg\}.\nonumber
\end{align}
The trace subtraction of the twist-2 operator give rise to the so-called
``kinematical'' target-mass corrections. For the forward-scattering case, 
contributions of this type have been considered by Nachtmann. 
For the vector and tensor operators, we obtain mass corrections 
which are proportional to $ x_\alpha$~\footnote{This kind
of mass corrections survives on the light-cone (see trace terms in
Section~\ref{meson}).} and terms proportional to $x^2$ which are resummed to
Gegenbauer polynomials (local operators) or Bessel functions
(nonlocal operators).

An analogous calculation using expression (\ref{XO3})
gives the local twist-3 matrix element
\begin{align}
\label{lmatrix_O_tw3}
\langle 0|O^{\rm tw3}_{\alpha n}(x)|\rho(P,\lambda)\rangle
=\frac{2}{(n+1)}\,f_\rho m_\rho\,\Phi_n^{(3)} x^\beta
&\Big\{
e^{(\lambda)}_{[\alpha} P_{\beta]} h^2_{n-1}(P|x)
+\frac{n+2}{2(n+1)}\,m_\rho^2 x_{[\alpha}e_{\beta]}^{(\lambda)} h^2_{n-2}(P|x)
\nonumber\\
&
+\frac{1}{n+1}\,m_\rho^2 x_{[\alpha} P_{\beta]} (e^{(\lambda)}x) h^3_{n-3}(P|x)
\Big\},
\end{align}
which is close to Wandzura's expression, 
$\langle PS|O^{\rm tw3}_{5\alpha n}(q)|PS\rangle$,
(Eq.~(C.9) in~\cite{Wandzura77}).
Additionally the resummed bilocal matrix element of twist-3 reads
\begin{align}
\label{nmatrix_O_tw3}
&\langle 0|O^{\rm tw3}_{\alpha}(x,-x)|\rho(P,\lambda)\rangle 
=f_\rho m_\rho\int_{-1}^1\d\xi\, \hat\Phi^{(3)}(\xi)
\left(2+\xi\pd_\xi\right) x^\beta
\bigg\{2(\ii\xi) e^{(\lambda)}_{[\alpha}P_{\beta]}{\cal H}_2(P\xi| x)
\nonumber\\
&\qquad\qquad
+m_\rho^2\int_0^1\d t\Big(
(\ii\xi t)^2  x_{[\alpha} e^{(\lambda)}_{\beta]}{\cal H}_2(P\xi| t x)
+(\ii\xi t)^3 x_{[\alpha}P_{\beta]} (e^{(\lambda)}x){\cal H}_3(P\xi| t x)
\Big)\bigg\}.
\end{align}
It is well-known that the twist-3 operator $O^{\rm tw3}_{\alpha}(x,-x)$ 
is related to other twist-3 operators containing total derivatives and 
operators of Shuryak-Vainshtein type by means of QCD equations of 
motion~\cite{BB88,ball99}. Therefore, the matrix 
elements~(\ref{lmatrix_O_tw3}) and (\ref{nmatrix_O_tw3}) include the 
contribution of the twist-3 operator containing total derivatives 
which is as large as those from twist-2 (see also~\cite{BL01}) and
also give rise to so-called ``dynamic'' mass corrections. 
Dynamical mass-corrections have so far only been considered for the 
exclusive case, Ref.~\cite{ball99}.
In addition the next higher twist contributions of the chiral-even vector 
operator are of twist-4 which also contain dynamical mass corrections. 
They act in the same direction as the mass corrections of twist-2 and twist-3
operators ($\sim x_\alpha$ and $\sim x^2$).

Now we consider the chiral-even {\em axial vector operator }
$O_{5\alpha n}(x)$ 
with the same twist projectors (\ref{OPROJ-x}) as for the chiral-even vector operator 
$O_{\alpha n}(x)$.
We define the corresponding moments of the
meson DAs $\Xi^{(\tau)}_n$ by
\begin{align}
\label{}
\langle 0|O^{(\tau)}_{5\alpha n}(x)|\rho(P,\lambda)\rangle 
= \frac{1}{2}\Big(f_\rho -f_\rho^{\rm T}\frac{m_u+m_d}{m_\rho}\Big)m_\rho\,
{\cal P}^{(\tau)\beta}_{\alpha n}
\left(\epsilon_\beta^{\ \,\gamma\mu\nu}
e^{(\lambda)}_\gamma P_\mu x_\nu (Px)^n \Xi^{(\tau)}_n\right),
\end{align} 
where $f^{\rm T}_\rho$ denotes the tensor decay constant. First, 
we observe that the twist-2 contribution vanishes. The nontrivial local 
vacuum-to-meson matrix elements of this  axial vector operator 
are of twist-3:
\begin{align}
\label{lmatrix_O5_tw3}
\langle 0|O^{\rm tw3}_{5\alpha n}(x)|\rho(P,\lambda)\rangle 
&=\frac{1}{2}\Big(f_\rho-f_\rho^{\rm T}\frac{m_u+m_d}{m_\rho}\Big)m_\rho\,
\epsilon_\alpha^{\ \,\beta\mu\nu}
e^{(\lambda)}_\beta P_\mu x_\nu\,\Xi^{(3)}_n h^1_n(P|x),
\end{align} 
and the bilocal matrix element of twist-3 reads
\begin{align}
\label{nmatrix_O5_tw3}
\langle 0|O^{\rm tw3}_{5\alpha}(x,-x)|\rho(P,\lambda)\rangle 
&=\frac{1}{2}
\Big(f_\rho-f_\rho^{\rm T}\frac{m_u+m_d}{m_\rho}\Big)m_\rho\,
\epsilon_\alpha^{\ \,\beta\mu\nu}
e^{(\lambda)}_\beta P_\mu x_\nu
\int_{-1}^1\d\xi\, \hat\Xi^{(3)}(\xi)
\left(1+\xi\pd_\xi\right)\nonumber\\
&\hspace{8cm}
\times{\cal H}_1(P\xi| x).
\end{align}
By the way the twist-4 matrix element also vanishes and the next higher 
twist contribution would be of twist-5. Also here, the matrix 
elements~(\ref{lmatrix_O5_tw3}) and (\ref{nmatrix_O5_tw3}) include the 
contribution of the twist-3 operator containing total derivatives.

The matrix element of the chiral-odd {\em scalar operator} (see Eq.~(\ref{NPROJ-x})) 
is defined as
\begin{align}
\label{}
\langle 0|N^{\rm tw3}_{n}(x)|\rho(P,\lambda)\rangle 
=-\ii\Big( f^{\rm T}_\rho-f_\rho \frac{m_u+m_d}{m_\rho}\Big)
m_\rho^2 {\cal P}^{(3)}_{n}
\left(\big(e^{(\lambda)} x\big) (Px)^n \Upsilon^{(3)}_n\right),
\end{align}
where $\Upsilon^{(3)}_n$ is the moment of a spin-independent twist-3 
distribution function. Using expression (\ref{XN3}) the local matrix 
element is given as
\begin{align}
\label{lmatrix_N_tw3}
\langle 0|N^{\rm tw3}_{n}(x)|\rho(P,\lambda)\rangle 
&=-\ii\Big( f^{\rm T}_\rho-f_\rho \frac{m_u+m_d}{m_\rho}\Big)
\big(e^{(\lambda)} x\big) m_\rho^2
\,\Upsilon^{(3)}_n  h^1_n(P|x),
\end{align}
and for the bilocal matrix element of twist-3 we obtain
\begin{align}
\label{nmatrix_N_tw3}
\langle 0|N^{\rm tw3}(x,-x)|\rho(P,\lambda)\rangle 
&=-\ii\Big( f^{\rm T}_\rho-f_\rho \frac{m_u+m_d}{m_\rho}\Big)
\big(e^{(\lambda)} x\big) m_\rho^2
\int_{-1}^1\d \xi\, \hat\Upsilon^{(3)}(\xi)
\left(1+\xi\pd_\xi\right){\cal H}_1(P\xi| x).
\end{align}
The next higher twist contributions of the scalar chiral-odd operator
are of order twist-5.

Now, we consider the matrix elements of the chiral-odd {\em  skew tensor 
operators}. 
Using Eq.~(\ref{MPROJ-x}),
the corresponding moments of the wave function may be  
introduced according to
\begin{align}
\label{}
\langle 0|M^{(\tau)}_{[\alpha\beta]n}(x)|\rho(P,\lambda)\rangle 
=\ii f^{\rm T}_\rho {\cal P}^{(\tau)\mu\nu}_{[\alpha\beta]n}
\left(\Big(e^{(\lambda)}_\mu P_\nu-e^{(\lambda)}_\nu P_\mu\Big) 
(Px)^n \Psi^{(\tau)}_n\right).
\end{align}
The local matrix element of the skew tensor operator of twist-2, 
performing the differentiations in the expression (\ref{XM2}), is given by
\begin{align}
\label{lmatrix_M_tw2}
&\langle 0|M^{\rm tw2}_{[\alpha\beta] n}(x)|\rho(P,\lambda)\rangle
=\ii f^{\rm T}_\rho\,\Psi_n^{(2)} 
\bigg\{
\frac{2}{n+1}\,
e^{(\lambda)}_{[\alpha}P_{\beta]} h^2_n(P|x)
+\frac{n+3}{(n+2)^2}\,m_\rho^2 x_{[\alpha} e^{(\lambda)}_{\beta]} h^2_{n-1}(P|x)
\nonumber\\
&\qquad+\frac{m_\rho^2}{(n+2)^2(n+1)}
\Big\{
4 x_{[\alpha} e^{(\lambda)}_{\beta]} h^3_{n-1}(P|x)
-\Big(2 x_{[\alpha} e^{(\lambda)}_{\beta]}(Px)
+4x_{[\alpha}P_{\beta]} (e^{(\lambda)}x)\Big)h^3_{n-2}(P|x)
\nonumber\\
&\qquad\qquad\qquad\qquad
+12x_{[\alpha}P_{\beta]} (e^{(\lambda)}x)h^4_{n-2}(P|x)
-6x_{[\alpha}P_{\beta]} (e^{(\lambda)}x)(Px) h^4_{n-3}(P|x)
\Big\}\bigg\},
\end{align}
and the resummed bilocal matrix element of twist-2 reads
\begin{align}
\label{nmatrix_M_tw2}
&\langle 0|M^{\rm tw2}_{[\alpha\beta]}(x,-x)|\rho(P,\lambda)\rangle 
=\ii\,f^{\rm T}_\rho \int_{-1}^1\d\xi \, \hat\Psi^{(2)}(\xi)
\left(3+\xi\pd_\xi\right)
\bigg\{2\left(2+\xi\pd_\xi\right) 
e^{(\lambda)}_{[\alpha}P_{\beta]}{\cal H}_2(P\xi| x)\nonumber\\
&
+m_\rho^2\int_0^1\d t\, t\bigg[
\left(1+\xi\pd_\xi\right)
(\ii\xi t)  x_{[\alpha} e^{(\lambda)}_{\beta]}{\cal H}_2(P\xi| t x)\\
&
+\Big(2\left(4+\xi\pd_\xi\right)
(\ii\xi t) x_{[\alpha}e^{(\lambda)}_{\beta]} 
-(\ii\xi t)^2\big( x_{[\alpha}e^{(\lambda)}_{\beta]}(Px)
+2 x_{[\alpha}P_{\beta]}(e^{(\lambda)}x)\big)\Big){\cal H}_3(P\xi| tx)
\nonumber\\
&
+\Big(2\left(5+\xi\pd_\xi\right)\left(4+\xi\pd_\xi\right)
(\ii\xi t)^2 x_{[\alpha}P_{\beta]}(e^{(\lambda)}x) 
-\left(4+\xi\pd_\xi\right)
(\ii\xi t)^3 x_{[\alpha}P_{\beta]}(e^{(\lambda)}x)(Px) \Big)
{\cal H}_4(P\xi| tx)
\bigg]\bigg\}\nonumber.
\end{align}
The local matrix element of the skew tensor operator of 
twist-3  is obtained from (\ref{XM3}) as follows:
\begin{align}
\label{lmatrix_M_tw3}
\langle 0|M^{\rm tw3}_{[\alpha\beta] n}(x)|\rho(P,\lambda)\rangle
&=-\frac{2}{(n+2)n}\,\ii f^{\rm T}_\rho\,\Psi_n^{(3)} m_\rho^2
\Big\{
x_{[\alpha}e^{(\lambda)}_{\beta]} h^2_{n-1}(P|x)
+2 x_{[\alpha} P_{\beta]} (e^{(\lambda)}x) h^3_{n-2}(P|x)
\Big\},
\end{align}
and the corresponding bilocal matrix element reads
\begin{align}
\label{nmatrix_M_tw3}
&\langle 0|M^{\rm tw3}_{[\alpha\beta]}(x,-x)|\rho(P,\lambda)\rangle 
=-2\ii f^{\rm T}_\rho m_\rho^2\int_{0}^1\frac{\d t}{t} \int_{-1}^1\d\xi\, \hat\Psi^{(3)}(\xi)
\left(1+\xi\pd_\xi\right)\nonumber\\
&\qquad\times\Big\{(\ii\xi t) x_{[\alpha}e^{(\lambda)}_{\beta]}{\cal H}_2(P\xi| t x)
+\left(3+\xi\pd_\xi\right)(\ii\xi t)^2  
x_{[\alpha}P_{\beta]}(e^{(\lambda)}x){\cal H}_3(P\xi| t x)\Big\}.
\end{align}
The next higher twist contributions of the skew tensor operator would
be of twist-4.
Again, the matrix elements~(\ref{lmatrix_N_tw3}), (\ref{nmatrix_N_tw3}),
(\ref{lmatrix_M_tw3}) and (\ref{nmatrix_M_tw3})
include the contribution of the twist-3 operator containing 
total derivatives.
Let us note, that the contribution of the additional twist-3 operator,
Eq.~(\ref{XtM3}), vanishes.

Finally, we remark that after projection onto the light-cone the 
matrix elements, which have been introduced in Section~(\ref{meson}), 
are recovered. This can be easily checked by using the expansion of the 
Gegenbauer polynomials or the Poisson integral representation of 
Bessel functions. 
Let us point to the fact that 
the $x^2-$dependence is contained in the polynomials $h^\nu_n(P|x)$ and the 
functions
${\cal H}_\nu(P|x)$ in the case of the moments and the wave
functions, respectively. Since the latter is given by
\begin{align}
2\Gamma(\nu){\cal H}_\nu(P|x)
= 4^\nu \int_{-1}^{+1}\d t (1-t^2)^{\nu-1} 
\e^{\ii (Px) \big( 1 + t \sqrt{1-m_\rho^2 x^2/(Px)^2}\big)/2}
\end{align}
one would obtain a fairly simple expansion in powers of $m_\rho^2 x^2/(Px)^2$.
Because the meson DAs of definite twist are independent from any 
coordinates or momenta, they coincide on-cone 
and off-cone. The whole information about the power corrections is
contained in twist projection of the operators from which the matrix 
elements are taken. 

Trace-subtractions of leading twist operators give
rise to the so-called ``kinematical'' (or rather geometric)
mass-corrections. 
They are formally analogous to Nachtmann corrections in inclusive 
processes.
Operators of higher twist can be decomposed into operators 
containing total derivatives and 
quark-quark-gluon operators of Shuryak-Vainshtein type 
entering by the EOM. The contributions of operators with total derivatives
is a specific feature in exclusive processes as well as in off-forward
scattering, which make the structure of these corrections much more complex.
Numerically, these corrections turned out to be relevant. We conclude
that, at least for exclusive vector-meson DAs, a resummation of
mass-corrections induced by trace-subtractions in the leading twist
matrix element and the higher twist operators containing total derivatives
are relevant for a good approximation.

\subsubsection{Mass corrections of pion distributions}
Amplitudes of processes involving pseudoscalar mesons (pions) can be
expressed in terms of matrix elements of two-particle operators
sandwiched between the vacuum and the meson state
$\langle 0|\bar u(x)\Gamma U(x,-x) d(-x)|\pi(P)\rangle$.
We are now in particular interested in pion-mass corrections.
In contrast to the vector mesons, the matrix elements of the pion
are just related to the pion momentum $P$ with $P^2=m_\pi^2$ where
$m_\pi$ denotes the pion mass.
Therefore, the DAs of pions are given by totally symmetric harmonic
operators. Because no asymmetric Young tableaux 
give a contribution we can carry out the classification of the DAs
by means of the totally symmetric scalar two particle operators of any twist.
Thus we are able to make a classification of pion DAs up to infinite twist.

We start with the two particle DAs of minimal twist.
The light-cone matrix-element of the pseudo scalar operator of twist-2
reads~\cite{chern84,braun89}
\begin{align}
\label{O_twn_sc-pi-lc}
\langle 0|O^{\rm tw2}_5(\lcx,-\lcx)|\pi(P)\rangle 
=\ii f_\pi (P\lcx)\int_{-1}^1\!\d \xi\, \hat\phi_\pi^{(2)}(\xi)\,
\e^{\ii\xi P\lcx},
\end{align}
where $\hat\phi_\pi^{(2)}(\xi)$ is the twist-2 DA. 

Using Eqs.~(\ref{O_twn_sc}) and (\ref{O^(2+2j)}),
we are able to give the following (infinite) twist 
decomposition of the corresponding operator and of the DAs as follows
\begin{align}
\label{O_twn_sc-pi}
\langle 0|O_5(x,-x)|\pi(P)\rangle&=\langle 0|O_5^{{\rm tw}2}(x,-x)|\pi(P)\rangle\nonumber\\
&\quad+\sum_{j=1}^{\infty}
\frac{x^{2j}}{4^j j!(j-1)!}
\int_0^1\!\d t\, t\, (1-t)^{j-1}
\langle 0|O_5^{{\rm tw}(2+2j)}(tx,-tx)|\pi(P)\rangle,
\end{align}
with 
\begin{align}
\label{O^(2+2j)-pi}
\langle 0|O_5^{{\rm tw}(2+2j)}(tx,-tx)|\pi(P)\rangle
&=\ii f_\pi \int_{-1}^1\!\d \xi\, \hat\phi^{(2+2j)}_\pi(\xi)
\left(- m_\pi^2\xi^2\right)^{j}
\bigg\{
\left(t(Px)(2+\xi\pd_\xi)
-\frac{1}{2}\, \ii\xi t^2 m_\pi^2 x^2 \right)\nonumber\\
&\qquad\times
\left(3+\xi\pd_\xi\right)
{\cal H}_2(P\xi| t x)
- 2j \frac{\ii}{\xi}  \left(1+\xi\pd_\xi\right)
{\cal H}_1(P\xi| t x)
\bigg\}.
\end{align}
Here $\hat\phi^{(2+2j)}_\pi(\xi)$ is a pion DA of twist $\tau=2+2j$.
The harmonic twist-2 operator is obtained from Eq.~(\ref{O^(2+2j)-pi}) 
by setting $t=1$ and $j=0$ and gives the pure kinematical target-mass 
corrections. 

The light-cone matrix-element of the pseudo scalar operator of twist-3
is given by~\cite{braun89}
\begin{align}
\label{N_twn_sc-pi-lc}
\langle 0|N^{\rm tw3}_5(\lcx,-\lcx)|\pi(P)\rangle 
=\frac{f_\pi m_\pi^2}{m_u+m_d}
\int_{-1}^1\!\d \xi\, \hat\phi_p^{(3)}(\xi)\, \e^{\ii\xi P\lcx}.
\end{align}

The (infinite) twist decomposition of the bilocal scalar operator 
$N_5(x,-x)=\bar u(x)\ii\gamma_5 d(-x)$  into harmonic operators 
$N^{{\rm tw}(3+2j)}(x,-x)$ of twist $\tau=3+2j,\, j= 0,1, 2, \ldots,$ 
is given by Eqs.~(\ref{N_twn_sc}) and (\ref{N^(3+2j)}) and 
reads
\begin{align}
\label{N_twn_sc-pi}
\langle 0|N_5(x,-x)|\pi(P)\rangle &=\langle 0| N_5^{{\rm tw}3}(x,-x)|\pi(P)\rangle\nonumber\\
&\quad+ \sum_{j=1}^{\infty}
\frac{x^{2j}}{4^j j!(j-1)!}
\int_0^1\!\d t\, t\,(1-t)^{j-1}
\langle 0|N_5^{{\rm tw}(3+2j)}(tx,-tx)|\pi(P)\rangle,
\end{align}
with 
\begin{align}
\label{N^(3+2j)-pi}
&\langle 0|N_5^{{\rm tw}(3+2j)}(tx,-tx)|\pi(P)\rangle
=\frac{f_\pi m_\pi^2}{m_u+m_d}
\int_{-1}^1\!\d \xi\, \hat\phi_p^{(3+2j)}(\xi)
\left(-m_\pi^2\xi^2\right)^{j}\left(1+\xi\pd_\xi\right)
{\cal H}_1(P\xi| t x).
\end{align}
Eq.~(\ref{N^(3+2j)-pi}) is the generalization of Ball's~\cite{ball99} 
pion mass corrections of leading twist for any twist. These higher twist
operators give also rise to dynamical target-mass corrections.

Another twist-3 light-cone matrix element is given as~\cite{braun89}
\begin{align}
\label{M-pi}
\langle 0|M^{{\rm tw3}}_{5[\alpha\beta]}(\lcx,-\lcx)|\pi(P)\rangle
=P_{[\alpha}\lcx_{\beta]} \frac{\ii f_\pi m^2_\pi}{3(m_u-m_d)}
\int_{-1}^1\!\!\d\xi\,\hat\phi_\sigma^{(3)}(\xi)\,\e^{\ii\xi P\lcx}.
\end{align}

Using the parametrization of Eq.~(\ref{M-pi}) together with Eqs.~(\ref{M_twn_sc})
and (\ref{M^(2+2j)}),
we finally get the
infinite twist decomposition:
\begin{align}
\label{M_twn_sc-pi}
\langle 0|M_5(x,-x)|\pi(P)\rangle&=\langle 0|M_5^{{\rm tw}3}(x,-x)|\pi(P)\rangle\\
&\quad+\sum_{j=1}^{\infty}\frac{x^{2j}}{4^j j!(j-1)!}
\int_0^1\d t\, t\, (1-t)^{j-1}\, 
\langle 0|M_5^{{\rm tw}(3+2j)}(tx,-tx)|\pi(P)\rangle,\nonumber
\end{align}
with 
\begin{align}
\label{M^(3+2j)-pi}
\langle 0|M_5^{{\rm tw}(3+2j)}(tx,-tx)|\pi(P)\rangle
=\frac{f_\pi m^2_\pi}{6(m_u-m_d)}&
\int_{-1}^1\!\d\xi\,\hat\phi_\sigma^{(3+2j)}(\xi)
\left(-m_\pi^2\xi^2\right)^{j}\, 
t \xi
\left((Px)^2-m_\pi^2x^2\right)
\nonumber\\
&\quad\times
\left(2+\xi\pd_\xi\right)\left(3+\xi\pd_\xi\right)
{\cal H}_2(P\xi| t x).
\end{align}
Again these higher twist operators contain also dynamical target-mass 
corrections.

Let me, additionally, note that the formal structure of Eqs.
(\ref{O^(2+2j)-pi}), (\ref{N^(3+2j)-pi}) and (\ref{M^(3+2j)-pi})
is, up to the factor $(-m_\pi^2\xi^2)^j$, the same for any twist.

\subsection{Conclusions}

In this Section, we introduced a general procedure of parametrizing
non-forward matrix elements of off-cone QCD operators
${\cal O}_\Gamma(\kappa x,-\kappa x)$ 
in terms of  two-variable distribution 
amplitudes $f_a^{(\tau)}({\mathbb Z})$ of well-defined geometric twist $\tau$, 
namely, single variable hadron distributions, double distributions etc..

The procedure relies on the unique twist decomposition of non-local
operators off the light-cone leading to an infinite series of 
operators with growing twist. This decomposition is completely of
group theoretical origin and is equivalent to the decomposition of
the local operators into irreducible tensor representations of the 
Lorentz group. 

Using these results we determined the off-cone power corrections
to various double distributions and the vector as well as scalar 
meson distribution amplitudes
being the inputs of the corresponding scattering amplitudes and
hadronic form factors, respectively. These power corrections, in the 
case of local operators, are expressed in terms of 
Gegenbauer polynomials being multiplied with the moments of the 
distribution amplitudes and, in the case of non-local operators,
in terms of Bessel functions now multiplied with the 
distribution amplitudes directly. Accordingly, the off-cone
expressions are obtained from the on-cone ones by harmonic extension.

Concerning the computation of the scattering amplitudes 
of physical relevance some Fourier transformation has to be carried out
whose result mainly depends on the (singular) coefficient functions as
well as on the Gegenbauer polynomials or Bessel functions 
(see Appendix~\ref{mass}). 
The easiest way to calculate the power corrections of scattering amplitudes
is to use the local operators in terms of Gegenbauer polynomials in 
the $q$-space. The currently most interesting 
example is the (deeply) virtual Compton scattering.



\chapter{The twist decomposition of nonlocal tensor operators}
In this Chapter we give the explicite twist decomposition of 
(non)local operators on the light cone as well as off-cone. 
In Section~\ref{tensor} we discuss the basics of irreducible
tensors with respect to the Lorentz and rotation group in $2h$-dimensional 
spacetime ($2h\ge 3$) in the framework of polynomial technique. 
We are using the harmonic irreducible tensor polynomials with $h=2$
in order to rewrite harmonic operators in terms of Gegenbauer polynomials 
and Bessel functions in Section~\ref{off-cone}. We have used these harmonic
operators in Section~\ref{harmonic} to calculate the power corrections of 
distribution amplitudes with definite geometric twist.
In Section~\ref{gluon} we decompose a general bilocal tensor operator
of second rank. In this connection we derive the twist decomposition of
the light-cone operators $O_\alpha(\kappa_1\lcx,\kappa_2\lcx)$ and
$M_{[\alpha\beta]}(\kappa_1\lcx,\kappa_2\lcx)$ which we used in 
Section~\ref{forward} and \ref{meson}.
In Section~\ref{trilocal} we discuss the twist decomposition of trilocal 
and multilocal operators on the light-cone.

\section{Irreducible tensor polynomials and harmonic extension in 
$2h$-dimensional spacetime -- polynomial technique}
\renewcommand{\theequation}{\thesection.\arabic{equation}}
\setcounter{equation}{0}
\label{tensor}

The objective of the present Section is to use the polynomial technique
for symmetric tensor and some asymmetric tensor representations of the
orthogonal group as well as the Lorentz group in $2h$-dimensions ($2h\ge 3$). 
By means of the polynomial technique we can use a $2h$-vector as a
device for writing tensors with special symmetries in an analytic form.

In order to construct irreducible vector and tensor polynomials 
corresponding to irreducible tensors with special Young symmetry,
we generalize the homogeneous polynomial technique which is well-known
for the scalar case in constructing irreducible symmetric tensor 
representations of the orthogonal group $SO(2h,{\Bbb C})$.

\noindent
(i)\hspace{.5cm}
Let us now consider a symmetric traceless tensor 
$\tl T_{\mu_1\ldots\mu_n}$ corresponding to the Young tableau
\unitlength0.3cm
\begin{picture}(4.5,1)
\linethickness{0.075mm}
\put(1,0){\framebox(3,1){$\SC n$}}
\end{picture}. 
Such tensors are, for the group $SO(2h,{\Bbb C})$,
in one-to-one correspondence with harmonic polynomials of the complex $2h$-vector
$\zeta$:
\begin{align}
\tl T_n(\zeta)\equiv \zeta^{\mu_1}\cdots\zeta^{\mu_n}\, \tl T_{\mu_1\ldots\mu_n},
\qquad\zeta\in{\Bbb C}^{2h}
\qquad
\Delta \tl T_n(\zeta)=0,
\end{align}
or, equivalently, with their restriction $T_n(z)$ on the
complex light cone $K_{2h}({\Bbb C})$:
\begin{align}
T_n(z)\equiv z^{\mu_1}\cdots z^{\mu_n}\,  T_{\mu_1\ldots\mu_n}
=\lim_{\zeta\rightarrow z, z^2=0} \tl T_n(\zeta).
\end{align} 
Then, the space of (complex) harmonic polynomials $\tl T_n(\zeta)$ and the
space of homogeneous polynomials $T_n(z)$ on the complex light cone provide
convenient realizations of the carrier space of
irreducible finite dimensional symmetric tensor representations of the 
group $SO(2h,{\Bbb C})$ (see Ref.~\cite{BT77}).

For symmetric tensor representations of the group $SO(1,2h-1)$ the elements 
of the carrier space are harmonic (more strictly: $\square$-harmonic) polynomials of degree $n$, 
$\tl T_n(x)$, with respect to the real $2h$-vector $x$, iff 
\begin{align}
\square \tl T_n(x)=0,\qquad x\in M^{2h}
\end{align}
or, alternatively, their restrictions  $T_n(\lcx)$ on the
real light cone $K_{2h}({\Bbb R})$.
The tensor representations $\tl T_n(\zeta)$ and $T_n(z)$ of 
$SO(2h,{\Bbb C})$ are equivalent to the analytic continuation of the representations
$\tl T_n(x)$ and $T_n(\lcx)$ of $SO(1,2h-1)$, respectively.
In the following we consider finite dimensional representations of the group 
$SO(1,2h-1)$. 

Let us now regard the equivalence between the representation
$\tl T_n(x)$ and $T_n(\lcx)$. 
For this we note the decomposition of any homogeneous polynomial in the
form
\begin{align}
T_n(x)= \tl T_{n}(x)+ x^2 T_{n-2}(x),
\quad
\tl T_{n}(x)\in {\cal H}^{2h,n},
\end{align}
where $T_n(x)\in {\cal R}^{2h,n}$ is a homogeneous polynomial of degree $n$ in $x$. 
On the light cone, the value of any polynomial coincides with the value
on the cone of a uniquely defined harmonic polynomial. 
This establishes an isomorphism between the coset space 
${\cal R}^{2h,n}/x^{2}{\cal R}^{2h,n-2}$ and the space ${\cal H}^{2h,n}$.

By means of the extension $T_n(x)$ for any polynomial $T_n(\lcx)$ on the light 
cone, the harmonic extension is defined by
\begin{align}
\square\tl T_n(x)=0,\quad \tl T_n(x)|_{x^2=0}\equiv T_n(x)|_{x=\lcx}.
\end{align}
Therefore, the harmonic extension 
establishes an one-to-one relation between homogeneous polynomials on the 
light-cone and corresponding harmonic polynomials off-cone which 
provide a convenient realization for symmetric tensor representations of 
the Lorentz group and the same is valid for tensor polynomials carrying more 
complicated tensor representations of the Lorentz group. It is important to
note that the unique harmonic extension must not destroy the type 
of Lorentz representation. 

The solutions of 
\begin{align}
\label{H0}
\square \tl T_n(x) = 0,
\end{align}
are {\em (scalar) 
harmonic polynomials} of order $n$ corresponding to symmetric 
traceless tensors of rank $n$ ($\tl T_n(x)\equiv T^{(n)}_n(x)$). They are given by
(see, e.g., \cite{VK}, Chapter IX) and \cite{BT77})\\

\begin{align}
\label{T_harm_d}
\tl T_n(x) 
&=
{\cal P}^{(n)}_n (x)T_n(x),
\end{align}
with the harmonic projection operator
\begin{align}
\label{P^[n]_n}
{\cal P}^{(n)}_n(x)
\equiv H^{(2h)}_n(x^2|\square)
&=
\sum_{k=0}^{[\frac{n}{2}]}
\frac{(-1)^k(h+n-k-2)!}{4^k k! (h+n-2)!}\,
x^{2k}\,\square^{k} \,.
\end{align}
Therefore,  the unique harmonic extension $\tl T_n(x)$
of $T_n(\lcx)$ is uniquely given by (\ref{T_harm_d}) and (\ref{P^[n]_n}). 
Additionally, the restriction of Eq.~(\ref{P^[n]_n}) to the light cone
is ${\cal P}^{(n)}_n(\lcx)=1$.
The condition of tracelessness (\ref{H0}) is trivially satisfied on the 
light cone by $\lcx^2=0$.
For $T_n(\lcx)=(a\lcx)^n$ the harmonic extension can be rewritten as
\begin{align}
\tl T_n(x)&=\frac{2^n n! (2h+n-3)!}{(2h+2n-3)!} \left(\sqrt{a^2 x^2}\right)^{n}
P_n^{\left(h-\frac{3}{2},\, h-\frac{3}{2}\right)}
\left(\frac{ax}{\sqrt{a^2 x^2}}\right)\nonumber\\
&=\frac{n!}{(h-1)_n} \left(\frac{1}{2}\sqrt{a^2 x^2}\right)^{n}
C_n^{h-1}\left(\frac{ax}{\sqrt{a^2 x^2}}\right),
\end{align}
with $(h-1)_n=\Gamma(h-1+n)/\Gamma(h-1)$.
Here $P_n^{\left(h-\frac{3}{2},\, h-\frac{3}{2}\right)}$ are 
the Jacobi polynomials and 
$C_n^{h-1}$ are the 
Gegenbauer polynomials of degree $n$.
Note that in three dimensions $C^{1/2}_n$ coincide with the 
Legendre polynomials.

A vector polynomial of degree $n-1$ in $x$  
for the Young tableau
\unitlength0.3cm
\begin{picture}(5.0,2)
\linethickness{0.075mm}
\put(1,0){\framebox(1,1){$\SC\alpha$}}
\put(2,0){\framebox(3,1){$\SC n-1$}}
\end{picture}
is recovered by straightforward differentiation as
\begin{align}
\label{T^[n]_alpha}
T^{(n)}_{\alpha n-1}(x)
&=\frac{1}{n}\, \pd_\alpha \tl T_{n}(x),\nonumber\\
&=\frac{1}{n}
\sum_{k=0}^{[\frac{n-1}{2}]}\frac{(-1)^k (h+n-k-3)!}{4^k k!(h+n-2)!}\, x^{2k}
\left\{(h+n-k-2)\pd_\alpha-\frac{1}{2}\, x_\alpha\square\right\}\square^k
T_{n}(x).
\end{align}
The conditions of tracelessness are now given by
\begin{align}
\label{}
\pd^\alpha T^{(n)}_{\alpha n-1}(x)=0,
\quad 
\square T^{(n)}_{\alpha n-1}(x)=0.
\end{align}
On the light-cone, 
the vector polynomial of degree $n-1$ in $\lcx$  
is obtained from $k=0$ in Eq.~(\ref{T^[n]_alpha}) as follows
\begin{align}
\label{T^[n]_alpha(lcx)}
T^{(n)}_{\alpha n-1}(\lcx)
=\frac{1}{n(h+n-2)}\, \d_\alpha T_{n}(\lcx),
\end{align}
where the inner derivative $\d_\alpha$, as a lowering operator 
on the light cone by construction, maps the scalar polynomial of degree $n$ to 
a vector polynomial of degree $n-1$.
The condition of tracelessness on the light-cone can be formulated 
by the help of the inner derivative: 
\begin{align}
\label{}
\d^\alpha T^{(n)}_{\alpha n-1}(\lcx)=0,
\end{align}
and is easily proved by the property $\d^2=0$.
The harmonic extension of (\ref{T^[n]_alpha(lcx)}) is given by
\begin{align}
\label{T^{[n]}_{alpha n-1}(x)-he}
T^{(n)}_{\alpha n-1}(x)
&=\frac{1} {n(h+n-2)}\, {\cal P}^{(n)}_{\alpha n-1}(x) T_{n}(x),
\end{align}
where the harmonic projection operator of vector polynomials of type $[n]=(n)$ 
is defined by
\begin{align}
{\cal P}^{(n)}_{\alpha n-1}(x)
&:=(h+n-2)\pd_\alpha {\cal P}^{(n)}_{n}(x)\nonumber\\
&=\sum_{k=0}^{[\frac{n-1}{2}]}\frac{(-1)^k (h+n-k-3)!}{4^k k!(h+n-3)!}
\,x^{2k} {\cal D}_\alpha(k)\, \square^k
\end{align}
and where the differential operator,
\begin{align}
{\cal D}_\alpha(k):=(h-1+k+x\pd)\pd_\alpha-\frac{1}{2}\, x_\alpha\square,
\end{align}
can be understood as a formal off-cone modification of the interior differentiation
$\d_\alpha$
which gives in the limit $k=0$ and $x\rightarrow\lcx$ the interior 
differential operator 
(${\cal D}_\alpha(0)=\d_\alpha$ and
${\cal P}^{(n)}_{\alpha n-1}(\lcx)=\d_\alpha$).
The {\it prescription for the harmonic extension} is:
Replace the interior derivative $\d_\alpha$ by the differential operator
${\cal D}_\alpha(k)$ which has to be sandwiched in the scalar harmonic 
projection operator  of the corresponding polynomial degree 
between $x^{2k}$ and $\square^k$ so that in the limit $k=0$ the
projector of the Lorentz spin on the light-cone is recovered.

The generalization of (\ref{T^[n]_alpha(lcx)}) is
\begin{align}
\label{T^[n]_mu_l (lcx)}
T^{(n)}_{\mu_1\cdots\mu_l n-l}(\lcx)
=\frac{(n-l)!}{n!}
\frac{ (h+n-2-l)!}{(h+n-2)!}\,
\d_{\mu_1}\ldots\d_{\mu_l}\, T_{n}(\lcx),
\end{align}
and its harmonic extension is given by
\begin{align}
\label{}
T^{(n)}_{\mu_1\cdots\mu_l n-l}(x)
&=\frac{(n-l)!}{n!} \frac{(h+n-2-l)!}{(h+n-2)!}\, 
{\cal P}^{(n)}_{\mu_1\cdots\mu_l n-l}(x) T_{n}(x),
\end{align}
where
\begin{align}
{\cal P}^{(n)}_{\mu_1\cdots\mu_l n-l}(x)
=\sum_{k=0}^{[\frac{n-1}{2}]}\frac{(-1)^k (h+n-k-2-l)!}{4^k k!(h+n-2-l)!}
\,x^{2k} {\cal D}_{\mu_1}(k)\ldots{\cal D}_{\mu_l}(k) \, \square^k.
\end{align}

\noindent
(ii)\hspace{.5cm}
Now we consider an irreducible tensor 
$T^{(n,1)}_{[\alpha\beta]\mu_1\cdots\mu_{n-1}}$ having the symmetry type of
the Young tableau
\unitlength0.3cm
\begin{picture}(5.0,2)
\linethickness{0.075mm}
\put(1,0){\framebox(1,1){$\SC\alpha$}}
\put(1,1){\framebox(1,1){$\SC\beta$}}
\put(2,1){\framebox(3,1){$\SC n-1$}}
\end{picture}.\\
In order to obtain an irreducible tensor, 
it has to be:\\
 (a) traceless in $\mu_1\cdots\mu_{n-1}$;\\
 (b) traceless in one antisymmetric and one symmetric index.\\
Using the polynomial technique we replace 
$T^{(n,1)}_{[\alpha\beta]\mu_1\cdots\mu_{n-1}}$ by a homogeneous polynomial 
of degree $n-1$ in $x$.
Then we can build the tensor polynomial carrying the required symmetry by
means of
\begin{align}
\label{as_ha_ve_po_0}
T^{(n,1)}_{[\alpha\beta] n-1}(x)=
\frac{2}{ n+1}\,
\pd_{[\beta} \tl T_{\alpha] n}(x),
\end{align}
where $\tl T_{\mu n}(x)$ has to be harmonic and traceless. 
Let me note that the condition of the correct symmetry type can be formulated 
as
\begin{align}
\label{n,1-cond}
\epsilon^{\alpha\beta\nu\mu_1\ldots\mu_{2h-3}} \pd_\nu T^{(n,1)}_{[\alpha\beta] n-1}(x)=0.
\end{align}
The conditions of tracelessness for the 
irreducible tensor polynomial are given by
\begin{align}
\label{H1ir}
\square T^{(n,1)}_{[\alpha\beta] n-1}(x) = 0,
\quad
\pd^\alpha T^{(n,1)}_{[\alpha\beta] n-1}(x) = 0,
\end{align}
and can be reduced to the conditions of tracelessness for $\tl T_{\mu n}(x)$:
\begin{align}
\label{H1}
\square \tl T_{\alpha n}(x) = 0,
\quad
\pd^\alpha \tl T_{\alpha n}(x) = 0,
\end{align}
resulting from (a) and and (b), respectively.
We make an
Ansatz with the special structure of traces of this tensor as follows:
\begin{equation}
\tl T_{\mu n}(x)
= 
\left\{\delta_{\mu}^{\nu}
-a_n x_\mu \pd^{\nu}
   -   b_n x^2\pd_\mu\pd^{\nu}
\right\}
{\cal P}^{(n)}_n(x)
T_{\nu  n}(x).
\end{equation}
From the conditions of tracelessness (\ref{H1}) we obtain two equations
for the coefficients $a_n,\ b_n$:
\begin{eqnarray}
1 &=&  ( 2h+n-1) a_n + 2( n - 1) b_n,\nonumber\\
0 &=&  a_n + 2 (h+n-2) b_n.
\nonumber
\end{eqnarray}
The unique solution is:
\begin{eqnarray}
a_n = \frac{ h+n-2}{( h+n-1)(2h+n-3)},
\qquad
b_n = - \frac{1}{2(h+n-1)(2h+n-3)}.
\nonumber
\end{eqnarray}
Then the harmonic and traceless vector polynomial is (not irreducible):
\begin{align}
\label{ha_ve_po}
\tl T_{\mu n}(x)
= 
\Big\{\delta_{\mu}^{\nu}
&-\frac{h+n-2 }{(h+n-1)(2h+n-3)}\, x_\mu \pd^{\nu}\nonumber\\
&+\frac{1}{2(h+n-1)(2h+n-3)}\, x^2\pd_\mu\pd^{\nu}
\Big\}
{\cal P}^{(n)}_n(x)
T_{\nu  n}(x).
\end{align}
Finally, using (\ref{as_ha_ve_po_0}) and (\ref{ha_ve_po}) we obtain for 
the irreducible representation of type $[n+1]=(n,1)$ as tensor polynomial 
of degree $n-1$:
\begin{align}
\label{[n,1](n-1)}
T^{(n,1)}_{[\alpha\beta] n-1}(x)
&=\frac{2}{n+1}
\bigg\{\delta^\mu_{[\alpha}\pd_{\beta]}-\frac{1}{2h+n-3}\, 
x_{[\alpha}\pd_{\beta]}\pd^{\mu}\bigg\}
\sum_{k=0}^{[\frac{n}{2}]}\frac{(-1)^k (h+n-k-2)!}{4^k k!(h+n-2)!}\, 
x^{2k}\square^k T_{\mu n}(x),\nonumber\\
&=\frac{2}{n+1}
\sum_{k=0}^{[\frac{n-1}{2}]}\frac{(-1)^k (h+n-k-3)!}{4^k k!(h+n-2)!}\, x^{2k}
\bigg\{\delta^\mu_{[\alpha}
\Big((h+n-k-2)\pd_{\beta]}-\frac{1}{2}\, x_{\beta]}\square\Big)\nonumber\\
&\qquad\qquad-\frac{1}{2h+n-3}\, x_{[\alpha}\pd_{\beta]}
\Big((h+n-k-2)\pd^{\mu}-\frac{1}{2}\, x^{\mu}\square\Big)
\bigg\}\square^k
T_{\mu n}(x),
\end{align}
and the irreducible representation of type $(n,1)$ as tensor polynomial of
degree $n-2$ is obtained from Eq.~(\ref{[n,1](n-1)}) by means of partial differentiation
with respect to the index $\gamma$
\begin{align}
\label{[n,1](n-2)}
&T^{(n,1)}_{[\alpha\beta] \gamma n-2}(x)
=\frac{1}{n-1}\,
\pd_\gamma T^{(n,1)}_{[\alpha\beta] n-1}(x),\nonumber\\
&\quad=\frac{2}{(n+1)(n-1)}
\bigg\{\delta^\mu_{[\alpha}\pd_{\beta]}\pd_\gamma-
\frac{1}{2h+n-3}
\left(g_{\gamma[\alpha}\pd_{\beta]}
+x_{[\alpha}\pd_{\beta]}\pd_\gamma\right)
\pd^{\mu}\bigg\}\\
&\hspace{7cm}\times
\sum_{k=0}^{[\frac{n}{2}]}\frac{(-1)^k (h+n-k-2)!}{4^k k!(h+n-2)!}\, 
x^{2k}\square^k T_{\mu n}(x),\nonumber\\
&\quad=\frac{2}{(n+1)(n-1)}
\sum_{k=0}^{[\frac{n-1}{2}]}\frac{(-1)^k (h+n-k-4)!}{4^k k!(h+n-2)!}\, x^{2k}
\bigg\{\delta^\mu_{[\alpha}
\Big((h+n-k-3)\pd_{\beta]}-\frac{1}{2}\, x_{\beta]}\square\Big)
\nonumber\\
&\hspace{7cm}
\times
\Big((h+n-k-2)\pd_{\gamma}-\frac{1}{2}\, x_{\gamma}\square\Big)\nonumber\\
&\quad
-\frac{1}{2h+n-3}\,\left\{ 
g_{\gamma[\alpha}
\Big((h+n-k-3)\pd_{\beta]}-\frac{1}{2}\, x_{\beta]}\square\Big)
+
x_{[\alpha}\pd_{\beta]}
\Big((h+n-k-3)\pd_{\gamma}-\frac{1}{2}\, x_{\gamma}\square\Big)
\right\}\nonumber\\
&\hspace*{7cm}
\times
\Big((h+n-k-2)\pd^{\mu}-\frac{1}{2}\, x^{\mu}\square\Big)
\bigg\}\square^k
T_{\mu n}(x),\nonumber
\end{align}
The irreducible tensor polynomial of degree $n-1$ on the light-cone 
is obtained from Eq.~(\ref{[n,1](n-1)}) for $k=0$ as
\begin{align}
\label{[n,1](n-1)lc}
T^{(n,1)}_{[\alpha\beta] n-1}(\lcx)
=\frac{2}{(n+1)(h+n-2)}
\bigg\{\delta^\mu_{[\alpha}\d_{\beta]}
+\frac{1}{2h+n-3}\, X_{[\alpha\beta]}
\d^{\mu}\bigg\}T_{\mu n}(\lcx),
\end{align}
where $X_{[\alpha\beta]}=-\lcx_{[\alpha}\lcd_{\beta]}$.
Now the condition of tracelessness is imposed as
\begin{align}
\d^\alpha T^{(n,1)}_{[\alpha\beta] n-1}(\lcx)=0,
\end{align}
and can be easily checked by means of the algebraic properties of
the interior derivative on the cone.
Additionally, the condition of the correct symmetry type is 
given on the cone by~\cite{Dobrev82}
\begin{align}
\label{}
\epsilon^{\alpha\beta\nu\mu_1\ldots\mu_{2h-3}} \d_\nu T^{(n,1)}_{[\alpha\beta] n-1}(\lcx)=0.
\end{align}
The harmonic extension of $T^{(n,1)}_{[\alpha\beta] n-1}(\lcx)$ is defined
by
\begin{align}
\label{[n,1](n-1)he}
T^{(n,1)}_{[\alpha\beta] n-1}(x)
&=\frac{2}{(n+1)(h+n-2)}\,
{\cal P}^{(n,1);\, \mu}_{[\alpha\beta] n-1}(x) T_{\mu n}(x),
\end{align}
with
\begin{align}
{\cal P}^{(n,1);\, \mu}_{[\alpha\beta] n-1}(x)
=
\sum_{k=0}^{[\frac{n-1}{2}]}\frac{(-1)^k (h+n-k-3)!}{4^k k!(h+n-3)!}\, x^{2k}
\bigg\{
\delta^\mu_{[\alpha}{\cal D}_{\beta]}(k)
+\frac{1}{2h+n-3}\, X_{[\alpha\beta]}{\cal D}^{\mu}(k)
\bigg\}\square^k,
\end{align}
where we used $X_{[\alpha\beta]}=-x_{[\alpha}\pd_{\beta]}$ in the 
off-cone expression.
Eventually, the restriction of the irreducible representation (\ref{[n,1](n-2)})
to the light cone is given by
\begin{align}
\label{[n,1](n-2)lc}
&T^{(n,1)}_{[\alpha\beta]\gamma n-2}(\lcx)
=\frac{1}{(n-1)(h+n-3)}\,
\d_\gamma T^{(n,1)}_{[\alpha\beta] n-1}(\lcx)\nonumber\\
&\quad=\frac{2}{(n+1)(n-1)(h+n-2)(h+n-3)}\nonumber\\
&\hspace{3cm}
\times
\bigg\{\delta^\mu_{[\alpha}\d_{\beta]}\d_\gamma
-\frac{1}{2h+n-3}
\left(g_{\gamma[\alpha}\d_{\beta]}
-X_{[\alpha\beta]}\d_\gamma\right)
\d^{\mu}\bigg\}T_{\mu n}(\lcx),
\end{align}
It satisfies the conditions of tracelessness
\begin{align}
\d^\alpha T^{(n,1)}_{[\alpha\beta]\gamma n-2}(\lcx)=0,
\quad
\d^\gamma T^{(n,1)}_{[\alpha\beta]\gamma n-2}(\lcx)=0.
\end{align}
It is important to note that the formulas (\ref{[n,1](n-1)}) and 
(\ref{[n,1](n-2)}) are the unique harmonic extensions
 (\ref{[n,1](n-1)lc}) and (\ref{[n,1](n-2)lc}), respectively.
A harmonic projector ${\cal P}^{(n,1);\, \mu}_{[\alpha\beta]\gamma n-2}(x)$
can be proven in the same way.

\noindent
(iii)\hspace{.5cm}
Now we regard an irreducible tensor 
$T^{(n,1,1)}_{[\alpha\beta\gamma]\mu_1\cdots\mu_{n-1}}$ having the symmetry type of
the Young tableau
\unitlength0.3cm
\begin{picture}(5.0,3)
\linethickness{0.075mm}
\put(1,0){\framebox(1,1){$\SC\alpha$}}
\put(1,1){\framebox(1,1){$\SC\beta$}}
\put(1,2){\framebox(1,1){$\SC\gamma$}}
\put(2,2){\framebox(3,1){$\SC n-1$}}
\end{picture} .
We are using the convention $T_{[\alpha\beta\gamma]} = \frac{1}{6}
(T_{\alpha\beta\gamma}-T_{\alpha\gamma\beta}
+T_{\beta\gamma\alpha}-T_{\beta\alpha\gamma}
+T_{\gamma\alpha\beta}-T_{\gamma\beta\alpha})$.\\
In order to obtain an irreducible tensor, 
it has to be:\\
 (a) traceless in $\mu_1\cdots\mu_{n-1}$;\\
 (b) traceless in one antisymmetric and one symmetric index.\\
Again, by means of the polynomial technique, the tensor 
$T^{(n,1,1)}_{[\alpha\beta\gamma]\mu_1\cdots\mu_{n-1}}$ can be replaced
by a homogeneous polynomial of degree $n-1$ in $x$.
Then we obtain the tensor polynomial carrying the required symmetry 
as follows
\begin{align}
\label{as_ha_ve_po_1}
T^{(n,1,1)}_{[\alpha\beta\gamma] n-1}(x)&=
\frac{3}{ n+2}\,
\delta^\mu_{[\alpha}\delta^\nu_\beta \pd_{\gamma]} 
\,\tl T_{\mu\nu n}(x),\nonumber\\
&=\frac{1}{ n+2}\,\Big(
\delta^\mu_{\alpha}\delta^\nu_\beta \pd_{\gamma}
+\delta^\mu_{\beta}\delta^\nu_\gamma \pd_{\alpha}
+\delta^\mu_{\gamma}\delta^\nu_\alpha \pd_{\beta}\Big)
\tl T_{[\mu\nu] n}(x),
\end{align}
where $\tl T_{[\mu\nu] n}(x)$ has to be traceless.
The (symmetry) condition of the Young tableau type is given, by means of the polynomial 
technique, by
\begin{align}
\label{n,1,1-cond}
\epsilon^{\alpha\beta\gamma\nu\mu_1\ldots\mu_{2h-4}} \pd_\nu T^{(n,1,1)}_{[\alpha\beta\gamma] n-1}(x)=0.
\end{align}
The conditions of tracelessness for the 
irreducible tensor polynomial are given by
\begin{align}
\label{H2t}
\square T^{(n,1,1)}_{[\alpha\beta\gamma] n-1}(x) = 0,
\quad
\pd^\alpha T^{(n,1,1)}_{[\alpha\beta\gamma] n-1}(x) = 0,
\end{align}
and give the following conditions of tracelessness for
$\tl T_{[\alpha \beta] n}(x)$:
\begin{align}
\label{H2}
\square \tl T_{[\alpha \beta] n}(x) = 0,
\quad
\pd^\alpha \tl T_{[\alpha \beta] n}(x) = 0.
\end{align}
Using the same procedure as before we may construct the harmonic
skew tensor polynomial $\tl T_{[\mu\nu] n}(x)$.
The solutions of Eqs.~(\ref{H2}) are the {\em antisymmetric 
harmonic tensor polynomials} $\tl T_{[\alpha\beta] n}(x)$ of 
order $n$.
The most general Ansatz for it has the following structure:
\begin{equation}
\tl T_{[\alpha\beta] n}
=
\left\{\delta_\alpha^\mu\delta_\beta^\nu
-a_n x_{[\alpha}\delta_{\beta]}^\nu\pd^\mu
-b_n x^2\pd^\mu\pd_{[\alpha}\delta_{\beta]}^\nu
-c_n x^\mu x_{[\alpha}\pd_{\beta]}\pd^\nu
\right\}
{\cal P}^{(n)}(x) 
T_{[\mu\nu] n}(x).
\end{equation}
The corresponding system of linear equations reads
\begin{eqnarray}
1&=&(2h+n-2)a_n/2 + (n-1)b_n \nonumber\\
0&=&2b_n+(2h+n-2)c_n      \nonumber\\
0&=&a_n+c_n+2(h+n-2)b_n\, .   \nonumber
\end{eqnarray}
It is solved by the following values:
\begin{eqnarray}
(a_n + c_n, \;b_n, \;c_n)
=\frac{1}{(h+n-1)(2h+n-4)}
\left(2(h+n-2),\; -1,\; \frac{2}{2h+n-2}\right).
\nonumber
\end{eqnarray}
The harmonic and traceless skew tensor polynomial is given by
\begin{align}
\label{T_skewtensor-tl}
\tl T_{[\alpha\beta] n}(x)
&=\Big\{\delta_{[\alpha}^\mu\delta_{\beta]}^\nu
+\frac{2}{(h+n-1)(2h+n-4)}
\left((h+n-2) x_{[\alpha}\delta_{\beta]}^{\mu}\pd^{\nu} 
-\frac{1}{2}\, x^2\pd_{[\alpha}\delta_{\beta]}^{\mu}\pd^{\nu}\right)\nonumber\\
&\qquad-\frac{2}{(h+n-1)(2h+n-4)(2h+n-2)}\,
 x_{[\alpha}\pd_{\beta]}x^{\mu}\pd^{\nu}\Big\}
{\cal P}^{(n)}_n\left(x^2|\square\right)T_{[\mu\nu]n}(x).
\end{align}
Finally, we obtain for the irreducible tensor polynomial of degree $n-1$ of
Young tableau $[n+2]=(n,1,1)$:
\begin{align}
\label{[n,1,1]}
T^{(n,1,1)}_{[\alpha\beta\gamma] n-1}(x)
&=\frac{3}{n+2}
\bigg\{\delta^\mu_{[\alpha}\delta^\nu_\beta\pd_{\gamma]}-
\frac{2}{2h+n-4}\, 
x_{[\alpha}\pd_{\beta}\delta^\mu_{\gamma]}\pd^{\nu}\bigg\}\\
&\hspace{5cm}\times
\sum_{k=0}^{[\frac{n}{2}]}\frac{(-1)^k (h+n-k-2)!}{4^k k!(h+n-2)!}\, 
x^{2k}\square^k T_{[\mu\nu] n}(x),\nonumber\\
&=\frac{3}{n+2}
\sum_{k=0}^{[\frac{n-1}{2}]}\frac{(-1)^k (h+n-k-3)!}{4^k k!(h+n-2)!}\, x^{2k}
\bigg\{\delta^\mu_{[\alpha}\delta^\nu_\beta
\Big((h+n-k-2)\pd_{\gamma]}-\frac{1}{2}\, x_{\gamma]}\square\Big)\nonumber\\
&\qquad\qquad-\frac{2}{2h+n-4}\, x_{[\alpha}\pd_{\beta}\delta^\mu_{\gamma]}
\Big((h+n-k-2)\pd^{\nu}-\frac{1}{2}\, x^{\nu}\square\Big)
\bigg\}\square^k
T_{[\mu\nu] n}(x),\nonumber
\end{align}
and for its restriction to the light cone:
\begin{align}
\label{[n,1,1]lc}
T^{(n,1,1)}_{[\alpha\beta\gamma] n-1}(\lcx)
&=\frac{3}{(n+2)(h-n-2)}
\bigg\{\delta^\mu_{[\alpha}\delta^\nu_\beta\d_{\gamma]}
+\frac{2}{2h+n-4}\, 
X_{[\alpha\beta}\delta^\mu_{\gamma]}\d^{\nu}\bigg\}
T_{[\mu\nu] n}(\lcx).
\end{align}
Eq.~(\ref{[n,1,1]}) is the unique harmonic extension of (\ref{[n,1,1]lc}).
On the light cone, 
the condition of tracelessness is given by
\begin{align}
\d^\alpha T^{(n,1,1)}_{[\alpha\beta\gamma] n-1}(\lcx)=0,
\end{align}
and the condition of the symmetry type reads
\begin{align}
\label{n,1,1-cond-lc}
\epsilon^{\alpha\beta\gamma\nu\mu_1\ldots\mu_{2h-4}} \d_\nu T^{(n,1,1)}_{[\alpha\beta\gamma] n-1}(\lcx)=0.
\end{align}
The harmonic extension of $T^{(n,1,1)}_{[\alpha\beta\gamma] n-1}(\lcx)$ is
defined by
\begin{align}
\label{[n,1,1](n-1)}
T^{(n,1,1)}_{[\alpha\beta\gamma] n-1}(x)
&=\frac{3}{(n+2)(h+n-2)}\,
{\cal P}^{(n,1,1);\, \mu\nu}_{[\alpha\beta\gamma] n-1}(x) T_{[\mu\nu] n}(x),
\end{align}
with
\begin{align}
{\cal P}^{(n,1,1);\, \mu\nu}_{[\alpha\beta\gamma] n-1}(x)
=
\sum_{k=0}^{[\frac{n-1}{2}]}\frac{(-1)^k (h+n-k-3)!}{4^k k!(h+n-3)!}\, x^{2k}
\bigg\{
\delta^\mu_{[\alpha}\delta^\nu_\beta {\cal D}_{\gamma]}(k)
+\frac{2}{2h+n-4}\, X_{[\alpha\beta}\delta^\mu_{\gamma]}{\cal D}^{\nu}(k)
\bigg\}\square^k.
\end{align}
By means of partial differentiation
with respect to the index $\delta$, one can also compute the 
tensor polynomial:
$T^{(n,1,1)}_{[\alpha\beta\gamma]\delta  n-2}(x)
=\frac{1}{n-1}\,
\pd_\delta T^{(n,1,1)}_{[\alpha\beta\gamma] n-1}(x)$.

\noindent
(iv)\hspace{.5cm} 
Now we want to determine the reducible {\em symmetric harmonic
tensor polynomials} $\tl T_{(\alpha\beta) n}(x)$ which we need for the
Young tableau $[n+2]=(n,2)$. 
We renounce to construct the proper irreducible tensor polynomial, 
\begin{align}
T^{(n,2)}_{[\alpha\rho][\beta\tau] n-2}(x)=\frac{8}{(n+1)n}\,
\delta^\mu_{[\alpha}\pd_{\lambda]}\delta^\nu_{[\beta}\pd_{\gamma]}
\delta^\lambda_{(\rho}\delta^\gamma_{\tau)}\tl T_{(\mu\nu) n}(x),
\end{align}
which has the symmetry type of
the Young tableau
\unitlength0.3cm
\begin{picture}(5.5,2)
\linethickness{0.075mm}
\put(1,0){\framebox(1,1){$\SC\alpha$}}
\put(1,1){\framebox(1,1){$\SC\rho$}}
\put(2,0){\framebox(1,1){$\SC\beta$}}
\put(3,1){\framebox(3,1){$\SC n-2$}}
\put(2,1){\framebox(1,1){$\SC\tau$}}
\end{picture} .
In Section~\ref{gluon-iv} we 
will consider the more easier tensor polynomial
$T^{(n,2)}_{(\alpha\beta) n}(x)= x^\rho x^\tau\, T^{(n,2)}_{[\alpha\rho][\beta\tau] n-2}(x)$
(see Eq.~(\ref{Gx_tw4_iv})).\\
The conditions of tracelessness for the reducible tensor polynomial are:
\begin{align}
\label{H3}
\square \tl T_{(\alpha \beta) n}(x) =0,
\quad
\pd^\alpha \tl T_{(\alpha \beta) n}(x) = 0,
\quad
g^{\alpha\beta}\tl T_{(\alpha \beta) n}(x) = 0.
\end{align}
Its general structure is as follows:
\begin{align}
\label{Proj6}
\tl T_{(\alpha\beta) n}(x)
=&\;
\Big\{\delta_\alpha^\mu\delta_\beta^\nu
+a_n g_{\alpha\beta} x^{(\nu}\pd^{\mu)}
+b_n x_{(\alpha}\delta_{\beta)}^{(\nu}\pd^{\mu)}
+c_n x_{(\alpha}x^{(\nu}\pd^{\mu)}\pd_{\beta)}
\nonumber\\
&+d_n x^2\delta_{(\alpha}^{(\nu}\pd_{\beta)}\pd^{\mu)}
+e_n x^2 x^{(\nu}\pd^{\mu)} \pd_{\alpha}\pd_{\beta}
+f_n x^2 x_{(\alpha}\pd_{\beta)} \pd^{\mu}\pd^{\nu}
\nonumber\\
&+g_n x_{\alpha}x_{\beta} \pd^{\mu}\pd^{\nu}
+k_n x^2 g_{\alpha\beta} \pd^{\mu}\pd^{\nu}
+l_n x^4 \pd_{\alpha}\pd_{\beta} \pd^{\mu}\pd^{\nu}
\Big\}
\breve{T}_{(\mu\nu) n}(x)
\end{align}
with the partially traceless polynomials
\begin{equation}
\breve{T}_{(\mu\nu) n}(x)=\left(\delta^\rho_\mu\delta^\sigma_\nu
-\frac{1}{2h}\, g_{\mu\nu}g^{\rho\sigma}\right)
{\cal P}^{(n)}_n(x)T_{(\rho\sigma) n}(x).
\end{equation}
It holds by construction 
\begin{equation}
\square \breve{T}_{(\mu\nu) n}(x)=0,\qquad 
g^{\mu\nu} \breve{T}_{(\mu\nu) n}(x)=0.
\end{equation}
From Eqs.~(\ref{H3}) the following system of linear equations
for the nine unknown coefficients results:
\begin{eqnarray*}
0 &=& (n-1)c_n+b_n+2h a_n\\
0 &=& d_n+g_n+(n-2) f_n+2h k_n\\
0 &=& 2(h+n-2)d_n+c_n+b_n\\
0 &=& 2(h+n-2)e_n+c_n\\
0 &=& 2(h+n-2)f_n+c_n+2g_n\\
0 &=& 2(h+n-2)k_n+a_n+g_n\\
0 &=& 4(h+n-3)l_n+f_n+e_n\\
0 &=& (2h+n)b_n+2(n-1)d_n +2a_n+2\\
0 &=& (2h+n)c_n+2d_n +2a_n+4(n-2)e_n\\
0 &=& (2h+n)f_n+2k_n+d_n+8(n-3)l_n+2e_n\\
0 &=& 2(2h+n-1)g_n+4k_n+c_n+2(n-2)f_n+b_n.
\end{eqnarray*}
Its unique solution is
\begin{eqnarray*}
a_n&=& \frac{h+n-2}{(h-1)(h+n)(2h+n-2)},  \\
b_n&=& - 2\,\frac{(h(h+n-1)-n+1)(h+n-2)}{(h-1)(h+n)(h+n-1)(2h+n-2)},   \\
c_n&=& - 2\,\frac{h+n-2}{(h-1)(h+n)(h+n-1)(2h+n-2)},  \\
d_n&=& \frac {h(h+n-1)-n+2}{(h-1)(h+n)(h+n-1)(2h+n-2)},  \\
e_n&=&\frac{1}{(h-1)(h+n)(h+n-1)(2h+n-2)},   \\
f_n&=& -\frac {h(h+n-4)-2(n-3)}{(h-1)(h+n)(h+n-1)(2h+n-2)(2h+n-3)},  \\
g_n&=&\frac{\big(h(h+n-2)-n+3\big)(h+n-2)}{(h-1)(h+n)(h+n-1)(2h+n-2)(2h+n-3)},\\ 
k_n&=&-\frac{h(h+n-3)-2n+3+\hbox{\large$\frac{1}{2}$}(h+n)(h+n-1)}{(h-1)(h+n)(h+n-1)(
2h+n-2)(2h+n-3)}, \\
l_n&=&\frac{h-3}{4(h-1)(h+n)(h+n-1)(2h+n-2)(2h+n-3)}.
\end{eqnarray*}
In the case $h=2$ these coefficients simplify; they are given as follows:
\begin{align*}
a_n&=\frac {n}{(n +2)^2}
&&b_n= - 2\, \frac{n(n + 3)}{(n + 1)(n + 2)^2},  \\
c_n&= - 2\, \frac {n}{(n + 1)(n + 2)^2}, 
&&d_n=\frac {4 + n}{(n + 1)(n + 2)^2}, \\
e_n&=\frac {1}{(n + 1)({n} + 2)^2}, 
&&f_n=-2\,\frac{1}{(n + 1)^2(n + 2)^2},\\ 
g_n&=\frac {n(n + 3)}{(n+ 1)^{2}(n + 2)^2},  
&&k_n=-\frac{1}{2}\frac{n^2+3n+4}{(n+1)^2(n +2)^2}, \\
l_n&=-\frac{1}{4}\frac {1}{({n} + 1)^2({n} + 2)^2}\,.  
\end{align*}
 
This finishes the determination of the harmonic tensor polynomials 
being necessary for the present study. In principle, the extension 
of the procedure to arbitrary tensors of higher order is obvious. 
However, the explicit computation,  without any additional information 
about their general properties, will be quite complicated. Let us 
remark that, to the best of our knowledge, such quantities have not 
been considered in the mathematical literature.

\section{Twist decomposition of a bilocal 2nd rank 
tensor operator}
\label{gluon}
\setcounter{equation}{0}
This Section is devoted to the twist decomposition of a general 
2nd rank light-cone tensor operator. 
Since the twist decomposition is independent from the 
chirality we demonstrate it only for the chiral-even gluon operator 
(\ref{Gtensor}). The twist decomposition of the chiral-odd gluon operator
Eq.~(\ref{Gtensor5}) is the same.
In addition,
we make the special choice $\ka = 0, \kb = \kappa$, thereby also 
simplifying the generalized
covariant derivatives, Eq.~(\ref{D_kappa}), to the usual ones: 
\begin{align}
\label{Gtensor0}
G_{\alpha\beta}(0,\kappa x)
=&\,
F_\alpha^{\ \rho}(0)U(0,\kappa x)F_{\beta\rho}(\kappa x)
=
\sum_{n=0}^\infty \frac{\kappa^n}{n!}x^{\mu_1} \ldots x^{\mu_n}
G_{\alpha\beta(\mu_1\ldots\mu_n)}\ ,
\\
\text{with}&\;\;\qquad G_{\alpha\beta(\mu_1\ldots\mu_n)}
=
F_\alpha^{\ \rho}(y) D^y_{(\mu_1}\ldots D^y_{\mu_n)}
F_{\beta\rho}(y)\big|_{y=0}\ ;
\nonumber
\end{align}
here the symbol $(\ldots)$ denotes symmetrization of the enclosed 
indices.

Physically, the forward matrix elements of the both gluon operators 
provide the gluon distribution functions~\cite{CS82,Manohar,Ji92}.  

The local tensor operators 
$G_{\alpha\beta(\mu_1\ldots\mu_n)} $ can be decomposed according to
the Young patterns (i) to (iv) and possible higher ones
(i.e.,~for $m_1+m_2 = n+2, m_2 \geq 3$):
\begin{align}
\label{G_loc}
G_{\alpha\beta(\mu_1\ldots\mu_n)}=
G_{\alpha\beta(\mu_1\ldots\mu_n)}^{\mathrm{(i)}}+
\alpha_{n+1}
G_{\alpha\beta(\mu_1\ldots\mu_n)}^{\mathrm{(ii)}}
+\beta_n
G_{\alpha\beta(\mu_1\ldots\mu_n)}^{\mathrm{(iii)}}
+\gamma_n
G_{\alpha\beta(\mu_1\ldots\mu_n)}^{\mathrm{(iv)}}
+ \ldots\, ,
\end{align}
with the (nontrivial) normalization coefficients 
$f_{[m]}/m!$ of the Young operators given by
$\alpha_{n+1} = 2(n+1)/(n+2)$ for $[m] = (n+1,1)$,
$\beta_n = 3n/(n+2)$ for $[m] = (n,1,1)$
and
$\gamma_n = 4(n-1)/(n+1)$ for $[m] = (n,2)$.
The corresponding Clebsch-Gordan series in terms of 
representations $(j_1,j_2)$ of the Lorentz group is
\begin{align}
&\hbox{\Large$\big(\frac{1}{2},\frac{1}{2}\big)$}
\otimes
\hbox{\Large$\big(\frac{1}{2},\frac{1}{2}\big)$}
\otimes
\left(
\hbox{\Large$\big(\frac{n}{2},\frac{n}{2}\big)$}
\oplus
\hbox{\Large$\big(\frac{n-2}{2},\frac{n-2}{2}\big)$}
\oplus
\ldots\right)\nonumber\\
&\qquad =
\hbox{\Large$\big(\frac{n+2}{2},\frac{n+2}{2}\big)$}
\oplus
\left(\hbox{\Large$\big(\frac{n+2}{2},\frac{n}{2}\big)$}
\oplus
\hbox{\Large$\big(\frac{n}{2},\frac{n+2}{2}\big)$}\right)
\oplus
\hbox{\Large$\big(\frac{n}{2},\frac{n}{2}\big)$}
\nonumber\\
\label{CGR1}
&\qquad\qquad 
\oplus
\left(\hbox{\Large$\big(\frac{n+2}{2},\frac{n-2}{2}\big)$}
\oplus
\hbox{\Large$\big(\frac{n-2}{2},\frac{n+2}{2}\big)$}\right)\nonumber\\
&\qquad\qquad
\oplus
\left(\hbox{\Large$\big(\frac{n}{2},\frac{n-2}{2}\big)$}
\oplus
\hbox{\Large$\big(\frac{n-2}{2},\frac{n}{2}\big)$}\right)
\oplus
\hbox{\Large$\big(\frac{n-2}{2},\frac{n-2}{2}\big)$}
\oplus\ldots\ ;
\end{align}
the corresponding tensor spaces will be denoted by ${\bf T} (j_1,j_2)$.
In the last line of Eq.~(\ref{CGR1}) such representations are listed down
which correspond to higher twist contributions contained in the trace
terms of symmetry classes (i) -- (iv).
The canonical (or scale) dimension $d$ of the local operator 
(\ref{G_loc}) is $n+4$, 
and the spin of the various contributions in (\ref{CGR1}) 
ranges from $n+2$ up to $1$ or $0$ if $n$ is even or odd,
respectively;
therefore, the local operators with well-defined twist are 
irreducible tensors of the Lorentz group or, equivalently,
of the group 
$SL(2,{\Bbb C})\times {\Bbb R}_+$, having scale dimension $d$.

In general, according to the spin content and the rank of the corresponding 
local tensor operators, the nonlocal operator (\ref{Gtensor0}) for arbitrary
$x$ contains contributions of twist ranging from $\tau = 2$
until $\tau=\infty$. 
However, after projection onto the light--cone,
$x \rightarrow {\tilde x}$, this infinite series terminates
at least at $\tau = 6$ since higher order terms are
proportional to $x^2$. 

In order to be more explicit, as well as for later use, we 
introduce the (anti)sym\-metrization with respect to $\alpha$ and
$\beta$ and we define the following nonlocal operators:
\begin{eqnarray}
\label{GT}
G^\pm_{\alpha\beta}(0,\kappa x) 
&:=& 
\hbox{$\frac{1}{2}$} \left(G_{\alpha\beta}(0,\kappa x) \pm 
G_{\beta\alpha}(0,\kappa x)\right)\ ,
\\
\label{GV}
G^\pm_{\alpha }(0,\kappa x)
 &:=& x^\beta\ G^\pm_{\alpha\beta}(0,\kappa x)
 \equiv G^\pm_{\alpha\bullet}(0,\kappa x)\ ,
\\
\label{GS}
G (0,\kappa x)
&:=&  x^\alpha x^\beta\ G^+_{\alpha\beta}(0,\kappa x)\ .
\end{eqnarray}
In fact, the twist decomposition of these bilocal light--ray 
operators reads:\footnote{
We use the notation
$A_{(\alpha\beta)}\equiv\frac{1}{2}(A_{\alpha\beta}+A_{\beta\alpha})$
and 
$A_{[\alpha\beta]}\equiv\frac{1}{2}(A_{\alpha\beta}-A_{\beta\alpha})$.
}
\begin{align} 
\label{G^+_t}
G^+_{\alpha\beta}(0,\kappa\tilde{x})
=&\phantom{+}
 G^{\mathrm{tw2}}_{(\alpha\beta)}(0,\kappa\tilde{x})
+G^{\mathrm{tw3}}_{(\alpha\beta)}(0,\kappa\tilde{x}) 
+G^{\mathrm{tw4}}_{(\alpha\beta)}(0,\kappa\tilde{x})
\nonumber\\
&\!+\!G^{\mathrm{tw5}}_{(\alpha\beta)}(0,\kappa\tilde{x}) 
+G^{\mathrm{tw6}}_{(\alpha\beta)}(0,\kappa\tilde{x})\,,
\\
\label{G^-_t}
G^-_{\alpha\beta}(0,\kappa\tilde{x})
=&\phantom{+}
G^{\mathrm{tw3}}_{[\alpha\beta]}(0,\kappa\tilde{x}) 
+G^{\mathrm{tw4}}_{[\alpha\beta]}(0,\kappa\tilde{x}) 
+G^{\mathrm{tw5}}_{[\alpha\beta]}(0,\kappa\tilde{x})\,, 
\\
\label{G^+_vec}
G^+_{\alpha}(0,\kappa\tilde{x})
=&\phantom{+}
 G^{\mathrm{tw2}}_{(\alpha\bullet)}(0,\kappa\tilde{x})
+G^{\mathrm{tw3}}_{(\alpha\bullet)}(0,\kappa\tilde{x}) 
+G^{\mathrm{tw4}}_{(\alpha\bullet)}(0,\kappa\tilde{x}) \,,
\\
\label{G^-_vec}
G^-_{\alpha}(0,\kappa\tilde{x})
=&\phantom{+}
 G^{\mathrm{tw3}}_{[\alpha\bullet]}(0,\kappa\tilde{x}) 
+G^{\mathrm{tw4}}_{[\alpha\bullet]}(0,\kappa\tilde{x})\,,
\\
\label{G_sca}
G(0,\kappa\tilde{x})
=&\phantom{+}
 G^{\mathrm{tw2}}(0,\kappa\tilde{x})\,.
\end{align}
The various twist contributions individually decompose further 
according to the symmetry classes which may contribute.
The explicit expressions for generic tensor operators
are given in the following subsections, where
it will be shown that Young patterns (i) and (ii) as well as (iv) 
contribute to the symmetric tensor operators and related vector
and scalar operators, and Young patterns (ii) and (iii) contribute to
the antisymmetric tensor operators and related vector 
operators. 

Fortunately, in the case of forward scattering one has to consider the 
symmetry type (i) in the polarized and unpolarized case, and additionally, 
the symmetry type (ii) for the polarized case. 
But in order to consider non-forward matrix elements all symmetry types are
involved.

\subsection{Tensor operators of symmetry class (i)}
\subsubsection{Construction of nonlocal symmetric class-(i) operators 
of definite twist}
Let us start with the simplest case of the totally symmetric traceless 
tensors, and their contractions with $x$, which have twist $\tau =2$ and
are contained in the tensor space ${\bf T} (\frac{n+2}{2},\frac{n+2}{2})$. 
They have symmetry class (i) and are uniquely characterized by the 
following standard tableau (with normalizing factor 1):
\\
\\
\unitlength0.5cm
\begin{picture}(30,1)
\linethickness{0.075mm}
\put(1,0){\framebox(1,1){$\alpha$}}
\put(2,0){\framebox(1,1){$\beta$}}
\put(3,0){\framebox(1,1){$\mu_1$}}
\put(4,0){\framebox(3,1){$\ldots$}}
\put(7,0){\framebox(1,1){$\mu_n$}}
\put(9.5,0){$\stackrel{\wedge}{=}$}
\put(11,0){$\relstack{\alpha\beta\mu_1\ldots\mu_n}{\cal S}
F_\alpha^{\ \rho}(0) D_{\mu_1}\ldots D_{\mu_n}F_{\beta\rho}(0)
- \mathrm{trace~terms~,}$}
\end{picture}
\\
\\
The irreducible local twist--2 operator reads
\begin{align}
G^{\mathrm{tw2(i)}}_{(\alpha\beta\mu_1\ldots\mu_n)}
=
F_{(\alpha}^{\, \rho}(0) D_{\mu_1}\ldots D_{\mu_n}F_{\beta)\rho}(0)
-\mathrm{trace~terms}\,.
\nonumber
\end{align}
Let us postpone the determination of the trace terms and make the 
resummation to the corresponding nonlocal operator in advance. 
This is obtained, at first, by contracting with $x^{\mu_1}\ldots x^{\mu_n}$ 
and rewriting in the following form:
\begin{eqnarray}
\label{2betan}
\hspace{-.2cm}
G^{\mathrm{tw2(i)}}_{(\alpha\beta) n}(x)
=
x^{\mu_1}\ldots x^{\mu_n}\
G^{\mathrm{tw2(i)}}_{(\alpha\beta\mu_1\ldots\mu_n)}
\label{O2_x}
=
\hbox{\Large$\frac{1}{(n+2)(n+1)}$}\,
\pd_\alpha\pd_\beta \tl G_{n+2}(x), 
\end{eqnarray}
with 
\begin{eqnarray}
\tl G_{n+2}(x) &=& G_{n+2}(x)-\mathrm{~trace~terms~,}
\nonumber\\
G_{n+2}(x) &\equiv& x^\mu x^\nu F_\mu^{\ \rho} (0) (xD)^n F_{\nu\rho}(0)\ .
\nonumber
\end{eqnarray}
Here, $\tl G_{n+2}(x)$ is a harmonic polynomial of order $n+2$, 
cf.~Eq.~(\ref{T_harm_d}):
\begin{align}
\label{Proj_tw2}
\hspace{-.3cm}
\tl G_{n+2} (x)
=
H_{n+2}^{(4)}\!\left(x^2|\square\right) G_{n+2}(x)
\equiv
\sum_{k=0}^{[\frac{n+2}{2}]}\frac{(n+2-k)!}{(n+2)!k!}
\!\left(\frac{-x^2}{4}\right)^{\!k}\!\square^k
G_{n+2}(x)\,.
\nonumber
\end{align}
Then, using $((n+2)(n+1))^{-1} = \int^1_0 \d\lambda 
(1-\lambda)\lambda^n$ and
the integral representation of Euler's beta function we obtain
the nonlocal {\em twist--2 tensor operator}:\footnote{
Here, and in the following, we omit the indication of the symmetry class
of the nonlocal twist--2 operators because only totally symmetric 
tensors contribute. However, the trace terms being of higher twist 
must be classified according to their symmetry type.}
\begin{equation}
\label{G_tw2_tens}
G^{\mathrm{tw2}}_{(\alpha\beta)} (0,\kappa x)
=
\sum_{n=0}^{\infty}
\frac{\kappa^n}{n!}\ 
G^{\mathrm{tw2(i)}}_{(\alpha\beta)n}(x)
=
\pd_\alpha\pd_\beta
\int_{0}^{1} \d\lambda (1-\lambda)\,
\tl G(0,\kappa\lambda x) ,
\end{equation}
with
\begin{equation}
\label{proj_tw2}
\tl G(0,\kappa x)
=
G(0,\kappa x)
+\sum_{k=1}^{\infty}\int_0^1\!\d t\, t
\left(\frac{-x^2}{4}\right)^{\!k}\!
\frac{\square^k}{k!(k-1)!}
\left(\frac{1-t}{t}\right)^{\! k-1}
\!G(0,\kappa tx).
\end{equation}
The operator $G^{\mathrm{tw2}}_{(\alpha\beta)} (0,\kappa x)$ 
satisfies the conditions of a harmonic tensor function:
\begin{equation}
\label{cond}
g^{\alpha\beta}G^{\mathrm{tw2}}_{(\alpha\beta)}(0,\kappa x)=0 ,
\quad
\square G^{\mathrm{tw2}}_{(\alpha\beta)}(0,\kappa x)=0 ,
\quad
\pd^\alpha G^{\mathrm{tw2}}_{(\alpha\beta)}(0,\kappa x)=0\, .
\end{equation}
\subsubsection{Reduction to vector and scalar operators}
Now, we specify to the nonlocal vector and scalar operators. 
Multiplying Eq.~(\ref{G_tw2_tens}) by $x^\beta$ and observing
the equality
$(x\pd_x) \tl G(0,\kappa\lambda x) = (\lambda \partial_\lambda +2) 
\tl G(0,\kappa\lambda x)$ gives the {\em twist--2 vector operator},
\begin{equation}
\label{G_tw2_v}
G^{\mathrm{tw2}}_{(\alpha\bullet)}(0,\kappa x)
=
\pd_\alpha\int_{0}^{1} \d\lambda\,\lambda\, \tl G(0,\kappa\lambda x),
\end{equation}
which satisfies the conditions
\begin{equation}
\label{cond2}
\square\ G^{\mathrm{tw2}}_{(\alpha\bullet)}(0,\kappa x)=0,
\quad
\pd^\alpha\ G^{\mathrm{tw2}}_{(\alpha\bullet)}(0,\kappa x)=0\,.
\end{equation}
Multiplying by $x^\alpha x^\beta$  gives the {\em twist--2 scalar 
operator},
\begin{equation}
\label{G_tw2_s}
G^{\mathrm{tw2}}(0,\kappa x) = \tl G(0,\kappa x)\,,
\end{equation}
which, by definition, satisfies the condition
\begin{equation}
\label{cond20}
\square G^{\mathrm{tw2}}(0,\kappa x)=0\,.
\end{equation}
The expression (\ref{proj_tw2}) for the scalar operator has already 
been given by Balitsky and Braun \cite{BB88}.

Comparing Eqs.~(\ref{G_tw2_tens}), (\ref{G_tw2_v}) and (\ref{G_tw2_s})
we may recognize that in the case of symmetry class (i) the tensor 
and vector operators are obtained from the scalar operator by very
simple operations. Furthermore, we observe how in the case of the
scalar operator the trace terms -- being proportional to $x^2$ --
are to be subtracted from $G(0,\kappa x)$ in order to make that
operator traceless. In the case of vector and tensor operators
such subtraction, because of the appearance of the derivatives,
is more complicated.
\subsubsection{Projection onto the light--cone}
Let us now project onto the light--cone and, at the same time,
also extend to the case of general values $(\ka, \kb)$.
Because of the derivatives $\pd_\alpha$ and $\pd_\beta$,
appearing in Eq.~(\ref{G_tw2_tens}), only the terms with $k=1,2$ 
in Eq.~(\ref{proj_tw2}) contribute. The final expression for the
{\em symmetric twist--2 light--cone tensor operator} is given by
\begin{align}
\hspace{-.3cm}
\label{Gtw2_gir}
G^{\mathrm{tw2}}_{(\alpha\beta)}(\ka\xx,\kb\xx)
=
&\,
\pd_\alpha\pd_\beta\!\int_{0}^{1}\!\d\lambda (1-\lambda)
G(\kappa_1\lambda x,\kappa_2\lambda x)\big|_{x=\tilde{x}}\!
-G_{(\alpha\beta)}^{\mathrm{>(i)}}(\kappa_1\lcx,\kappa_2\lcx),
\\
\hspace{-.3cm}
\label{G-high(i)}
G_{(\alpha\beta)}^{\mathrm{>(i)}}(\kappa_1\lcx,\kappa_2\lcx)
=&
\int_0^1\d\lambda\Big\{(1-\lambda+\lambda\ln\lambda)
\left(\hbox{\Large$\frac{1}{2}$}
g_{\alpha\beta}+x_{(\alpha}\pd_{\beta)}\right)\square
\\
\hspace{-.3cm}
&+\hbox{\Large$\frac{1}{4}$}\big(2(1-\lambda)+(1+\lambda)\ln\lambda\big)
x_\alpha x_\beta\square^2\Big\}
G(\kappa_1\lambda x,\kappa_2\lambda x)\big|_{x=\tilde{x}}.\nonumber
\end{align}
The operator $G_{(\alpha\beta)}^{\mathrm{>(i)}}(\kappa_1\xx,\kappa_2\xx)$
contains the {\em higher twist contributions} which have to be 
subtracted from the first term, Eq.~(\ref{Gtw2_gir}),
in order to make the whole expression traceless. 
In order to disentangle the different terms of well-defined twist
we observe that the tensor operator (\ref{G-high(i)}) 
consists of scalar and vector parts. 
Here, an operator being contained in the trace terms is 
called a scalar and vector part, if the expression 
multiplied by $g_{\alpha\beta}$, $x_\alpha$ or $x_\beta$ 
is a scalar, like $\square G(\ka x,\kb x)$ or $\pd^\mu G_\mu(\ka x,\kb x)$, 
and a vector, like $\pd_\beta G(\ka x,\kb x)$ or 
$\pd_\beta\pd^\mu G_\mu(\ka x,\kb x)$, respectively.\footnote{
Strictly speaking, $x_\alpha$ and $x_\beta$ as well as $g_{\alpha\beta}$ 
which are necessary for the dimension and tensor structure of the whole
expression do not belong to the higher twist operator itself.}
The scalar operators, $\square G(\ka x,\kb x)|_{x=\tilde{x}}$ and 
$\square^2 G(\ka x,\kb x)|_{x=\tilde{x}}$,
occurring in Eq.~(\ref{G-high(i)}) 
have already well-defined twist $\tau =4$ and $\tau =6$, respectively. 
To ensure that the vector operator 
$\pd_\beta \square G(\ka x,\kb x)|_{x=\tilde{x}}$ also obtains
well-defined twist, it is necessary to subtract its own trace terms 
which are of twist  $\tau=6$.

The higher twist operators contained in the trace terms of the 
twist--2 tensor operator are the following:
\begin{align}
G^{\mathrm{>(i)}}_{(\alpha\beta)}
(\kappa_1\tilde{x},\kappa_2\tilde{x})
&=
G^{\mathrm{tw4(i)a}}_{(\alpha\beta)}(\kappa_1\tilde{x},\kappa_2\tilde{x})+
G^{\mathrm{tw4(i)b}}_{(\alpha\beta)}(\kappa_1\tilde{x},\kappa_2\tilde{x})
\nonumber\\
&\qquad+ G^{\mathrm{tw6(i)a}}_{(\alpha\beta)}(\kappa_1\tilde{x},\kappa_2\tilde{x})
+G^{\mathrm{tw6(i)b}}_{(\alpha\beta)}(\kappa_1\tilde{x},\kappa_2\tilde{x}),
\end{align}
with
\begin{align}
G_{(\alpha\beta)}^{\mathrm{tw4(i)a}}(\kappa_1\lcx,\kappa_2\lcx)
=&\,\hbox{\Large$\frac{1}{2}$}\,g_{\alpha\beta}\square
\int_0^1\d\lambda(1-\lambda+\lambda\ln\lambda)
G(\kappa_1\lambda x,\kappa_2\lambda x)\big |_{x=\tilde{x}}\,,
\nonumber\\
G_{(\alpha\beta)}^{\mathrm{tw4(i)b}}(\kappa_1\lcx,\kappa_2\lcx)
=&\, x_{(\alpha}\pd_{\beta)}\square
\int_0^1\d\lambda(1-\lambda+\lambda\ln\lambda)
G(\kappa_1\lambda x,\kappa_2\lambda x)\big |_{x=\tilde{x}}
-G_{(\alpha\beta)}^{\mathrm{tw6(i)b}}(\kappa_1\lcx,\kappa_2\lcx),
\nonumber
\end{align}
and
\begin{align}
G_{(\alpha\beta)}^{\mathrm{tw6(i)a}}(\kappa_1\lcx,\kappa_2\lcx)
=&
\hbox{\Large$\frac{1}{4}$}\,x_\alpha x_\beta\square^2\!
\int_0^1\!\d\lambda
\big( 2(1\!-\!\lambda)+(1\!+\!\lambda)\ln\lambda\big)
G(\kappa_1\lambda x,\kappa_2\lambda x)\big |_{x=\tilde{x}}\,,
\nonumber\\
G_{(\alpha\beta)}^{\mathrm{tw6(i)b}}(\kappa_1\lcx,\kappa_2\lcx)
=&\,
\hbox{\Large$\frac{1}{4}$}\,x_\alpha x_\beta\square^2\!
\int_0^1\!\d\lambda
\Big( \hbox{\Large$\frac{(1-\lambda)^2}{\lambda}$}
-\hbox{\Large$\frac{1-\lambda^2}{2\lambda}$}-\lambda\ln\lambda\Big)
G(\kappa_1\lambda x,\kappa_2\lambda x)\big|_{x=\tilde{x}}\,.
\nonumber
\end{align}
Obviously, by inspection of the spin content of the symmetry type (i),
the local twist--4 and twist--6 operators carry the 
representation ${\bf T} (\frac{n}{2},\frac{n}{2})$ and 
${\bf T} (\frac{n-2}{2},\frac{n-2}{2})$, respectively. 

Finally,
completing the twist decomposition of the symmetric nonlocal tensor 
operator, we write down the symmetric twist--4 light--cone
tensor operator contained in the expression (\ref{G-high(i)}) 
\begin{align}
G_{(\alpha\beta)}^{\mathrm{tw4(i)}}(\kappa_1\lcx,\kappa_2\lcx)
&=G_{(\alpha\beta)}^{\mathrm{tw4(i)a}}(\kappa_1\lcx,\kappa_2\lcx)
+G_{(\alpha\beta)}^{\mathrm{tw4(i)b}}(\kappa_1\lcx,\kappa_2\lcx)
\\
&=
\int_0^1\d\lambda\Big\{(1-\lambda+\lambda\ln\lambda)
\left(\hbox{\Large$\frac{1}{2}$}\,
g_{\alpha\beta}+x_{(\alpha}\pd_{\beta)}\right)\square\nonumber\\
&\qquad
-\hbox{\Large$\frac{1}{4}$}
\Big(\hbox{\Large$\frac{(1-\lambda)^2}{\lambda}$}
-\hbox{\Large$\frac{1-\lambda^2}{2\lambda}$}-\lambda\ln\lambda\Big)
x_\alpha x_\beta\square^2\Big\}
\left.G(\kappa_1\lambda x,\kappa_2\lambda x)\right|_{x=\tilde{x}}\,,
\nonumber
\end{align}
whereas the symmetric twist--6 light--cone operator contained 
in (\ref{G-high(i)}) reads
\begin{align}
G_{(\alpha\beta)}^{\mathrm{tw6(i)}}(\kappa_1\lcx,\kappa_2\lcx)
&=G_{(\alpha\beta)}^{\mathrm{tw6(i)a}}(\kappa_1\lcx,\kappa_2\lcx)
+G_{(\alpha\beta)}^{\mathrm{tw6(i)b}}(\kappa_1\lcx,\kappa_2\lcx)\\
&=
\hbox{\Large$\frac{1}{4}$}\,x_\alpha x_\beta\square^2
\int_0^1\d\lambda
\Big( \hbox{\Large$\frac{1-\lambda^2}{2\lambda}$}+\ln\lambda\Big)
\left.G(\kappa_1\lambda x,\kappa_2\lambda x)\right|_{x=\tilde{x}}\nonumber\, .
\end{align}
\subsubsection{Vector and scalar light--ray operators}
Contracting Eq.~(\ref{Gtw2_gir}) with $\lcx^\beta$, making use of formula
$(x\pd_x)f(\lambda x)=\lambda\pd_\lambda f(\lambda x)$ and performing the 
partial integrations we obtain the final version of the {\em twist--2 
light--cone vector operator},
\begin{equation}
\label{G_tw2_vec}
G^{\mathrm{tw2}}_{(\alpha\bullet)}(\kappa_1\lcx,\kappa_2\lcx)
=\int_0^1\d\lambda\, \lambda
\Big[\pd_\alpha+\hbox{\Large$\frac{1}{2}$}(\ln\lambda)x_\alpha \square\Big]
G(\kappa_1\lambda x,\kappa_2\lambda x)\big |_{x=\tilde{x}}\,,
\end{equation}
and the twist--4 light--cone vector operator,
\begin{equation}
\label{G4i}
G^{\mathrm{tw4(i)}}_{(\alpha\bullet)}(\kappa_1\lcx,\kappa_2\lcx)=
-\hbox{\Large$\frac{1}{2}$}\,x_\alpha \square
\int_0^1\d\lambda\, \lambda(\ln\lambda)
G(\kappa_1\lambda x,\kappa_2\lambda x)\big |_{x=\tilde{x}}\, .
\end{equation}

In order to obtain the {\em scalar twist--2 light--ray operator} 
we multiply (\ref{G_tw2_vec}) by $\lcx^\alpha$. Then, 
the twist--4 part vanishes and the remaining twist--2 operator
restores the scalar operator (compare Eq.~(\ref{G_sca})), 
\begin{equation}
G^{\mathrm{tw2}}(\kappa_1\lcx,\kappa_2\lcx)
=
G(\kappa_1\lcx,\kappa_2\lcx)\,.
\end{equation}
Let us point to the fact that the trace of the original gluon
tensor, $g^{\alpha\beta} G_{\alpha\beta}(\ka\xx,\kb\xx)$, is a
twist--4 scalar operator. It is contained in 
$G^{\rm tw4(i)a}_{(\alpha\beta)}(\ka\xx,\kb\xx)$
as well as similar expressions occurring below, cf.~also 
Eq.~(\ref{G_trace}).
\subsubsection{Condition of tracelessness on the cone}
Finally, it should be remarked, that the conditions (\ref{cond}),
(\ref{cond2}) and (\ref{cond20}),
if translated into the corresponding ones containing derivatives 
with respect to $\xx$, 
no longer hold for the light--cone operators. 
This is clear because, by projecting onto the light--cone, part 
of the original structure of the operators has been lost.
Nevertheless, the conditions of tracelessness of the 
light--cone operators may be formulated by using the 
{\em interior derivative} on the light--cone \cite{BT77,Dobrev77,GL99c}
which has been extensively used for the construction of local
conformal operators \cite{Dobrev76a,Dobrev82}.
In four dimensions it is given by
\begin{align}
\d_\alpha\equiv
\big(1+\lcx\lcd\big)\lcd_\alpha
-\hbox{\Large$\frac{1}{2}$}\,
\lcx_\alpha\lcd^2\quad\text{with}\quad
\lcd_\alpha\equiv\frac{\pd}{\pd\lcx^\alpha}\,,
\nonumber
\end{align}
and has the following properties (see Section~\ref{tensor})
\begin{align}
\d^2 = 0\, ,\quad
[\d_\alpha,\d_\beta]=0
\quad\text{and}\quad
\d_\alpha\lcx^2=\lcx^2\big(\d_\alpha+2\lcd_\alpha\big)\,.
\nonumber
\end{align}
Then, the conditions of tracelessness simplify, namely, they read:
\begin{align}
g^{\alpha\beta} 
G^{\mathrm{tw2}}_{(\alpha\beta)}(\ka\xx,\kb\xx)=0\,,
\qquad
\d^\alpha G^{\mathrm{tw2}}_{(\alpha\beta)}(\ka\xx,\kb\xx)=0\,,
\nonumber
\end{align}
as well as
\begin{align}
\d^\alpha G^{\mathrm{tw2}}_{(\alpha\bullet)}(\ka\xx,\kb\xx)=0\,.
\nonumber
\end{align}

Analogous conditions hold for the light--cone operators of
definite twist in the case of symmetry classes (ii) -- (iv)
obtained below. 

We remark that the interior derivative
may be used in defining more directly totally symmetric
local light--cone operators. For example it holds 
(see (\ref{T^[n]_alpha(lcx)}) and (\ref{T^[n]_mu_l (lcx)}))
\begin{align}
G^{\mathrm{tw2(i)}}_{(\alpha\beta)n} (\xx)
&=
\hbox{\Large$\frac{1}{(n+2)^2(n+1)^2}$}\,
\d_\alpha\d_\beta G_{n+2}(\xx)\,,
\nonumber\\
G^{\mathrm{tw2(i)}}_{(\alpha\bullet)n} (\xx)
&=
\hbox{\Large$\frac{1}{(n+2)^2}$}\,
\d_\alpha G_{n+2}(\xx)\,. 
\nonumber
\end{align}
However, in case of symmetry types (ii) -- (iv)  
the corresponding expressions should be more complicated 
(see Section~\ref{tensor}).

\subsection{Tensor operators of symmetry class (ii)}
\subsubsection{Construction of nonlocal (anti)symmetric class-(ii)
operators of definite twist: Young tableau A}
Now we consider tensor operators, and their contractions with $x$,
having symmetry class (ii) and whose local twist--3 parts are contained
in ${\bf T}\hbox{$(\frac{n+2}{2},\frac{n}{2})
\oplus {\bf T}(\frac{n}{2},\frac{n+2}{2})$}$.  Contrary to the 
totally symmetric case we have different possibilities to put 
the tensor indices into the corresponding Young pattern. Without
presupposing any symmetry of indices $\alpha$ and
$\beta$ we should start with the following Young tableau:
\\
\\
\unitlength0.5cm
\begin{picture}(30,1)
\linethickness{0.075mm}
\put(1,-1){\framebox(1,1){$\alpha$}}
\put(1,0){\framebox(1,1){$\mu_1$}}
\put(2,0){\framebox(1,1){$\mu_2$}}
\put(3,0){\framebox(3,1){$\ldots$}}
\put(6,0){\framebox(1,1){$\mu_n$}}
\put(7,0){\framebox(1,1){$\beta$}}
\put(8.5,0){$\stackrel{\wedge}{=}$}
\put(9.5,0)
{${\hbox{\Large$\frac{2(n+1)}{n+2}$}}
\relstack{\alpha\mu_1}{\cal A}\;
\relstack{\beta\mu_1\ldots\mu_n}{\cal S} \!
F_\alpha^{\ \rho}(0)D_{\mu_1}\ldots D_{\mu_n}F_{\beta\rho}(0)
\! - \!\mathrm{trace~terms,} $}
\end{picture}
\\ \\
with normalizing factor $\alpha_{n+1}$. (The tableau with 
$\alpha\leftrightarrow\beta$ will be considered thereafter.)
Denoting this
symmetry behaviour by (iiA) we may write the local twist-3 tensor
operator as follows:
\begin{align}
G^{\mathrm{tw3(iiA)}}_{\alpha\beta\mu_1\ldots\mu_n}
&=\;
\hbox{\Large$\frac{1}{n+2}$}
\Big\{
F_\alpha^{\ \rho}(0)D_{(\mu_1}\ldots D_{\mu_n)}F_{\beta\rho}(0)
+
F_{\alpha}^{\ \rho}(0)D_{\beta} D_{(\mu_2}\ldots
D_{\mu_n}F_{\mu_1)\rho}(0)
\nonumber\\
& \quad
+\!\sum\limits_{l=2}^{n}\!
F_\alpha^{\ \rho}(0)
D_{(\mu_1}\ldots D_{\mu_{l-1}}D_{|\beta|} D_{\mu_{l+1}}
\ldots D_{\mu_n } F_{\mu_l)\rho}(0)
- (\alpha \leftrightarrow \mu_1)
\Big\}-\text{trace terms}.
\nonumber
\end{align}
Proceeding in the same manner as in the last Subsection we 
multiply by $x^{\mu_1}\ldots x^{\mu_n}$ and obtain:
\begin{align}
G^{\mathrm{tw3(iiA)}}_{\alpha\beta n}(x)
&=\hbox{\Large$\frac{2}{(n+2)n}$}\, x^\nu
\delta^\mu_{[\alpha} \pd_{\nu]}\pd_\beta \tl G_{\mu\bullet|n+1}(x) 
\\
&=\hbox{\Large$\frac{1}{(n+2)n}$}
\Big(\delta_\alpha^\mu (x\pd)-x^\mu\pd_\alpha\Big)
\pd_\beta \tl G_{\mu\bullet|n+1}(x) 
\nonumber
\end{align}
with
\begin{align}
\tl G_{\mu\bullet|n+1}(x)
\equiv x^\nu \tl G_{\mu\nu|n}(x)
= \Big\{\delta^\alpha_\mu - 
\hbox{\Large$\frac{1}{(n+2)^2}$}
\Big[x_\mu\pd^\alpha (x\pd) -
\hbox{\Large$\frac{1}{2}$}x^2\pd_\mu\pd^\alpha\Big]\Big\}
H^{(4)}_{n+1}(x^2|\square)G_{\alpha\bullet|n+1}(x) \,.
\nonumber
\end{align}
Here, $\tl G_{\mu\bullet|n+1}(x)$ is the harmonic vector polynomial
of order $n+1$
(see, Eq.~(\ref{ha_ve_po}), Section~\ref{tensor})
 and the symmetry behaviour of class (ii) is obtained 
through the differential operator in front of it.
Now, using 
$((n+2)n)^{-1}=\int_0^1\d\lambda(1-\lambda^2)\lambda^n/(2\lambda)$
and Euler's beta function we sum up to obtain the following nonlocal 
twist--3 tensor operator:
\begin{align}
\label{G_tw3_nl_1}
G^{\mathrm{tw3(iiA)}}_{\alpha\beta} (0,\kappa x)
&=\int_{0}^{1} \d\lambda\,\hbox{\Large$\frac{1-\lambda^2}{2\lambda}$}
\Big(\delta_\alpha^\mu(x\pd)-x^\mu\pd_\alpha\Big)
\pd_\beta \tl G_{\mu\bullet}(0,\kappa \lambda x)\,,
\end{align}
with the nonlocal traceless (vector) operator
\begin{align}
\label{Gv_nl_tl}
\tl G_{\alpha\bullet}(0,\kappa x)
&=G_{\alpha\bullet}(0,\kappa x)
+\sum_{k=1}^\infty\int_0^1\!\d t
\left(\frac{-x^2}{4}\right)^{\!k}\!
\frac{\square^k}{k!(k-1)!}
\left(\frac{1-t}{t}\right)^{\!k-1}\!
G_{\alpha\bullet}(0,\kappa t x)\\
\label{Gv_nl_tl}
&-\big[ x_\alpha\pd^\mu(x\pd)
-\hbox{\Large$\frac{1}{2}$} x^2\pd_\alpha\pd^\mu\big]
\sum_{k=0}^\infty
\int_0^1\!\d\tau\tau \int_0^1\!\d t\,t
\left(\frac{-x^2}{4}\right)^{\!k}
\frac{\square^k}{k!k!}\left(\frac{1-t}{t}\right)^{\!k}\!
G_{\mu\bullet}(0,\kappa\tau t x).\nonumber
\end{align}

The following decomposition into symmetric and antisymmetric 
part
is useful for the further calculations
\begin{equation}
\label{G3A}
G^{\mathrm{tw3(iiA)}}_{\alpha\beta} (0,\kappa x)=
G^{\mathrm{tw3(iiA)}}_{[\alpha\beta]} (0,\kappa x)+
G^{\mathrm{tw3(iiA)}}_{(\alpha\beta)} (0,\kappa x)\, ,
\end{equation}
with
\begin{align}
\label{G3as}
G^{\mathrm{tw3(iiA)}}_{[\alpha\beta]} (0,\kappa x)
&=
\int_0^1\d\lambda\, \lambda\delta^\mu_{[\alpha}\pd_{\beta]} 
\tl G_{\mu\bullet}(0,\kappa \lambda x)\,,
\\
\label{G3sym}
G^{\mathrm{tw3(iiA)}}_{(\alpha\beta)} (0,\kappa x)
&=
\int_0^1\d\lambda\, \hbox{\Large$\frac{1}{\lambda}$}
\Big(\delta^\mu_{(\alpha}\pd_{\beta)}
-\hbox{\Large$\frac{1-\lambda^2}{2}$}
\pd_\alpha\pd_\beta x^\mu\Big) 
\tl G_{\mu\bullet}(0,\kappa \lambda x)\\
&=
\int_{0}^{1} \d\lambda\,\hbox{\Large$\frac{1-\lambda^2}{2\lambda}$}
\Big(\delta_{(\alpha}^\mu(x\pd)-x^\mu\pd_{(\alpha}\Big)
\pd_{\beta)} \tl G_{\mu\bullet}(0,\kappa \lambda x)\,.\nonumber
\end{align}

Again, these operators are harmonic tensor functions.
The conditions of tracelessness for the tensor operator are
\begin{eqnarray}
\label{M2harm}
g^{\alpha\beta} G^{\mathrm{tw3(iiA)}}_{\alpha\beta}(0,\kappa x) = 0\,,
&\quad&
\square G^{\mathrm{tw3(iiA)}}_{\alpha\beta}(0,\kappa x) = 0\,,
\nonumber\\
\pd^\alpha G^{\mathrm{tw3(iiA)}}_{\alpha\beta}(0,\kappa x) =0\,,
&\quad&
\pd^\beta G^{\mathrm{tw3(iiA)}}_{\alpha\beta}(0,\kappa x) =0\,.
\end{eqnarray}
\subsubsection{Reduction to vector operators}
The corresponding twist--3 vector operator is obtained 
from Eq.~(\ref{G_tw3_nl_1}) by multiplication by $x^\beta$: 
\begin{equation}
\label{Mab2}
G^{\mathrm{tw3(iiA)}}_{\alpha\bullet}(0,\kappa x)
=
 \int_{0}^{1} \d\lambda\,\lambda
\Big(\delta_\alpha^\mu(x\pd)-x^\mu\pd_\alpha\Big)
\tl G_{\mu\bullet}(0,\kappa\lambda x).
\end{equation}
Obviously, a corresponding scalar operator does not exist.

This vector operator fulfils the following conditions of tracelessness
\begin{eqnarray}
\square G_{\alpha\bullet}^{\rm tw3(iiA)}(0,\kappa x) =0\,,
\qquad
\pd^\alpha G_{\alpha\bullet}^{\rm tw3(iiA)}(0,\kappa x)=0\,.
\end{eqnarray}
\subsubsection{Projection onto the light--cone}
(a)~~The calculation of the {\em antisymmetric} tensor operator
$G^{\mathrm{tw3(iiA)}}_{[\alpha\beta]}(\ka\xx,\kb\xx)$ 
on the light--cone is similar to 
that of $M^{\mathrm{tw2}}_{[\alpha\beta]}(\ka\xx,\kb\xx)$
in~\cite{Lazar98,GLR99}; for the details we refer to it. The resulting expression is:
\begin{eqnarray}
\label{M_tw2_ir}
\hspace{-.5cm}
G^{\mathrm{tw3(iiA)}}_{[\alpha\beta]} (\ka\xx,\kb\xx)
\!&=&\!
 \int_{0}^{1}\!\d\lambda\,\lambda
\delta^\mu_{[\alpha}\pd_{\beta]}
G_{\mu\bullet}(\ka\lambda\xx,\kb\lambda\xx)
\big|_{x=\tilde{x}}
\!-\!
G^{\mathrm{>(iiA)}}_{[\alpha\beta]} (\ka\xx,\kb\xx),
\\
\label{M_hi.tw_a}
\hspace{-.5cm}
G^{\mathrm{>(iiA)}}_{[\alpha\beta]} (\ka\xx,\kb\xx)
\!&=&\!\
\hbox{\Large$\frac{1}{2}$}
 \int_{0}^{1}\!\d\lambda\,(1-\lambda)
\Big\{\big(2 x_{[\alpha}\pd_{\beta]}\pd^\mu
- 
x_{[\alpha}\delta_{\beta]}^\mu\square\big)
G_{\mu\bullet}(\ka\lambda x,\kb\lambda x) \big|_{x=\tilde{x}}
\nonumber\\
&& \!
-(1-\lambda+\lambda\ln\lambda)x_{[\alpha}\pd_{\beta]}\square 
G(\ka\lambda x,\kb\lambda x)\big|_{x=\tilde{x}}\Big\}\,.
\end{eqnarray}
$G^{\mathrm{>(iiA)}}_{[\alpha\beta]} (\ka\xx,\kb\xx)$ 
contains twist--4 and twist--5 contributions, but
twist--6 contributions do not appear due to $x_{[\alpha}x_{\beta]}=0$.
The higher twist operator (\ref{M_hi.tw_a}) contains two
vector operators multiplied by $x_\alpha$ and $x_\beta$, respectively.
For their twist decomposition one has to take into account Young pattern (i) 
as well as (ii). The procedure is analogous to the decomposition 
of the vector operator $O_\alpha(\ka\xx,\kb\xx)$ made
in~\cite{Lazar98,GLR99}. After a straightforward calculation we obtain
\begin{align}
G^{\mathrm{>(iiA)}}_{[\alpha\beta]}
(\kappa_1\tilde{x},\kappa_2\tilde{x})
=
G^{\mathrm{tw4(iiA)a}}_{[\alpha\beta]}(\kappa_1\tilde{x},\kappa_2\tilde{x})+
G^{\mathrm{tw4(iiA)b}}_{[\alpha\beta]}(\kappa_1\tilde{x},\kappa_2\tilde{x})
+G^{\mathrm{tw5(iiA)a}}_{[\alpha\beta]}(\kappa_1\tilde{x},\kappa_2\tilde{x})
\end{align}
with
\begin{eqnarray}
\hspace{-.5cm}
\label{G_tw4_iiA}
G^{\mathrm{tw4(iiA)a}}_{[\alpha\beta]}
(\kappa_1\tilde{x},\kappa_2\tilde{x})
\!&=&\!
\hbox{\Large$\frac{1}{2}$}\,
x_{[\alpha}\pd_{\beta]}\pd^{\mu} 
\int_{0}^{1}\d\lambda\,
\hbox{\Large$\frac{1-\lambda^2}{\lambda}$}
G_{\mu\bullet}(\kappa_1\lambda x,\kappa_2\lambda x)
\big|_{x=\xx}\,,
\nonumber\\
\hspace{-.5cm}
G^{\mathrm{tw4(iiA)b}}_{[\alpha\beta]}
(\kappa_1\tilde{x},\kappa_2\tilde{x})
\!&=&\!
- \hbox{\Large$\frac{1}{2}$}\,x_{[\alpha}\pd_{\beta]}\square
\int_{0}^{1}\d\lambda
\Big(
\hbox{\Large$\frac{1-\lambda^2}{2\lambda}$}+\lambda\ln\lambda
\Big)
G(\kappa_1\lambda x,\kappa_2\lambda x)
\big|_{x=\xx}\,,
\nonumber\\
\hspace{-.5cm}
G^{\mathrm{tw5(iiA)a}}_{[\alpha\beta]}
(\kappa_1\tilde{x},\kappa_2\tilde{x})
\!&=&\!
\label{M_tw5_ii}
- 
x_{[\alpha}\big(\delta_{\beta]}^{\mu}(x\pd)
-x^{\mu}\pd_{\beta]}\big)\square\! 
\int_{0}^{1}\!\d\lambda \,
\hbox{\Large$\frac{(1-\lambda)^2}{4\lambda}$}
G_{\mu\bullet}(\kappa_1\lambda x,\kappa_2\lambda x)
\big|_{x=\xx}\,.
\nonumber
\end{eqnarray}
(b)~~Now, we determine the {\em symmetric} tensor operator
$G^{\mathrm{tw3(iiA)}}_{(\alpha\beta)}(\ka\xx,\kb\xx)$
on the light--cone. Putting
Eq.~(\ref{Gv_nl_tl}) into (\ref{G3sym}), after some lengthy 
but straightforward calculation (taking into account  only the 
relevant terms of the $k$-summation and performing some
partial integrations), we get the following result:
\begin{align}
G^{\mathrm{tw3(iiA)}}_{(\alpha\beta)} (\ka\xx,\kb\xx)
=
\int_0^1\d\lambda \,
\hbox{\Large$\frac{1-\lambda^2}{2\lambda}$}
\Big(\delta^\mu_{(\alpha}(x\pd)
-x^\mu\pd_{(\alpha}\Big)\pd_{\beta)} 
G_{\mu\bullet} (\lambda x,\kb \lambda x)\big|_{x=\xx}
-G^{\mathrm{>(iiA)}}_{(\alpha\beta)} (\ka\xx,\kb\xx)\, ,
\end{align}
with the higher twist contributions of the trace terms
\begin{align}
\label{G_high_(ii)_sy}
G^{\mathrm{>(iiA)}}_{(\alpha\beta)}& (\ka\xx,\kb\xx)
=\int_{0}^{1}\!\d\lambda\Big\{\Big[
\hbox{\Large$\frac{1-\lambda^2}{2\lambda}$}
g_{\alpha\beta}\pd^\mu
+\hbox{\Large$\frac{1-\lambda}{2\lambda}$}
\delta^\mu_{(\alpha}x_{\beta)}\square
+(1-\lambda)x_{(\alpha}\pd_{\beta)}\pd^\mu\nonumber\\
&
-\hbox{\Large$\frac{(1-\lambda)^2}
{4\lambda}$}x_\alpha x_\beta\pd^\mu\square
\Big] G_{\mu\bullet} (\ka\lambda\xx,\kb\lambda\xx)\nonumber\\
&-\Big[\hbox{\Large$\frac{1}{2}$}\Big(
\hbox{\Large$\frac{1-\lambda^2}{2\lambda}$}
+\lambda\ln\lambda\Big)g_{\alpha\beta}\square
+\Big(\hbox{\Large$\frac{1}{2}$}(1-\lambda)+
\hbox{\Large$\frac{1-\lambda^2}{4\lambda}$}
+\lambda\ln\lambda\Big)
x_{(\alpha}\pd_{\beta)}\square\nonumber\\
&+\hbox{\Large$\frac{1}{4}$}\Big(
\hbox{\Large$\frac{1-\lambda^2}{2\lambda}$}
-\hbox{\Large$\frac{(1-\lambda)^2}{\lambda}$}+\lambda\ln\lambda\Big)
x_\alpha x_\beta\square^2\Big] G(\ka\lambda x,\kb\lambda x)
\Big\}\Big|_{x=\tilde{x}}.
\end{align}
It is obvious that Eq.~(\ref{G_high_(ii)_sy}) contains scalar and vector
operators.
Again, using Young patterns (i) and (ii) and subtracting the trace terms, 
we can decompose the vector part appearing in Eq.~(\ref{G_high_(ii)_sy}) 
into twist--4, twist--5 and twist--6 operators. This procedure is analogous
to the twist decomposition of the (vector) quark operator 
$O_\alpha(\ka\lcx,\kb\lcx)$ in~\cite{Lazar98,GLR99}I.
From the twist--4 and twist--5 vector part we have to subtract their trace terms 
being of twist--6  and add it to the other twist--6 scalar operator.
Let us recall that the scalar twist--4 and twist--6 operators are already 
traceless on the cone.
In that manner we get the following decomposition:
\begin{align}
G^{\mathrm{>(iiA)}}_{(\alpha\beta)} (\ka\xx,\kb\xx)
&=\!
G^{\mathrm{tw4(iiA)a}}_{(\alpha\beta)} (\ka\xx,\kb\xx)\!+\!
G^{\mathrm{tw4(iiA)b}}_{(\alpha\beta)} (\ka\xx,\kb\xx)\!+\!
G^{\mathrm{tw4(iiA)c}}_{(\alpha\beta)} (\ka\xx,\kb\xx)
\nonumber\\
&\!\!\!\!\!\!\!\!\!+
G^{\mathrm{tw4(iiA)d1}}_{(\alpha\beta)} (\ka\xx,\kb\xx)+
G^{\mathrm{tw4(iiA)d2}}_{(\alpha\beta)} (\ka\xx,\kb\xx)+
G^{\mathrm{tw4(iiA)e}}_{(\alpha\beta)} (\ka\xx,\kb\xx)
\nonumber\\
&\!\!\!\!\!\!\!\!\!+
G^{\mathrm{tw5(iiA)d}}_{(\alpha\beta)} (\ka\xx,\kb\xx)+
G^{\mathrm{tw6(iiA)a}}_{(\alpha\beta)} (\ka\xx,\kb\xx)+
G^{\mathrm{tw6(iiA)b}}_{(\alpha\beta)} (\ka\xx,\kb\xx)
\nonumber\\
&\!\!\!\!\!\!\!\!\!+
G^{\mathrm{tw6(iiA)c}}_{(\alpha\beta)}(\ka\xx,\kb\xx)+
G^{\mathrm{tw6(iiA)d1}}_{(\alpha\beta)} (\ka\xx,\kb\xx)+
G^{\mathrm{tw6(iiA)d2}}_{(\alpha\beta)} (\ka\xx,\kb\xx)
\nonumber\\
&\!\!\!\!\!\!\!\!\!+
G^{\mathrm{tw6(iiA)d3}}_{(\alpha\beta)} (\ka\xx,\kb\xx)+
G^{\mathrm{tw6(iiA)d4}}_{(\alpha\beta)} (\ka\xx,\kb\xx)+
G^{\mathrm{tw6(iiA)e}}_{(\alpha\beta)} (\ka\xx,\kb\xx),
\end{align}
with  
\begin{align}
G_{(\alpha\beta)}^{\mathrm{tw4(iiA)a}}(\ka\xx,\kb\xx)
=&-\hbox{\Large$\frac{1}{2}$} g_{\alpha\beta}\square
\int_0^1\d\lambda
\Big(
\hbox{\Large$\frac{1-\lambda^2}{2\lambda}$}
+\lambda\ln\lambda\Big)
G(\ka\lambda x,\kb\lambda x)\big|_{x=\tilde{x}}\,,
\nonumber\\
G_{(\alpha\beta)}^{\mathrm{tw4(iiA)b}}(\ka\xx,\kb\xx)
=&\,- \hbox{\Large$\frac{1}{2}$} x_{(\alpha}\pd_{\beta)}\square
\int_0^1\d\lambda\Big(
(1-\lambda)+\hbox{\Large$\frac{1-\lambda^2}{2\lambda}$}
+2\lambda\ln\lambda\Big)
G(\ka\lambda x,\kb\lambda x)\big|_{x=\tilde{x}}
\nonumber\\&
-G_{(\alpha\beta)}^{\mathrm{tw6(iiA)b}}(\ka\xx,\kb\xx)\,,
\nonumber\\
G_{(\alpha\beta)}^{\mathrm{tw4(iiA)d1}}(\ka\xx,\kb\xx)
=&\,- \hbox{\Large$\frac{1}{2}$} x_{(\alpha}\pd_{\beta)}\square
\int_0^1\d\lambda\Big(
\hbox{\Large$\frac{1-\lambda}{\lambda}$}
+\hbox{\Large$\frac{\ln\lambda}{\lambda}$}\Big)
G(\ka\lambda x,\kb\lambda x)\big|_{x=\tilde{x}}
\nonumber\\&
-G_{(\alpha\beta)}^{\mathrm{tw6(iiA)d1}}(\ka\xx,\kb\xx)\,,
\nonumber
\end{align}
and
\begin{align}
G_{(\alpha\beta)}^{\mathrm{tw6(iiA)a}}(\ka\xx,\kb\xx)
=&
\hbox{\Large$\frac{1}{4}$}x_\alpha x_\beta\square^2\!
\int_0^1\!\d\lambda
\Big( 
\hbox{\Large$\frac{(1-\lambda)^2}{\lambda}$}
- \hbox{\Large$\frac{1-\lambda^2}{2\lambda}$}
- \lambda\ln\lambda\Big)\!
G(\ka\lambda x,\kb\lambda x)\big|_{x=\tilde{x}}\,,
\nonumber\\
G_{(\alpha\beta)}^{\mathrm{tw6(iiA)b}}(\ka\xx,\kb\xx)
=&
-\hbox{\Large$\frac{1}{4}$}x_\alpha x_\beta\square^2
\int_0^1\d\lambda
\Big( \hbox{\Large$\frac{(1-\lambda)^2}{2\lambda}$}
-\hbox{\Large$\frac{3}{2}$}
\hbox{\Large$\frac{1-\lambda^2}{2\lambda}$}
-\lambda\ln\lambda
-\hbox{\Large$\frac{\ln\lambda}{2\lambda}$}\Big)
G(\ka\lambda x,\kb\lambda x)\big|_{x=\tilde{x}}\,,
\nonumber\\
G_{(\alpha\beta)}^{\mathrm{tw6(iiA)d1}}(\ka\xx,\kb\xx)
=&
\hbox{\Large$\frac{1}{4}$}x_\alpha x_\beta\square^2
\int_0^1\d\lambda
\Big( \hbox{\Large$\frac{1-\lambda}{\lambda}$}
+\hbox{\Large$\frac{\ln\lambda}{\lambda}$}
+\hbox{\Large$\frac{\ln^2\lambda}{2\lambda}$}\Big)
G(\ka\lambda x,\kb\lambda x)\big|_{x=\tilde{x}}\,,
\nonumber\\
G_{(\alpha\beta)}^{\mathrm{tw6(iiA)d3}}(\ka\xx,\kb\xx)
=&
-\hbox{\Large$\frac{1}{2}$}x_\alpha x_\beta\square^2
\int_0^1\d\lambda
\Big( \hbox{\Large$\frac{1-\lambda}{\lambda}$}
+\hbox{\Large$\frac{\ln\lambda}{\lambda}$}
+\hbox{\Large$\frac{\ln^2\lambda}{2\lambda}$}\Big)
G(\ka\lambda x,\kb\lambda x)\big|_{x=\tilde{x}}\,,
\nonumber
\end{align}
being related to the scalar operator $G(\ka\xx,\kb\xx)$, 
as well as
\begin{align}
G_{(\alpha\beta)}^{\mathrm{tw4(iiA)c}}(\ka\xx,\kb\xx)
=&\hbox{\Large$\frac{1}{2}$} g_{\alpha\beta}\pd^\mu 
\int_0^1\d\lambda
\hbox{\Large$\frac{1-\lambda^2}{\lambda}$}
G_{\mu\bullet}(\ka\lambda x,\kb\lambda x)\big|_{x=\tilde{x}},
\nonumber\\
G_{(\alpha\beta)}^{\mathrm{tw4(iiA)d2}}(\ka\xx,\kb\xx)
=&\,x_{(\alpha}\pd_{\beta)}\pd^\mu 
\int_0^1\d\lambda\Big(
\hbox{\Large$\frac{1-\lambda}{\lambda}$}
+\hbox{\Large$\frac{\ln\lambda}{\lambda}$}\Big)
G_{\mu\bullet}(\ka\lambda x,\kb\lambda x)\big|_{x=\tilde{x}}
\nonumber\\&
-G_{(\alpha\beta)}^{\mathrm{tw6(iiA)d2}}(\ka\xx,\kb\xx)\,,
\nonumber\\
G_{(\alpha\beta)}^{\mathrm{tw4(iiA)e}}(\ka\xx,\kb\xx)
=&\,x_{(\alpha}\pd_{\beta)}\pd^\mu 
\int_0^1\d\lambda(1-\lambda)
G_{\mu\bullet}(\ka\lambda x,\kb\lambda x)\big|_{x=\tilde{x}}
-G_{(\alpha\beta)}^{\mathrm{tw6(iiA)e}}(\ka\xx,\kb\xx)\,,
\nonumber
\end{align}
and
\begin{align}
G_{(\alpha\beta)}^{\mathrm{tw5(iiA)}}(\ka\xx,\kb\xx)
=&
-\hbox{\Large$\frac{1}{2}$}
x_{(\alpha}\big(\delta^\mu_{\beta)}(x\pd)-x^\mu\pd_{\beta)}\big)\square 
\int_0^1\d\lambda\Big(
\hbox{\Large$\frac{1-\lambda}{\lambda}$}
+\hbox{\Large$\frac{\ln\lambda}{\lambda}$}\Big)
G_{\mu\bullet}(\ka\lambda x,\kb\lambda x) \big |_{x=\tilde{x}}\,,
\nonumber\\
&-G_{(\alpha\beta)}^{\mathrm{tw6(iiA)d3}}(\ka\xx,\kb\xx)
-G_{(\alpha\beta)}^{\mathrm{tw6(iiA)d4}}(\ka\xx,\kb\xx)\,,
\nonumber
\end{align}
and
\begin{align}
G_{(\alpha\beta)}^{\mathrm{tw6(iiA)c}}(\ka\xx,\kb\xx)
\!=&\,
-\hbox{\Large$\frac{1}{4}$}x_\alpha x_\beta\square\pd^\mu 
\int_0^1\d\lambda
\hbox{\Large$\frac{(1-\lambda)^2}{\lambda}$}
G_{\mu\bullet}(\ka\lambda x,\kb\lambda x)\big|_{x=\tilde{x}}\,,
\nonumber\\
G_{(\alpha\beta)}^{\mathrm{tw6(iiA)d2}}(\ka\xx,\kb\xx)
\!=&
-\!\hbox{\Large$\frac{1}{2}$}x_\alpha x_\beta\square\pd^\mu 
\!\int_0^1\!\d\lambda\Big(
\hbox{\Large$\frac{1-\lambda}{\lambda}$}
+\hbox{\Large$\frac{\ln\lambda}{\lambda}$}
+\hbox{\Large$\frac{\ln^2\lambda}{2\lambda}$}\!\Big)
G_{\mu\bullet}(\ka\lambda x,\kb\lambda x)\big|_{x=\tilde{x}}\,,
\nonumber\\
G_{(\alpha\beta)}^{\mathrm{tw6(iiA)d4}}(\ka\xx,\kb\xx)
\!=\,&\hbox{\Large$\frac{1}{2}$}x_\alpha x_\beta\square\pd^\mu 
\!\int_0^1\!\d\lambda\Big(
\hbox{\Large$\frac{1-\lambda}{\lambda}$}
+\hbox{\Large$\frac{\ln\lambda}{\lambda}$}
+\hbox{\Large$\frac{\ln^2\lambda}{\lambda}$}\!\Big)
G_{\mu\bullet}(\ka\lambda x,\kb\lambda x)\big|_{x=\tilde{x}}\,,
\nonumber\\
G_{(\alpha\beta)}^{\mathrm{tw6(iiA)e}}(\ka\xx,\kb\xx)
\!=&
\hbox{\Large$\frac{1}{4}$}
x_\alpha x_\beta\square\pd^\mu  
\int_0^1\d\lambda\,
\hbox{\Large$\frac{(1-\lambda)^2}{\lambda}$}
G_{\mu\bullet}(\ka\lambda x,\kb\lambda x)\big|_{x=\tilde{x}}\,,
\nonumber
\end{align}
being related to the vector operator 
$G_{\mu\bullet}(\ka\xx,\kb\xx)$.
\subsubsection{Contributions of Young tableau B}
Let us now consider the other Young tableau which is obtained from the 
former one by exchanging $\alpha$ and $\beta$:
\\
\\
\unitlength0.5cm
\begin{picture}(30,1)
\linethickness{0.075mm}
\put(1,-1){\framebox(1,1){$\beta$}}
\put(1,0){\framebox(1,1){$\mu_1$}}
\put(2,0){\framebox(1,1){$\mu_2$}}
\put(3,0){\framebox(3,1){$\ldots$}}
\put(6,0){\framebox(1,1){$\mu_n$}}
\put(7,0){\framebox(1,1){$\alpha$}}
\put(8.5,0){$\stackrel{\wedge}{=}$}
\put(9.5,0)
{${\hbox{\Large$\frac{2(n+1)}{n+2}$}}\!
\relstack{\beta\mu_1}{\cal A}\;
\relstack{\alpha\mu_1\ldots\mu_n}{\cal S} \!\!
F_\alpha^{\ \rho}(0)D_{\mu_1}\ldots D_{\mu_n}\!F_{\beta\rho}(0)
\! - \!\mathrm{trace~terms.} $}
\end{picture}
\\ \\
This symmetry behaviour will be denoted by (iiB). The corresponding 
nonlocal twist--3 operator which we obtain after analogous 
calculations is (with self-explaining denotations):
\begin{equation}
\hspace{-1cm}
\label{M_tw2_nl}
G^{\mathrm{tw3(iiB)}}_{\alpha\beta} (0,\kappa x)
=
\int_{0}^{1} \d\lambda\,\hbox{\Large$\frac{1-\lambda^2}{2\lambda}$}
\Big(\delta_\beta^\mu(x\pd)-x^\mu\pd_\beta\Big)
\pd_\alpha \tl G_{\bullet\mu}(0,\kappa \lambda x)\,.
\end{equation}
Its decomposition into symmetric and antisymmetric part yields
\begin{equation}
\label{G3B}
G^{\mathrm{tw3(iiB)}}_{\alpha\beta} (0,\kappa x)=
G^{\mathrm{tw3(iiB)}}_{[\alpha\beta]} (0,\kappa x)+
G^{\mathrm{tw3(iiB)}}_{(\alpha\beta)} (0,\kappa x)\, ,
\end{equation}
with
\begin{align}
G^{\mathrm{tw3(iiB)}}_{[\alpha\beta]} (0,\kappa x)
&=
\int_0^1\d\lambda\, \lambda\delta^\mu_{[\beta}\pd_{\alpha]}
\tl G_{\bullet\mu} (0,\kappa \lambda x)\,,
\\
G^{\mathrm{tw3(iiB)}}_{(\alpha\beta)} (0,\kappa x)
&=
\int_{0}^{1} \d\lambda\,\hbox{\Large$\frac{1-\lambda^2}{2\lambda}$}
\Big(\delta_{(\alpha}^\mu(x\pd)-x^\mu\pd_{(\alpha}\Big)
\pd_{\beta)} 
\tl G_{\bullet\mu}(0,\kappa \lambda x)\,.
\end{align}
The projection onto the light--cone and the
calculation of the higher twist operators contained in the trace terms
is completely analogous to case (A) and should be omitted here.
\subsubsection{Construction of the complete twist--3 tensor operators
on the light--cone}
In order to obtain the complete twist--3 operator it is necessary to add 
both twist--3 operators (\ref{G3A}) and (\ref{G3B}) resulting from 
the Young patterns (iiA) and (iiB).
Since no further Young pattern contributes to twist--3
operators we omit the index (ii).
After projection onto the light--cone the final result is
\begin{equation}
\label{G3full}
G^{\mathrm{tw3}}_{\alpha\beta} (\ka\xx,\kb\xx) =
G^{\mathrm{tw3}}_{[\alpha\beta]} (\ka\xx,\kb\xx) +
G^{\mathrm{tw3}}_{(\alpha\beta)} (\ka\xx,\kb\xx) \, ,
\end{equation}
with the {\em antisymmetric twist--3 light--cone tensor 
operator}\footnote
{
Here, we used the abbreviation $G^-_{\mu}\equiv x^\nu G_{[\mu\nu]}$ 
instead of $G_{[\mu\bullet]}$ 
in order not to come into conflict with the antisymmetrization
of the indices $\alpha$ and $\beta$.} 
\begin{align}
\label{G3_anti}
G^{\mathrm{tw3}}_{[\alpha\beta]} (\ka\xx,\kb\xx)
=&
2\!\int_0^1\!\d\lambda\, \lambda\pd_{[\beta}
G^-_{\alpha]} (\ka\lambda x,\kb\lambda x)\big|_{x=\xx} 
-G^{\mathrm{tw4(ii)}}_{[\alpha\beta]} (\ka\xx,\kb\xx) 
-G^{\mathrm{tw5(ii)}}_{[\alpha\beta]} (\ka\xx,\kb\xx),
\end{align}
where
\begin{align}
\label{G_tw4_ii}
G^{\mathrm{tw4(ii)}}_{[\alpha\beta]}(\ka\xx,\kb\xx) 
&=
x_{[\alpha}\pd_{\beta]}\pd^{\mu}
\int_{0}^{1}\d\lambda\,
\hbox{\Large$\frac{1-\lambda^2}{\lambda}$}\
G^-_{\mu} (\ka\lambda x,\kb\lambda x) 
\big|_{x=\xx}\,,\\
G^{\mathrm{tw5(ii)}}_{[\alpha\beta]}(\ka\xx,\kb\xx) 
&=
\label{G_tw5_ii}
-x_{[\alpha}\big(\delta_{\beta]}^{\mu}(x\pd)
-x^{\mu}\pd_{\beta]}\big)\square 
\int_{0}^{1}\d\lambda\
\hbox{\Large$\frac{(1-\lambda)^2}{2\lambda}$}
G^-_{\mu}(\ka\lambda x,\kb\lambda x)
\big|_{x=\xx}.
\end{align}
The  local twist--4 operators are contained in the 
tensor space ${\bf T} (\frac{n}{2},\frac{n}{2})$
and the local twist--5 operators are contained in the 
tensor space ${\bf T} (\frac{n}{2},\frac{n-2}{2})\oplus
{\bf T} (\frac{n-2}{2},\frac{n}{2})$. 

In turn, let us remark that the bilocal light--ray operator 
$G^{\mathrm{tw3}}_{[\alpha\beta]} (\ka\xx,\kb\xx)$ 
is the same as the antisymmetric tensor operator which obtains
from the Young tableau
\unitlength0.35cm
\begin{picture}(8.5,2)
\linethickness{0.05mm}
\put(1,0){\framebox(1,1){$\SC\alpha$}}
\put(1,1){\framebox(1,1){$\SC\beta$}}
\put(2,1){\framebox(1,1){$\SC\mu_1$}}
\put(3,1){\framebox(1,1){$\SC\mu_2$}}
\put(4,1){\framebox(3,1){$\SC\ldots$}}
\put(7,1){\framebox(1,1){$\SC\mu_n$}}
\end{picture}\,.
That tableau has been used for the computation of 
$M^{\mathrm{tw2}}_{[\alpha\beta]} (\ka\xx,\kb\xx)$ in Ref.~\cite{Lazar98}.
Furthermore, let us note that the local operators contained in
$M^{\mathrm{tw2}}_{[\alpha\beta]} (\ka\xx,\kb\xx)$ and in
$G^{\mathrm{tw3}}_{[\alpha\beta]} (\ka\xx,\kb\xx)$ are in 
complete agreement with the local operators on the light--cone 
determined by Dobrev and Ganchev~\cite{Dobrev82} by means of the
interior derivative.

From the above point of view our result for the antisymmetric 
part of twist--3 operator (\ref{G3full}) could be obtained much easier.
However, the {\em symmetric twist--3 light--cone tensor operator}
 is much more involved. It is given 
by\footnote
{
Similarly, we use the notation $G^+_\mu$ instead of $G_{(\mu\bullet)}$.
}
\begin{align}
\label{G3_symm}
G^{\mathrm{tw3}}_{(\alpha\beta)} (\ka\xx,\kb\xx)
=&
\int_{0}^{1} \d\lambda\,\hbox{\Large$\frac{1-\lambda^2}{2\lambda}$}
\Big(\delta_{(\alpha}^\mu(x\pd)-x^\mu\pd_{(\alpha}\Big)
\pd_{\beta)} 
G^+_{\mu} (\ka\lambda x,\kb\lambda x) \big|_{x=\xx}
\nonumber\\
&-G^{\mathrm{>(ii)}}_{(\alpha\beta)} (\ka\xx,\kb\xx)\, ,
\end{align}
where $G^{\mathrm{>(ii)}}_{(\alpha\beta)} (\ka\xx,\kb\xx)$ includes
all the trace terms having higher twist; these operators are given by
\begin{align}
G_{(\alpha\beta)}^{\mathrm{tw4(ii)a}} (\ka\xx,\kb\xx)
=&
-g_{\alpha\beta}\square
\int_0^1\d\lambda
\Big(
\hbox{\Large$\frac{1-\lambda^2}{2\lambda}$}
+\lambda\ln\lambda\Big)
G(\ka\lambda x,\kb \lambda x)\big|_{x=\tilde{x}}\,,
\nonumber\\
G_{(\alpha\beta)}^{\mathrm{tw4(ii)b}} (\ka\xx,\kb\xx)
=&
-x_{(\alpha}\pd_{\beta)}\square\!
\int_0^1\!\d\lambda
\Big(\!(1-\lambda)\!+\!
\hbox{\Large$\frac{1\!-\!\lambda^2}{2\lambda}$}
\!+\!2\lambda\!\ln\lambda\!
\Big)
G(\ka\lambda x,\kb \lambda x)\big|_{x=\tilde{x}}
\nonumber\\
&
-G_{(\alpha\beta)}^{\mathrm{tw6(ii)b}} (\ka\xx,\kb\xx)\,,
\nonumber\\
G_{(\alpha\beta)}^{\mathrm{tw4(ii)d1}} (\ka\xx,\kb\xx)
=&
-x_{(\alpha}\pd_{\beta)}\square\!
\int_0^1\!\d\lambda
\Big(\!
 \hbox{\Large$\frac{1\!-\!\lambda}{\lambda}$}
+\hbox{\Large$\frac{\ln\lambda}{\lambda}$}\Big)
G(\ka\lambda x,\kb \lambda x)\big|_{x=\tilde{x}}
\nonumber\\
&
-G_{(\alpha\beta)}^{\mathrm{tw6(ii)d1}} (\ka\xx,\kb\xx)\,,
\nonumber\\
\intertext{and}
G_{(\alpha\beta)}^{\mathrm{tw6(ii)a}} (\ka\xx,\kb\xx)
=&
\hbox{\Large$\frac{1}{2}$}x_\alpha x_\beta\square^2\!
\int_0^1\!\d\lambda
\Big( 
 \hbox{\Large$\frac{(1-\lambda)^2}{\lambda}$}
- \hbox{\Large$\frac{1-\lambda^2}{2\lambda}$}
- \lambda\ln\lambda
\Big)
G(\ka\lambda x,\kb\lambda x)\big|_{x=\tilde{x}}\,,
\nonumber\\
G_{(\alpha\beta)}^{\mathrm{tw6(ii)b}} (\ka\xx,\kb\xx)
=&
-\hbox{\Large$\frac{1}{2}$}x_\alpha x_\beta\square^2
\int_0^1\d\lambda
\Big( 
 \hbox{\Large$\frac{(1-\lambda)^2}{2\lambda}$}
-\hbox{\Large$\frac{3}{2}$}
 \hbox{\Large$\frac{1-\lambda^2}{2\lambda}$}
\nonumber\\
&\qquad\qquad\qquad\qquad
-\lambda\ln\lambda
-\hbox{\Large$\frac{\ln\lambda}{2\lambda}$}
\Big)
G(\ka\lambda x,\kb \lambda x)\big|_{x=\tilde{x}}\,,
\nonumber\\
G_{(\alpha\beta)}^{\mathrm{tw6(ii)d1}} (\ka\xx,\kb\xx)
=&
\hbox{\Large$\frac{1}{2}$}x_\alpha x_\beta\square^2
\int_0^1\d\lambda
\Big( 
 \hbox{\Large$\frac{1-\lambda}{\lambda}$}
+\hbox{\Large$\frac{\ln\lambda}{\lambda}$}
+\hbox{\Large$\frac{\ln^2\lambda}{2\lambda}$}
\Big)
G(\ka\lambda x,\kb \lambda x)\big|_{x=\tilde{x}}\,,
\nonumber\\
G_{(\alpha\beta)}^{\mathrm{tw6(ii)d3}}(\ka\xx,\kb\xx)
=&
-x_\alpha x_\beta\square^2
\int_0^1\d\lambda
\Big( \hbox{\Large$\frac{1-\lambda}{\lambda}$}
+\hbox{\Large$\frac{\ln\lambda}{\lambda}$}
+\hbox{\Large$\frac{\ln^2\lambda}{2\lambda}$}\Big)
G(\ka\lambda x,\kb\lambda x)\big|_{x=\tilde{x}}\,,
\nonumber
\end{align}
being related to the scalar operator $G(\ka x, \kb x)$, 
as well as
\begin{align}
G_{(\alpha\beta)}^{\mathrm{tw4(ii)c}} (\ka\xx,\kb\xx)
=& g_{\alpha\beta}\pd^\mu 
\int_0^1\d\lambda
\hbox{\Large$\frac{1-\lambda^2}{\lambda}$}
G^{\mathrm{+}}_{\mu} (\ka\lambda x,\kb \lambda x)
\big|_{x=\tilde{x}}\,,
\nonumber\\
G_{(\alpha\beta)}^{\mathrm{tw4(ii)d2}} (\ka\xx,\kb\xx)
=&\,2 x_{(\alpha}\pd_{\beta)}\pd^\mu 
\int_0^1\d\lambda\Big(
\hbox{\Large$\frac{1-\lambda}{\lambda}$}
+\hbox{\Large$\frac{\ln\lambda}{\lambda}$}\Big)
G^{\mathrm{+}}_{\mu} (\ka\lambda x,\kb \lambda x)
\big|_{x=\tilde{x}}
\nonumber\\
&-G_{(\alpha\beta)}^{\mathrm{tw6(ii)d2}} (\ka\xx,\kb\xx)\,,
\nonumber\\
G_{(\alpha\beta)}^{\mathrm{tw4(ii)e}} (\ka\xx,\kb\xx)
=&\,2 x_{(\alpha}\pd_{\beta)}\pd^\mu 
\int_0^1\d\lambda(1-\lambda)
G^{\mathrm{+}}_{\mu} (\ka\lambda x,\kb \lambda x)
\big|_{x=\tilde{x}}
\nonumber\\
&-G_{(\alpha\beta)}^{\mathrm{tw6(ii)e}} (\ka\xx,\kb\xx)\,,
\nonumber
\end{align}
and
\begin{align}
G_{(\alpha\beta)}^{\mathrm{tw5(ii)}} (\ka\xx,\kb\xx)
=&
-x_{(\alpha}\big(\delta^\mu_{\beta)}(x\pd)\!-\!x^\mu\pd_{\beta)}\big)
\square 
\nonumber\\
&\qquad
\times
\int_0^1\!\d\lambda\Big(
 \hbox{\Large$\frac{1-\lambda}{\lambda}$}
+\hbox{\Large$\frac{\ln\lambda}{\lambda}$}\Big)
G^+_{\mu} (\ka\lambda x,\kb \lambda x)
\big|_{x=\tilde{x}}\,,
\nonumber\\
&-G_{(\alpha\beta)}^{\mathrm{tw6(ii)d3}}(\ka\xx,\kb\xx)
-G_{(\alpha\beta)}^{\mathrm{tw6(ii)d4}}(\ka\xx,\kb\xx)\,,
\nonumber
\end{align}
and
\begin{align}
G_{(\alpha\beta)}^{\mathrm{tw6(ii)c}} (\ka\xx,\kb\xx)
=&
-\hbox{\Large$\frac{1}{2}$}x_\alpha x_\beta\square\pd^\mu 
\int_0^1\d\lambda\,
\hbox{\Large$\frac{(1-\lambda)^2}{\lambda}$}
G^{\mathrm{+}}_{\mu} (\ka\lambda x,\kb \lambda x)
\big|_{x=\tilde{x}}\,,
\nonumber\\
G_{(\alpha\beta)}^{\mathrm{tw6(ii)d2}} (\ka\xx,\kb\xx)
=&
-\!x_\alpha x_\beta\square\pd^\mu 
\!\int_0^1\!\d\lambda\Big(
 \hbox{\Large$\frac{1-\lambda}{\lambda}$}
+\hbox{\Large$\frac{\ln\lambda}{\lambda}$}
+\hbox{\Large$\frac{\ln^2\!\lambda}{2\lambda}$}\!\Big)
G^{\mathrm{+}}_{\mu} (\ka\lambda x,\kb \lambda x)
\big|_{x=\tilde{x}}\,,
\nonumber\\
G_{(\alpha\beta)}^{\mathrm{tw6(ii)d4}} (\ka\xx,\kb\xx)
=&\, x_\alpha x_\beta\square\pd^\mu 
\!\int_0^1\!\d\lambda\Big(
 \hbox{\Large$\frac{1-\lambda}{\lambda}$}
+\hbox{\Large$\frac{\ln\lambda}{\lambda}$}
+\hbox{\Large$\frac{\ln^2\!\lambda}{\lambda}$}\!\Big)
G^{\mathrm{+}}_{\mu} (\ka\lambda x,\kb \lambda x)
\big|_{x=\tilde{x}}\,,
\nonumber\\
G_{(\alpha\beta)}^{\mathrm{tw6(ii)e}} (\ka\xx,\kb\xx)
=&
\hbox{\Large$\frac{1}{2}$}
x_\alpha x_\beta\square\pd^\mu\!
\int_0^1\!\d\lambda
\hbox{\Large$\frac{(1-\lambda)^2}{\lambda}$}
G^{\mathrm{+}}_{\mu} (\ka\lambda x,\kb \lambda x)
\big|_{x=\tilde{x}}\,,
\nonumber
\end{align}
being related to the (symmetric) vector operator
$G^+_{\mu} (\ka x,\kb x)$.
\subsubsection{Determination of complete light--cone vector operators}
Now we are able to determine the twist--3 vector operator resulting
from the tensor operator (\ref{G3full}) by multiplying by $x^\beta$.
Because neither symmetry type (iii) nor (iv) may contribute to the
vector operator this will be the final expression.
The {\em  twist--3 light--cone vector operator} consists of
two parts, one originating from the symmetric  and the
other from the antisymmetric tensor operator, 
\begin{align}
\label{Gtw3-vec}
&G^{\mathrm{tw3}}_\alpha(\kappa_1\lcx,\kappa_2\lcx)
=
G^{\mathrm{tw3}}_{(\alpha\bullet)}(\kappa_1\lcx,\kappa_2\lcx)
+
G^{\mathrm{tw3}}_{[\alpha\bullet]}(\kappa_1\lcx,\kappa_2\lcx)\,,
\end{align}
where
\begin{align}
\hspace{-.3cm}
\label{G_tw3_vec_sy}
G^{\mathrm{tw3}}_{(\alpha\bullet)}(\kappa_1\lcx,\!\kappa_2\lcx)
&=\!\int_0^1\!\d\lambda \lambda
\Big[\delta_\alpha^\mu(x\pd)\!-\!x^\mu\pd_\alpha
\!-\!x_\alpha\big(\pd^\mu\!+\!\ln\lambda \square x^\mu\big)\!\Big]
G^+_\mu(\kappa_1\lambda x,\!\kappa_2\lambda x)
\big|_{x=\tilde{x}}\,,
\\
\hspace{-.3cm}
G^{\mathrm{tw3}}_{[\alpha\bullet]}(\kappa_1\lcx,\!\kappa_2\lcx)
\label{G_tw3_vec_asy}
&=\!\int_0^1\!\d\lambda\, \lambda
\Big[\delta_\alpha^\mu(x\pd)\!-\!x^\mu\pd_\alpha
\!-\!x_\alpha\pd^\mu\Big]
G^-_\mu(\kappa_1\lambda x,\kappa_2\lambda x)
\big|_{x=\tilde{x}}\,,
\end{align}
and the twist--4 vector operator which is contained in the 
trace terms of that twist--3 vector operator is given by
\begin{align}
\label{G4ii}
G^{\mathrm{tw4(ii)}}_\alpha(\kappa_1\lcx,\kappa_2\lcx)
&=
G^{\mathrm{tw4(ii)}}_{(\alpha\bullet)}(\kappa_1\lcx,\kappa_2\lcx)
+
G^{\mathrm{tw4}}_{[\alpha\bullet]}(\kappa_1\lcx,\kappa_2\lcx)\,,
\end{align}
where
\begin{align}
\label{G_tw4_vec_sy(ii)}
G^{\mathrm{tw4(ii)}}_{(\alpha\bullet)}(\kappa_1\lcx,\kappa_2\lcx)
&=x_\alpha 
\int_0^1\d\lambda\,\lambda\big[\pd^\mu+(\ln\lambda)\square x^\mu\big]
G^+_\mu(\kappa_1\lambda x,\kappa_2\lambda x)\big|_{x=\tilde{x}}\,,
\\
G^{\mathrm{tw4}}_{[\alpha\bullet]}(\kappa_1\lcx,\kappa_2\lcx)
\label{G_tw4_vec_asy}
&=x_\alpha 
\int_0^1\d\lambda\,\lambda\,\pd^\mu
G^-_\mu(\kappa_1\lambda x,\kappa_2\lambda x)\big|_{x=\tilde{x}}\,.
\end{align}
The antisymmetric part of the twist--4 vector operator is already
complete. The complete {\em twist--4 light--cone vector operator} 
is given by 
\begin{align}
\label{G_tw4_vec}
G^{\mathrm{tw4}}_\alpha(\kappa_1\lcx,\kappa_2\lcx)
&=
G^{\mathrm{tw4}}_{(\alpha\bullet)}(\kappa_1\lcx,\kappa_2\lcx)+
G^{\mathrm{tw4}}_{[\alpha\bullet]}(\kappa_1\lcx,\kappa_2\lcx)
\nonumber\\
&= x_\alpha 
\int_0^1\d\lambda\,\lambda\big[\pd^\mu+
\hbox{\Large$\frac{\ln\lambda}{2}$}\square x^\mu\big]
G_\mu(\kappa_1\lambda x,\kappa_2\lambda x)\big|_{x=\tilde{x}}\,,
\end{align}
where the symmetric twist--4 vector operator obtains by adding together
expressions (\ref{G4i}) and (\ref{G4ii}),
\begin{align}
\label{G_tw4_vec_sy}
G^{\mathrm{tw4}}_{(\alpha\bullet)}(\kappa_1\lcx,\kappa_2\lcx)
&=
G^{\mathrm{tw4(i)}}_{(\alpha\bullet)}(\kappa_1\lcx,\kappa_2\lcx)+
G^{\mathrm{tw4(ii)}}_{(\alpha\bullet)}(\kappa_1\lcx,\kappa_2\lcx)
\\
&=x_\alpha 
\int_0^1\d\lambda\,\lambda\big[\pd^\mu+
\hbox{\Large$\frac{\ln\lambda}{2}$}\square x^\mu\big]
G^+_\mu(\kappa_1\lambda x,\kappa_2\lambda x)\big|_{x=\tilde{x}}\,.\nonumber
\end{align}
Furthermore, putting together expressions 
(\ref{G_tw2_vec}), (\ref{G_tw3_vec_sy}), (\ref{G_tw4_vec_sy}) 
and (\ref{G_tw3_vec_asy}), (\ref{G_tw4_vec_asy})
we obtain for the complete decomposition of the {\em light--cone vector operator} 
the following results anticipated by Eqs.~(\ref{G^+_vec}) and (\ref{G^-_vec}),
respectively:
\begin{align}
G_{(\alpha\bullet)}(\kappa_1\lcx,\kappa_2\lcx)
&=
G^{\mathrm{tw2}}_{(\alpha\bullet)}(\kappa_1\lcx,\kappa_2\lcx)+
G^{\mathrm{tw3}}_{(\alpha\bullet)}(\kappa_1\lcx,\kappa_2\lcx)+
G^{\mathrm{tw4}}_{(\alpha\bullet)}(\kappa_1\lcx,\kappa_2\lcx)\\
G_{[\alpha\bullet]}(\kappa_1\lcx,\kappa_2\lcx)
&=G^{\mathrm{tw3}}_{[\alpha\bullet]}(\kappa_1\lcx,\kappa_2\lcx)+
G^{\mathrm{tw4(ii)}}_{[\alpha\bullet]}(\kappa_1\lcx,\kappa_2\lcx)
\end{align}
\subsection{Tensor operators of symmetry class (iii)}
\subsubsection{Construction of nonlocal antisymmetric class-(iii)
operators of definite twist}
Now we consider the twist--4 and twist--5 contributions originating
from the (unique) Young tableau related to the symmetry class (iii)
(with the normalizing factor $\beta_n$):\\
\\
\unitlength0.5cm
\begin{picture}(30,1)
\linethickness{0.075mm}
\put(1,-2){\framebox(1,1){$\alpha$}}
\put(1,-1){\framebox(1,1){$\beta$}}
\put(1,0){\framebox(1,1){$\mu_1$}}
\put(2,0){\framebox(1,1){$\mu_2$}}
\put(3,0){\framebox(1,1){$\mu_3$}}
\put(4,0){\framebox(3,1){$\ldots$}}
\put(7,0){\framebox(1,1){$\mu_n$}}
\put(8.5,0){$\stackrel{\wedge}{=}$}
\put(9.5,0)
{$\hbox{\Large$\frac{3n}{n+2}$}\!
\relstack{\alpha\beta\mu_1}{\cal A}\,
\relstack{\mu_1\ldots\mu_n}{\cal S}\!
F_\alpha^{\ \rho}(0)D_{\mu_1}\ldots
D_{\mu_n}F_{\beta\rho}(0)\! -\mathrm{trace~terms\,.} $}
\end{picture}
\\ \\ \\
The corresponding traceless local operator having twist $\tau = 4$ 
and being contained in the tensor space
${\bf T}\!\left(\frac{n}{2},\frac{n}{2}\right)$ is
given by:
\begin{eqnarray}
\label{M3}
G^{\mathrm{tw4(iii)}}_{\alpha\beta\mu_1\ldots\mu_n}
\!&=&\!
\hbox{\Large$\frac{n}{n+2}$}
\Big(
F_{[\alpha|}^{\ \ \rho}(0)D_{(\mu_1}\ldots D_{\mu_n)} F_{|\beta]\rho}(0)- 
F_{[\alpha|}^{\ \ \rho}(0)D_{(\beta} D_{\mu_2}
\ldots D_{\mu_n)} F_{|\mu_1]\rho}(0)
\nonumber\\
\label{M_tw3_l}
& & \qquad
-F_{[\mu_1|}^{\ \ \ \rho}(0) D_{(\alpha}D_{\mu_2}
\ldots D_{\mu_n) } F_{|\beta]\rho}(0)
\Big)
-\mathrm{trace~terms}.
\nonumber
\end{eqnarray}
Now, contracting this expression  
with $x^{\mu_1}\ldots x^{\mu_n}$ we obtain:
\begin{align}
\label{Mx_tw3_l}
G^{\mathrm{tw4(iii)}}_{[\alpha\beta] n}(x)
&=
x^{\mu_1}\ldots x^{\mu_n}
G^{\mathrm{tw4(iii)}}_{[\alpha\beta]\mu_1\ldots\mu_n}  
\nonumber\\
&=\hbox{\Large$\frac{3}{n+2}$}\, x^\gamma\,
\delta^\mu_{[\alpha}\delta^\nu_{\beta}\pd_{\gamma]}
\tl G_{[\mu\nu] n}(x)\\
&=
\hbox{\Large$\frac{1}{n+2}$}
\left(
(x\pd)\delta^\nu_{[\beta}
- 2 x^\nu\pd_{[\beta}\right)\delta_{\alpha]}^\mu
\tl G_{[\mu\nu] n}(x)\,.
\nonumber
\end{align}
with the (antisymmetric) tensorial harmonic polynomials of order $n$
(see~(\ref{T_skewtensor-tl})):
\begin{align}
\tl G_{[\alpha\beta] n}(x)
=&
\bigg\{\delta_\alpha^\mu\delta_\beta^\nu
+\hbox{\Large$\frac{1}{(n+1)n}$}
\left(
2 x_{[\alpha}\delta_{\beta]}^{[\mu}\pd^{\nu]}(x\pd)
-
x^2\pd_{[\alpha}\delta_{\beta]}^{[\mu}\pd^{\nu]}\right)\nonumber\\
&
-\hbox{\Large$\frac{2}{(n+2)(n+1)n}$}\,
x_{[\alpha}\pd_{\beta]}x^{[\mu}\pd^{\nu]}\bigg\}
H^{(4)}_n \!\big(x^2|\square\big)
G_{[\mu\nu] n}(x).
\nonumber
\end{align}

Resumming these local terms gives the nonlocal twist--4 operator 
as follows
\begin{eqnarray}
\label{M_tw3_nl_ir}
G^{\mathrm{tw4(iii)}}_{[\alpha\beta]}(0,\kappa x)
=
\int_0^1\d \lambda\,\lambda
\delta^\mu_{[\alpha}\left(
(x\pd)\delta^\nu_{\beta]}
- 2 x^\nu\pd_{\beta]}\right)
\tl G_{[\mu\nu]}(0,\kappa\lambda x)\, ,
\end{eqnarray}
where, using the integral representation of 
the additional factor $1/n$ (the remaining factors of the 
denominator are taken together with $1/n!$ to get $1/(n+2)!$) 
and of the beta function, we introduced:
\begin{eqnarray}
\label{G0anti}
\hspace{-.6cm}
&&\tl G_{[\alpha\beta]}(0,\kappa x)
=
G_{[\alpha\beta]}(0,\kappa x)
+\sum_{k=1}^\infty\int_0^1\!
\frac{\d t}{t}
\!\left(\!\frac{-x^2}{4}\right)^{\!k}\!
\frac{\square^k}{k!(k-1)!}
\left(\!\frac{1-t}{t}\right)^{\!k-1}\!
G_{[\alpha\beta]}(0,\kappa t x)
\nonumber\\
\hspace{-.6cm}
&&\quad
+\int_0^1\!\frac{\d\tau}{\tau}\bigg\{\!
\left(
2x_{[\alpha}\delta_{\beta]}^{[\mu}\pd^{\nu]}(x\pd)
- x^2\pd_{[\alpha}
\delta_{\beta]}^{[\mu}\pd^{\nu]}\right)
\!\sum_{k=0}^\infty
\int_0^1\!\d t
\left(\!\frac{-x^2}{4}\right)^{\!k}\!
\frac{\square^k}{k!k!}\!
\left(\!\frac{1-t}{t}\right)^{\!k}
\nonumber\\
\hspace{-.8cm}
&&\quad
-\,2x_{[\alpha}\pd_{\beta]}x^{[\mu}\pd^{\nu]}
\sum_{k=0}^\infty\!
\int_0^1\!\d t\,t\!
\left(\!\frac{-x^2}{4}\right)^{\!k}\!
\frac{\square^k}{(k+1)!k!}
\left(\!\frac{1-t}{t}\right)^{\!k+1}
\!\bigg\}
G_{[\mu\nu]}(0,\kappa\tau t x).\qquad\quad
\end{eqnarray}

Then, by construction, Eq.~(\ref{M_tw3_nl_ir}) defines 
a nonlocal operator of twist--4. As is easily seen by partial 
integration it fulfils the following relation
\begin{equation}
\label{M_tw3}
G^{\rm tw4(iii)}_{[\alpha\beta]}(0,\kappa x)
=\tl G_{[\alpha\beta]}(0,\kappa x)
- G^{\rm tw3}_{[\alpha\beta]}(0,\kappa x)\, 
\end{equation}
and, because of Eq.~(\ref{M2harm}) and the properties of 
(\ref{G0anti}), it is a harmonic tensor operator: 
\begin{equation}
\label{M0harm}
\square 
G^{\rm tw4(iii)}_{[\alpha\beta]}(0,\kappa x)=0\ ,
\qquad
\pd^\alpha G^{\rm tw4(iii)}_{[\alpha\beta]}(0,\kappa x)
=0\,.
\end{equation}
\subsubsection{Projection onto the light--cone}
Let us now project onto the light--cone. After the same 
calculations as has been carried out for the operator
$M^{\rm tw3}_{[\alpha\beta]}(0,\kappa\lcx)$ in~\cite{GLR99}
we obtain 
\begin{align}
\label{G_tw4_iii}
G^{\rm tw4(iii)}_{[\alpha\beta]}(\ka\xx,\kb\xx)
=&
\int_{0}^{1}\d\lambda\, \lambda
\delta^\mu_{[\alpha}\left( (x\pd)\delta_{\beta]}^\nu
- 2 x^\nu\pd_{\beta]}\right)
G_{[\mu\nu]}(\ka\lambda x,\kb\lambda x)\big|_{x=\xx}
\\
& 
- G^{\rm tw5(iii)}_{[\alpha\beta]}(\ka\xx,\kb\xx),
\nonumber
\end{align}
where the twist--5 part is determined by the trace
terms, namely
\begin{align}
\label{G_tw5_iii}
G^{\rm tw5(iii)}_{[\alpha\beta]}(\ka\xx,\kb\xx)
= 
& - \int_{0}^{1}\d\lambda 
\hbox{\Large$\frac{1-\lambda^2}{\lambda}$}\Big\{
x_{[\alpha}\big(\delta_{\beta]}^{[\mu}(x\pd)
-x^{[\mu}\pd_{\beta]}\big)\pd^{\nu]}
\nonumber\\
&\qquad\qquad 
- x_{[\alpha}\delta_{\beta]}^{[\mu}x^{\nu]}\square\Big\}
G_{[\mu\nu]}(\ka\lambda x,\kb\lambda x)
\big|_{x=\xx}\,. 
\end{align}
Obviously, there exist no vector (and scalar) operators of
symmetry type (iii).
\subsubsection{Determination of the complete antisymmetric
light--cone tensor operator}
This finishes the computation of twist contributions of the 
antisymmetric light--cone tensor operator. The complete 
{\em antisymmetric twist--4 light--cone tensor operator} obtains from
Eq.~(\ref{G_tw4_iii}) and the twist--4 trace terms of the 
twist--3 operator, Eq.~(\ref{G_tw4_ii}):
\begin{eqnarray}
\label{G_tw4_as}
\lefteqn{G^{\rm tw4}_{[\alpha\beta]}(\ka\xx,\kb\xx)
=
G^{\rm tw4(ii)}_{[\alpha\beta]}(\ka\xx,\kb\xx)
+
G^{\rm tw4(iii)}_{[\alpha\beta]}(\ka\xx,\kb\xx)}
\nonumber\\
&&=\!
\int_{0}^{1}\!\d\lambda\Big\{
\lambda\left(
(x\pd)\delta_{[\beta}^\nu
- 2 x^\nu\pd_{[\beta}\right)\delta_{\alpha]}^{\mu}
+\hbox{\Large$\frac{1-\lambda^2}{\lambda}$}\Big(
  x_{[\alpha}\big(\delta_{\beta]}^{[\mu}(x\pd)
  - x^{[\mu}\pd_{\beta]}\big)\pd^{\nu]}
\\
& &\quad
- x_{[\alpha}\delta_{\beta]}^{[\mu}x^{\nu]}\square
- x_{[\alpha}\pd_{\beta]}x^{[\mu}\pd^{\nu]}\Big)
\Big\}G_{[\mu\nu]}(\kappa_1\lambda x,\kappa_2\lambda x)
\big|_{x=\xx}.
\nonumber
\end{eqnarray}
Furthermore, the complete {\em  antisymmetric twist--5 
light--cone tensor operator} obtains from 
Eqs. (\ref{G_tw5_ii}) and (\ref{G_tw5_iii}) as follows:
\begin{eqnarray}
\label{G_tw5_as}
 \lefteqn{
G^{\rm tw5}_{[\alpha\beta]}(\ka\xx,\kb\xx)
=
G^{\rm tw5(ii)}_{[\alpha\beta]}(\ka\xx,\kb\xx)
+
G^{\rm tw5(iii)}_{[\alpha\beta]}(\ka\xx,\kb\xx)}
\\
&&=\!
 \int_{0}^{1}\d\lambda 
\hbox{\Large$\frac{1-\lambda}{\lambda}$}\Big\{
x_{[\alpha}\delta_{\beta]}^{[\mu}x^{\nu]}\square
-2x_{[\alpha}\big(\delta_{\beta]}^{[\mu}(x\pd)
-x^{[\mu}\pd_{\beta]}\big)\pd^{\nu]}
\Big\}G_{[\mu\nu]}(\kappa_1\lambda x,\kappa_2\lambda x)
\big|_{x=\xx}.\nonumber
\end{eqnarray}
Together with the twist-3 part, Eq.~(\ref{G3_anti}), we finally 
obtain the complete decomposition of the {\em antisymmetric 
light--cone tensor operator} (compare Eq.~(\ref{G^-_t})):
\begin{equation}
G_{[\alpha\beta]}(\ka\xx,\kb\xx)=
G^{\rm tw3}_{[\alpha\beta]}(\ka\xx,\kb\xx)+
G^{\rm tw4}_{[\alpha\beta]}(\ka\xx,\kb\xx)+
G^{\rm tw5}_{[\alpha\beta]}(\ka\xx,\kb\xx)\, .
\end{equation}

\subsection{Tensor operators of symmetry class (iv)}
\label{gluon-iv}
\subsubsection{Construction of the nonlocal symmetric class-(iv) 
operator of definite twist}
Now let us investigate the symmetric tensor operators 
of the symmetry class (iv) which are given by the following Young tableau
(with normalizing factor $\gamma_{n}$):
\\
\\
\unitlength0.5cm
\begin{picture}(30,1)
\linethickness{0.075mm}
\put(1,-1){\framebox(1,1){$\alpha$}}
\put(2,-1){\framebox(1,1){$\beta$}}
\put(1,0){\framebox(1,1){$\mu_1$}}
\put(2,0){\framebox(1,1){$\mu_2$}}
\put(3,0){\framebox(1,1){$\mu_3$}}
\put(4,0){\framebox(3,1){$\ldots$}}
\put(7,0){\framebox(1,1){$\mu_n$}}
\put(8.5,0){$\stackrel{\wedge}{=}$}
\put(9.5,0)
{${\hbox{\Large$\frac{4(n-1)}{n+1}$}}\!
\relstack{\alpha\mu_1}{\cal A}\,
\relstack{\beta\mu_2}{\cal A}\,
\relstack{\alpha\beta}{\cal S}\,
\relstack{\mu_1\ldots\mu_n}{\cal S} \!
F_\alpha^{\ \rho}(0)D_{\mu_1}\ldots D_{\mu_n}F_{\beta\rho}(0)
\! - \!\mathrm{traces\,.} $}
\end{picture}
\\ \\
These local tensor operators have twist--4 and transform according to
the representation\linebreak
${\bf T}\hbox{$(\frac{n+2}{2},\frac{n-2}{2})\oplus
{\bf T}(\frac{n-2}{2},\frac{n+2}{2})$}$. 
They are given by:
\begin{align}
\hspace{-1cm}
G^{\mathrm{tw4(iv)}}_{\alpha\beta\mu_1\ldots\mu_n}
=&\;
\hbox{\Large$\frac{1}{(n+1)n}$}
\Big(G_{(\alpha\beta)(\mu_1\ldots\mu_n)}
-G_{(\alpha\mu_2)(\mu_1\beta\mu_3\ldots\mu_n)}
-G_{(\mu_1\beta)(\alpha\mu_2\ldots\mu_n)}
\nonumber\\
\hspace{-1cm}
&\qquad\qquad
+G_{(\mu_1\mu_2)(\alpha\beta\mu_3\ldots\mu_n)}\Big)
-\mathrm{trace~terms}\,,
\nonumber
\end{align}
with $G_{(\alpha\beta)(\mu_1\ldots\mu_n)}\equiv
F_{(\alpha|}^{\ \ \rho}(0)D_{(\mu_1}\ldots D_{\mu_n)} F_{|\beta)\rho}(0)$.
If we multiply this expression by $x^{\mu_1}\ldots x^{\mu_n}$, we obtain:
\begin{align}
\label{Gx_tw4_iv}
G^{\mathrm{tw4(iv)}}_{(\alpha\beta) n}(x)
&=x^{\mu_1}\ldots x^{\mu_n}G^{\mathrm{tw4(iv)}}_{\alpha\beta\mu_1\ldots\mu_n}
\nonumber\\
&=\hbox{\Large$\frac{4}{(n+1)n}$}\, x^\rho x^\tau\,
\delta_{[\alpha}^\mu \pd_{\rho]} \delta_{[\beta}^\nu  \pd_{\tau]}\,
\tl G_{(\mu\nu) n}(x)\\
&=\hbox{\Large$\frac{1}{(n+1)n}$}
\Big(\delta_\alpha^\mu\delta_\beta^\nu x^\rho (x\pd)\pd_\rho
-2x^\mu(x\pd) \delta_{(\alpha}^\nu \pd_{\beta)}
+x^\mu x^\nu\pd_\alpha\pd_\beta\Big)
\tl G_{(\mu\nu) n}(x)\, .
\nonumber
\end{align}
Here, $\tl G_{(\mu\nu) n}(x)$ is the harmonic symmetric tensor
polynomial of order $n$ (cf. Eq.~(\ref{Proj6}), Section~\ref{tensor}):
\begin{align}
&\tl G_{(\alpha\beta) n}(x)
=\Big\{ 
\delta_\alpha^\mu\delta_\beta^\nu
+\hbox{\Large$\frac{n}{(n+2)^2}$} g_{\alpha\beta} x^{(\nu}\pd^{\mu)}
-\hbox{\Large$\frac{1}{(n+2)(n+1)}$} 
\Big(2n x_{(\alpha}\delta_{\beta)}^{(\nu}\pd^{\mu)}
-x^2\delta_{(\alpha}^{(\nu}\pd_{\beta)}\pd^{\nu)}\Big)
\nonumber\\
&\qquad-\hbox{\Large$\frac{1}{(n+2)^2(n+1)}$}
\Big(2n x_{(\alpha}\pd_{\beta)}x^{(\nu}\pd^{\mu)}
-x^2 \pd_{\alpha}\pd_{\beta} x^{(\nu}\pd^{\mu)}\Big)
+\hbox{\Large$\frac{(n+3)n}{(n+2)^2(n+1)^2}$} 
\Big(x_{\alpha}x_{\beta} 
-\hbox{\Large$\frac{1}{2}$}x^2 g_{\alpha\beta}\Big)\pd^{\mu}\pd^{\nu}
\nonumber\\
&\qquad-\hbox{\Large$\frac{1}{(n+2)^2(n+1)^2}$}
\Big(2 x^2 \pd_{(\alpha}x_{\beta)} +\hbox{\Large$\frac{1}{4}$}
x^4 \pd_{\alpha}\pd_{\beta}\Big) \pd^{\mu}\pd^{\nu}
\Big\}
\breve{G}_{(\mu\nu) n}(x)\,.
\nonumber
\end{align}
Resumming all local operators of symmetry class (iv) 
and using the integral representation 
$((n+1)n)^{-1}=\int_0^1\d\lambda \lambda^n(1-\lambda)/\lambda$ 
gives
\begin{align}
\label{G_tw4_1}
G^{\mathrm{tw4(iv)}}_{(\alpha\beta)}(0,\kappa x)
=
\Big(\delta_\alpha^\mu\delta_\beta^\nu x^\rho (x\pd)\pd_\rho
-2 x^\mu (x\pd)\delta_{(\beta}^\nu\pd_{\alpha)}
+x^\mu x^\nu\pd_\alpha\pd_\beta\Big)
\int_0^1\d\lambda\,\hbox{\Large$\frac{1-\lambda}{\lambda}$}
\,\tl G_{(\mu\nu)}(0,\kappa\lambda x),
\end{align}
where $\tl G_{(\mu\nu)}(0,\kappa x)$ is given by
\begin{eqnarray}
\lefteqn{\tl G_{(\alpha\beta)}(0,\kappa x)
=
\breve{G}_{(\alpha\beta)}(0,\kappa x)
+\sum_{k=1}^\infty\int_0^1\!
\frac{\d t}{t}
\!\left(\!\frac{-x^2}{4}\right)^{\!k}\!
\frac{\square^k}{k!(k-1)!}\!
\left(\!\frac{1-t}{t}\right)^{\!k-1}\!
\breve{G}_{(\alpha\beta)}(0,\kappa t x)}
\nonumber\\
&&\!-\big[2x_{(\alpha}\delta_{\beta)}^{(\mu}\pd^{\nu)}(x\pd)
- x^2\pd_{(\alpha}\delta_{\beta)}^{(\mu}\pd^{\nu)}\big]
\sum_{k=0}^\infty
\int_0^1\!\d t\, t
\left(\!\frac{-x^2}{4}\right)^{\!k}\!
\frac{\square^k}{k!(k+1)!}
\left(\!\frac{1-t}{t}\right)^{\!k+1}\!
\breve{G}_{(\mu\nu)}(0,\kappa t x)
\nonumber\\
& &\!-\bigg\{\big[2x_{(\alpha}\pd_{\beta)}x^{(\mu}\pd^{\nu)}(x\pd)
-g_{\alpha\beta}x^{(\mu}\pd^{\nu)}(x\pd)\big(x\pd+1\big)
-x^2\pd_\alpha\pd_\beta x^{(\mu}\pd^{\nu)}\big]
\!\int_0^1\!\d \lambda\,\lambda\nonumber\\
& &\!-\big[\big(x_\alpha x_\beta-\hbox{\Large$\frac{x^2}{2}$}
g_{\alpha\beta}\big)
\pd^{\mu}\pd^{\nu}(x\pd)\big(x\pd+3\big)
-\big(2x^2\pd_{(\alpha}x_{\beta)}+\hbox{\Large$\frac{x^4}{4}$} 
\pd_\alpha\pd_\beta\big)
\pd^\mu\pd^\nu\big]
\int_0^1\d \lambda\,(1-\lambda)\bigg\}
\nonumber\\
&&\qquad
\times
\sum_{k=0}^\infty\!
\int_0^1\!\d t\,t
\left(\!\frac{-x^2}{4}\right)^{\!k}\!
\frac{\square^k}{(k+1)!k!}
\left(\!\frac{1-t}{t}\right)^{\!k+1}
\breve{G}_{(\mu\nu)}(0,\kappa\lambda t x).\qquad\quad
\end{eqnarray}
Of course, Eq.~(\ref{G_tw4_1}) may be rewritten as follows
\begin{align}
\hspace{-.3cm}
G^{\rm tw4(iv)}_{(\alpha\beta)}(0,\kappa x)
\label{Gx_tw_2}
=&\;\tl G_{(\alpha\beta)}(0,\kappa x)
-2\!\int_0^1\!\d\lambda\,\hbox{\Large$\frac{1}{\lambda}$}
\pd_{(\alpha}
\TLL {G^+_{\beta)}}(0,\kappa\lambda x)
+\!\int_0^1\!\d\lambda\, \hbox{\Large$\frac{1-\lambda}{\lambda}$}
\pd_\alpha \pd_{\beta}
\tl G(0,\kappa\lambda x),
\nonumber
\end{align}
which will be used in the further considerations.
Contrary to any of the formerly obtained symmetric twist--4 tensor 
operators, this operator is the `true tensor part', i.e.,~not being
proportional to $g_{\alpha\beta}$ or $x_\alpha$ and $x_\beta$. 
Obviously, this twist--4 operator satisfies the conditions of tracelessness
\begin{equation}
\square G^{\rm tw4(iv)}_{(\alpha\beta)}(0,\kappa x)=0,\quad
\pd^\alpha G^{\rm tw4(iv)}_{(\alpha\beta)}(0,\kappa x)=0, \quad
g^{\alpha\beta} G^{\rm tw4(iv)}_{(\alpha\beta)}(0,\kappa x).
\end{equation}
\subsubsection{Projection onto the light--cone}
The symmetric twist--4 tensor operator of symmetry type (iv) 
on the light--cone is obtained as follows
\begin{align}
\label{G_tw4_iv}
G^{\mathrm{tw4(iv)}}_{(\alpha\beta)} (\ka\xx,\kb\xx)
=&
\Big(\delta_\alpha^\mu\delta_\beta^\nu x^\rho (x\pd)\pd_\rho
-2x^\mu (x\pd)\delta_{(\beta}^\nu\pd_{\alpha)}
+x^\mu x^\nu\pd_\alpha\pd_\beta\Big)\nonumber\\
&\times
\int_0^1\d\lambda\hbox{\Large$\frac{1-\lambda}{\lambda}$}
G_{(\mu\nu)}(\ka\lambda x,\kb\lambda x)
\big|_{x=\xx}
-G^{\mathrm{>(iv)}}_{(\alpha\beta)} (\ka\xx,\kb\xx),
\end{align}
where the trace terms of higher twist are given by
\begin{eqnarray}
\lefteqn{
\label{G_high_(iv)_sy}
G^{\mathrm{>(iv)}}_{(\alpha\beta)} (\ka\xx,\kb\xx)
=
\int_{0}^{1}\!\d\lambda\Big\{
\Big(2\lambda x_{(\alpha}\delta_{\beta)}^\nu\pd^\mu
-(1-\lambda)x_\alpha x_\beta\pd^\mu\pd^\nu\Big) 
G_{(\mu\nu)}(\ka\lambda x,\kb\lambda x)
}\nonumber\\
&&-\Big(
\hbox{\Large$\frac{1}{\lambda}$}g_{\alpha\beta}\pd^\mu
+\hbox{\Large$\frac{1-\lambda^2}{\lambda}$}x_{(\alpha}\delta^\mu_{\beta)}\square
-\hbox{\Large$\frac{(1-\lambda)^2}{2\lambda}$}x_\alpha x_\beta\square\pd^\mu
\Big) G_{\mu}^{+}(\ka\lambda x,\kb\lambda x) \nonumber\\
&&+\Big(\hbox{\Large$\frac{1-\lambda}{2\lambda}$}g_{\alpha\beta}\square
+\hbox{\Large$\frac{(1-\lambda)^2}{2\lambda}$}
x_{(\alpha}\pd_{\beta)}\square
-\hbox{\Large$\frac{1}{2}$}\Big(
\hbox{\Large$\frac{1-\lambda^2}{2\lambda}$}+\ln\lambda\Big)
x_\alpha x_\beta\square^2\Big)G(\ka\lambda x,\kb\lambda x)\nonumber\\
&&+
\Big(
\hbox{\Large$\frac{1}{2\lambda}$}g_{\alpha\beta}(x\pd)
-\lambda x_{(\alpha}\pd_{\beta)}
+\hbox{\Large$\frac{1}{2}$} (1-\lambda)
x_\alpha x_\beta\square\Big) G^\rho_{\ \rho}(\ka\lambda x,\kb\lambda x)
\Big\}\Big|_{x=\tilde{x}}.
\end{eqnarray}
The explicit twist decomposition of this operator 
by means of the Young tableaux (i) and (ii) gives
\begin{align}
G^{\mathrm{>(iv)}}_{(\alpha\beta)}
(\ka\xx,\kb\xx)
&=
G^{\mathrm{tw4(iv)a}}_{(\alpha\beta)}(\ka\xx,\kb\xx)+
G^{\mathrm{tw4(iv)b}}_{(\alpha\beta)}(\ka\xx,\kb\xx)+
G^{\mathrm{tw4(iv)c}}_{(\alpha\beta)}(\ka\xx,\kb\xx)
\nonumber\\
&+
G^{\mathrm{tw4(iv)d1}}_{(\alpha\beta)}(\ka\xx,\kb\xx)+
G^{\mathrm{tw4(iv)d2}}_{(\alpha\beta)}(\ka\xx,\kb\xx)+
G^{\mathrm{tw4(iv)e1}}_{(\alpha\beta)}(\ka\xx,\kb\xx)
\nonumber\\
&+
G^{\mathrm{tw4(iv)e2}}_{(\alpha\beta)}(\ka\xx,\kb\xx)+
G^{\mathrm{tw4(iv)f}}_{(\alpha\beta)}(\ka\xx,\kb\xx)+
G^{\mathrm{tw4(iv)h}}_{(\alpha\beta)}(\ka\xx,\kb\xx)
\nonumber\\
&+
G^{\mathrm{tw5(iv)d}}_{(\alpha\beta)}(\ka\xx,\kb\xx)+
G^{\mathrm{tw5(iv)e}}_{(\alpha\beta)}(\ka\xx,\kb\xx)
\nonumber\\
&+
G^{\mathrm{tw6(iv)a}}_{(\alpha\beta)}(\ka\xx,\kb\xx)+
G^{\mathrm{tw6(iv)b}}_{(\alpha\beta)}(\ka\xx,\kb\xx)+
G^{\mathrm{tw6(iv)c}}_{(\alpha\beta)}(\ka\xx,\kb\xx)
\nonumber\\
&+
G^{\mathrm{tw6(iv)d1}}_{(\alpha\beta)}(\ka\xx,\kb\xx)+
G^{\mathrm{tw6(iv)d2}}_{(\alpha\beta)}(\ka\xx,\kb\xx)+
G^{\mathrm{tw6(iv)d3}}_{(\alpha\beta)}(\ka\xx,\kb\xx)
\nonumber\\
&+
G^{\mathrm{tw6(iv)d4}}_{(\alpha\beta)}(\ka\xx,\kb\xx)+
G^{\mathrm{tw6(iv)e1}}_{(\alpha\beta)}(\ka\xx,\kb\xx)+
G^{\mathrm{tw6(iv)e2}}_{(\alpha\beta)}(\ka\xx,\kb\xx)
\nonumber\\
&+G^{\mathrm{tw6(iv)e3}}_{(\alpha\beta)}(\ka\xx,\kb\xx)+
G^{\mathrm{tw6(iv)e4}}_{(\alpha\beta)}(\ka\xx,\kb\xx)+
G^{\mathrm{tw6(iv)e5}}_{(\alpha\beta)}(\ka\xx,\kb\xx)
\nonumber\\
&+
G^{\mathrm{tw6(iv)f}}_{(\alpha\beta)}(\ka\xx,\kb\xx)+
G^{\mathrm{tw6(iv)h}}_{(\alpha\beta)}(\ka\xx,\kb\xx)+
G^{\mathrm{tw6(iv)k}}_{(\alpha\beta)}(\ka\xx,\kb\xx)\,,
\end{align}
with the following operators of well-defined twist
\begin{align}
G_{(\alpha\beta)}^{\mathrm{tw4(iv)a}}(\ka\xx,\kb\xx)
=& g_{\alpha\beta}\square
\int_0^1\d\lambda
\hbox{\Large$\frac{1-\lambda}{2\lambda}$}
G(\ka\lambda x,\kb\lambda x)\big|_{x=\tilde{x}}\,,
\nonumber\\
G_{(\alpha\beta)}^{\mathrm{tw4(iv)b}}(\ka\xx,\kb\xx)
=&\,  x_{(\alpha}\pd_{\beta)}\square
\int_0^1\d\lambda
\hbox{\Large$\frac{(1-\lambda)^2}{2\lambda}$}
G(\ka\lambda x,\kb\lambda x)\big|_{x=\tilde{x}}
-G_{(\alpha\beta)}^{\mathrm{tw6(iv)b}}(\ka\xx,\kb\xx)\,,
\nonumber\\
G_{(\alpha\beta)}^{\mathrm{tw4(iv)d1}}(\ka\xx,\kb\xx)
=&\,  x_{(\alpha}\pd_{\beta)}\square
\int_0^1\d\lambda\Big(
\hbox{\Large$\frac{1-\lambda^2}{2\lambda}$}
+\hbox{\Large$\frac{\ln\lambda}{\lambda}$}\Big)
G(\ka\lambda x,\kb\lambda x)\big|_{x=\tilde{x}}
\nonumber\\
&-G_{(\alpha\beta)}^{\mathrm{tw6(iv)d1}}(\ka\xx,\kb\xx)\,,
\nonumber\\
\intertext{and}
G_{(\alpha\beta)}^{\mathrm{tw6(iv)a}}(\ka\xx,\kb\xx)
=&\,
-\hbox{\Large$\frac{1}{2}$}x_\alpha x_\beta\square^2\!
\int_0^1\!\d\lambda
\Big( \hbox{\Large$\frac{1-\lambda^2}{2\lambda}$}+\ln\lambda\Big)
G(\ka\lambda x,\kb\lambda x)\big|_{x=\tilde{x}}\,,
\nonumber\\
G_{(\alpha\beta)}^{\mathrm{tw6(iv)b}}(\ka\xx,\kb\xx)
=&
-\hbox{\Large$\frac{1}{2}$}x_\alpha x_\beta\square^2\!
\int_0^1\!\d\lambda
\Big( \hbox{\Large$\frac{1-\lambda}{\lambda}$}
-\hbox{\Large$\frac{1-\lambda^2}{4\lambda}$}
+\hbox{\Large$\frac{\ln\lambda}{2\lambda}$}\Big)
G(\ka\lambda x,\kb\lambda x)\big|_{x=\tilde{x}}\,,
\nonumber\\
G_{(\alpha\beta)}^{\mathrm{tw6(iv)d1}}(\ka\xx,\kb\xx)
=&\,
-\hbox{\Large$\frac{1}{4}$}x_\alpha x_\beta\square^2\!
\int_0^1\!\d\lambda
\Big( \hbox{\Large$\frac{1-\lambda^2}{2\lambda}$}
+\hbox{\Large$\frac{\ln\lambda}{\lambda}$}
+\hbox{\Large$\frac{\ln^2\lambda}{\lambda}$}\Big)
G(\ka\lambda x,\kb\lambda x)\big|_{x=\tilde{x}}\,,
\nonumber\\
G_{(\alpha\beta)}^{\mathrm{tw6(iv)d3}}(\ka\xx,\kb\xx)
=&\,
\hbox{\Large$\frac{1}{2}$}x_\alpha x_\beta\square^2\!
\int_0^1\!\d\lambda
\Big( \hbox{\Large$\frac{1-\lambda^2}{2\lambda}$}
+\hbox{\Large$\frac{\ln\lambda}{\lambda}$}
+\hbox{\Large$\frac{\ln^2\lambda}{\lambda}$}\Big)
G(\ka\lambda x,\kb\lambda x)\big|_{x=\tilde{x}}\,,
\nonumber
\end{align}
being related to the scalar operator,
as well as 
\begin{align}
G_{(\alpha\beta)}^{\mathrm{tw4(iv)c}}(\ka\xx,\kb\xx)
=& - g_{\alpha\beta}\pd^\mu 
\int_0^1\d\lambda
\hbox{\Large$\frac{1}{\lambda}$}
G^{+}_{\mu}(\ka\lambda x,\kb\lambda x)\big|_{x=\tilde{x}}\,,
\nonumber\\
G_{(\alpha\beta)}^{\mathrm{tw4(iv)d2}}(\ka\xx,\kb\xx)
=&\,-2 x_{(\alpha}\pd_{\beta)}\pd^\mu 
\int_0^1\d\lambda\Big(
\hbox{\Large$\frac{1-\lambda^2}{2\lambda}$}
+\hbox{\Large$\frac{\ln\lambda}{\lambda}$}\Big)
G^{\mathrm{+}}_{\mu}(\ka\lambda x,\kb\lambda x)\big|_{x=\tilde{x}}
\nonumber\\&
-G_{(\alpha\beta)}^{\mathrm{tw6(iv)d2}}(\ka\xx,\kb\xx)\,,
\nonumber\\
G_{(\alpha\beta)}^{\mathrm{tw4(iv)e1}}(\ka\xx,\kb\xx)
=&\, x_{(\alpha}\pd_{\beta)}\pd^\mu 
\int_0^1\d\lambda \hbox{\Large$\frac{1-\lambda^2}{\lambda}$}
G^{\mathrm{+}}_{\mu}(\ka\lambda x,\kb\lambda x)\big|_{x=\tilde{x}}
-G_{(\alpha\beta)}^{\mathrm{tw6(iv)e1}}(\ka\xx,\kb\xx)\,,
\nonumber
\end{align}
and
\begin{align}
G_{(\alpha\beta)}^{\mathrm{tw5(iv)d}}(\ka\xx,\kb\xx)
=&\,
x_{(\alpha}\big(\delta^\mu_{\beta)}(x\pd)-x^\mu\pd_{\beta)}\big)\square 
\int_0^1\!\d\lambda\Big(
\hbox{\Large$\frac{1-\lambda^2}{2\lambda}$}
+\hbox{\Large$\frac{\ln\lambda}{\lambda}$}\Big)
G^{\mathrm{+}}_{\mu}(\ka\lambda x,\kb\lambda x)\big|_{x=\tilde{x}}\,,
\nonumber\\
&-G_{(\alpha\beta)}^{\mathrm{tw6(iv)d3}}(\ka\xx,\kb\xx)
-G_{(\alpha\beta)}^{\mathrm{tw6(iv)d4}}(\ka\xx,\kb\xx)\,,\nonumber
\end{align}
and
\begin{align}
G_{(\alpha\beta)}^{\mathrm{tw6(iv)c}}(\ka\xx,\kb\xx)
=&
\hbox{\Large$\frac{1}{2}$}x_\alpha x_\beta\square\pd^\mu 
\int_0^1\d\lambda
\hbox{\Large$\frac{(1-\lambda)^2}{\lambda}$}
G^{\mathrm{+}}_{\mu}(\ka\lambda x,\kb\lambda x)\big|_{x=\tilde{x}}\,,
\nonumber\\
G_{(\alpha\beta)}^{\mathrm{tw6(iv)d2}}(\ka\xx,\kb\xx)
=&\,\hbox{\Large$\frac{1}{2}$}
x_\alpha x_\beta\square\pd^\mu \!
\int_0^1\!\d\lambda\Big(
\hbox{\Large$\frac{1-\lambda^2}{2\lambda}$}
+\hbox{\Large$\frac{\ln\lambda}{\lambda}$}
+\hbox{\Large$\frac{\ln^2\lambda}{\lambda}$}\Big)
G^{\mathrm{+}}_{\mu}(\ka\lambda x,\kb\lambda x)\big|_{x=\tilde{x}}\,,
\nonumber\\
G_{(\alpha\beta)}^{\mathrm{tw6(iv)d4}}(\ka\xx,\kb\xx)
=&
-x_\alpha x_\beta\square\pd^\mu \!
\int_0^1\!\d\lambda\,
\hbox{\Large$\frac{\ln^2\lambda}{\lambda}$}\,
G^{\mathrm{+}}_{\mu}(\ka\lambda x,\kb\lambda x)\big|_{x=\tilde{x}}\,,
\nonumber\\
G_{(\alpha\beta)}^{\mathrm{tw6(iv)e1}}(\ka\xx,\kb\xx)
=&\,
-\hbox{\Large$\frac{1}{2}$}
x_\alpha x_\beta\square\pd^\mu 
\int_0^1\d\lambda\Big(
\hbox{\Large$\frac{1-\lambda^2}{2\lambda}$}
+\hbox{\Large$\frac{\ln\lambda}{\lambda}$}\Big)
G^{\mathrm{+}}_{\mu}(\ka\lambda x,\kb\lambda x)\big|_{x=\tilde{x}}\,,
\nonumber\\
G_{(\alpha\beta)}^{\mathrm{tw6(iv)e3}}(\ka\xx,\kb\xx)
=&\,
x_\alpha x_\beta\square\pd^\mu 
\int_0^1\d\lambda\Big(
\hbox{\Large$\frac{1-\lambda^2}{2\lambda}$}
+\hbox{\Large$\frac{\ln\lambda}{\lambda}$}\Big)
G^{\mathrm{+}}_{\mu}(\ka\lambda x,\kb\lambda x)\big|_{x=\tilde{x}}\,,
\nonumber
\end{align}
being related to the symmetric vector operator,
\begin{align}
G_{(\alpha\beta)}^{\mathrm{tw5(iv)e}}(\ka\xx,\kb\xx)
=&\,
x_{(\alpha}\big(\delta^\nu_{\beta)}(x\pd)-x^\nu\pd_{\beta)}\big)
\pd^\mu 
\int_0^1\d\lambda\hbox{\Large$\frac{1-\lambda^2}{\lambda}$}
\left.G_{(\mu\nu)}(\ka\lambda x,\kb\lambda x)\right|_{x=\tilde{x}}\,,
\nonumber\\
&\!\!\!\!-G_{(\alpha\beta)}^{\mathrm{tw6(iv)e3}}(\ka\xx,\kb\xx)
-G_{(\alpha\beta)}^{\mathrm{tw6(iv)e4}}(\ka\xx,\kb\xx)
-G_{(\alpha\beta)}^{\mathrm{tw6(iv)e5}}(\ka\xx,\kb\xx)
\nonumber\\
G_{(\alpha\beta)}^{\mathrm{tw6(iv)e4}}(\ka\xx,\kb\xx)
=&\, x_\alpha x_\beta\pd^\mu\pd^\nu
\int_0^1\d\lambda \hbox{\Large$\frac{1-\lambda^2}{\lambda}$}
G_{(\mu\nu)}(\ka\lambda x,\kb\lambda x)\big|_{x=\tilde{x}}\,,
\nonumber\\
G_{(\alpha\beta)}^{\mathrm{tw6(iv)f}}(\ka\xx,\kb\xx)
=&\,- x_\alpha x_\beta\pd^\mu\pd^\nu
\int_0^1\d\lambda(1-\lambda)
G_{(\mu\nu)}(\ka\lambda x,\kb\lambda x)\big|_{x=\tilde{x}}\,,
\nonumber
\end{align}
being related to the symmetric tensor operator and, furthermore,
\begin{align}
G_{(\alpha\beta)}^{\mathrm{tw4(iv)f}}(\ka\xx,\kb\xx)
=&\, g_{\alpha\beta}(x\pd)\int_0^1\d\lambda
\hbox{\Large$\frac{1}{2\lambda}$}
G^\rho_{\ \rho}(\ka\lambda x,\kb\lambda x)\big|_{x=\tilde{x}}\,,
\nonumber\\
G_{(\alpha\beta)}^{\mathrm{tw4(iv)h}}(\ka\xx,\kb\xx)
=& -x_{(\alpha}\pd_{\beta)}\!\int_0^1\!\d\lambda\,\lambda
G^\rho_{\ \rho}(\ka\lambda x,\kb\lambda x)\big|_{x=\tilde{x}}\!
-G_{(\alpha\beta)}^{\mathrm{tw6(iv)h}}(\ka\xx,\kb\xx)\,,
\nonumber\\
G_{(\alpha\beta)}^{\mathrm{tw4(iv)e2}}(\ka\xx,\kb\xx)
=& -x_{(\alpha}\pd_{\beta)}
\!\int_0^1\!\d\lambda
\hbox{\Large$\frac{1-\lambda^2}{\lambda}$}
G^\rho_{\ \rho}(\ka\lambda x,\kb\lambda x)\big|_{x=\tilde{x}}\!
-G_{(\alpha\beta)}^{\mathrm{tw6(iv)e2}}(\ka\xx,\kb\xx)\,,
\nonumber\\
\intertext{and}
G_{(\alpha\beta)}^{\mathrm{tw6(iv)h}}(\ka\xx,\kb\xx)
=&\,-\hbox{\Large$\frac{1}{4}$}x_\alpha x_\beta\square
\int_0^1\d\lambda \hbox{\Large$\frac{1-\lambda^2}{\lambda}$}
G^\rho_{\ \rho}(\ka\lambda x,\kb\lambda x)\big|_{x=\tilde{x}}\,,
\nonumber\\
G_{(\alpha\beta)}^{\mathrm{tw6(iv)e2}}(\ka\xx,\kb\xx)
=&\,
\hbox{\Large$\frac{1}{2}$}x_\alpha x_\beta\square
\int_0^1\d\lambda
\Big( \hbox{\Large$\frac{1-\lambda^2}{2\lambda}$}
+\hbox{\Large$\frac{\ln\lambda}{\lambda}$}\Big)
G^\rho_{\ \rho}(\ka\lambda x,\kb\lambda x)\big|_{x=\tilde{x}}\,,
\nonumber\\
G_{(\alpha\beta)}^{\mathrm{tw6(iv)k}}(\ka\xx,\kb\xx)
=&\,\hbox{\Large$\frac{1}{2}$}x_\alpha x_\beta\square
\int_0^1\d\lambda\,(1-\lambda)\,
G^\rho_{\ \rho}(\ka\lambda x,\kb\lambda x)\big|_{x=\tilde{x}}\,,
\nonumber\\
G_{(\alpha\beta)}^{\mathrm{tw6(iv)e5}}(\ka\xx,\kb\xx)
=&\,-
x_\alpha x_\beta\square
\int_0^1\d\lambda
\Big( \hbox{\Large$\frac{1-\lambda^2}{2\lambda}$}
+\hbox{\Large$\frac{\ln\lambda}{\lambda}$}\Big)
G^\rho_{\ \rho}(\ka\lambda x,\kb\lambda x)\big|_{x=\tilde{x}}\,,
\nonumber
\end{align}
being related to the trace of the symmetric tensor operator.

Here, it should be remarked that, contrary to the former cases (i)
-- (iii), the trace terms (\ref{G_high_(iv)_sy}) contain operators
having the same twist, $\tau = 4$, as the primary operator of symmetry 
class (iv). However, these operators being multiplied by
either $g_{\alpha\beta}$ or $x_\alpha$ and $x_\beta$ are scalar or
vector operators corresponding to symmetry type (i). Whereas
the local operators resulting from the twist--4 tensor operator of 
symmetry class (iv) are contained in the tensor space
${\bf T}\hbox{$(\frac{n+2}{2},\frac{n-2}{2})\oplus
{\bf T}(\frac{n-2}{2},\frac{n+2}{2})$}$, 
the local operators of the twist--4 scalar and vector operators 
are totally symmetric traceless tensors and, therefore, contained
in the tensor space ${\bf T}\hbox{$(\frac{n}{2},\frac{n}{2})$}$.
\subsubsection{Determination of the complete symmetric 
light--cone tensor operators}
Now, having finished the decomposition of a general 2nd rank tensor
operator, let us sum up the remaining symmetric tensor operators of
twist greater than 3, which appear in
the trace terms of the symmetric twist operators with Young symmetry (i), (ii)
as well as (iv), to  complete twist--4, twist--5 and twist--6 operators. 

The `scalar part' of the twist--4 light--ray tensor operator is given by
\begin{align}
\label{G_tw4sc_i}
\lefteqn{\hspace{-2cm}
G_{(\alpha\beta)}^{\mathrm{tw4,s}}(\ka\xx,\kb\xx)=-g_{\alpha\beta}
\bigg\{
\hbox{\Large$\frac{1}{2}$}\square\int_0^1\d\lambda\,\lambda(\ln\lambda)
G(\ka\lambda x,\kb\lambda x)
}\\
&-(x\pd)\int_0^1\d\lambda\hbox{\Large$\frac{1}{2\lambda}$}
G^\rho_{\ \rho}(\ka\lambda x,\kb\lambda x)
+\pd^\mu\int_0^1\d\lambda\,\lambda
G^+_{\mu}(\ka\lambda x,\kb\lambda x)
\bigg\}\bigg|_{x=\tilde{x}}\,,
\nonumber
\end{align}
and the `vector part' of the twist--4 light--ray tensor operator reads
\begin{eqnarray}
\label{G_tw4ve_i}
\lefteqn{
G_{(\alpha\beta)}^{\mathrm{tw4,v}}(\ka\xx,\kb\xx)
=-\lcx_{(\alpha}\times\nonumber}\\
&&\!\bigg\{\!
\int_0^1\!\d\lambda\Big\{
\Big(\hbox{\Large$\frac{1-\lambda^2}{2\lambda}$}+\lambda\ln\lambda\Big)
\pd_{\beta)}+\hbox{\Large$\frac{1}{4}$}x_{\beta)}
\Big(\hbox{\Large$\frac{1-\lambda^2}{\lambda}$}+\lambda\ln\lambda
+\hbox{\Large$\frac{\ln\lambda}{\lambda}$}\Big)\square\Big\}\square
G(\ka\lambda x,\kb\lambda x)\nonumber\\
&&\!+\int_0^1\!\hbox{\Large$\frac{\d\lambda}{\lambda}$}
\Big\{\pd_{\beta)}+\hbox{\Large$\frac{1}{2}$}x_{\beta)}
(\ln\lambda)\square\Big\}
G^\rho_{\ \rho}(\ka\lambda x,\kb\lambda x)\\
&&\!-\int_0^1\!\hbox{\Large$\frac{\d\lambda}{\lambda}$}
\Big\{2\big(1-\lambda^2\big)\pd_{\beta)}
+\hbox{\Large$\frac{1}{2}$}x_{\beta)}
\big(1-\lambda^2+2\ln\lambda\big)\square\Big\}\pd^\mu
G^+_{\mu}(\ka\lambda x,\kb\lambda x)
\bigg\}\bigg|_{x=\tilde{x}}.
\nonumber
\end{eqnarray}
Furthermore, the contributions to the complete `vector part' of the 
twist--5 light--ray tensor operator are 
constructed by means of the Young tableau (ii), and their 
local operators are antisymmetric\footnote{
Note that antisymmetry is not in $\alpha$ and $\beta$ but results
from the differential operator!} traceless tensors contained
in the space ${\bf T}\hbox{$(\frac{n}{2},\frac{n-2}{2})\oplus
{\bf T}(\frac{n-2}{2},\frac{n}{2})$}$. 
This complete twist--5 tensor operator is given by
\begin{align}
\label{G_tw5_ii_sy}
&G_{(\alpha\beta)}^{\mathrm{tw5,v}}(\ka\xx,\kb\xx)
=-\lcx_{(\alpha}
\bigg\{
\int_0^1\d\lambda\Big\{
\hbox{\Large$\frac{(1-\lambda)^2}{2\lambda}$}\,
\big(\delta^\nu_{\beta)}(x\pd)-x^\nu\pd_{\beta)}\big)\square
+\Big(\hbox{\Large$\frac{1-\lambda}{\lambda}$}
+\hbox{\Large$\frac{\ln\lambda}{\lambda}$}\Big)
x_{\beta)}\pd^\nu\square\nonumber\\
&\qquad\qquad\qquad\qquad
-\Big(\hbox{\Large$\frac{1-\lambda}{\lambda}$}
-\hbox{\Large$\frac{1-\lambda^2}{4\lambda}$}
+\hbox{\Large$\frac{1}{2\lambda}$}\ln\lambda\Big)
x_{\beta)}\square^2 x^\nu\Big\}
G^+_\nu(\ka\lambda x,\kb\lambda x)\nonumber\\
&\qquad-\int_0^1\d\lambda\Big\{
\hbox{\Large$\frac{1-\lambda^2}{\lambda}$}\,
\big(\delta^\nu_{\beta)}(x\pd)-x^\nu\pd_{\beta)}-x_{\beta)}\pd^\nu\big)\pd^\mu 
-\Big(\hbox{\Large$\frac{1-\lambda^2}{2\lambda}$}
+\hbox{\Large$\frac{\ln\lambda}{\lambda}$}\Big)
x_{\beta)}\square
\big(\pd^\mu x^\nu-g^{\mu\nu}\big)
\Big\}\nonumber\\
&\hspace{5cm}\times G_{(\mu\nu)}(\ka\lambda x,\kb\lambda x)
\bigg\}\bigg|_{x=\tilde{x}}.
\end{align}
The `scalar part' of the twist--6 light--ray tensor operator has 
Young symmetry (i). 
Thus, the local twist--6 operators are totally symmetric traceless tensors 
lying in the space ${\bf T}\hbox{$(\frac{n-2}{2},\frac{n-2}{2})$}$. 
This nonlocal twist--6 operator is given by
\begin{align}
\label{G_tw6_i}
&G_{(\alpha\beta)}^{\mathrm{tw6,s}}(\ka\xx,\kb\xx)
=-\lcx_\alpha\lcx_\beta\bigg\{
\hbox{\Large$\frac{1}{2}$}\,\square^2
\int_0^1\d\lambda
\Big(\hbox{\Large$\frac{1-\lambda}{\lambda}$}
+\hbox{\Large$\frac{1+\lambda}{2\lambda}$}\ln\lambda\Big)
G(\ka\lambda x,\kb\lambda x)\nonumber\\
&\qquad
+\hbox{\Large$\frac{1}{2}$}\,\square\int_0^1\d\lambda\,
\Big(\hbox{\Large$\frac{1-\lambda}{\lambda}$}
+\hbox{\Large$\frac{\ln\lambda}{\lambda}$}\Big)
\Big( G^\rho_{\ \rho}(\ka\lambda x,\kb\lambda x)
-2\pd^\mu G^+_\mu(\ka\lambda x,\kb\lambda x)\Big)\nonumber\\
&\qquad
-\pd^\mu\pd^\nu\int_0^1\d\lambda\,\hbox{\Large$\frac{1-\lambda}{\lambda}$}\,
G_{(\mu\nu)}(\ka\lambda x,\kb\lambda x)
\bigg\}\bigg|_{x=\tilde{x}}.
\end{align}

Now, we may pick up all the contributions to the symmetric
(traceless) tensor operator.
Together with the twist--2 part, Eq.~(\ref{Gtw2_gir}), the twist--3 
part, Eq.~(\ref{G3_symm}), and the twist--4 part, Eq.~(\ref{G_tw4_iv}), 
we finally obtain the following complete decomposition
of the {\em symmetric light--cone tensor operator} (compare
Eq.~(\ref{G^+_t})):
\begin{align}
\label{complete_sym}
G_{(\alpha\beta)}(\ka\xx,\kb\xx)
&=
G^{\rm tw2(i)}_{(\alpha\beta)}(\ka\xx,\kb\xx)+
G^{\rm tw3(ii)}_{(\alpha\beta)}(\ka\xx,\kb\xx)+
G^{\rm tw4(iv)}_{(\alpha\beta)}(\ka\xx,\kb\xx)\quad\nonumber\\
&+
G^{\rm tw4,s}_{(\alpha\beta)}(\ka\xx,\kb\xx)+
G^{\rm tw4,v}_{(\alpha\beta)}(\ka\xx,\kb\xx)\quad\nonumber\\
&+
G^{\rm tw5,v}_{(\alpha\beta)}(\ka\xx,\kb\xx)+
G^{\rm tw6,s}_{(\alpha\beta)}(\ka\xx,\kb\xx)\,.
\end{align}

Let us remark that by construction the terms of the first line are
traceless, whereas the traces of the third line vanish because of
antisymmetry and $\xx^2 = 0$ on the light--cone, respectively.
However, the traces of the second line do not vanish but restore
the trace of the tensor operator,
\begin{eqnarray}
g^{\alpha\beta} G_{\alpha\beta} (\ka\xx,\kb\xx)
=
g^{\alpha\beta} G^{\rm tw4,s}_{(\alpha\beta)} (\ka\xx,\kb\xx)
+
g^{\alpha\beta} G^{\rm tw4,v}_{(\alpha\beta)} ((\ka\xx,\kb\xx)).
\end{eqnarray}
This may be proven explicitly by taking the trace of 
Eqs.~(\ref{G_tw4sc_i}) and (\ref{G_tw4ve_i}), using 
$(x\pd) \cong \lambda \pd_\lambda$,
performing partial integrations and observing the equality
\begin{eqnarray}
\label{G_trace}
\big(g^{\alpha\beta} - 2 \pd^\alpha x^\beta 
+ (1/2)\square x^\alpha x^\beta \big)
G_{(\alpha\beta)} (\ka\lambda x,\kb\lambda x)\big|_{\lambda =0} = 0\,.
\end{eqnarray}

Herewith, the twist decomposition of a generic bilocal 2nd rank
light--ray tensor operator is completed. In the next Section we 
show how this general formalism may be applied also to tri-- and
multi--local light--ray operators.

\subsection{Conclusions}
Let us shortly summarize our results.  
We determined in a systematic way the various contributions of definite twist 
for a generic bilocal 2nd rank tensor operator 
$G_{\alpha\beta}(\ka x, \kb x)$ as well as its symmetric 
and antisymmetric parts $G_{(\alpha\beta)}(\ka x, \kb x)$ and 
$G_{[\alpha\beta]}(\ka x, \kb x)$, respectively. In
addition, we determined the related vector and scalar operators which occur 
by truncating with $x^\beta$ and $x^\alpha x^\beta$ or $g^{\alpha\beta}$, 
respectively. 
By projection onto the light--cone 
we obtained the decomposition of a generic
bilocal light--ray tensor operator of 2nd rank -- and its reductions to 
(anti)symmetric tensors as
well as the related vector and scalar operators -- into operators of definite 
twist. In order to make the main results quite obvious we summarize them in 
Table~\ref{tab:gluon-ten} and \ref{tab:gluon-vec}.\footnote{
After substitution $\tau\rightarrow\tau-1$ in all formulas for $G_{[\alpha\beta]}(\ka\xx,\kb\xx)$
in Table~\ref{tab:gluon-ten} we get the corresponding twist decomposition for
the bilocal operator $M_{[\alpha\beta]}(\ka\xx,\kb\xx)$. 
If we substitute $\d\lambda\,\lambda\rightarrow\d\lambda$ in the formulas for
the vector operator $G_{\alpha}(\ka\xx,\kb\xx)$ in Table~\ref{tab:gluon-vec},
one obtains the twist decomposition for the operator 
$O_{\alpha}(\ka\xx,\kb\xx)$.} 
There, we indicate the different twist contributions 
to $G_{\alpha\beta}(\ka \xx, \kb \xx) =
G_{(\alpha\beta)}(\ka \xx, \kb \xx)+G_{[\alpha\beta]}(\ka \xx, \kb \xx)$
by their symmetry type (i) up to (iv) and the number of the
corresponding equations; in the second half of the Tables 
the contributions from the trace terms are shown. 
\begin{table}[h]
\begin{center}
\begin{tabular}{|c|c|c|c|}
\hline
Eq.& YT &   
$G_{(\alpha\beta)}(\kappa_1\lcx,\kappa_2\lcx)$
& 
$G_{[\alpha\beta]}(\kappa_1\lcx,\kappa_2\lcx)$\\
\hline
\ref{Gtw2_gir} & (i)    &  $\tau=2$ & --  \\
\ref{G3_symm}  & (ii)   & $\tau=3$ & -- \\
\ref{G3_anti}  & (ii)   & -- & $\tau=3$  \\
\ref{G_tw4_iii} & (iii) & -- & $\tau=4$  \\
\ref{G_tw4_iv} & (iv)   & $\tau=4$  &  -- \\
\ref{G_tw4sc_i} & (i)   & $\tau=4$ &  -- \\
\ref{G_tw4ve_i} & (i)   &  $\tau=4$ &  -- \\
\ref{G_tw4_as} & (i)    & -- & $\tau=4$ \\
\ref{G_tw5_ii_sy} & (ii) & $\tau=5$ &  -- \\
\ref{G_tw5_as} & (ii)   & -- & $\tau=5$  \\
\ref{G_tw6_i}  & (i)    & $\tau=6$ &  -- \\
\hline
\end{tabular}
\end{center}
\caption{Twist decomposition of general tensor operators
$G_{\alpha\beta} = G_{(\alpha\beta)} + G_{[\alpha\beta]}$.}
\label{tab:gluon-ten}
\end{table}
\begin{table}[h]
\begin{center}
\begin{tabular}{|c|c|c|c|}
\hline
Eq.& YT &  
$G_{(\alpha\bullet)}(\kappa_1\lcx,\kappa_2\lcx)$
& 
$G_{[\alpha\bullet]}(\kappa_1\lcx,\kappa_2\lcx)$\\
\hline
\ref{G_tw2_vec} & (i)  & $\tau=2$   &  -- \\
\ref{G_tw3_vec_sy} & (ii) & $\tau=3$  & -- \\
\ref{G_tw3_vec_asy} & (ii) & -- & $\tau=3$  \\
\ref{G_tw4_vec_sy} & (i)  & $\tau=4$   &  -- \\
\ref{G_tw4_vec_asy} & (i) & -- & $\tau=4$  \\
\hline
\end{tabular}
\end{center}
\caption{Twist decomposition of general vector operators
$G_\alpha = G_{(\alpha\bullet)} + G_{[\alpha\bullet]}$.}
\label{tab:gluon-vec}
\end{table}

The present study made also obvious how difficult an exact treatment of
higher twist contributions will be. There is a complicate interplay
between the higher twists resulting from different symmetry types
and the corresponding trace terms. The results also suffer from 
the equations of motion (EOM) which are frequently used in the 
applications
because they do not contribute between physical states \cite{P}. 
The twist decomposition of these EOM operators
\begin{align}
{}^{\text{EOM}\!}\OO_\Gamma(\ka\lcx,\kb\lcx)&=
\bar{\psi}(\ka\lcx) \Gamma
U(\ka\lcx,\kb\lcx)\Big[\ii\RD D\Dslas-m\Big]\psi(\kb\lcx)\nonumber\\
&-\bar{\psi}(\ka\lcx) \Big[\ii\LD D\Dslas+m\Big]\Gamma
U(\ka\lcx,\kb\lcx)\psi(\kb\lcx),
\nonumber
\end{align}
is analogous to the corresponding quark operators
$\OO_\Gamma(\ka\lcx,\kb\lcx)$ which is given in~\cite{Lazar98,GLR99}.
The only difference is that for the equation of motion operators
the canonical dimension and the twist raises by one unit.

\section{Extension to trilocal 
tensor operators}
\label{trilocal}
\setcounter{equation}{0}
The procedure discussed in the previous Section for bilocal gluon operators may be 
extended to trilocal operators, too. These operators occur in the higher 
twist contributions 
to light--cone dominated processes and, additionally, as counterparts in the 
renormalization of such higher twist operators which have been obtained
in Section~\ref{gluon}. In the following we consider various trilocal operators of 
minimal twist--3 and twist--4. 
The local versions of such higher twist operators have been considered
already in the early days of QCD in systematic studies of deep inelastic
scattering \cite{G} and on behalf of giving some parton interpretation
for the distribution amplitudes
\cite{P,EFP,SV,jaffe83,BFKL}. Various studies of local twist--4 operators determined
their anomalous dimension matrices and the behaviour of their structure 
functions \cite{Wada,Okawa,Luttrell81,Luttrell82,JS82}. The relevance of local 
twist--3 operators for
the structure function $g_2$ of polarized deep inelastic scattering is
also well-known~\cite{AR,SV,BKL,Ratcliffe,Kodaira95a,Kodaira95b,Kodaira96,Kodaira99}. 
If written nonlocal all these 
operators are necessarily multilocal. However, nonlocal higher twist 
operators have been extensively used in the study of the structure 
function $g_2$ \cite{BB88,Geyer96a,Geyer96b} and of the photon and 
vector meson wave functions \cite{BBK89,BB91,ball98,ball99}.

Here, our main interest is not to give an exhaustive study of the various
scattering processes but to present the twist decomposition of some 
characteristic multilocal operators.

\subsection{General trilocal
tensor operators: 
     Quark--antiquark--gluon, four--quark and four--gluon operators}

This Subsection is
devoted to the consideration of trilocal operators 
which share the same twist decomposition as the gluon operators 
$G_{\alpha\beta}$ (or their (anti)symmetric parts). We call them 
unconstrained because their truncation with $x^\alpha x^\beta$ does
not vanish identically. In order to compactify notations we introduce 
the following abbreviations:
\begin{align}
\Gamma^i 
= \{ 1, \gamma_5\}, \quad
\Gamma^i_\alpha 
= \{\gamma_\alpha, \gamma_5\gamma_\alpha\}, \quad
\Gamma^i_{\alpha\beta} 
= \{\sigma_{\alpha\beta}, \gamma_5\sigma_{\alpha\beta}\}, \quad
F^i_{\mu\nu} = \{F_{\mu\nu}, \widetilde F_{\mu\nu}\}\,.
\nonumber
\end{align}
\subsubsection{Quark--antiquark--gluon operators}
First, we consider the following quark--antiquark--gluon 
operators,\footnote{
For simplicity we restrict to the nonsinglet case and suppress flavour 
matrices.}
\begin{align}
V_{\alpha\beta}^{ij}(\ka x,\tau x,\kb  x)
&=
\bar{\psi}(\ka x)U(\ka x,\tau x)\Gamma_\alpha^{i\,\rho} 
F^j_{\beta\rho}(\tau x)U(\tau x,\kb x)\psi(\kb x)\,,
\\
\intertext{and}
W^{ij}_{[\alpha\beta]}(\ka x,\tau x,\kb  x)
&=
\bar{\psi}(\ka x) U(\ka x,\tau x) \Gamma^i 
F^j_{\alpha\beta}(\tau x) U(\tau x,\kb x)\psi(\kb x)\,,
\end{align}
whose minimal twist is $\tau_{\rm min} = 3$ and $4$, respectively; 
they have the same symmetry as the gluon tensor operators 
$G_{\alpha\beta}$ and $G_{[\alpha\beta]}$, respectively. 
The trilocal operators $V_{\alpha\beta}^{ij}(\ka x,\tau x,\kb  x)$
are inalienable for a systematic study of the chiral-odd twist-3
distribution functions $h_L(x)$ and $e(x)$~\cite{JJ92,Koike95,Koike97,ball96} which
give information about the quark-gluon correlations in the nucleon.

In order to be able to apply the procedure of twist decomposition
as well as the results of Section~\ref{gluon} we have to verify the local structure 
of these trilocal operators.
Let us study, for example, the first of the above quark--gluon
operators,
$V^{11}_{\alpha\beta}$, in some detail 
and Taylor expand its three local fields around
$y=0$:
\begin{align}
\lefteqn{
\hspace{-.9cm}
V^{11}_{\alpha\beta}(\ka x,\tau x,\kb x)=
}\nonumber\\
&=\
\bar{\psi}(\ka x)U(\ka x,0)U(0,\tau x)\sigma_\alpha^{\ \rho}
F_{\beta\rho}(\tau x)U(\tau x,0)U(0,\kb x)\psi(\kb x)
\nonumber\\
&=
\!\sum_{n_1,m,n_2 =0}^\infty\!
\hbox{\Large$\frac{\ka^{n_1}\tau^m\kb^{n_2}}{n_1!m!n_2!}$}
\Big[\bar\psi(y)\!\left
(\!\LD D_y x\right)^{\!\!n_1}\!\Big]
\sigma_\alpha^{\ \rho}
\Big[\!\left(x D_z\right)^m\! F_{\beta\rho}(z)\Big]\!
\Big[\!\!\left(x \RD D_y \right)^{\!\!n_2}\!\!\!
\psi(y)\Big]\Big|_{y=z=0}
\nonumber\\
&=
\sum_{N=0}^\infty
\hbox{\Large$\frac{1}{N!}$}\, \bar\psi(0)\!
\left(
\sum_{n=0}^N \!\sum_{\ell=0}^n
\!
\hbox{\Large$\binom{N}{n}  \binom{n}{\ell}$}
\!\left(\!\ka \!\LD D x \right)^{\!\!n-\ell}\!
\!\!\!\sigma_\alpha^{\ \rho}
\Big[\!\left(\tau x D \right)^{N-n}\!
F_{\beta\rho}(0) \!\Big]\!\!
\left(\!\kb x\! \RD D \right)^{\!\!\ell}\!\right)\!
\psi(0)\,,
\nonumber
\end{align}
where the left and right derivatives are given by
Eqs.~(\ref{D_kappa}), 
and
\begin{equation}
D_\mu F_{\alpha\beta} 
= \pd_\mu F_{\alpha\beta} + \ii g [A_\mu, F_{\alpha\beta}]\,.
\end{equation}
This leads to the
following expansion into local tensor operators
\begin{equation}
\label{Vij}
V^{11}_{\alpha\beta}(\ka x,\tau x,\kb  x)
=
\sum_{N=0}^\infty \frac{1}{N!} x^{\mu_1}\ldots x^{\mu_N}
V^{11}_{\alpha\beta\mu_1\ldots\mu_N}(\ka,\tau,\kb)\,,
\end{equation}
 where
\begin{equation}
\label{V_local3}
V^{11}_{\alpha\beta\mu_1\ldots\mu_N}(\ka,\tau,\kb)
\equiv 
\bar\psi(0) \sigma_\alpha^{\ \rho}\
{\bf D}_{[\beta\rho]\mu_1\ldots\mu_N}(\ka,\tau,\kb) \psi(0)\,,
\end{equation}
and
the field strength--dependent generalized covariant derivative 
of $N$th order is given by
\begin{align}
\label{D_F}
&{\bf D}_{[\beta\rho]\mu_1\ldots\mu_N}(\ka,\tau,\kb)=
\\
&\quad= N!\sum_{n=0}^N \sum_{\ell=0}^n
\hbox{\Large$\frac{\ka^{n-\ell}}{(n-\ell)!}$}
\LD D_{\mu_1}\! \ldots \!\LD D_{\mu_{n-\ell}}
\hbox{\Large$\frac{\tau^{N-n}}{(N-n)!}$}
\Big[D_{\mu_{n+1}}\!\ldots\!D_{\mu_{N}} 
F_{\beta\rho}(0)\Big]
\hbox{\Large$\frac{\kb^\ell}{\ell!}$}
\RD D_{\mu_{n-\ell+1}}\!\ldots \!\RD D_{\mu_n}.
\nonumber
\end{align}
Obviously, if the field
strength were not present this operation 
would reduce to the product 
$\Tensor D_{\mu_1}(\ka,\kb)\ldots \Tensor D_{\mu_N}(\ka,\kb)$
of generalized derivatives
(\ref{D_kappa}).
Furthermore, for $\tau =0$ we are left with a more
simple expression 
which, however, cannot be reduced to the $N$th power of some 
extended
derivative because $F$ is equipped with some matrix structure. 
Anyway, the local tensor
operators (\ref{V_local3}) of rank $N+2$
with canonical dimension $d=N+5$, which are given
as a sum of
 $\frac{1}{2} (N+1)(N+2)$ terms, have to be decomposed 
according to their
geometric twist. In principle, this has to be done 
term-by-term. But, because of the linearity of
that procedure we are 
not required to do this for any term explicitly.

Due to the same general local structure of $V^{ij}_{\alpha\beta}$ and 
$W^{ij}_{\alpha\beta}$ -- which is governed by Taylor expansions 
completely analogous to Eq.~(\ref{Vij}) -- their decomposition into 
terms of definite twist leads to the same expressions as we obtained 
in Section~\ref{gluon} but
with $G_{\alpha\beta}$ and $G_{[\alpha\beta]}$ exchanged by 
$V^{ij}_{\alpha\beta}$ and $W^{ij}_{\alpha\beta}$, respectively. 
The only difference consists in a shift of any twist by one unit, 
$\tau \rightarrow \tau + 1$, in the various inputs of Table~\ref{tab:gluon-ten} and 
\ref{tab:gluon-vec}.
 
In contrast to the operators $V$ and $W$ the quark--antiquark--gluon
operators
\begin{align}
\label{OO}
\OO^{ij}_{\alpha\beta}(\ka x,\tau x,\kb x)
&= x^\rho\,\bar{\psi}(\ka x)U(\ka x,\tau x)\Gamma_\alpha^i 
F^j_{\beta\rho}(\tau x)U(\tau x,\kb x)\psi(\kb x)
\\ 
\intertext{and}
\label{OO'}
{\OO'}^{ij}_{\alpha\beta}(\ka x,\tau x,\kb x)
&=x^\rho x^\sigma \bar{\psi}(\ka x)U(\ka x,\tau x)\Gamma^i_{\alpha\rho} 
F^j_{\beta\sigma}(\tau x)U(\tau x,\kb x)\psi(\kb x)
\end{align}
are special, namely, despite of their arbitrary symmetry with respect to
$\alpha$ and $\beta$ they vanish identically if multiplied by $x^\beta$
and $x^\alpha$ or $x^\beta$, respectively. Therefore, they show some
peculiarities which will be treated in Subsection 4.2.
\subsubsection{Four--quark and four--gluon operators}
Let us now consider trilocal operators which are built up from 
four quark or four gluon fields. We denote them by 
${}^{I}\!\QQ^{ij}_{\alpha\beta}$ with 
$I=0$ and $I=2$, respectively (here, $I$ counts the
number of external $x$'s of the operators):
\begin{align}
{}^{I=0}\!\QQ^{ij}_{\alpha\beta}(\ka x,\tau x,\kb x)
=
\big(\,\bar\psi(\ka x)\Gamma^i_\alpha U(\ka x,\tau x)\psi(\tau x)\big)
\big(\,\bar\psi(\tau x) U(\tau x,\kb x)\Gamma^j_\beta\psi(\kb x)\big)
\end{align}
as well
as
\begin{align}
{}^{I=2}\!\QQ^{ij}_{\alpha\beta}&(\ka x,\tau x,\kb x)=
x^\rho x^\sigma\,
{}^{I=2}\!\QQ^{ij}_{\alpha\rho\sigma\beta}(\ka x,\tau x,\kb x)\,
\\
&= x^\rho x^\sigma
\Big(F_{\alpha\mu}(\ka x)U(\ka x,\tau x)F_\rho^{i\mu}(\tau
x)\Big)
\Big(F_\sigma^{j\nu}(\tau x)U(\tau x,\kb x)F_{\beta\nu}(\kb x)\Big)\,.
\nonumber
\end{align}

Both sets of operators have minimal twist $\tau_{\rm min} =4$. 
Their local versions have been already considered by various authors, 
see, e.g.,~\cite{JS82}. The explicit form of the local operators
will be given below. Because
there are no restrictions concerning the free indices $\alpha$ and $\beta$
the Young patterns (i) -- (iv) are involved. The twist decomposition 
may be performed along the lines of Section~\ref{gluon} and, as in the
foregoing cases, the outcome will be almost the same as for 
$G_{\alpha\beta}$. Again, the difference is that now the value of 
twist raises by two units, $\tau \rightarrow \tau  2$, relative to the 
gluon operators. 
In addition, because of the change in the external $x$--factors 
(not being accompanied by $\lambda$)
the measures of the $\lambda$--integrations 
have to be changed according to (see also Subsection 4.2)
\begin{equation}
\d\lambda
\longrightarrow  \d\lambda\,\lambda^{I}\,.
\qquad 
\end{equation}

Open, up to now, is the structure of the local operators. Let us study 
in detail the first of the four--quark operators,
${}^{I=0}\! \QQ^{11}_{\alpha\beta}$, whose 
Taylor expansion may be written in two equivalent ways:
\begin{align}
{}^{I=0}\!
\QQ^{11}_{\alpha\beta}&(\ka x,\tau x,\kb x)=\\
&=
\left(\,\bar\psi(\ka x)\gamma_\alpha U(\ka x,0)
U(0,\tau x)\psi(\tau x)\right)
\left(\bar\psi(\tau x) U(\tau x,0)
U(0,\kb x)\gamma_\beta\psi(\kb x)\right) 
\nonumber\\
&= \!\sum_{n_1,m,n_2
=0}^\infty\! \!\!
\hbox{\Large$\frac{\ka^{n_1}\tau^m\kb^{n_2}}{n_1!m!n_2!}$}
\Big[\bar\psi(y)\!\left(\!\LD D_y x\right)^{\!\!n_1}\!\Big]\!
\gamma_\alpha 
\!\Big[\!\left(xD_z\right)^m\! \left(\psi(z)\bar\psi(z)\right)\Big]\!
\gamma_\beta
\!\Big[\!\!\left(x \RD D_y \right)^{\!\!n_2}\!\!\! \psi(y)\Big]\Big|_{y=z=0}
\nonumber\\
&=
\!\sum_{n_1=0}^\infty\!
\hbox{\Large$\frac{1}{n_1!}$}
\Big(\bar\psi(y)\!\left(x\Tensor D_y (\ka,\tau)\right)^{\!\!n_1}
\!\!\!\gamma_\alpha \psi(y)\!\Big)\Big|_{y=0}
\!\sum_{n_2=0}^\infty\! 
\hbox{\Large$\frac{1}{n_2!}$}
\Big(\bar\psi(z)\!\left(x\Tensor D_z (\tau,\kb)\right)^{\!\!n_2}
\!\!\!\gamma_\beta \psi(z)\!\Big)\Big|_{z=0},
\nonumber
\end{align}
where the two possibilities also use different generalized covariant 
derivatives. The left, right and left-right derivatives are given
by Eqs.~(\ref{D_kappa}), and, treating $\psi(0)\bar\psi(0)$ as a 
matrix in the group algebra, another form of the derivative obtains:
\begin{eqnarray}
D_\mu \left(\psi(0)\bar\psi(0)\right) 
&= 
\pd_\mu \left(\psi(0)\bar\psi(0)\right) 
+ \ii g 
[A_\mu, \left(\psi(0)\bar\psi(0)\right)]
\nonumber\\
&=
\left(\RD D_\mu
\psi(0)\right)\bar\psi(0) 
+
\psi(0)\left(\bar\psi(0)\LD D_\mu\right)
\,.
\nonumber
\end{eqnarray}
This leads to the following expansion into local
tensor operators 
\begin{equation}
{}^{I=0}\! \QQ^{ij}_{\alpha\beta}(\ka x,\tau x,\kb  x)
=
\sum_{N=0}^\infty \frac{1}{N!} x^{\mu_1}\ldots x^{\mu_N} \,
{}^{I=0}\!
\QQ^{ij}_{\alpha\beta\mu_1\ldots\mu_N}(\ka,\tau,\kb)\,,
\end{equation}
{\rm where}
\begin{equation}
\label{O_loc4}
{}^{I=0}\!
\QQ^{ij}_{\alpha\beta\mu_1\ldots\mu_N}(\ka,\tau,\kb)
\equiv 
\bar\psi(0)\gamma_\alpha
{\bf D}_{\mu_1\ldots\mu_N}(\ka,\tau,\kb) 
\gamma_\beta\psi(0)\,,
\end{equation}
and the quark field--dependent generalized covariant derivative 
of $N$th
order is given by
\begin{align}
\label{D_psi}
\lefteqn{
\hspace{-2cm}
{\bf D}_{\mu_1\ldots\mu_N}(\ka,\tau,\kb)
=
\sum_{n=0}^N
\hbox{\Large$\binom{N}{n}$}
\Tensor D_{\mu_1}\ldots \Tensor D_{\mu_{n}}
\psi(0)\times
\bar\psi(0)
\Tensor D_{\mu_{n+1}}\ldots \Tensor D_{\mu_N}
\qquad\qquad\qquad}
\\
&\quad=
\sum_{n=0}^N N!
\sum_{\ell=0}^n 
\LD D_{\mu_1}\ldots\LD D_{\mu_{n-\ell}}
\hbox{\Large$\frac{\ka^{n-\ell}\tau^\ell}{(n-\ell)!\ell!}$}
\RD D_{\mu_{n-\ell+1}}\ldots\RD D_{\mu_{n}} \psi(0)
\nonumber\\
&\quad\qquad\quad
\times
\sum_k^{N-n} 
\bar\psi(0) \LD D_{\mu_{n+1}}\ldots\LD D_{\mu_{n+k}}
\hbox{\Large$\frac{\tau^{k}\kb^{N-n-k}}{k![N-n-k)!}$}
\RD D_{\mu_{n+k+1}}\ldots\RD D_{\mu_{N}}\,.
\nonumber
\end{align}
Analogous generalized covariant derivatives occur for the four--gluon
operators. The only difference will be that instead of the quark fields
corresponding gluon field strengths appear in (\ref{D_psi}), and
that the derivatives are to be taken in the adjoint representation.
Despite having a complicated structure this does not matter in the
twist decomposition of the trilinear operators 
${}^{I}\! \QQ^{ij}_{\alpha\beta}(\ka x,\tau x,\kb  x)$.

Let us finish this Subsection with a short remark concerning so-called 
{\em four-particle} operators which also have been considered in the 
literature for local operators. These operators have the general structure,
cf. \cite{P,EFP}
\begin{equation}
(\bar\psi U \psi)(\bar\psi U\psi), \quad
(\bar\psi U
F U F U \psi) \quad {\rm or}
\quad
(F U F U F U F), 
\nonumber
\end{equation}
with any field being located at a different point
$\kappa_i x, i = 1,\ldots,4$.
The expansion into local tensors is obtained in the same way as above,
leading to a general expression of the following form:
\begin{eqnarray}
{}^{I}\! {\cal  R}_{\alpha\beta}(\kappa_i x)
=\sum_{N=0}^\infty \frac{1}{N!} x^{\mu_1}\ldots x^{\mu_N}
x^\rho \ldots x^\sigma\,
{}^{I}\!{\cal R}_{\alpha\beta\rho\ldots\sigma\mu_1\ldots\mu_N}(\kappa_i)\,, 
\end{eqnarray}
where the terms with equal value of $N$ are given as well-defined sums
of local tensor operators. Any of these operators decompose in the
same manner into irreducible representations of the Lorentz group.
The latter result is immediately related to the fact that all indices
$\mu_1\ldots\mu_N$ are to be symmetrized, i.e.,~lying in the first row
of any relevant Young tableau! Different symmetry classifications solely 
depend on the distribution of the remaining indices $\alpha\beta
\rho\ldots\sigma$ -- the others being somehow truncated -- 
to the various Young tableaux. As long as only
up to two free indices $\alpha\beta$ are relevant the twist
decomposition takes place according to the results of Section~\ref{gluon},
eventually modified by the power $\lambda^I$ which is related to the
number of external $x$'s in ${}^{I}\!{\cal R}$. In addition, the twist 
of the various components may be shifted by some (equal) amount. 

\subsection{Constrained trilocal operators:
Shuryak-Vainshtein-- and three--gluon operators}

Now we consider the special quark--antiquark--gluon operator
$\OO^{ij}_{\alpha\beta}$, Eq.~(\ref{OO}), and related three--gluon 
operators both having minimal twist--3; they will be denoted by
${}^{I}\! \OO^{ij}_{\alpha\beta}(\ka x,\tau x,\kb x)$ with $I=0$ and
$I=1$, respectively. The quark--antiquark--gluon operators are given by
\begin{align}
{}^{I=0}\!
\OO^{ij}_{\alpha\beta}(\ka x,\!\tau x,\!\kb x)
&= x^\rho \,{}^{I=0}\!
\OO^{ij}_{\alpha\beta\rho}(\ka x,\!\tau x,\!\kb x)\\
&=\! x^\rho\,
\bar{\psi}(\ka x)U(\ka
x,\tau x)\Gamma^i_\alpha 
F^j_{\beta\rho}(\tau x)U(\tau x,\kb x)\psi(\kb
x)\,,
\nonumber
\end{align}
where again some possible flavour structure has been suppressed. These
operators are related to the following generalizations of the so-called 
Shuryak-Vainshtein operators \cite{SV,BB88},
\begin{eqnarray}
S^\pm_\beta(\ka x,\tau x,\kb x)
=
x^\alpha \big(
{}^{I=1}\! \OO_{\alpha\beta}^{11}(\ka x,\tau x,\kb x) \pm \ii\, 
{}^{I=1}\! \OO_{\alpha\beta}^{22}(\ka x,\tau x,\kb x) 
\big)x^\alpha , \nonumber
\end{eqnarray}
which, in the flavour singlet case, mix with the 
following three--gluon operators:
\begin{align}
&
{}^{I=1}\! \OO^{i}_{\alpha\beta}(\ka x,\tau x,\kb x)
=
x^\rho
x^\sigma\,
{}^{I=1}\! \OO^{i}_{\alpha\beta\rho\sigma}(\ka x,\tau x,\kb x)\\
&\qquad\qquad
=\!x^\rho x^\sigma\,
F^a_{\alpha\nu}(\ka x)U^{ab}(\ka x,\tau x) 
F^{i\,bc}_{\beta\rho}(\tau
x)U^{cd}(\tau x,\kb x)F^{d\nu}_\sigma(\kb x)\,,
\nonumber
\end{align}
where $F^{ab}_{\alpha\beta}(x)\equiv f^{acb}F^c_{\alpha\beta}(x)$
and the phase factors are taken in the adjoint representation.
 
Because of their construction these special
trilocal operators have the property
\begin{eqnarray}
\label{prop0}
x^\beta\, {}^{I}\! \OO^{ij}_{\alpha\beta}(\ka x,\tau
x,\kb x) \equiv 0\,.
\end{eqnarray} 
They contain twist--3 up to twist--6 contributions which, because of 
the antisymmetry of the gluon field strength, have to be determined by
means of the Young tableaux (ii) -- (iv). 
Because of (\ref{prop0}) the scalar operators
$x^\alpha x^\beta\, {}^{I}\! \OO^{ij}_{\alpha\beta}(\ka x,\tau x,\kb x)$ 
vanish identically.

The Taylor expansion of the operator 
${}^{I=0}\! \OO^{11}_{\alpha\beta}(\ka x,\tau x,\kb x)$ reads
\begin{equation}
\label{Oij}
{}^{I=0}\! \OO^{11}_{\alpha\beta}(\ka x,\tau x,\kb  x)
=
\sum_{N=0}^\infty \frac{1}{N!} x^{\mu_1}\ldots x^{\mu_N} x^\rho \,
{}^{I=0}\! \OO^{11}_{\alpha\beta\rho\mu_1\ldots\mu_N}(\ka,\tau,\kb)\,,
\end{equation}
 with
\begin{equation}
\label{O_local3}
{}^{I=0}\!\OO^{11}_{\alpha\beta\rho\mu_1\ldots\mu_N}(\ka,\tau,\kb)
\equiv 
\bar\psi(0)
\gamma_\alpha
{\bf D}_{[\beta\rho]\mu_1\ldots\mu_N}(\ka,\tau,\kb) \psi(0)\,,
\end{equation}
where the field strength--dependent derivative already has been given
by (\ref{D_F}). The Taylor expansion of the related three--gluon 
operator obtains as
follows:
\begin{align}
{}^{I=1}\! \OO^{11}_{\alpha\beta}(\ka x,\tau x,\kb x)
=&
\sum_{N=0}^\infty \frac{1}{N!} x^{\mu_1}\ldots x^{\mu_N} x^\rho x^\sigma
\;{}^{I=1}\!\OO^A_{\alpha\beta\rho\sigma\mu_1\ldots\mu_N}(\ka,\tau,\kb)
\end{align}
where
\begin{align}
\label{O_local4}
{}^{I=1}\!
\OO^{11}_{\alpha\beta\rho\sigma\mu_1\ldots\mu_N}(\ka,\tau,\kb)
\equiv
&
F^a_{\alpha\nu}(0)
{\bf D}^{ab}_{[\beta\rho]\mu_1\ldots\mu_N}(\ka,\tau,\kb)
F_{\sigma}^{b\nu}(0)\,,
\end{align}
with the same field strength dependent generalized covariant derivative
(\ref{D_F}), but now in the adjoint representation. The generalization
arbitrary values $i$ and $j$ is obvious. 

As it became obvious by the above considerations all the nonlocal 
three--particle operators ${}^{I}\!\OO^{ij}_{\alpha\beta}$ have the same 
general local structure and, therefore, decompose according to the same symmetry 
patterns. They only differ in the rank of the local tensor operators
which is due to the external powers of $x$. Again, this 
must be taken into account by a change of the integration 
measure according to\footnote{
Observe, that because of notational simplicity in the explicit 
expressions below we shifted the variable $I \rightarrow I-1$.} 
\begin{equation*}
\d\lambda \longrightarrow
\d\lambda\,\lambda^{I+1}\,.
\end{equation*}
From now on we omit the indices $i$ and $j$.

In order to determine the twist--3 light--cone operators 
${}^{I}\! \OO_{\alpha\beta}(\ka\xx,\tau\xx,\kb\xx)$ for $I=0$ and $I=1$,
let us consider the following Young tableaux of symmetry pattern (iiB)
\noindent
\unitlength0.40cm
\begin{picture}(8.5,2)
\linethickness{0.05mm}
\put(1,0){\framebox(1,1){$\SC\beta$}}
\put(1,1){\framebox(1,1){$\SC\rho$}}
\put(2,1){\framebox(1,1){$\SC\mu_1$}}
\put(3,1){\framebox(3,1){$\SC\ldots$}}
\put(6,1){\framebox(1,1){$\SC\mu_N$}}
\put(7,1){\framebox(1,1){$\SC\alpha$}}
\end{picture}
and
\unitlength0.4cm
\begin{picture}(8.5,2)
\linethickness{0.05mm}
\put(1,0){\framebox(1,1){$\SC\beta$}}
\put(1,1){\framebox(1,1){$\SC\rho$}}
\put(2,1){\framebox(1,1){$\SC\sigma$}}
\put(3,1){\framebox(1,1){$\SC\mu_1$}}
\put(4,1){\framebox(3,1){$\SC\ldots$}}
\put(7,1){\framebox(1,1){$\SC\mu_N$}}
\put(8,1){\framebox(1,1){$\SC\alpha$}}
\end{picture}
\hspace{.2cm}
, respectively:
\begin{align}
\label{O_tw3_1}
{}^{I}\!\OO^{\mathrm{tw}3}_{\alpha\beta}(\ka x,\tau x,\kb x)
&=\hbox{\Large$\frac{1}{2}$}\!\!Π\int_{0}^{1}\!\!\!
\d\lambda\,\lambda^I
\Big[\big(1\!+\!\lambda^2\big)\delta_\beta^\mu\pd_\alpha
\!+\!\big(1\!-\!\lambda^2\big)\delta_\alpha^\mu\pd_\beta\Big]
{}^{I}\!\tl \OO_{\mu}(\ka\lambda x,\!\tau\lambda x,\!\kb\lambda x)
\nonumber\\
&={}^{I}\!\OO^{\mathrm{tw}3}_{[\alpha\beta]} (\ka x,\tau x,\kb x)
+{}^{I}\!\OO^{\mathrm{tw}3}_{(\alpha\beta)} (\ka x,\tau x,\kb x)\,;
\end{align}
here, and in the following, we write 
$\OO_\mu \equiv x^\rho\OO_{\rho\mu}$.
Let us remark that, first, because of property (\ref{prop0}), only the
above mentioned Young tableaux contribute and, second, the difference
between the expressions (\ref{O_tw3_1}) and (\ref{M_tw2_nl}) result 
also from that property, namely the truncation by $x^\rho$ in the second 
term of the integrand, $-x^\mu \pd_\alpha\pd_\beta
{}^{I}\!\tl \OO_{\mu}(\ka x,\tau x,\kb x) = 
2\delta_{(\alpha}^\mu\pd_{\beta)} {}^{I}\!\tl\OO_{\mu}(\ka x,\tau x,\kb x)$,
after partial integrations leads to the above expression (\ref{O_tw3_1}).
In the case $I=0$ this operator is in agreement with the expression 
given by Balitsky and Braun (cf.~Eq.~(5.14) of \cite{BB88}). However,
their operator lacks to be really of twist--3 since it is not traceless; 
therefore it contains also twist--4, twist--5 as well as twist--6 operators 
resulting from the trace terms. 

Now, let us to project onto the light--cone. As in the case of the
gluon tensor only the first terms, $k=0,1,2$, in the expansion
of ${}^{I}\!\tl \OO_{\mu}(\ka\lambda x,\!\tau\lambda x,\!\kb\lambda x)$ 
are to be taken into account.
For the antisymmetric part of the twist--3 operator one gets
\begin{equation}
\label{OO_tw3_iiasy}
{}^{I}\!\OO^{\mathrm{tw}3}_{[\alpha\beta]}(\ka\lcx,\tau\lcx,\kb\lcx)
=
\int_0^1\d\lambda\,
\lambda^{2+I}\pd_{[\alpha}
{}^{I}\!\OO_{\beta]}(\ka\lambda\lcx,\tau\lambda\lcx,\kb\lambda\lcx)
-{}^{I}\!\OO^{\mathrm{>(ii)}}_{[\alpha\beta]}(\ka\lcx,\tau\lcx,\kb\lcx)
\end{equation}
where
\begin{align}
\label{}{}^{I}\!\OO^{\mathrm{>(ii)}}_{[\alpha\beta]}
(\ka\lcx,\tau\lcx,\kb\lcx)
=\hbox{\Large$\frac{1}{2}$}
&\int_{0}^{1}\d\lambda\,\lambda^{1+I}
(1-\lambda)
\big(2x_{[\alpha}\pd_{\beta]}\pd^\mu-x_{[\alpha}\delta_{\beta]}^\mu\square
\big)\nonumber\\
&\times{}^{I}\! \OO_{\mu}(\ka\lambda x,\tau\lambda x,\kb\lambda x)
\big|_{x=\xx}.
\end{align}
The decomposition of
${}^{I}\!\OO^{\mathrm{>(ii)}}_{[\alpha\beta]}
(\ka\lcx,\tau\lcx,\kb\lcx)$
arises as
\begin{align}
{}^{I}\!\OO^{\mathrm{>(ii)}}_{[\alpha\beta]}
(\kappa_1\tilde{x},\tau\lcx,\kappa_2\tilde{x})
=
{}^{I}\!\OO^{\mathrm{tw4(ii)}}_{[\alpha\beta]}
(\kappa_1\tilde{x},\tau\lcx,\kappa_2\tilde{x})+
{}^{I}\!\OO^{\mathrm{tw5(ii)}}_{[\alpha\beta]}
(\kappa_1\tilde{x},\tau\lcx,\kappa_2\tilde{x})\,,
\end{align}
with
\begin{eqnarray}
\hspace{-.5cm}
\label{OO_tw4_iiasy}
{}^{I}\!\OO^{\mathrm{tw4(ii)}}_{[\alpha\beta]}
(\kappa_1\tilde{x},\tau\lcx,\kappa_2\tilde{x})
\!\!\!&=&\!\!\!
\hbox{\Large$\frac{1}{2}$}x_{[\alpha}\pd_{\beta]}\pd^{\mu}
\int_{0}^{1}\d\lambda\,\lambda^I
(1-\lambda^2)\,
{}^{I}\!\OO_{\mu}(\kappa_1\lambda x,\tau\lambda x,\kappa_2\lambda x)
\big|_{x=\xx}\,,
\nonumber\\
\hspace{-.5cm}
{}^{I}\!\OO^{\mathrm{tw5(ii)}}_{[\alpha\beta]}
(\kappa_1\tilde{x},\tau\lcx,\kappa_2\tilde{x})
\!\!\!&=&\!\!\!
\label{OO_tw5_iiasy}
-
\hbox{\Large$\frac{1}{4}$}
x_{[\alpha}\big(\delta_{\beta]}^{\mu}(x\pd)
-x^{\mu}\pd_{\beta]}\big)\square\nonumber\\
&&\qquad\qquad\times
\int_{0}^{1}\!\!\!\d\lambda\,\lambda^I
(1\!-\!\lambda)^2\,
{}^{I}\!\OO_{\mu}(\kappa_1\lambda x,\tau\lambda x,\kappa_2\lambda x)
\big|_{x=\xx}\,.
\nonumber
\end{eqnarray}
The symmetric part reads
\begin{equation}
\label{OO_tw3_iisy}
{}^{I}\!\OO^{\mathrm{tw}3}_{(\alpha\beta)}
(\ka\lcx,\tau\lcx,\kb\lcx)
=
\int_0^1\d\lambda\,\lambda^{I}\pd_{(\alpha}
{}^{I}\!\OO_{\beta)}(\ka\lambda\lcx,\tau\lambda\lcx,\kb\lambda\lcx)
-{}^{I}\!\OO^{\mathrm{>(ii)}}_{(\alpha\beta)}(\ka\lcx,\tau\lcx,\kb\lcx)
\end{equation}
where
\begin{eqnarray}
\label{OO_high_(ii)}
\lefteqn{{}^{I}\!\OO^{\mathrm{>(ii)}}_{(\alpha\beta)}
(\ka\lcx,\tau\lcx,\kb\lcx)
=
\int_{0}^{1}\!\d\lambda\,\lambda
^I\Big\{
\hbox{\Large$\frac{1}{2}$}(1-\lambda^2)
g_{\alpha\beta}\pd^\mu
+\hbox{\Large$\frac{1}{2}$}(1-\lambda)\delta^\mu_{(\alpha}x_{\beta)}\square}\qquad\\
&&+\lambda(1-\lambda)x_{(\alpha}\pd_{\beta)}\pd^\mu
-\hbox{\Large$\frac{1}{4}$}(1-\lambda)^ 2x_\alpha x_\beta\pd^\mu\square
\Big\}
{}^{I}\!\OO_{\mu}(\ka\lambda x,\tau\lambda x,\kb \lambda x)
\big|_{x=\xx}.\nonumber
\end{eqnarray}
Let us remark that the conditions of tracelessness
for the light-cone operators are:
\begin{eqnarray}
g^{\alpha\beta}\,{}^{I}\!\OO^{\mathrm{tw}3}_{(\alpha\beta)}
(\ka\xx,\tau\xx,\kb\xx)
=
0\,,\qquad\d^\alpha\,
{}^{I}\!\OO^{\mathrm{tw}3}_{(\alpha\beta)}(\ka\xx,\tau\xx,\kb\xx)
&=&0
\,,
\nonumber\\
\d^\alpha\,
{}^{I}\!\OO^{\mathrm{tw}3}_{[\alpha\beta]}(\ka\xx,\tau\xx,\kb\xx)
&=&0\,.
\nonumber
\end{eqnarray}

Now we can calculate the
twist--3 vector operator from the twist--3 tensor 
operator.
The twist--3 light--cone vector
operator
reads
\begin{align}
\label{OO_tw3_vec}
&{}^{I}\!\OO^{\mathrm{tw3}}_\alpha(\ka\lcx,\tau\lcx,\kb\lcx)
=\int_0^1\d\lambda\,
\lambda^{2+I}
\Big[\delta_\alpha^\mu(x\pd)-x^\mu\pd_\alpha -x_\alpha\pd^\mu\Big]
{}^{I}\!\OO_\mu(\ka\lambda x,\tau\lambda x,\kb\lambda x)\big|_{x=\tilde{x}}
\nonumber\\
&\qquad={}^{I}\!\OO_\alpha(\ka\lcx,\tau\lcx,\kb\lcx)
-\lcx_\alpha\int_0^1\d\lambda\, \lambda^{2+I}\pd^\mu
\,{}^{I}\!\OO_\mu(\ka\lambda x,\tau\lambda x,\kb\lambda x)
\big|_{x=\tilde{x}},
\end{align}
and the twist--4 vector operator
which is contained in the trace terms
of the twist--3 vector operator
\begin{equation}
\label{OO_tw4_ivec}
{}^{I}\!\OO^{\mathrm{tw4(ii)}}_\alpha(\ka\lcx,\tau\lcx,\kb\lcx)=
\lcx_\alpha\int_0^1\d\lambda\,\lambda^{2+I}\pd^\mu\,
{}^{I}\!\OO_\mu(\ka\lambda x,\tau\lambda x,\kb\lambda x)
\big|_{x=\tilde{x}}\, .
\end{equation}

The determination of the antisymmetric twist--4 operator having symmetry 
type (iii) obtains from Young tableaux 
\unitlength0.4cm
\begin{picture}(8,3)
\linethickness{0.05mm}
\put(1,0){\framebox(1,1){$\SC\alpha$}}
\put(1,1){\framebox(1,1){$\SC\beta$}}
\put(1,2){\framebox(1,1){$\SC\rho$}}
\put(2,2){\framebox(1,1){$\SC\mu_1$}}
\put(3,2){\framebox(3,1){$\SC\ldots$}}
\put(6,2){\framebox(1,1){$\SC\mu_N$}}
\end{picture}
and
\unitlength0.4cm
\begin{picture}(9,3)
\linethickness{0.05mm}
\put(1,0){\framebox(1,1){$\SC\alpha$}}
\put(1,1){\framebox(1,1){$\SC\beta$}}
\put(1,2){\framebox(1,1){$\SC\rho$}}
\put(2,2){\framebox(1,1){$\SC\sigma$}}
\put(3,2){\framebox(1,1){$\SC\mu_1$}}
\put(4,2){\framebox(3,1){$\SC\ldots$}}
\put(7,2){\framebox(1,1){$\SC\mu_N$}}
\end{picture},
respectively.
Again, making use of property (\ref{prop0}), the result is
\begin{align}
\label{OO_tw4_iiiasy}
{}^{I}\!\OO^{\rm tw4(iii)}_{[\alpha\beta]}(\ka\lcx,\tau\lcx,\kb\lcx)
=&\!
\int_{0}^{1}\!\!\!\d\lambda\lambda^{2+I}
\delta^\mu_{[\alpha}\!\left( (x\pd)\delta_{\beta]}^\nu
\!-\! 2x^\nu\pd_{\beta]}\right)\!\!
{}^{I}\!\OO_{[\mu\nu]}(\ka\lambda x,\!\tau\lambda x,\!\kb\lambda x)
\big|_{x=\xx}\nonumber
\\
& 
- {}^{I}\!\OO^{\rm tw5(iii)}_{[\alpha\beta]}(\ka\lcx,\tau\lcx,\kb\lcx),
\end{align}
twist--5 part is determined by the trace namely with
\begin{align}
{}^{I}\!\OO^
{\rm tw5(iii)}_{[\alpha\beta]}(\ka\lcx,\tau\lcx,\kappa\lcx)
= 
&
-
\int_{0}^{1}\d\lambda\,\lambda^I
(1-\lambda^2)\Big\{x_{[\alpha}\big(\delta_{\beta]}^{[\mu}(x\pd)
-x^{[\mu}\pd_{\beta]}\big)\pd^{\nu]}
\nonumber\\
&\ -
x_{[\alpha}\delta_{\beta]}^{[\mu}x^{\nu]}\square\Big\}
{}^{I}\!\OO_{[\mu\nu]}(\ka\lambda x,\tau\lambda x,\kb\lambda x)
\big|_{x=\xx}.
\end{align}
The antisymmetric twist--5 tensor operator is given by:
\begin{align}
\label{OO_tw5_ii}
\lefteqn{
\hspace{-1cm}
{}^{I}\!\OO^{\rm tw5}_{[\alpha\beta]}(\ka\lcx,\tau\lcx,\kappa\lcx)
= 
{}^{I}\!\OO^{\rm tw5(ii)}_{[\alpha\beta]}(\ka\lcx,\tau\lcx,\kappa\lcx)+
{}^{I}\!\OO^{\rm tw5(iii)}_{[\alpha\beta]}(\ka\lcx,\tau\lcx,\kappa\lcx)
}\nonumber\\
&
=-\hbox{\Large$\frac{1}{4}$}
x_{[\alpha}\big(\delta_{\beta]}^{\mu}(x\pd)
-x^{\mu}\pd_{\beta]}\big)\square\!\!
\int_{0}^{1}\!\!\!\d\lambda\,\lambda^I
(1\!-\!\lambda)^2\,
{}^{I}\!\OO_{\mu}(\kappa_1\lambda x,\tau\lambda x,\kappa_2\lambda x)
\big|_{x=\xx}\nonumber\\
&
-\int_{0}^{1}\d\lambda\,\lambda^I
(1-\lambda^2)\Big\{
x_{[\alpha}\big(\delta_{\beta]}^{[\mu}(x\pd)
-x^{[\mu}\pd_{\beta]}\big)\pd^{\nu]}
\nonumber\\
&\qquad\qquad\qquad\qquad\qquad
-x_{[\alpha}\delta_{\beta]}^{[\mu}x^{\nu]}\square\Big\}
{}^{I}\!\OO_{[\mu\nu]}(\ka\lambda
x,\tau\lambda x,\kb\lambda x)
\big|_{x=\xx}.
\end{align}
Finally, we obtain the complete decomposition of the antisymmetric 
tensor operator:
\begin{align}
{}^{I}\!\OO_{[\alpha\beta]}(\ka\lcx,\tau\lcx,\kb\lcx)&=
{}^{I}\!\OO^{\rm tw3(ii)}_{[\alpha\beta]}(\ka\lcx,\tau\lcx,\kb\lcx)+
{}^{I}\!\OO^{\rm tw4(iii)}_{[\alpha\beta]}(\ka\lcx,\tau\lcx,\kb\lcx)
\nonumber\\
&+{}^{I}\!\OO^{\rm tw4(ii)}_{[\alpha\beta]}(\ka\lcx,\tau\lcx,\kb\lcx)+
{}^{I}\!\OO^{\rm tw5}_{[\alpha\beta]}(\ka\lcx,\tau\lcx,\kb\lcx).
\end{align}

The determination of the symmetric
twist--4 operator having symmetry 
type (iv) obtains from Young
tableaux
\unitlength0.4cm
\begin{picture}(9,2)
\linethickness{0.05mm}
\put(1,0){\framebox(1,1){$\SC\beta$}}
\put(2,0){\framebox(1,1){$\SC\alpha$}}
\put(1,1){\framebox(1,1){$\SC\rho$}}
\put(2,1){\framebox(1,1){$\SC\mu_1$}}
\put(3,1){\framebox(1,1){$\SC\mu_2$}}
\put(4,1){\framebox(3,1){$\SC\ldots$}}
\put(7,1){\framebox(1,1){$\SC\mu_N$}}
\end{picture}
and
\unitlength0.4cm
\begin{picture}(9,2)
\linethickness{0.05mm}
\put(1,0){\framebox(1,1){$\SC\beta$}}
\put(2,0){\framebox(1,1){$\SC\alpha$}}
\put(1,1){\framebox(1,1){$\SC\rho$}}
\put(2,1){\framebox(1,1){$\SC\sigma$}}
\put(3,1){\framebox(1,1){$\SC\mu_1$}}
\put(4,1){\framebox(3,1){$\SC\ldots$}}
\put(7,1){\framebox(1,1){$\SC\mu_N$}}
\end{picture},
respectively.
Making use of property (\ref{prop0}) the result is 
\begin{eqnarray}
\label{OO_tw4_ivsy}
\lefteqn{{}^{I}\!\OO^{\mathrm{tw4(iv)}
}_{(\alpha\beta)} 
(\ka\lcx,\tau\lcx,\kb\lcx)
=}\nonumber\\
&&\int_0^1\!\!\d\lambda\,\lambda^{I}(1-\lambda)
\Big(\delta_\alpha^\mu\delta_\beta^\nu x^\rho (x\pd)\pd_\rho
-2x^\mu(x\pd)\delta_{(\beta}^\nu\pd_{\alpha)}
\Big) {}^{I}\!\OO_{(\mu\nu)}(\ka\lambda x,\tau\lambda x,\kb\lambda
x)
\big|_{x=\xx}\quad\nonumber\\
&&-{}^{I}\!\OO^{\mathrm{>(iv)}}_{(\alpha\beta)}(\ka\lcx,\tau\lcx,\kb\lcx)\,,
\end{eqnarray}
where
\begin{align}
\label{OO_high_(iv)_sy}
&{}^{I}\!\OO^{\mathrm{>(iv)}}_{(\alpha\beta)}(\ka\lcx,\tau\lcx,\kb\lcx)
=\\
&
\int_{0}^{1}\!\!\d\lambda\Big\{\Big(2\lambda^{2+I}
x_{(\alpha}\delta_{\beta)}^\nu\pd^\mu
-\lambda^{1+I}(1-\lambda)x_\alpha
x_\beta\pd^\mu\pd^\nu\Big)
{}^{I}\!\OO_{(\mu\nu)}(\ka\lambda x,\tau\lambda x,\kb\lambda x)
\nonumber\\
&-
\hbox{\Large$\frac{1}{2}$}
\Big( \lambda^{I} g_{\alpha\beta}\pd^\mu
+\lambda^{I}(1-\lambda^2)x_{(\alpha}\delta^\mu_{\beta)}\square
-\hbox{\Large$\frac{1}{2}$}\lambda^{I}(1-\lambda)^2 
x_\alpha x_\beta\square\pd^\mu \Big)
{}^{I}\!\OO_{\mu}(\ka\lambda x,\tau\lambda x,\kb\lambda x) \nonumber\\
&
+
\Big(
\hbox{\Large$\frac{1}{2}$}\lambda^I g_{\alpha\beta}(x\pd)
-\lambda^{2+I}
x_{(\alpha}\pd_{\beta)}
+\hbox{\Large$\frac{1}{2}$} \lambda^{1+I} (1-\lambda)
x_\alpha x_\beta\square\Big) {}^{I}\!\OO^\rho_{\ \rho}
(\ka\lambda x,\tau\lambda x,\kb\lambda x)
\Big\}\Big|_{x=\tilde{x}}\!.\nonumber
\end{align}

Now, we use the property (\ref{prop0}) and sum up the symmetric 
{\em higher twist} operators, which appear in the trace terms of the 
symmetric twist operators with Young symmetry (ii)
and (iv), to a complete twist--4, twist--5 and twist--6 operator. The
complete twist--4 operators are  constructed by means of the Young tableau 
(i). The `scalar part' of the twist--4 tensor operator is given by
\begin{align}
\label{OO_tw4_sc}
{}^{I}\!\OO_{(\alpha\beta)}^{\mathrm{tw4,s}}(\ka\lcx,\tau\lcx,\kb\lcx)
=\hbox{\Large$\frac{1}{2}$}
g_{\alpha\beta}\bigg\{
&(x\pd)\int_0^1\d\lambda\,\lambda^I\,
{}^{I}\!\OO^\rho_{\ \rho}(\ka\lambda x,\tau\lambda x,\kb\lambda x)\\
&-\pd^\mu\int_0^1\d\lambda\,\lambda^{2+I}\,
{}^{I}\!\OO_{\mu}(\ka\lambda
x,\tau\lambda x,\kb\lambda x)
\bigg\}\bigg|_{x=\tilde{x}}\nonumber
\end{align}
and the twist--4 `vector part' reads
\begin{align}
\label{OO_tw4_vec}
&{}^{I}\!\OO_{(\alpha\beta)}^{\mathrm{tw4,v}}(\ka\lcx,\tau\lcx,\kb\lcx)
=-\lcx_{(\alpha}
\bigg\{\int_0^1\!\!\d\lambda\,\lambda^I\,
\Big\{\pd_{\beta)}+\hbox{\Large$\frac{1}{2}$}x_{\beta)}
(\ln\lambda)\square\Big\}
{}^{I}\!\OO^\rho_{\ \rho}(\ka\lambda x,\tau\lambda x,\kb\lambda x)\nonumber\\
&-\int_0^1\!\!\d\lambda\,\lambda^I\,
\Big\{\big(1-\lambda^2\big)\pd_{\beta)
}+\hbox{\Large$\frac{1}{4}$}x_{\beta)}
\big(1-\lambda^2+2\ln\lambda\big)\square\Big\}\pd^\mu\,
{}^{I}\!\OO_{\mu}(\ka\lambda x,\tau\lambda x,\kb\lambda x)
\bigg\}\bigg|_{x=\tilde{x}}\!\!.
\end{align}
Furthermore, the complete twist--5 vector operator is again constructed by means of the 
Young tableau (ii). The `vector part' of the twist--5 tensor operator is 
given by
\begin{align}
\label{OO_tw5_vec}
&{}^{I}\!\OO_{(\alpha\beta)}^{\mathrm{tw5,v}}
(\ka\xx,\tau\xx,\kb\xx)=
-\lcx_{(\alpha}\times\\
&\bigg\{\hbox{\Large$\frac{1}{4}$}
\int_0^1\d\lambda\,\lambda^I\Big(\big(1-\lambda\big)^2
\big(\delta^\nu_{\beta)}(x\pd)-x^\nu\pd_{\beta)}\big)\square
+2\big(1-\lambda+\ln\lambda)x_{\beta)}\pd^\nu\square\Big)
\,{}^{I}\!\OO_\nu(\ka\lambda x,\tau\lambda x,\kb\lambda x)\nonumber\\
&-\!\!\int_0^1\!\!\d\lambda\,\lambda^I\Big\{
\big(1-\lambda^2\big)
\big(\delta^\nu_{\beta)}(x\pd)-x^\nu\pd_{\beta)}-x_{\beta)}\pd^\nu\big)\pd^\mu 
-\Big(\hbox{\Large$\frac{1}{2}$}\big(1-\lambda^2\big)
+\ln\lambda\Big)
x_{\beta)}\square
\big(\pd^\mu x^\nu-g^{\mu\nu}\big)\Big\}\nonumber\\
&\hspace{8cm}\times
\,{}^{I}\!\OO_{(\mu\nu)}(\ka\lambda x,\tau\lambda x,\kb\lambda x)
\bigg\}\bigg|_{x=\tilde{x}}.\nonumber
\end{align}
The full twist--6 scalar operator has Young symmetry (i).  
The `scalar part' of the twist--6 tensor operator is given by
\begin{align}
\label{OO_tw6_sc}
&{}^{I}\!\OO_{(\alpha\beta)}^{\mathrm{tw6,s}}(\ka\lcx,\tau\lcx,\kb\lcx)\nonumber\\
&\
=\lcx_\alpha\lcx_\beta\bigg\{\hbox{\Large$\frac{1}{2}$}\square
\int_0^1\d\lambda\,\lambda^I
\big(1-\lambda+\ln\lambda\big)
\big(
\pd^\mu {}^{I}\!\OO_\mu(\ka\lambda x,\tau\lambda x,\kb\lambda x)-
{}^{I}\!\OO^\rho_{\ \rho}(\ka\lambda x,\tau\lambda x,\kb\lambda x)
\big)
\nonumber\\
&\qquad\qquad\qquad
+\pd^\mu\pd^\nu\int_0^1\d\lambda\,\lambda^{I}(1-\lambda)\,
{}^{I}\!\OO_{(\mu\nu)}(\ka\lambda x,\tau\lambda x,\kb\lambda x)
\bigg\}\bigg|_{x=\tilde{x}}.
\end{align}

Thus, together with the twist--3 part, Eq.~(\ref{OO_tw3_iisy}), and the 
twist--4 part, Eq.~(\ref{OO_tw4_ivsy}), 
we finally obtain the complete decomposition of the
symmetric tensor operator:
\begin{eqnarray}
\lefteqn{{}^{I}\!\OO_{(\alpha\beta)}(\ka\lcx,\tau\lcx,\kb\lcx)=
{}^{I}\!\OO^{\rm tw3(ii)}_{(\alpha\beta)}(\ka\lcx,\tau\lcx,\kb\lcx)+
{}^{I}\!\OO^{\rm tw4(iv)}_{(\alpha\beta)}(\ka\lcx,\tau\lcx,\kb\lcx)}
\nonumber\\
&&+{}^{I}\!\OO^{\rm tw4,s}_{(\alpha\beta)}(\ka\lcx,\tau\lcx,\kb\lcx)+
{}^{I}\!\OO^{\rm tw4,v}_{(\alpha\beta)}(\ka\lcx,\tau\lcx,\kb\lcx)+
{}^{I}\!\OO^{\rm tw5,v}_{(\alpha\beta)}(\ka\lcx,\tau\lcx,\kb\lcx)\nonumber\\
&&+{}^{I}\!\OO^{\rm tw6,s}_{(\alpha\beta)}(\ka\lcx,\tau\lcx,\kb\lcx).
\end{eqnarray}

\subsection{Conclusions}
In this Section we have discussed the twist decomposition
of trilocal light--ray operators. 
The only difference is that the respective values of twist 
increase by a definite amount which is equal for any given
operator and possibly, depending on the number $I$ of external 
$x$--factors, by a change in the integration measure, 
$\d\lambda \rightarrow \d\lambda\ \lambda^I$. 
The applicability of our procedure  
for bilocal operators rests upon the fact that the general form
of the Taylor expansion (around $y = 0$) for any of the nonlocal operators 
is the same: It is given by an infinite sum of terms of $N$th 
order being multiplied by $x^{\mu_1} x^{\mu_2} \ldots x^{\mu_N}$ where each 
term consists of a finite sum of local operators having exactly the same 
tensor structure. In principle, this works
also for multi-local operators of higher order.

However, there are special trilocal operators which do not immediately 
fit into the general scheme of Section~\ref{gluon}. 
Its twist decomposition is obtained by restricting 
the expressions of the generic case to the  special conditions under 
consideration. The results are classified in Table~\ref{tab:O-ten} and 
\ref{tab:O-vec}, again by 
indicating their symmetry type and the number of the corresponding 
equation. In the same manner one should proceed for other kinds of 
special multilocal operators like that of Eq.~(\ref{OO'}).
\begin{table}[h]
\begin{center}
\begin{tabular}{|c|c|c|c|}
\hline
Eq.& YT &  type & 
${}^{I}\!\OO_{\alpha\beta}(\ka\lcx,\tau\lcx,\kb\lcx)$
\\
\hline
\ref{OO_tw3_iisy} & (ii) & sym  & $\tau=3$  \\
\ref{OO_tw3_iiasy} & (ii) & asym & $\tau=3$  \\
\ref{OO_tw4_iiiasy} & (iii) & asym & $\tau=4$  \\
\ref{OO_tw4_ivsy} & (iv) & sym  & $\tau=4$  \\
\ref{OO_tw4_sc} & (i)   & sym & $\tau=4$  \\
\ref{OO_tw4_vec} & (i)   & sym &  $\tau=4$  \\
\ref{OO_tw4_iiasy} & (i)   & asym & $\tau=4$  \\
\ref{OO_tw5_vec} & (ii) & sym  & $\tau=5$  \\
\ref{OO_tw5_ii} & (ii) & asym & $\tau=5$  \\
\ref{OO_tw6_sc} & (i)   & sym   & $\tau=6$  \\
\hline
\end{tabular}
\end{center}
\caption{Twist decomposition of special tensor operators 
${}^{I}\! \OO_{\alpha\beta}$
 with the property $x^\beta\, {}^{I}\! \OO_{\alpha\beta} \equiv 0$}
\label{tab:O-ten}
\end{table}

\begin{table}[h]
\begin{center}
\begin{tabular}{|c|c|c|}
\hline
Eq.& YT &  
${}^{I}\!\OO_{\alpha}(\ka\lcx,\tau\lcx,\kb\lcx)$
\\
\hline
\ref{OO_tw3_vec} & (ii) & $\tau=3$  \\
\ref{OO_tw4_ivec} & (i) & $\tau=4$   \\
\hline
\end{tabular}
\end{center}
\caption{Twist decomposition of special vector operators 
${}^{I}\! \OO_{\alpha}$  with the property
$x^\alpha\, {}^{I}\! \OO_{\alpha} \equiv 0$}
\label{tab:O-vec}
\end{table}

The various tensor operators which we considered here are nonlocal 
generalizations of local operators partly considered earlier in the 
literature for twist--3 and twist--4. There, as far as possible an explicit twist 
decomposition has been circumvented by truncating {\em any} of the 
(constant) totally symmetrized tensor indices of the local operators by 
some light--like vector $n^\mu$, i.e., by reducing to the scalar 
case. To the best of our knowledge the only work where the twist 
decomposition of (scalar) bilocal operators has been considered is that 
of Balitsky and Braun \cite{BB88}. However, their work is based on the
external field formalism and does not have an obvious
group theoretical systematics. 

In principle, our procedure may be extended also to tensors of
arbitrary high rank. Despite being defined by an obvious algorithm, its 
application to the next step, the twist decomposition of a generic 
3rd rank tensor operator, will be very cumbersome. First of all, it would
be necessary to determine the projection operators onto all traceless 
3rd rank tensorial harmonic polynomials and, secondly, also additional 
Young patterns had to be taken into account. Fortunately, such kind 
of nonlocal operators  -- at least in the near future --  may not be of 
physical relevance. Therefore, any further study in that direction seems 
to be reasonable only from a group theoretical point of view.

\section{Harmonic operators of definite geometric twist in terms of Gegenbauer polynomials 
and Bessel functions}
\label{off-cone}
\subsection{Totally symmetric harmonic operators of any 
geometric twist}
\setcounter{equation}{0}
Let us now discuss the twist decomposition of a scalar operator 
of degree $n$ into all its twist parts by means of the polynomial technique.
In this way, the scalar operators of degree $n$ 
are the generating polynomials of symmetric tensors of rank $n$.
In this case, the infinite twist decomposition of non-local (scalar)
operators into the infinite tower of harmonic operators of definite twist 
is possible. 
The corresponding group theoretical background has been formulated 
and proven by Bargmann and Todorov~\cite{BT77}. 
In order to demonstrate the efficiency of the polynomial 
technique and because of their physical relevance we apply it to the complete 
twist decomposition of the scalar operators $N(x,-x),\, M(x,-x)$ and 
$O(x,-x)$ 
having minimal twist $\tau = 3$ and $\tau = 2$, respectively, as well as to 
the related vector and tensor operators. 

\subsubsection{Scalar harmonic operators without external operations}
To begin with the simplest case let us study the local operators $N_n(x)$ 
of degree $n$ which are generating polynomials of symmetric tensors of 
degree $n$. The general formula for the decomposition of the 
local scalar operator $N_{n}(x)$ which can be considered as a homogeneous 
polynomial of degree $n$ into its harmonic polynomials of definite
 twist $\tau=3+2j,\, j= 0, 1, \ldots ,$ is given by~\cite{BT77}
\begin{align}
\label{N_tw_sc}
N_{n}(x)=
\sum_{j=0}^{[\frac{n}{2}]}
\frac{(n+1-2j)!}{4^j j!(n+1-j)!}\, x^{2j}\, 
N^{{\rm tw}(3+2j)}_{n-2j}(x),
\end{align}
where the harmonic operators of definite twist are defined by 
(see also Ref.~\cite{VK}, Eq.~9.3.2(3))
\begin{align}
\label{Ntg}
N^{{\rm tw}(3+2j)}_{n-2j}(x)=
\sum_{k=0}^{[\frac{n-2j}{2}]}
\frac{(-1)^k (n-2j-k)!}{4^k k!(n-2j)!}\, x^{2k}\square^k
\left(\square^{j} N_{n}(x)\right).
\end{align}
They satisfy the condition of tracelessness 
\begin{align}
\square N^{{\rm tw}(3+2j)}_{n-2j}(x)=0.
\nonumber
\end{align}
Therefore, the polynomials of Eq.~(\ref{Ntg}) span
the space of homogeneous harmonic polynomials of degree $n-2j$.

After the substitution $N_n(x) = \int \d^4 p\, N(p) (xp)^n$
into (\ref{Ntg}), and using 
\begin{align}
\square^{k+j} \,(xp)^n = 
\frac{n!}{(n-2k-2j)!} (p^2)^{k+j} (xp)^{n-2k-2j},
\nonumber
\end{align}
we rewrite
these harmonic operators in terms of Gegenbauer polynomials as 
follows:\footnote{Let me note that the Chebyshev polynomials of the 
second kind, $U_n(z)$, used
in Refs.~\cite{ball99b,BM01} are defined in terms of Gegenbauer 
polynomials as follows: $U_n(z)=C^1_n(z)$ (see~\cite{VK}).}
\begin{align}
\label{Nttw_GP}
N^{\text{tw}(3+2j)}_{n-2j}(x) =
\frac{n!}{(n-2j)!}
\int\!\d^4 p\, N(p)\, 
p^{2j}\left(\frac{1}{2}\sqrt{p^2 x^2}\right)^{n-2j}
\!C_{n-2j}^1\left(\frac{px}{\sqrt{p^2 x^2}}\right).
\end{align}
From the group theoretical point of view, Eq.~(\ref{Nttw_GP}) is another form
of the harmonic extension (\ref{Ntg}). 
Here, the $p-$integration introduces a superposition
of the Fourier transforms $N(p)$ times $p^{2j}$ with coefficients 
being (two-sided) harmonic polynomials, 
\begin{align}
h^1_{n-2j}(p|x) = \Big(\frac{1}{2}
\sqrt{p^2 x^2}\Big)^{n-2j}
\!C_{n-2j}^1\big({px}/{\sqrt{p^2 x^2}}\big)
\end{align}
with
\begin{align}
\square_x\, h^1_{n-2j}(p|x) = 0 = \square_p\, h^1_{n-2j}(p|x),\qquad
h^1_{n-2j}(p|x)=h^1_{n-2j}(x|p).
\end{align}

The resummation of the local expressions $N_n(x)$ according to 
$N(x,-x) = \sum_{n=0}^\infty ({\ii^n}/{n!}) N_n(x)$ is obtained by using the 
well-known integral representation of Euler's beta function,
\begin{align}
\label{beta}
B(n,m)=\frac{\Gamma(n)\Gamma(m)}{\Gamma(n+m)}=
\int_0^1\d t\, t^{n-1} (1-t)^{m-1}.
\end{align}
Then the (infinite) twist decomposition of the nonlocal scalar operator 
$N(x,-x)$  into nonlocal harmonic operators $N^{{\rm tw}(3+2j)}(x,-x)$
of twist $\tau=3+2j,\, j= 0,1, 2, \ldots,$ reads
\begin{align}
\label{N_twn_sc}
N(x,-x)= N^{{\rm tw}3}(x,-x)
+ \sum_{j=1}^{\infty}
\frac{x^{2j}}{4^j j!(j-1)!}
\int_0^1\d t\, t\,(1-t)^{j-1}\, 
N^{{\rm tw}(3+2j)}(tx,-tx),
\end{align}
with 
\begin{align}
\label{N^(3+2j)}
N^{{\rm tw}(3+2j)}(x,-x)
&=\sqrt{\pi}\int\!\d^4 p\, N(p)\, 
(- p^2)^{j}\left(1+p\pd_p\right)
\left(\sqrt{(px)^2-p^2 x^2}\right)^{-1/2}\nonumber\\
&\hspace{5cm}\times
J_{1/2}\left(\frac{1}{2}\sqrt{(px)^2-p^2 x^2}\right)
\e^{\ii  px/2}.
\end{align}
Let us remark that, for notational simplicity, in Eq.~(\ref{N_twn_sc})
we wrote the nonlocal harmonic operators for $j \geq 1$ 
without including the integration over $t$  
which is due to the normalization of the local harmonic operators,
cf.~Eqs.~(\ref{N_tw_sc}) and (\ref{Nttw_GP}). The normalization
of the nonlocal harmonic operators may be read off from the terms
in front of the $t-$integral in Eq.~(\ref{N_twn_sc}).
Let us point also to the remarkable fact that the property of
harmonicity is independent of $j$; this obtains immediately from the 
fact that the following expressions are (two-sided) harmonic functions,
\begin{align}
\label{harm}
\hspace{-.5cm}
{\cal H}_1(p|x) = \sqrt{\pi}\left(\sqrt{(px)^2-p^2 x^2}\right)^{-1/2}
\!J_{1/2}\left(\frac{1}{2}\sqrt{(px)^2-p^2 x^2}\right)
\e^{\ii  px/2}
\end{align}
with
\begin{align}
\square_x\, {\cal H}_1(p|x) = 0 = \square_p\,{\cal H}_1(p|x),\qquad
{\cal H}_1(p|x)={\cal H}_1(x|p).
\end{align}
Furthermore, we observe that the factor $(1+p\pd/\pd p)$ in 
Eq.~(\ref{N^(3+2j)}) may be changed into $(1+x\pd/\pd x)$ after which it
can be taken outside the $p-$integration and, in Eq.~(\ref{N_twn_sc}),
could be changed into $(1+t\pd/\pd t)$; if desired, this could
be used for a partial integration.

The resulting expression (\ref{N^(3+2j)}) can be rewritten in
a form which, in the case $j=0$, has been already introduced 
by \cite{BB91,ball99}. Namely, using
\begin{align}
z^{-n-1/2} J_{n+1/2}(z) = (-1)^n \sqrt{\frac{2}{\pi}} 
\Big(\frac{1}{z} \frac{\d}{\d z}\Big)^n
\Big(\frac{\sin z}{z}\Big),
\end{align}
we obtain
\begin{align}
\label{N(3+2j)}
&N^{{\rm tw}(3+2j)}(x,-x) 
=
\int \d^4 p\, N(p)\, (-p^2)^j\,
\frac{2}{\sqrt{(px)^2-p^2 x^2}}\, p\pd_p \,
\Big( \sin \Big(\frac{1}{2}
\sqrt{(px)^2-p^2 x^2}\Big) \, \e^{\ii px/2} \Big)
\nonumber\\
&\qquad=
\int \d^4 p\, N(p)\, (-p^2)^j\,
\frac{1}{2\sqrt{(px)^2-p^2 x^2}}
\bigg( 
\big(px + \sqrt{(px)^2-p^2 x^2}\big)\,
\e^{(\ii/2)\big(px + \sqrt{(px)^2-p^2 x^2}\big)}
\nonumber\\
&\qquad\qquad\qquad\qquad
-\big(px - \sqrt{(px)^2-p^2 x^2}\big)\,
\e^{(\ii/2)\big(px - \sqrt{(px)^2-p^2 x^2}\big)}
\bigg)\\
&=
\int \d^4 p\, N(p)\, (-p^2)^j\,
\Big\{
\cos \Big(\frac{1}{2}\sqrt{(px)^2-p^2 x^2}\Big)
+\frac{\ii (px)}{\sqrt{(px)^2-p^2 x^2}}\, 
\sin \Big(\frac{1}{2} \sqrt{(px)^2-p^2 x^2}\Big) 
\Big\} \e^{\ii px/2}.\nonumber
\end{align}
If we use the abbreviation, which was introduced by Balitsky and Braun
(see Eq.~(5.15) in Ref.~\cite{BB91}),
\begin{align}
\left[\e^{\ii px}\right]_{\rm{lt}}=
\bigg\{
\cos \Big(\frac{1}{2}\sqrt{(px)^2-p^2 x^2}\Big)
+\frac{\ii (px)}{\sqrt{(px)^2-p^2 x^2}}\, 
\sin \Big(\frac{1}{2} \sqrt{(px)^2-p^2 x^2}\Big) 
\bigg\} \e^{\ii px/2},
\end{align}
we may rewrite Eq.~(\ref{N(3+2j)}) according to
\begin{align}
N^{{\rm tw}(3+2j)}(x,-x) 
=
\int \d^4 p\, N(p)\, (-p^2)^j\,\left[\e^{\ii px}\right]_{\rm{lt}}.
\end{align}
Analogous results may be obtained also in the more
complicated cases which will be considered below.

\subsubsection{Scalar harmonic operators with external operations}

Now, let us consider the case of the operators 
$O(x,-x) = x^\alpha O_\alpha(x,-x)$, 
where an additional power of $x$ occurs through contraction of the vector
operator with $x^\alpha$. Therefore, the formulas (\ref{N_tw_sc}) and 
(\ref{Ntg}) are to be written down for the scalar operator 
$O_{n+1}(x)= \int \d^4 p\, x_\mu O^\mu(p) (xp)^n$, 
i.e.,~by replacing $n$ by $n+1$ within these expressions for the 
decomposition into harmonic operators of any twist $\tau=2+2j$.
Then the decomposition of the scalar local operators 
$O_{n+1}(x)$ reads:
\begin{align}
\label{O_loc_i}
O_{n+1}(x)=\sum_{j=0}^{[\frac{n+1}{2}]}
\frac{(n+2-2j)!}{4^j j!(n+2-j)!}\, x^{2j}\, 
O^{{\rm tw}(2+2j)}_{n+1-2j}(x),
\end{align}
where the local harmonic operators of twist $\tau = 2 + 2j$ are given by 
\begin{align}
\label{Otg}
O^{{\rm tw}(2+2j)}_{n+1-2j}(x)=
\sum_{k=0}^{[\frac{n+1-2j}{2}]}\frac{(-1)^k (n+1-2j-k)!}{4^k k!(n+1-2j)!}\, 
x^{2k}\square^k\left(\square^{j} O_{n+1}(x)\right). 
\end{align}
After performing the differentiations $\square^{k+j} O_{n+1}(x)$ 
we can rewrite
these harmonic operators in terms of Gegenbauer polynomials as follows:
\begin{align}
\label{Ottw_GP}
&O^{\text{tw}(2+2j)}_{n+1-2j}(x)=
\frac{n!}{(n+1-2j)!}
\int\!\d^4 p\, O^\mu(p)\, p^{2j}\,
\bigg\{x_\mu \left(\frac{1}{2}\sqrt{p^2 x^2}\right)^{n-2j}
\!C_{n-2j}^2\bigg(\frac{px}{\sqrt{p^2 x^2}}\bigg)\\
&
-\frac{1}{2} p_\mu  x^2
\left(\frac{1}{2}\sqrt{p^2 x^2}\right)^{n-1-2j}
\!C_{n-1-2j}^2\bigg(\frac{px}{\sqrt{p^2 x^2}}\bigg)
+2j \, \frac{p_\mu}{p^2}
\left(\frac{1}{2}\sqrt{p^2 x^2}\right)^{n+1-2j}
\!C_{n+1-2j}^1\bigg(\frac{px}{\sqrt{p^2 x^2}}\bigg)
\bigg\}, \nonumber
\end{align}
where the additional terms proportional to $p_\mu$ occur 
because of the appearance of derivatives with respect to Gamma structure 
$(x\gamma)$.
The change in the order of the Gegenbauer polynomials and, consequently,
in the order of the Bessel functions and the accompanying factors
$\sqrt{(px)^2-p^2 x^2}$ occurring below, is also due to that fact. 
Of course, for $j=0$ we re-obtain the expression (\ref{O2_n+1})

Resumming with respect to $n\ (\,\geq 2j \pm 1$ or $2j,$ respectively) 
and, again, using the representation (\ref{beta}) of Euler's beta function
we obtain the following (infinite) twist decomposition
\begin{align}
\label{O_twn_sc}
O(x,-x)=O^{{\rm tw}2}(x,-x)
+\sum_{j=1}^{\infty}
\frac{x^{2j}}{4^j j!(j-1)!}
\int_0^1\d t\, t\, (1-t)^{j-1}\, 
O^{{\rm tw}(2+2j)}(tx,-tx),
\end{align}
with 
\begin{align}
\label{O^(2+2j)}
O^{{\rm tw}(2+2j)}(x,-x)
&=\sqrt{\pi}\int\!\d^4 p\, O^\mu(p)\,(- p^2)^{j}
\bigg\{
\left( x_\mu(2+p\pd_p)
-\frac{1}{2}\, \ii p_\mu x^2  
\right)\\
&\qquad\qquad\times
\left(3+p\pd_p\right)\left(\sqrt{(px)^2-p^2 x^2}\right)^{-3/2}
\!J_{3/2}\left(\frac{1}{2}\sqrt{(px)^2-p^2 x^2}\right)
\nonumber\\
&\quad\; - 2j\, \frac{ \ii p_\mu}{ p^{2}} \left(1+p\pd_p\right)
\left(\sqrt{(px)^2-p^2 x^2}\right)^{-1/2}
\!J_{1/2}\left(\frac{1}{2}\sqrt{(px)^2-p^2 x^2}\right)
\bigg\}\,\e^{\ii px/2}.\nonumber
\end{align}
Again, by convention, the resummed nonlocal harmonic operators 
$O^{{\rm tw}(2+2j)}(tx,-tx)$ of twist $\tau=2+2j, \, j = 0,1,2,\cdots,$ 
are obtained by changing any $x \rightarrow tx$. 
Also here analogous comments are in order as for the simpler case of the 
operator $N(x,-x)$. Harmonicity of the nonlocal operators (\ref{O^(2+2j)}), 
$\square O^{{\rm tw}(2+2j)}(x,-x) = 0$, holds by construction.
Obviously, for $j=0$ we recover the expression (\ref{XYZ}) .

Finally, we consider the scalar operator
$M(x,-x) =  x^\nu \pd^\mu M_{[\mu\nu]} (x,-x)$ resulting from the 
skew tensor operator $M_{[\mu\nu]} (x,-x)$. This operator is
governed by totally symmetric tensors since the skew tensors  
$M_{[\mu\nu]n}$ having symmetry type $[n+2]=(n+1,1)$ are composed with another 
skew tensor $x^{[\nu}\pd^{\mu]}$ having symmetry type $[2]=(1,1)$ and the 
Clebsch--Gordan series of their direct product consists only of one
symmetry type, $[n]=(n)$.

The local operators are decomposed according to formulas
(\ref{N_tw_sc}) and (\ref{Ntg}). However, because of the
special `external' structure the resulting expressions
simplify considerably. Namely, after performing the differentiations 
\begin{align*}
 \square^{k+j} \pd^\mu\sigma_{\mu\nu} x^\nu (xp)^n
= \frac{n!}{(n-1-2k-2j)!}\, \big(p^\mu\sigma_{\mu\nu} x^\nu\big)
 p^{2j} (p^2)^k (px)^{n-1-2k-2j},
\end{align*}
we arrive at the following harmonic 
operators in terms of Gegenbauer polynomials:
\begin{align}
\label{Mttw_GP}
&M^{\text{tw}(3+2j)}_{n-2j}(x) =
\frac{n!}{(n-2j)!}
\int\!\d^4 p\, M_{[\mu\nu]}(p)\, p^\mu x^\nu\,p^{2j}\,
\left(\frac{1}{2}\sqrt{p^2 x^2}\right)^{n-1-2j}
\!C_{n-1-2j}^2\left(\frac{px}{\sqrt{p^2 x^2}}\right).
\end{align}
Now, resumming with respect to $n\, (\,\geq 2j+1)$ we finally get the 
infinite twist decomposition:
\begin{align}
\label{M_twn_sc}
M(x,-x)=M^{{\rm tw}3}(x,-x)
+\sum_{j=1}^{\infty}
\frac{x^{2j}}{4^j j!(j-1)!}
\int_0^1\d t\, t\, (1-t)^{j-1}\, 
M^{{\rm tw}(3+2j)}(tx,-tx),
\end{align}
with 
\begin{align}
\label{M^(2+2j)}
&M^{{\rm tw}(3+2j)}(x,-x)
=\sqrt{\pi}\int\!\d^4 p\, 
M_{[\mu\nu]}(p)\,(- p^2)^{j}\,\ii p^\mu x^\nu
\nonumber\\
&\qquad\qquad\times
\left(2+p\pd_p\right)\left(3+p\pd_p\right)
\left(\sqrt{(px)^2-p^2 x^2}\right)^{-3/2}
\!J_{3/2}\left(\frac{1}{2}\sqrt{(px)^2-p^2 x^2}\right)
\e^{\ii px/2}.
\end{align}
Again, harmonicity of the  nonlocal operators 
$M^{{\rm tw}(3+2j)}(x,-x)$ of twist $\tau=3+2j, \, j = 0,1,2,\cdots,$ 
is fulfilled without taking care of $j$.


\subsubsection{Harmonic vector and tensor operators
related to totally symmetric local operators}


The prescribed procedure may be also used for the twist decomposition 
of any totally symmetric (tensor) operator 
with an arbitrary number of free tensor indices which, 
in the framework of the polynomial technique, are
recovered by straightforward differentiation  with respect to $x$ 
of the corresponding scalar operators. 

The simplest of these operators, whose tower of infinite twist part starts 
with $\tau =2$ and which contains the leading contributions to virtual 
Compton scattering is the operator $O^{\sf S}_\alpha (x,-x)$. Its symmetry 
type $\sf S$ is characterized by the Young frames
$(n+1),\ 1\leq n\leq\infty$ (cf.~also Ref.~\cite{GLR99} where we denoted
that symmetry type by (i)).
The local expressions, $O^{\sf S}_{\alpha n}(x)$,  are obtained from the local 
scalar operators (\ref{O_loc_i}) of degree $n+1$ by applying 
$(1/(n+1)) \pd_\alpha$.  
Eventually, we get
\begin{align}
\pd_\alpha O_{n+1}(x)
&=\sum_{j-1=0}^{[\frac{n+1}{2}]}
\frac{(n+2-2j)!}{4^j (j-1)!(n+2-j)!}\, 2x_\alpha\, x^{2(j-1)}\, 
O^{{\rm tw}(2+2j)}_{n+1-2j}(x)
\nonumber\\
&+
\sum_{j=0}^{[\frac{n+1}{2}]}
\frac{(n+2-2j)!}{4^j j!(n+2-j)!}\, x^{2j} 
\left(\pd_\alpha O^{{\rm tw}(2+2j)}_{n+1-2j}(x)\right).
\nonumber
\end{align}
and the representation (\ref{beta}) of the beta function,
both sums may be rewritten as
\begin{align}
&\sum_{j=0}^{[\frac{n+1}{2}]}
\frac{(n-2j)!}{4^j j!(n+1-j)!}\,  x^{2j}\, 
O^{{\rm tw}(4+2j)}_{n-1-2j}(x)
=\sum_{j=0}^{[\frac{n-1}{2}]}
\frac{x^{2j}}{4^j (j!)^2}  
\int_0^1\d t\, t\, (1-t)^j\,
O^{{\rm tw}(4+2j)}_{n-1-2j}(tx),
\nonumber\\
&\sum_{j=0}^{[\frac{n+1}{2}]}
\frac{(n+2-2j)!}{4^j j!(n+2-j)!}\, x^{2j} 
\pd_\alpha O^{{\rm tw}(2+2j)}_{n+1-2j}(x)
=
\pd_\alpha O^{{\rm tw}\ 2}_{n+1}(x)\nonumber\\
&\hspace{5cm}
+\sum_{j=1}^{[\frac{n+1}{2}]}
\frac{x^{2j}}{4^j j!(j-1)!}  
\int_0^1\d t\, t\, (1-t)^{j-1}
\left(\pd_\alpha O^{{\rm tw}(2+2j)}_{n+1-2j}(tx)\right),
\nonumber
\end{align}
respectively, where
$O^{{\rm tw}(4+2j)}_{n-1-2j}(x)$ is given by Eq.~(\ref{Ottw_GP})
with $j \rightarrow j+1$. From this it becomes obvious that, 
beginning with twist-4, one obtains two different contributions
of the same twist,
namely a vector and a scalar part.

Now, putting together these terms and 
resumming over $n$, thereby representing the normalizing coefficient
$1/(n+1)$ as $\int_0^1 \d \lambda\, \lambda^n$, we obtain
\begin{align}
\label{O_v_sym}
O^{\sf S}_\alpha(x, -x)
=\,& 
\int_0^1\,\d\lambda\,
\bigg\{\big(\pd_\alpha\,O^{\rm tw2}\big)(\lambda x,-\lambda x)\nonumber\\
&+
\sum_{j=1}^\infty
\frac{x^{2j}\lambda^{2j}}{4^j j!(j-1)!}\,  
\int_0^1\d t\, t\, (1-t)^{j-1}\,
\big(\pd_\alpha\, O^{{\rm tw}(2+2j)}\big)(\lambda tx, -\lambda tx)
\nonumber\\
&
-\frac{1}{2}\,x_\alpha
\sum_{j=0}^\infty
\frac{x^{2j}\lambda^{2j+1}}{4^j (j!)^2}\,  
\int_0^1\d t\, t\, (1-t)^{j}\,
O^{{\rm tw}(4+2j)}(\lambda tx, -\lambda tx)
\bigg\},
\end{align} 
where 
$O^{{\rm tw}(4+2j)}(x, -x)\equiv O^{{\rm tw}(2+2(j+1))}(x, -x)$
and $O^{{\rm tw}(2+2j)}(x, -x)$ are given by Eq.~(\ref{O^(2+2j)}).
Both the integrals over $\lambda$ and $t$ are due to normalizations
and, therefore, it depends on the personal taste if they are to be
included into the definition of the harmonic operators of twist $\tau$
or if they are considered as part of the twist decomposition.
It is interesting to note that in Eq.~(\ref{O_v_sym}) only the twist-2 
vector and the twist-4 scalar operator survives on the light-cone.
All other higher twist operators for $j\geq1$ are cancelled due to the 
factor $x^2$.  

In the same manner we are able to define the twist decomposition of 
(completely symmetric) tensor operators which result from the expression
\begin{align}
O^{\sf S}_{(\alpha_1\alpha_2\cdots\alpha_r)n}(x)
=
\frac{\pd_{\alpha_1}}{(n+1)} \cdots
\frac{\pd_{\alpha_r}}{(n+r)}
\sum_{j=0}^{[\frac{n+r}{2}]}
\frac{(n+1+r-2j)!}{4^j j!(n+1+r-j)!}\, x^{2j}\, 
O^{{\rm tw}(2+2j)}_{n+r-2j}(x)\ .
\nonumber
\end{align}
The general procedure is obvious from the consideration of the vector
operator. Completely analogous to that case with only one derivative we 
would obtain a finite number of different operators of a given twist $\tau$.

\subsection{Harmonic quark-antiquark operators of geometric twist 2 and 3}

In this Section we determine the harmonic quark-antiquark operators
up to twist $\tau = 2$ and $3$ which are relevant for the generalization
of the parton distributions of definite (geometric) twist \cite{GL01}
to the double distributions appearing in the parametrization in the
non-forward matrix elements of the corresponding non-local operators
off the light-cone.
According to the framework of polynomial technique, we start with the
local operators. 

\subsubsection{Local harmonic operators 
}

Now we discuss the local harmonic operators of geometric 
twist-2 and twist-3 for both the chiral-even vector operators,
\begin{align}
O_{\alpha n}(x)&=\bar{\psi}(0)\gamma_{\alpha}(\ii x\Tensor D)^n \psi(0),\\
O_{5\alpha n}(x)&
=\bar{\psi}(0)\gamma_{\alpha}\gamma_5 (\ii x\Tensor D)^n \psi(0),
\end{align}
and the chiral-odd scalar and skew tensor operators,
\begin{align}
N_{n}(x)&=\bar{\psi}(0)(\ii x\Tensor D)^n \psi(0),\\
M_{[\alpha\beta]n}(x)&
=\bar{\psi}(0)\sigma_{\alpha\beta}(\ii x\Tensor D)^n \psi(0),
\end{align}
together with their related scalar and vector operators
\begin{align}
O_{n+1}(x)=x^\alpha O_{\alpha n}(x),
\qquad
M_{\mu n+1}(x)=x^\nu M_{[\mu\nu] n}(x),
\qquad
M_{n}(x)=\pd^\mu M_{\mu n+1}(x).
\end{align}

Now we consider the decomposition the vector operators 
$O_{\alpha n}(x)$ into its harmonic twist-2 and twist-3 part. 
The irreducible vector operators $O^{\rm tw2}_{\alpha n}(x)$
of Lorentz type $(\frac{n+1}{2},\frac{n+1}{2})$ have symmetry types
$[n+1]=(n+1)$ and can be taken over from the previous Section. 
On the other hand, the twist-3 vector operators transform according to  
$(\frac{n+1}{2},\frac{n-1}{2})\oplus(\frac{n-1}{2},\frac{n+1}{2})$
and have symmetry types $[n+1]=(n,1)$. They may be summed up to the lowest
term of another infinite tower of non-local operators of twist
$\tau = 3, 4, \ldots.$
The conditions of tracelessness for harmonic vector operators are as 
follows: 
\begin{align}
\label{traceless}
\square O^{\rm tw\,\tau}_{\alpha n}(x)=0,
\qquad 
\pd^\alpha O^{\rm tw\,\tau}_{\alpha n}(x)=0
\qquad
{\rm for}
\qquad
\tau=2,3.
\end{align}

Eventually, the scalar twist-2 operator is given by 
(see Eqs.~(\ref{T_harm_d}) and (\ref{P^[n]_n}) with $h=2$)
\begin{align}
O^{{\rm tw} 2}_{n+1}(x)=
\sum_{k=0}^{[\frac{n+1}{2}]}\frac{(-1)^k (n+1-k)!}{4^k k!(n+1)!}\, 
x^{2k}\,\square^k O_{n+1}(x). 
\nonumber
\end{align}
by means of a partial differentiation, $\pd_\alpha$, together with 
the normalization factor $1/(n+1)$ as follows (see Eq.~(\ref{T^[n]_alpha})):
\begin{align}
\label{O_tw2}
O^{\rm tw2}_{\alpha n}(x)
&=\frac{1}{n+1}\, \pd_\alpha O^{\rm tw2}_{n+1}(x)
\nonumber\\
&=\frac{1}{n+1}
\sum_{k=0}^{[\frac{n}{2}]}\frac{(-1)^k (n-k)!}{4^k k!(n+1)!}\, x^{2k}\,
{\cal D}_\alpha(k)\, \square^k O_{n+1}(x).
\end{align}
Here, we introduced the operation
\begin{align}
\label{D}
{\cal D}_\alpha(k) = (k+1+x\pd)\pd_\alpha
-\hbox{\large$\frac{1}{2}$}\,x_\alpha\square
\end{align}
which is a generalization of the inner derivative 
off the light-cone and its extension to arbitrary 
values of $k\geq 0$ (see Section~\ref{tensor}).

The twist-3 vector operator which satisfies the conditions 
(\ref{traceless}) is\footnote{We use 
the convention $A_{[\mu\nu]}\equiv\frac{1}{2}\left(A_{\mu\nu}-A_{\nu\mu}\right)$.}
(see Eq.~(\ref{[n,1](n-1)}) with $h=2$)
\begin{align}
\label{O_tw3}
O^{\rm tw3}_{\alpha n}(x)
&=\frac{2}{n+1}\, x^\beta
\bigg\{\delta^\mu_{[\alpha}\pd_{\beta]}-\frac{1}{n+1}\, 
x_{[\alpha}\pd_{\beta]}\pd^{\mu}\bigg\}
\sum_{k=0}^{[\frac{n}{2}]}\frac{(-1)^k (n-k)!}{4^k k!n!}\, 
x^{2k}\,\square^k O_{\mu n}(x),\\
&=\frac{2}{n+1}
\sum_{k=0}^{[\frac{n-1}{2}]}
\frac{(-1)^k (n-1-k)!}{4^k k!n!}\, x^{2k} x^\beta
\Big\{\delta^\mu_{[\alpha}
{\cal D}_{\beta]}(k)
-\frac{1}{n+1}\, x_{[\alpha}\pd_{\beta]}
{\cal D}^\mu(k)
\Big\}\,\square^k
O_{\mu n}(x).\nonumber
\end{align}

The proper scalar twist-3 operator is of Lorentz type 
$(\frac{n}{2},\frac{n}{2})$ and reads
\begin{align}
\label{N_tw3}
N^{\rm tw3}_{n}(x)=
\sum_{k=0}^{[\frac{n}{2}]}\frac{(-1)^k (n-k)!}{4^k k!n!}\, x^{2k}
\square^k N_{n}(x).
\end{align}

Here I give the decomposition of the skew tensor operator 
$M_{[\alpha\beta] n}(x)$ into twist-2 and twist-3.
The twist-2 tensor operator $M^{\rm tw2}_{[\alpha\beta] n}(x)$
transforms according to 
$(\frac{n+2}{2},\frac{n}{2})\oplus(\frac{n}{2},\frac{n+2}{2})$, whereas
the twist-3 operator which is obtained
from the trace terms of $M^{\rm tw2}_{[\alpha\beta] n}(x)$
is of Lorentz type $(\frac{n}{2},\frac{n}{2})$.
They have symmetry types $[n+2]=(n+1,1)$ and $[n]=(n)$, respectively,
and are given by (see Eq.~(\ref{[n,1](n-1)}))
\begin{align}
\label{M_tw2}
M^{\rm tw2}_{[\alpha\beta] n}(x)
&=\frac{2}{n+2}
\bigg\{\delta^\mu_{[\alpha}\pd_{\beta]}-\frac{1}{n+2}\, 
x_{[\alpha}\pd_{\beta]}\pd^{\mu}\bigg\}
\sum_{k=0}^{[\frac{n+1}{2}]}\frac{(-1)^k (n+1-k)!}{4^k k!(n+1)!}\, 
x^{2k}\,\square^k M_{\mu n+1}(x)
\nonumber\\
&=\frac{2}{n+2}
\sum_{k=0}^{[\frac{n}{2}]}\frac{(-1)^k (n-k)!}{4^k k!(n+1)!}\, x^{2k}
\Big\{\delta^\mu_{[\alpha} {\cal D}_{\beta]}(k)
-\frac{1}{n+2}\, x_{[\alpha}\pd_{\beta]}{\cal D}^\mu(k)
\Big\}\,\square^k M_{\mu n+1}(x),
\end{align}
and
\begin{align}
\label{M_tw3a}
M^{\rm tw3}_{[\alpha\beta] n}(x)&=\frac{2}{(n+2)}\,
x_{[\alpha}\widehat{M}^{\rm tw3}_{\beta] n-1}(x),
\end{align}
with
\begin{align}
\label{vM_tw3}
\widehat{M}^{\rm tw3}_{\beta n-1}(x)
&=\frac{1}{n}\, \pd_\beta 
\sum_{k=0}^{[\frac{n}{2}]}\frac{(-1)^k (n-k)!}{4^k k!n!}\, x^{2k}
\square^k M_{n}(x)
\nonumber\\
&=\frac{1}{n}
\sum_{k=0}^{[\frac{n-1}{2}]}\frac{(-1)^k (n-1-k)!}{4^k k!n!}\, x^{2k}\,
{\cal D}_\beta(k)\,\square^k M_{n}(x)
\end{align}

Besides the twist-3 operator (\ref{M_tw3a}) resulting from the trace
terms of the twist-2 operator (\ref{M_tw2}) there exists also a
genuine twist-3 operator having Lorentz type 
$\big(\frac{n}{2},\frac{n}{2}\big)$ and being governed
by the symmetry type $[n+2]=(n,1,1)$. It is given by the following 
expression\footnote{We use the convention $A_{\alpha\beta\gamma}\equiv \frac{1}{3} 
(A_{\alpha[\beta\gamma]}+A_{\beta[\gamma\alpha]}+A_{\gamma[\alpha\beta]})$.}
(see Eq.~(\ref{[n,1,1]}) with $h=2$)
\begin{align}
\label{M_tw3b}
\widetilde M^{\rm tw3}_{[\alpha\beta] n}(x)
&=\frac{3}{n+2}\, x^\gamma
\bigg\{\delta^\mu_{[\alpha}\delta^\nu_{\beta}\pd_{\gamma]}
-\frac{2}{n}\, 
\delta^\mu_{[\alpha}x_\beta \pd_{\gamma]}\pd^{\nu}\bigg\}
\sum_{k=0}^{[\frac{n}{2}]}\frac{(-1)^k (n-k)!}{4^k k!n!}\, 
x^{2k}\,\square^k M_{[\mu\nu] n}(x)\\
&=\frac{3}{n+2}
\sum_{k=0}^{[\frac{n}{2}]}\frac{(-1)^k (n-1-k)!}{4^k k!n!}\,
x^{2k}\,x^\gamma
\Big\{\delta^\mu_{[\alpha}\delta^\nu_{\beta}{\cal D}_{\gamma]}(k)
-\frac{2}{n}\, 
\delta^\mu_{[\alpha}x_\beta \pd_{\gamma]}{\cal D}^{\nu}(k)\Big\}\, 
\square^k M_{[\mu\nu] n}(x).\nonumber
\end{align}
By construction, Eqs.~(\ref{M_tw2}), (\ref{vM_tw3}) and (\ref{M_tw3b}) are 
(traceless) harmonic polynomials obeying
\begin{align}
\square M^{\rm tw\,\tau}_{[\alpha\beta] n}(x)=0,\qquad 
\pd^\alpha M^{\rm tw\,\tau}_{[\alpha\beta] n}(x)=0=
\pd^\beta M^{\rm tw\,\tau}_{[\alpha\beta] n}(x),
\qquad
\tau = 2,\tilde 3,
\end{align}
and
\begin{align}
\square \widehat{M}^{\rm tw3}_{\alpha n-1}(x)=0,\qquad 
\pd^\alpha \widehat{M}^{\rm tw3}_{\alpha n-1}(x)=0.
\end{align}

Let us point to the structural similarities of the operators 
(\ref{O_tw2}) and (\ref{vM_tw3}) on the one hand and 
(\ref{O_tw3}) and (\ref{M_tw2}) on the other hand. --
Furthermore, we observe that when organizing the various expressions such
that everywhere the partial derivatives appear to the right of the $k$th
power of $x^2$ the differential operations ensuring the symmetry type 
are transformed according to $\pd_\alpha\rightarrow {\cal D}_\alpha(k)$
whereas the generalized (first order) inner derivative 
$x_{[\alpha}\pd_{\beta]}$ remains 
unchanged because it commutes with $x^2$. 
Namely, a short look on our results, 
Eqs.~(\ref{O_tw2}), (\ref{O_tw3}), (\ref{N_tw3}), (\ref{M_tw2}), 
(\ref{M_tw3a}) and (\ref{M_tw3b}), shows that the expressions on-cone 
are obtained by restricting the sums to those terms with $k=0$ and replacing 
$x\rightarrow \lcx,\;\pd_\alpha\rightarrow{\pd}/{\pd\lcx^\alpha}$.
Let us mention that this is nothing else but another form of the 
harmonic extension which is well-known in group theory 
\cite{BT77,Dobrev77,Dobrev82}. 
The harmonic extension gives an one-to-one relation between homogeneous 
polynomials on the light-cone and the corresponding harmonic polynomials 
off-cone. It is important to note that the unique harmonic extension 
must not destroy the type of the Lorentz representation. 

According to Eqs.~(\ref{O_tw2}), (\ref{O_tw3}), (\ref{N_tw3}), 
(\ref{M_tw2}), (\ref{M_tw3a}) and (\ref{M_tw3b}) we may introduce the 
orthogonal projection operators onto the subspaces with definite spin and, 
therefore, with definite geometric twist $\tau = 2,3$ as follows:
\begin{align}
\label{OPROJ-x}
O^{(\tau)}_{(5)\alpha n}(x) 
&= {\cal P}^{(\tau)\,\mu}_{\alpha n} O_{(5)\mu n}(x),\\
\label{NPROJ-x}
N^{(\tau)}_{n}(x) &= {\cal P}^{(\tau)}_n N_n(x),\\
\label{MPROJ-x}
M^{(\tau)}_{[\alpha\beta]n}(x) &= 
{\cal P}^{(\tau)\,[\mu\nu]}_{[\alpha\beta]n}M_{[\mu\nu]n}(x).
\end{align}
The corresponding twist projectors onto the light-cone are discussed 
in more detail in Section~\ref{tensor} (see also Ref.~\cite{GL01}).

\subsubsection{Resummed harmonic operators with twist $\tau = 2,\,3$}

Now we are able to rewrite the local harmonic operators of definite
twist $\tau = 2, 3$ in terms of Gegenbauer polynomials in the $x$-space
analogous to the scalar harmonic operators. 
In a second step we resum the infinite series
(for $n$) of Gegenbauer polynomials to the related Bessel functions  
thereby obtaining nonlocal harmonic operators of (the same) definite twist.

For notational simplicity we introduce the following abbreviations
for the homogeneous polynomials and their related functions:
\begin{align}
\label{h-n}
h_n^\nu(p|x)&=
\left(\frac{1}{2}\sqrt{p^2 x^2}\right)^n
\!C_n^\nu\bigg(\frac{px}{\sqrt{p^2 x^2}}\bigg),
\\
\label{H-n}
{\cal H}_\nu(p|x) &= 
\sqrt{\pi}
\left(\sqrt{(px)^2-p^2 x^2}\right)^{1/2-\nu}
\!J_{\nu-1/2}\left(\frac{1}{2}
\sqrt{(px)^2-p^2 x^2}\right)
\e^{\ii px/2}.
\end{align}
Using the following functional relations of the Gegenbauer and
Bessel functions \cite{PBM}, 
\begin{gather*}
m C_m^\nu(z) 
= 2\nu\left(zC_{m-1}^{\nu+1}(z) - C_{m-2}^{\nu+1}(z)\right),
\\
\frac{\d}{\d z} \left( z^{-\lambda} J_\lambda(z) \right)
= - z^{-\lambda} J_{\lambda+1}(z) ,
\\
z \frac{\d}{\d z}  J_\lambda(z)
= -\lambda J_{\lambda}(z) + z J_{\lambda-1}(z),
\end{gather*}
respectively, one obtains the following results for the partial derivations
of these functions:
\begin{align}
\pd_\alpha h_n^\nu(p|x)
&= \nu \left( p_\alpha h_{n-1}^{\nu+1}(p|x)
 - \frac{1}{2} p^2 x_\alpha h_{n-2}^{\nu+1}(p|x) \right),
\\
\pd_\alpha {\cal H}_\nu(p|x) 
&= \left( \ii p_\alpha (2\nu+1+x\pd) 
- \frac{1}{2}(\ii p)^2 x_\alpha \right){\cal H}_{\nu+1}(p|x).
\end{align}

The local twist-2 vector operator\footnote{The corresponding local
tensor operator of 2nd rank is given by: $O^{\text{tw2}}_{\alpha\beta n-1}(x)
=1/n\,\pd_\beta O^{\text{tw2}}_{\alpha n}(x)$.} 
is obtained through derivation with 
respect to $x^\alpha$ of the local twist-2 scalar operator as follows,
cf.,~Eq.~(\ref{O_tw2}),
\begin{align}
\label{XO2}
O^{\text{tw2}}_{\alpha n}(x) 
&=
\frac{1}{(n+1)^2}\,\pd_\alpha
\int\!\d^4 p\, \big(\bar{\psi}\gamma_\mu\psi\big)(p)
\left\{ x^\mu\, h_n^2(p|x)
-\frac{1}{2}\,  p^\mu x^2\, h_{n-1}^2(p|x)
\right\}
\nonumber\\
&=
\frac{1}{(n+1)^2}
\int\!\d^4 p\, \big(\bar{\psi}\gamma_\mu\psi\big)(p)
\bigg\{ 
\delta_\alpha^\mu \,h_n^2(p|x)
- p^\mu x_\alpha \,h_{n-1}^2(p|x)  
\\
&\qquad
+2 x^\mu p_\alpha \,h_{n-1}^3(p|x)
-\big(x^\mu x_\alpha p^2+p^\mu p_\alpha x^2\big) \,h_{n-2}^3(p|x)
+\frac{1}{2}\, p^\mu x_\alpha x^2 p^2 \,h_{n-3}^3(p|x)
\bigg\},
\nonumber
\end{align}
and the corresponding resummed nonlocal twist-2 vector 
operator reads
\begin{align}
\label{nl_O2}
&O^{\rm tw2}_\alpha (x,-x)
= \pd_\alpha \int_0^1\d t
\int\!\d^4 p\, \big(\bar{\psi}\gamma_\mu\psi\big)(p)
\left\{ x^\mu \left(2+p\pd_p\right)
-\frac{1}{2}\,\ii\,t  p^\mu\, x^2\right\}
\left(3+p\pd_p\right)\, {\cal H}_2(p|tx)
\nonumber\\
&\quad=\int\d^4 p\, \big(\bar{\psi}\gamma_\mu\psi\big)(p)
\left(2+p\pd_p\right)\int_0^1\d t
\bigg\{
\Big[\left(3+p\pd_p\right)\delta^\mu_{\alpha}
-\ii t p^\mu x_\alpha\Big]
{\cal H}_2(p|tx)
\\
&\qquad+\Big[\left(3+p\pd_p\right)
\left(\left(4+p\pd_p\right)\ii t p_\alpha  x^\mu 
-\frac{1}{2}(\ii t)^2 \big(p^2 x^\mu x_\alpha
+ x^2 p^\mu p_\alpha \big) 
\right)
+\frac{1}{4}
(\ii t)^3 p^\mu p^2 x_{\alpha}x^2 \Big]
{\cal H}_3(p|tx)
\bigg\}.
\nonumber
\end{align}
Here, the equality $1/(n+1) = \int_0^1\d t\, t^n$ has been used;
analogous equalities will be used in the following.

The resummed local twist-3 vector operator reads, cf.,~Eq.~(\ref{O_tw3}):
\begin{align}
\label{XO3}
&O^{\rm tw3}_{\alpha n}(x)
=\frac{2}{n+1}\, x^\beta
\Big\{\delta^\mu_{[\alpha}\pd_{\beta]}-\frac{1}{n+1}\, 
x_{[\alpha}\pd_{\beta]}\pd^{\mu}\Big\}
\int\d^4 p\, \big(\bar{\psi}\gamma_\mu\psi\big)(p) \,h_{n}^1(p|x)
\nonumber\\
&\qquad=\frac{1}{(n+1)^2}
\int\d^4 p\, \big(\bar{\psi}\gamma_\mu\psi\big)(p) x^\beta
\bigg\{ 2(n+1)
\delta_{[\alpha}^\mu p_{\beta]}  \,h_{n-1}^2(p|x)
+(n+2)\, x_{[\alpha}\delta_{\beta]}^\mu p^2 \,h_{n-2}^2(p|x)   
\nonumber\\
&\qquad\qquad\qquad\qquad\qquad
-4 x_{[\alpha}p_{\beta]} p^\mu \,h_{n-2}^3(p|x) 
+2 x_{[\alpha} p_{\beta]} x^\mu p^2 \,h_{n-3}^3(p|x)
\bigg\},
\end{align}
and the bilocal twist-3 vector operator is given by
\begin{align}
\label{nl_O3}
&O^{\rm tw3}_\alpha (x,-x)
=2\int\d^4 p\, \big(\bar{\psi}\gamma_\mu\psi\big)(p)\,
\,x^\beta\Big\{
\delta^\mu_{[\alpha}\pd_{\beta]}\,(1+p\pd_p)
-x_{[\alpha}\pd_{\beta]}\pd^\mu\Big\}
\int_0^1\d t\,
{\cal H}_1(p|tx)
\nonumber\\
&\qquad=
\int\d^4 p\, \big(\bar{\psi}\gamma^\mu\psi\big)(p)
\int_0^1\d t \, \left(2+p\pd_p\right)\, x^\beta
\bigg\{\Big[
2\,\ii t\,\delta^\mu_{[\alpha}p_{\beta]}(2+p\pd_p)\,
-(\ii t)^2 \delta^\mu_{[\alpha}x_{\beta]} p^2 \Big] {\cal H}_2(p|tx)
\nonumber\\
&\qquad\qquad\qquad\qquad\qquad
- \ii t\, x_{[\alpha} p_{\beta]} 
\Big[2\, \ii t\, p^\mu  \left(5+p\pd_p\right)
-(\ii t)^2 x^\mu  p^2  \Big]\,{\cal H}_3(p|tx)
\bigg\}.
\end{align}

The local twist-3 chiral-odd scalar operator will be given only
for completeness, cf.,~Eq.~(\ref{Nttw_GP}):
\begin{align}
\label{XN3}
N^{\rm tw3}_n(x)
&=
\int\d^4 p\, \big(\bar{\psi}\psi\big)(p) \,h_{n}^1(p|x),
\end{align}
and for the corresponding bilocal one, one gets
\begin{align}
\label{nl_N3}
N^{\rm tw3}(x,-x)
&=\int\d^4 p\, \big(\bar{\psi}\psi\big)(p)
\left(1+p\pd_p\right)\,{\cal H}_1(p|x).
\end{align}

The local resummed twist-2 chiral-odd skew tensor operator is given by,
cf.,~Eq.~(\ref{M_tw2}),
\begin{align}
\label{XM2}
M^{\rm tw2}_{[\alpha\beta] n}(x)
=&\frac{2}{(n+2)(n+1)}\,
\Big\{\delta^\mu_{[\alpha}\pd_{\beta]}-\frac{1}{n+2}\, 
x_{[\alpha}\pd_{\beta]}\pd^{\mu}\Big\}\nonumber\\
&\qquad\times\int\d^4 p\, \big(\bar{\psi}\sigma_{\mu\nu}\psi\big)(p)
\Big\{
x^\nu \,h_{n}^2(p|x)
-\frac{1}{2}\,p^\nu\,x^2 \,h_{n-1}^2(p|x)
\Big\},
\end{align}
and the related bilocal twist-2 skew tensor operator reads
\begin{align}
\label{nl_M2}
M^{\rm tw2}_{[\alpha\beta]} (x,-x)
&=2\int\!\d^4 p\, \big(\bar{\psi}\sigma_{\mu\nu}\psi\big)(p)
\int_0^1\!\d t\,t
\Big\{(2+p\pd_p)\,\delta^\mu_{[\alpha}\pd_{\beta]} 
-x_{[\alpha}\pd_{\beta]}\pd^\mu \Big\} \nonumber\\
&\qquad\qquad\qquad\qquad
\times
\Big[  x^\nu \left(3+p\pd_p\right)
-\frac{1}{2}\,\ii\,t p^\nu x^2 \Big]
{\cal H}_2(p|tx).
\end{align}
For the related, resummed local twist-3 skew tensor operator 
resulting from the trace terms we obtain
\begin{align}
\label{XM3}
{M}^{\rm tw3}_{[\alpha\beta] n}(x)
&=\frac{2}{(n+2)n}\,x_{[\alpha} \pd_{\beta]}
\int\d^4 p\, \big(\bar{\psi}\sigma_{\mu\nu}\psi\big)(p)\, p^\mu x^\nu
\,h_{n-1}^2(p|x),
\end{align}
and get the following twist-3 bilocal skew tensor
\begin{align}
\label{nl_M3}
M^{\rm tw3}_{[\alpha\beta]} (x,-x)
&=2 x_{[\alpha}\pd_{\beta]}
\int\d^4 p\, 
\big(\bar{\psi}\sigma_{\mu\nu}\psi\big)(p)\, \ii\, p^\mu x^\nu
(2+p\pd_p)
\int_0^1\d t\,{\cal H}_2(p|tx).
\end{align}

In addition, the independent local twist-3 skew tensor having symmetry type 
$[n+2]=(n,1,1)$ with Lorentz structure $\big(\frac{n}{2},\frac{n}{2}\big)$
reads
\begin{align}
\label{XtM3}
\widetilde M^{\rm tw3}_{[\alpha\beta] n}(x)
&=\frac{3}{n+2}\, x^\gamma
\bigg\{\delta^\mu_{[\alpha}\delta^\nu_{\beta}\pd_{\gamma]}
-\frac{2}{n}\, \delta^\mu_{[\alpha}x_{\beta}\pd_{\gamma]}\pd^{\nu}\bigg\}
\int\d^4 p\, \big(\bar{\psi}\sigma_{\mu\nu}\psi\big)(p) 
\,h_{n}^1(p|x)
\end{align}
and leads to the following expression for 
the twist-3 bilocal skew tensor
\begin{align}
\label{nl_tM3}
\widetilde M^{\rm tw3}_{[\alpha\beta]} (x,-x)
 =3 x^\gamma 
\bigg\{\delta^\mu_{[\alpha}\delta^\nu_{\beta}\pd_{\gamma]} x\pd
-  2\, \delta^\mu_{[\alpha}x_{\beta}\pd_{\gamma]}\pd^{\nu}\bigg\}
&\int\d^4 p\, 
\big(\bar{\psi}\sigma_{\mu\nu}\psi\big)(p)\,
(1+p\pd_p)\nonumber\\
&\times\int_0^1 \d t\, \frac{1-t^2}{2t}\,
{\cal H}_1(p|tx).
\end{align}

Let me note that we can use the formulas~(\ref{XO2}) and (\ref {XO3}) 
to calculate the target mass corrections of
the Compton amplitude in the $q$-space. In order to do this one has to use
the operators $O^{\rm tw2}_{(5)\alpha n}(q)$ and $O^{\rm tw3}_{(5)\alpha n}(q)$
which are multiplied by the coefficient functions in the $q$-space. 
In the forward case one finds 
the target mass corrections for the Bjorken sum rules and the Nachtmann 
moments (see Refs.~\cite{BE76,Wandzura77,MU80}). 
Additionally, one is able to
calculate the power corrections for the non-forward case, too.

Is also possible to rewrite all harmonic operators in terms of Gegenbauer 
polynomials and Bessel functions for $2h$-dimensional spacetime.

\newpage
\chapter{Conclusions and Outlook}

A unique group theoretical procedure was introduced which allows a
decomposition of nonlocal tensor-valued 
light-ray operators and off-cone operators into (tensorial) operators and 
harmonic operators, respectively, with definite geometric twist. 
This decomposition is completely of group theoretical origin and is
equivalent to the decomposition of local operators with respect to
irreducible tensor representations of the Lorentz group.
The explicite
twist decomposition in $1+3$-spacetime dimensions
has been given for all physically relevant bilocal 
quark and gluon tensor operators which appear, e.g., in the virtual
Compton scattering in the generalized Bjorken region 
as well as for multilocal
operators containing many quark and gluon fields, e.g., Shuryak-Vainshtein type
operators, four-quark and four gluon operators, and three-gluon
operators.
In the case of light-ray operators 
the twist projectors are given in terms of 
interior differential operators, which were introduced by Bargmann and
Todorov~\cite{BT77}. 
By the help of these interior differential operators the conditions of 
tracelessness can be easily formulated and satisfied. 
In the case of off-cone operators the local operators are given in terms of
Gegenbauer polynomials and the nonlocal ones are expressed in terms of Bessel
functions.
There are one-to-one relations between the light-ray and off-cone operators
with definite geometric twist. The light-ray operators are obtained 
through the projection of the off-cone operators
onto the light-cone.
In the other direction, the off-cone operators
can be understood as the harmonic extension of the light-ray operators.
Because we discussed the relations for arbitrary $2h$-spacetime 
dimensions one can also construct twist operators in higher or 
lower dimensions in straightforward manner. Perhaps this fact can
be useful in order to renormalize the QCD operators by the help
of dimensional regularization. 
Another interesting fact, from the mathematical point of view, 
may be the generalization of the twist 
decomposition for more complicated symmetry types. Perhaps such 
(higher) twist operators will be needed from the phenomenological point 
of view in future. 
Additionally,
one can use the polynomial technique to construct a general mathematical
theory of $SO(2h)$ or $SO(1,2h-1)$ irreducible tensor polynomials which 
do not exist in the mathematical literature. Then one will be able to compute the 
infinite twist decomposition of any multilocal tensor operator.

We calculated the forward and vacuum-to-vector meson matrix elements 
of bilocal light-ray operators with definite geometric twist. In this doing
we were able to classify the corresponding quark distribution and 
vector meson distribution amplitudes with respect to definite geometric twist
in a unique manner.
Concerning higher twists we introduced new parton distribution functions 
and $\rho$-meson distribution amplitudes, as wells as their moments, 
having definite geometric twist. We were able to derive one-to-one relations between
these new distribution amplitudes and the conventional ones with dynamical 
twist, which were earlier introduced by Jaffe and Ji~\cite{JJ92} and 
by Ball {\it et al.}~\cite{ball98}. Thereby, Wandzura-Wilczek type relations
and Burkhardt-Cottingham type sum rules for the dynamical distribution
amplitudes, which should be tested experimentally, have been obtained for
higher twists.

By calculating the (general) non-forward matrix elements of off-cone
operators with definite geometric twist we introduced for the first time
all geometric twist-2 and twist-3 double distribution amplitudes 
which are used from the phenomenological point of view. 
In the case of operators whose moments are related to totally symmetric 
tensors we were able to determine the complete twist decomposition 
up to infinite geometric twist. Using these bilocal operators we gave the 
classification of the (pseudo) scalar meson distribution amplitudes 
for all twists. 
An important result of our approach is that we were able to separate
all kinematical factors like $x^2$, $xP$, and $xS$ from the the distribution
amplitudes and to resum the power corrections to all orders. 
This is based on the group-theoretical content of the procedure of
trace subtraction from the local operators with definite symmetry type.
Because the distribution amplitudes of definite geometric twist 
are already determined by the nonlocal matrix elements on the light-cone 
their renormalization properties are determined by the light-cone operators
only. 
The nonlocal matrix elements off-cone contain 
supplementary kinematical structures which 
correspond to the structure of the harmonic extension of the matrix 
element on-cone.

Additionally, by means of these harmonic QCD operators one is able to 
calculate target and quark mass corrections in the Compton amplitude
of non-forward scattering processes, like deeply virtual Compton scattering 
and diffractive scattering. Hopefully, this will be useful for predictions 
of target mass effects in physical cross sections of these processes.
\\ \\ \\ 

{\Large{\bf{Acknowledgments}}}
\\ \\ \\
The author is grateful to B.~Geyer, 
P.~Ball, J.~Bl\"umlein, V.K.~Dobrev, J.~Eilers,
D.~M\"uller, S.~Neumeier, D.~Robaschik and C.~Weiss 
for various stimulating and useful discussions.
The author thanks J.~Eilers for the possibility to compare the decomposition 
of the 2nd rank tensor operator with his result 
which was very useful to find some mistakes. 
In addition, he gratefully acknowledges the Graduate College
``Quantum field theory'' at the Center for Theoretical Studies of Leipzig
University for financial support.

\newpage
\begin{appendix}
\chapter{Symmetry classes and tensor representations }
\renewcommand{\theequation}{\thechapter.\arabic{equation}}
\setcounter{equation}{0}
\label{young}

In this Appendix we collect some facts about tensor representations
of the classical matrix groups $G$ and their relation to the
symmetric group $S_n$ which are of relevance in our 
consideration (see, e.g.,~\cite{BR, Ham62}). 
Given any irreducible matrix representation, 
\begin{eqnarray}
G \ni g \mapsto {\hat g} \in {\cal B}(V): 
\qquad {\hat g} e_i = e_j\, {a^j}_i,
\end{eqnarray}
on some vector space $V$ (of dimension ${\rm dim}\,V$)
with base $\{e_i,\, i = 1,2,\ldots,{\rm dim}\,V \}$. 
The direct products
of these representations act on the {\em tensor space}
$T^n\, V$ which is build as tensor product of $n$ 
copies of the vector space $V$:
\begin{eqnarray}
T^n\,V
&=&
\left(V \otimes\ldots\otimes V\right).
\nonumber
\end{eqnarray}
The elements of the tensor space $T^n\,V$ are the $n$--times
contravariant 
{\it tensors}
${\bf t} = t^{i_1\ldots i_n} e_{i_1\ldots i_n}$, 
where $t^{i_1\ldots i_n}$ 
are the {\em components} of ${\bf t}$ in the base 
${e_{i_1\ldots i_n}}
\equiv
\big(e_{i_1}\otimes\ldots\otimes e_{i_n}\big),\;
i_k \in {[1,\ldots , {\rm dim\,}V ]}$.~\footnote{
Here, neither covariant tensors which 
act on direct products of the dual space $V^*$ nor mixed
ones are considered since, for the case under consideration, 
the indices may be raised or lowered by the metric tensor.
}
The components of the tensors of rank $n$ transform under
$G$ according to
\begin{eqnarray}
\label{tensortrf}
(t')^{i_1\ldots i_n}
\equiv
t^{i'_1\ldots i'_n}
=
{a^{i'_1}}_{i_1}
\ldots 
{a^{i'_n}}_{i_n}
t^{i_1\ldots i_n}.
\end{eqnarray}

In the following we shall comment on the connection between the
irreducible representations of the general linear group 
$GL(N, {\Bbb C})$ (together with
its subgroups) and of the symmetric group $S_n$ on 
these tensor spaces.
Let us, at the moment consider only the groups 
$G = GL(N,{\Bbb C}),SL(N,{\Bbb C}), SU(N)$.
Given any (irreducible) representation of $G$ on  
$V$, then its $n$--fold tensor product defines a reducible
representation $A^{(n)}$ which, together with the 
reducible representation $\hat \pi$ of $S_n$, is 
determined by:
\begin{eqnarray}
{\hat A}^{(n)}(g){\bf t}
\equiv \;
{\hat A}^{(n)}(g)(t^{i_1\ldots i_n}e_{i_1\ldots i_n})
&:=&
t^{i_1\ldots i_n}
({\hat g}e)_{i_1}\otimes\ldots\otimes ({\hat g}e)_{i_n},
\quad\forall g \in G,
\nonumber
\\
{\hat \pi}{\bf t}
\equiv 
~~\qquad
{\hat \pi}(t^{i_1\ldots i_n}e_{i_1\ldots i_n})
&:=&
t^{\pi(i_1\ldots i_n)}
e_{i_1}\otimes\ldots\otimes e_{i_n},
\quad \forall \pi \in S_n;
\nonumber
\end{eqnarray}
therefore the action of both groups on the tensors 
$\bf t$ commutes~\footnote{
This results from the fact that the product of the 
matrix elements $a^j_i$ in Eq.~(\ref{tensortrf})
is {\em bisymmetric} under $S_n$.
}
\begin{eqnarray}
\label{Schur}
\big[ {\hat A}^{(n)}(g), {\hat \pi} \big] {\bf t} = 0.
\end{eqnarray}
Of course, the same conclusion holds for the elements 
$\sum_\pi \alpha(\pi) {\hat \pi}, \alpha(\pi) \in {\Bbb R}$, 
of the group algebra ${\cal R}[S_n]$ of the symmetric group. 
This algebra, considered as a vector space, carries the 
{\em regular representation} of $S_n$ which is known to be
fully reducible. 

The irreducible representations $\Delta^{[m]}$ of the 
symmetric group are uniquely determined by the idempotent 
(normalized) Young operators
\begin{eqnarray}
\label{Young1}
{\cal Y}_{[m]} &=& \frac{f_{[m]}}{n!}{\cal P}{\cal Q}
\quad {\rm with} \quad
{\cal P} =\!\!\! \sum_{p \in H_{[m]}} p,
\quad 
{\cal Q} ~= \sum_{q \in V_{[m]}} \delta_q\,q,
\\
\label{Young2}
{\cal Y}_{[m]}{\cal Y}_{[m']}
&=&
\delta_{[m][m']}{\cal Y}_{[m]},
\end{eqnarray}
which are related to corresponding Young tableaux
being  denoted by $[m]$.
If
\begin{eqnarray}
{\underline m} = (m_1, m_2, \ldots m_r) 
\quad
{\rm with}
\quad
m_1 \geq m_2\geq \ldots \geq m_r,
\quad
\sum^r_{i=1}\, m_i = n,
\end{eqnarray}
defines a Young pattern (Fig.~\ref{Y-Rahmen}) then a {\em 
Young tableau} $[m]$ is obtained by putting in (without
repetition) 
the indices $i_1, \ldots i_n$ -- corresponding to different
``places''
within the direct product -- and $H_{[m]}$ and $V_{[m]}$ denotes 
their horizontal and vertical permutations with respect to
$[m]$.
A {\em standard tableau} is obtained when the indices
$i_1, \ldots, i_n$ are ordered lexicographically. There are
\begin{eqnarray}
\label{f}
f_{[m]} = n! \, 
\frac{\prod_{i<j} (l_i - l_j)}{\prod_{i=1}^r l_i!}
\quad
{\rm with}
\quad
l_i = m_i + r - i,
\quad
\sum_{[m]}f_{[m]}^2 = n!,
\end{eqnarray}
different standard tableaux which correspond to $f_{[m]}$ 
different, but equivalent, irreducible representations of $S_n$
whose dimension is given also by $f_{[m]}$.
The (normalized) Young operators ${\cal Y}_{[m]}$ 
according to (\ref{Young2}) project onto 
mutually orthogonal irreducible left ideals of ${\cal R}[S_n]$.
\begin{figure}[h]
\unitlength0.5cm
\begin{center}
\begin{picture}(15,5)
\linethickness{0.15mm}
\multiput(3,4)(1,0){5}{\line(0,1){1}}
\multiput(1,-1)(1,0){2}{\line(0,1){6}}
\multiput(3,1)(3,0){1}{\line(0,1){4}}
\multiput(4,3)(4,0){1}{\line(0,1){2}}
\put(1,5){\line(1,0){6}}
\put(1,4){\line(1,0){6}}
\put(1,3){\line(1,0){3}}
\put(1,2){\line(1,0){2}}
\put(1,1){\line(1,0){2}}
\put(1,0){\line(1,0){1}}
\put(1,-1){\line(1,0){1}}
\put(10,4){$m_1$ boxes}
\put(10,3){$m_2$ boxes}
\put(10,1){$\vdots$ }
\put(10,-1){$m_r$ boxes}
\end{picture}
\end{center}
\caption{\label{Y-Rahmen} Young pattern $\underline m$} 
\end{figure}
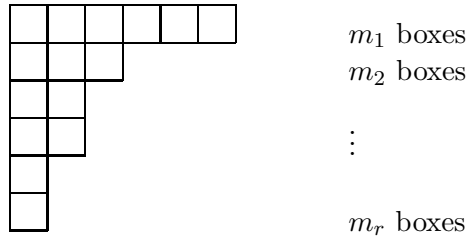

In this way the tensor product of $n$ spaces $V$ carries the 
regular representation of the group $S_n$ and decomposes into 
$\sum_{[m]} f_{[m]}$ irreducible subspaces containing tensors 
of symmetry class $[m]$. However, the tensors actually are
specified by taking for the indices $i_1, \ldots, i_n$ 
arbitrary values from the range $[1, \ldots, {\rm dim}\,V]$.
Therefore, with respect to the group $G$ the standard tableau 
is defined by putting into the Young pattern the {\em values of
the indices} $i_k$ such that they are non--decreasing from
left to right and increasing from top to bottom!
Because of Eq.~(\ref{Schur}) by Schur's Lemma it follows that
the representation ${\hat A}^{(n)}(g)$ of $G$ is reducible and 
it decomposes into as many irreducible representations as there
are (in general reducible) representations of $S_n$, i.e., any
tensor
representation of the group $G$ is characterized by a
standard tableau,
\begin{eqnarray}
{\hat A}^{(n)}(g) 
=
\underset{[m]}{\oplus}
R^{[m]}(g) \otimes E_{l_{[m]}},
\quad
{\hat \pi}
=
\underset{[m]}{\oplus}
 E_{r_{[m]}} \otimes \Delta^{[m]}(\pi),
\end{eqnarray}
where $E_{l_{[m]}}$ is the unit matrix; 
 more explicitly we have
\begin{eqnarray}
\label{tenrep}
{\hat A}^{(n)}(g)_{[m] \kappa \alpha, [m'] \kappa' \alpha'}
&=&
\delta_{[m][m']} 
\delta_{\alpha \alpha'}
R^{[m]}_{\kappa \kappa'}(g),
\end{eqnarray}
where $\kappa = 1, \ldots , r_{[m]}$ counts different
equivalent irreducible representations of $S_n$ and 
$ \alpha = 1, \ldots , l_{[m]} \leq f_{[m]}$ counts the 
basis elements of these representations. The  
representations $R^{[m]}(g)$ of $G$ in $T^n\,V$ are 
characterized by equivalent Young operators 
${\widehat{\cal Y}}_{[m]} = {\cal Q}{\cal P}$:~\footnote{
Note, that the symmetrizations $\cal P$ and the
antisymmetrizations $\cal Q$ with respect to $[m]$
are interchanged. This corresponds, in the terminology
of quantum mechanics, to ``quantum number permutations''
instead of the ``place permutations'' above. And, contrary
to ${\cal Y}_{[m]}$ which defines a left ideal in $\cal R$,
${\widehat{\cal Y}}_{[m]}$ defines a right ideal.
}
\begin{equation}
\label{dsum}
T^n\,V
~=~
\underset{[m]}{\oplus}
\underset{\kappa}{\oplus}
{\widehat{\cal Y}}_{[m]\kappa}(T^n\,V).
\end{equation}
The invariant subspaces projected out by
${\widehat{\cal Y}}_{[m]\kappa}(T^n\,V)$ are irreducible with 
respect to $GL(N,{\Bbb C})$, $SL(N,{\Bbb C})$ and $SU(N)$. 
After restriction onto these subgroups some of 
the irreducible representations become equivalent ones. 
 However, for $O(N,{\Bbb C}), SO(N,{\Bbb C})$ and
$Sp(2\nu)$ these representations, in general, are not
irreducible and decompose further.  

Concerning the orthogonal groups from the subspaces of
the above introduced symmetry classes only the 
completely antisymmetric ones (for $n \leq N$) remain 
irreducible. The reason is that, because of the very definition 
of the orthogonal groups,
$\delta_{ij} {a^i}_k {a^j}_l 
~=~ \delta_{kl}, \;\forall a \in O(N)$, 
the operation of taking the trace of a tensor 
commutes with the orthogonal transformations of that tensor:
\begin{eqnarray}
\label{decomp}
{\rm tr}\; {\bf T}'
~=~ \delta_{ij} {T'}^{ij}
~=~ \delta_{ij} {a^i}_k {a^j}_l T^{kl}
~=~ {\rm tr}\; {\bf T}.
\end{eqnarray}
Again, by Schur's Lemma, irreducible subspaces 
of $O(N)$ are spanned by {\em traceless}
tensors having definite symmetry class.
This decomposition is obtained as follows:
\begin{eqnarray}
\label{TL}
T^{i_1 i_2 \ldots i_n}_{[m]}
&=&
\TL {T_{[m]}^{i_1 i_2 \ldots i_n}}
+ \sum\limits_{1 \leq r,s \leq n} \delta^{i_r i_s}\, 
T^{i_1 \ldots i_{r-1}i_{r+1}\ldots 
i_{s-1}i_{s+1}\ldots i_n}_{[m-2]}.
\end{eqnarray}
The tensors which appear under the sum have degree $n-2$ and a
 Young pattern ${[m-2]}\subset {[m]}$ obtained by removing 
two boxes from the (right) border without destroying the property 
(\ref{even}) and (\ref{odd}) to be a pattern. They
 may be decomposed again into traceless ones plus some
remainder, and so on. Therefore, a traceless tensor is obtained
from the original one by successively subtracting the traces.

\chapter{Irreducible tensor representations of
the Lorentz group}
\renewcommand{\theequation}{\thechapter.\arabic{equation}}
\setcounter{equation}{0}
\label{lorentz}

In this Appendix the group theoretical background will be given
which is used for the construction of the  nonlocal 
operators whose local parts -- which are obtained by Taylor expansion --
transform according to irreducible tensor representations
 under the Lorentz group,
i.e.,~have definite spin and therefore also well defined
twist. These tensor representations are characterized by a 
specific symmetry class with respect to the symmetric group
and are determined by a few types of Young tableaux.

The Lie algebra of the Lorentz group ${\cal L}^\uparrow_+ $ 
is characterized by the generators of the three spatial
rotations $\boldsymbol{M}$ and the three boosts $\boldsymbol{N}$; from it the 
two (complex) linear combinations ${\boldsymbol{M}}_\pm$ may be build
which define two independent $SO(3)$ groups:
\begin{eqnarray}
[M_\pm^i, M_\pm^j ] =\ii \epsilon^{ijk} M_\pm^k,
\qquad
[ M_+^i, M_-^j ] ~=~ 0
\quad {\rm with} \quad
{\boldsymbol  M}_\pm = {\boldsymbol  M} \pm \ii {\boldsymbol  N}.
\nonumber
\end{eqnarray}
Therefore, the (Lie algebra of the) complex Lorentz group
is isomorphic to the (Lie algebra of the) direct product
$SO(3, {\Bbb C}) \otimes SO(3,{\Bbb C)} \simeq SO(4,{\Bbb C})$.
This characterization makes use of the fact that the irreducible 
(finite) representations 
\begin{eqnarray}
{\cal D}^{(j_+,j_-)}({\boldsymbol  \varphi}, {\boldsymbol  \vartheta})
~=~ 
{\cal D}^{(j_+)}({\boldsymbol  \varphi} - \ii {\boldsymbol  \vartheta}) 
\otimes
{\cal D}^{(j_-)}({\boldsymbol  \varphi} + \ii {\boldsymbol  \vartheta})
\end{eqnarray} 
of the restricted orthochronous Lorentz group 
${\cal L}^\uparrow_+ \simeq SO(1,3;{\Bbb R})$, 
where $\boldsymbol  \varphi$ and $\boldsymbol  \vartheta$ are the angle of rotation 
and the rapidity of the boost transformations, 
are determined by two numbers 
$(j_+, j_-), j_\pm = 0,1/2, 1, \cdots $, which define the
spin $j = j_+ + j_-$ of the representation.

Let us now state some general results concerning finite
dimensional representations of the (complex)
orthogonal groups (see, e.g.,~Chapters 8 and 10 of
Ref.~\cite{BR}):

\noindent $(1)~~$
The group $SO(N, {\Bbb C})$ has two series of 
complex--analytic\footnote{
A representation is complex--analytic if it depends
analytically on the group parameters.}
irreducible representations. Every representation of the first
(resp. the second) series determines and is in turn determined 
by a highest weight ${\underline m} = (m_1, m_2, \ldots ,
m_\nu)$ 
whose components $m_i$ are
integers (resp. half--odd integers) and satisfy the conditions
\begin{eqnarray}
\label{even}
m_1 \geq m_2 \geq \ldots \geq m_{\nu -1} \geq|m_\nu|
&{\rm for}& N = 2\nu,
\\
\label{odd}
m_1 \geq m_2 \geq \ldots \geq m_{\nu -1} \geq m_\nu \geq 0
&{\rm for}& N = 2\nu +1.
\end{eqnarray}
The first series determines the tensor representations,
 whereas the second series determines the spinor
representations.

\noindent$(2)~~$
A tensor representation of $SO(N; {\Bbb C})$, for either
$N = 2\nu$ or $N = 2\nu +1$, being determined
by the highest weight ${\underline m} = 
(m_1, m_2, \ldots , m_\nu), \;m_i$ integer, is 
equivalent to another tensor representation
$T_{i_1 i_2 \ldots i_n}, \; n = \sum m_i,$ which is
realized in the space of traceless tensors whose 
symmetry class is characterized by a 
Young pattern defined by the partition 
$[m]=  (m_1, m_2, \ldots , m_\nu)$. 

\noindent$(3)~~$
Furthermore, if ${\cal T}_{G_c}$ is a complex--analytic
representation of a (complex) semi-simple Lie group $G_c$
and ${\cal T}_{G_r}$ is the restriction of ${\cal T}_{G_c}$
to a real form $G_r$ of $G_c$, then, ${\cal T}_{G_c}$
is irreducible (fully reducible) iff ${\cal T}_{G_r}$ is
irreducible (fully reducible).
%
As a consequence,
any irreducible representation of the complex group remains
irreducible if restricted to a real subgroup and, 
on the other hand,
from any irreducible representation of the real group by 
analytic continuation in the group parameters 
an irreducible representation of its complexification is
obtained.

Therefore, in order to investigate irreducible tensor 
representations of the Lorentz group we may consider equally
well irreducible tensor representations of the complex
group $SO(4,{\Bbb C})$. Even more, any irreducible 
representation of the orthogonal group $SO(4)$ by analytic 
continuation induces an irreducible representation of 
$SO(4,{\Bbb C})$, which by restriction to their real subgroup 
${\cal L}^\uparrow_+$ subduces
an irreducible representation of the Lorentz group. Since
tensor representations of the orthogonal group are uniquely
determined by the symmetry class $[m]$ of their tensors 
we may study the irreducible tensor representations of 
the Lorentz group through a study of the corresponding 
Young tableaux which determine the irreducible representations
of the symmetric group $S_n$ of permutations.

This characterization of tensor representations through their 
symmetry class holds for any of the classical matrix groups, 
$GL(N,{\Bbb C})$ and their various subgroups 
 (see also Appendix \ref{young}).
However, for $O(N,{\Bbb C})$ and $SO(N,{\Bbb C})$ these 
representations, in general, are not irreducible. 
The reason is that taking the trace of a tensor 
commutes with the orthogonal transformations.
Therefore, by Schur's Lemma, irreducible subspaces 
of $SO(N,{\Bbb C})$ are spanned by {\em traceless}
tensors having definite symmetry class.

From the requirement of tracelessness the following 
restrictions obtain (see, e.g.,~\cite{Ham62}):
The only Young tableaux being relevant are those whose first
two columns are restricted to have length 
$\mu_1 + \mu_2 \leq N$ (see (i) -- (iv) below). 
Two representations $R$ and $R'$ 
whose first columns are related by $\mu'_1 = N - \mu_1$, 
where $\mu_1 \leq N/2$, are called associated; 
if $\mu_1 = \mu'_1 = N/2$ this representation
is called selfassociated. After restriction to the subgroup
$SO(N) \subset O(N)$ associated representations are equivalent,
whereas selfassociated representations decompose into two
nonequivalent irreducible representations. 

Let us illustrate this for the Lorentz group by the 
representations which will be of interest in the following.
First of all, vector and axial vector representations 
$V_\mu$ and $A_\mu = \epsilon_{\mu\nu\kappa\lambda}
A^{\nu\kappa\lambda}$, respectively, are associated ones, 
and antisymmetric tensor representations,
$A_{\mu\nu} = T_{\mu\nu} - T_{\nu\mu}$, are selfassociated
which, if restricted to $SO(1,3)$ decompose
into the (anti-)selfdual tensors
$A^\pm_{\mu\nu} = \frac{1}{2}( A_{\mu\nu} \mp \frac{1}{2} \, 
\ii \epsilon_{\mu\nu\kappa\lambda} A^{\kappa\lambda})$.
Since later on only representations of
the orthochronous Lorentz group containing the parity operation
are considered, this distinction is of no relevance.
Therefore, any tensor of second order, $T_{\mu\nu}$, may be 
decomposed according to
\begin{eqnarray}
\label{2tensor}
T_{\mu\nu}
&=&
\underbrace{
\hbox{\large $\frac{1}{2}$}
\left(T_{\mu\nu}+T_{\nu\mu}\right)
-
\hbox{\large $\frac{1}{4}$}
g_{\mu\nu}\, T^{\ \rho}_\rho }_{S_{\mu\nu}} 
+ 
\underbrace{
\hbox{\large $\frac{1}{2}$}
\left(T_{\mu\nu}-T_{\nu\mu}\right)  }_{A_{\mu\nu}}
+
\hbox{\large $\frac{1}{4}$}\,
g_{\mu\nu} T^{\ \rho}_\rho .
\end{eqnarray}
Let us denote by ${\bf T}(j_+, j_-),\, j_+ + j_-$ integer, 
the space of tensors
which carry an irreducible representation ${\cal D}^{(j_+,j_-)}$
of the Lorentz group. Then from Eq.~(\ref{2tensor}) we read off:
$S_{\mu\nu} \in {\bf T}(1,1)$ is a symmetric traceless tensor, 
$A_{\mu\nu} = A^+_{\mu\nu} + A^-_{\mu\nu} \in 
{\bf T}(1,0) \oplus {\bf T}(0,1)$ are the selfdual
and the anti-selfdual antisymmetric tensors, and
$T^{\ \rho}_\rho \in {\bf T}(0,0)$ corresponds to the trivial representation defined
through the unit tensor. This decomposition corresponds to the
Clebsch-Gordan decomposition of 
the direct product of two vector representations:
\begin{eqnarray}
\Big(
\hbox{\large $\frac{1}{2}$}, \hbox{\large $\frac{1}{2}$}
\Big)\otimes\Big(
\hbox{\large $\frac{1}{2}$}, \hbox{\large $\frac{1}{2}$}
\Big)
&=&
(1,1) \oplus\Big( (1,0)\oplus(0,1) \Big) \oplus (0,0) .
\end{eqnarray}

Now we consider the decomposition of the space of tensors of
rank $n$ into irreducible representation spaces of 
$SO(1,3; {\Bbb R})$. The different symmetry classes are strongly
restricted by the requirement that only such Young patterns
$[m]$
are allowed for which the sum of the first two columns is lower
or equal to four. Therefore, only the following Young patterns
correspond to nontrivial irreducible representations
by traceless tensors:
\begin{enumerate}
\item[i.] \unitlength0.4cm
\begin{picture}(30,1)
\linethickness{0.1mm}
\multiput(1,0)(1,0){13}{\line(0,1){1}}
\put(1,1){\line(1,0){12}}
\put(1,0){\line(1,0){12}}
\put(15,0){$j=n,n-2,n-4,\ldots$}
\end{picture}
\item[ii.]\unitlength0.4cm
\begin{picture}(5,1)
\linethickness{0.1mm}
\multiput(3,0)(1,0){10}{\line(0,1){1}}
\multiput(1,-1)(1,0){2}{\line(0,1){2}}
\put(1,1){\line(1,0){11}}
\put(1,0){\line(1,0){11}}
\put(1,-1){\line(1,0){1}}
\put(15,0){$j=n-1,n-2,n-3,\ldots$}
\end{picture}
\item[iii.] \unitlength0.4cm
\begin{picture}(5,2)
\linethickness{0.1mm}
\multiput(3,0)(1,0){9}{\line(0,1){1}}
\multiput(1,-2)(1,0){2}{\line(0,1){3}}
\put(1,1){\line(1,0){10}}
\put(1,0){\line(1,0){10}}
\put(1,-1){\line(1,0){1}}
\put(1,-2){\line(1,0){1}}
\put(15,0){$j=n-2,n-3,\ldots$}
\end{picture}
\item[iv.] \unitlength0.4cm
\begin{picture}(5,3)
\linethickness{0.1mm}
\multiput(4,0)(1,0){8}{\line(0,1){1}}
\multiput(1,-1)(1,0){3}{\line(0,1){2}}
\put(1,1){\line(1,0){10}}
\put(1,0){\line(1,0){10}}
\put(1,-1){\line(1,0){2}}
\put(15,0){$j=n-2,n-3,\ldots$}
\end{picture}
\end{enumerate}
\vspace*{3mm}
In addition, for $n = 4$, also the completely antisymmetric 
tensor of rank 4 which is proportional to 
$\epsilon_{\mu\nu\kappa\lambda}$, and therefore equivalent to
the trivial representation is allowed. 

For the cases (i) -- (iv)
the minimal spin $j$ -- depending on $n$ being either even or
odd --
will be zero or one. In the case of symmetry type (iv) we have
given only one special Young pattern; in principle the length
of the second row  may contain up to 
$m_2 = [\frac{n}{2}]$ boxes, and then the maximal spin is
given by $j = n - m_2$. ---
The representations corresponding to symmetry class (i) are
associated to representations of the symmetry class (iii) with
$n+2$ boxes. The representations corresponding to symmetry class 
(ii) and (iv) are selfassociated. The symmetry class (ii) 
contains two non-equivalent parts being related to  
$(\frac{n}{2},\frac{n}{2}-1)$ and 
$(\frac{n}{2}-1,\frac{n}{2})$; the symmetry class (iv) 
contains three nonequivalent parts related to
$(\frac{n}{2},\frac{n}{2}-2)$, $(\frac{n}{2}-1,\frac{n}{2}-1)$ 
and $(\frac{n}{2}-2,\frac{n}{2})$; and so on.
Any tensor whose symmetry class does not coincide with one 
of the above classes vanishes identically due to the requirement
of tracelessness.

There are two possible ways to construct the non-vanishing 
tensors. Either one symmetrizes the indices according to the
corresponding (standard) Young tableaux and afterwards subtracts 
the traces, or one starts from tensors being already traceless 
and finally symmetrizes because this does not destroy the
tracelessness. For practical reasons the latter procedure seems
to be preferable and will be used in the construction of
irreducible light--cone operators of definite twist. 

\chapter{The interior derivative on the light-cone and light-cone operators 
with definite geometric twist}
\label{inner}
\renewcommand{\theequation}{\thechapter.\arabic{equation}}
\setcounter{equation}{0}

In this Appendix we would like to discuss some properties of the so-called 
internal derivatives on the light--cone.

Let us now introduce the notion of {\it interior differential operators} 
on the complex light-cone 
\begin{align}
K_{2h}({\Bbb C})=\left\{z\in{\Bbb C}^{2h};\ z^2=z_1^2+\ldots+z_{2h}^2=0\right\},
\end{align}
and on the real light-cone
\begin{align}
K_{2h}({\Bbb R})=\left\{\lcx\in{\Bbb M}^{2h};\ \lcx^2=\lcx_1^2-\lcx_2^2-\ldots-\lcx_{2h}^2=0\right\},
\end{align}
respectively. The $2h$-dimensional Minkowski space ${\Bbb M}^{2h}$ has
the signature $(+-\cdots-)$.

For the first time, an interior differential operator has been used in order 
to characterize (irreducible)
symmetric tensor representations $U^n$ of $SO(2h)$ on the complex light-cone. 
In this turn one may consider the {\it graded algebra}
\begin{align} 
P\equiv P(K^n_{2h})=\bigoplus_{n=0}^\infty K^n_{2h}, 
\end{align}
where 
$K^n_{2h}$ is the space of homogeneous polynomials $T_n (\lcx)$ of degree $n$ 
on the cone~\cite{BT77,Dobrev77}. 
The reducible representation
\begin{align}
U(\Lambda) f(z)=f(\Lambda^{-1} z),\qquad\Lambda\in SO(2h),\quad f\in P,
\end{align}
of $SO(2h)$ generates all symmetric tensor representations $U^n$ 
\begin{align}
U=\bigoplus_{n=0}^\infty U^n,
\end{align}
which are given by the restriction of $U$ to $K^n_{2h}$~\cite{BT77}.

The real light-cone provides a 
convenient realization of the carrier space for the symmetric tensor
representations of the group $SO(1,2h-1)$. In the following we consider
the real light-cone.

A differential operator $Q$ is said to be an {\em interior} differential 
operator iff 
\begin{equation}
Q\big(x^2 T_n (x)\big){\big|}_{x^2\equiv \lcx^2=0}=0.
\end{equation}
For example, the generators of dilation $X=h-1+\lcx\lcd$ 
and rotations $X_{\mu\nu}=\lcx_\nu\lcd_\mu-\lcx_\mu\lcd_\nu$
are first order interior differential operators on the light-cone.
Obviously, they leave the space $K^n_{2h}$ {\em invariant}. 
The interior operators on $P$ form a (complex) algebra under addition and 
multiplication~\cite{BT77}.

A further 
interior differential operator $\d_\mu$ of second order may be introduced
by the following requirements:\\
(a) it should be a {\em lowering operator}, i.e.~mapping a homogeneous 
polynomial of degree $n$ (of $\lcx$) into a homogeneous polynomial of 
degree $n-1$, $K_{2h}^{n}\rightarrow K_{2h}^{n-1}$;\\
(b) it should behave as a vector under rotations,
\begin{equation}
\label{lie_1}
[X_{\mu\nu},\d_\lambda]=\delta_{\mu\lambda}\d_\nu-\delta_{\nu\lambda}\d_\mu;
\end{equation}
(c) it should be the lowest order differential operator satisfying 
(a) and (b).\\
Choosing the normalization of this interior derivative 
so that $2\d_\mu$ is the generator
of {\it special conformal transformations} of 
{\it massless} 0-helicity representations of the conformal Lie algebra 
${\mathfrak{so}}(2,2h)$, the interior derivative is given as
\begin{align}
\d_\mu f(\lcx)\equiv
\Big\{\big(h-1+x\pd\big)\pd_\mu
-\hbox{\large$\frac{1}{2}$}
x_\mu\square\Big\}f(x)\big|_{x=\lcx},
\end{align}
with the following properties
\begin{align}
\d^2 = 0\, ,\quad
[\d_\mu,\d_\nu]=0
\quad\text{and}\quad
\d_\mu\lcx^2=\lcx^2\big(\d_\mu+2\lcd_\mu\big)\,.
\end{align}
Obviously, $\d_\mu$ is a second order interior differential operator
which together with $\lcx_\mu$, $X$ and $X_{\mu\nu}$ satisfies 
the conformal {\em Lie algebra} ${\mathfrak{so}}(2,2h)$ in $\lcx$-space;
especially there hold the commutator relations
\begin{align}
\label{lie_3}
[\d_\mu,\lcx_\nu]=\delta_{\mu\nu} X+X_{\mu\nu},\qquad
[\d_\mu,X]=\d_\mu.
\end{align}
In that terminology $\lcx$ is a 
{\em raising operator} which plays the role of ``momentum''
in the conformal algebra:
\begin{align}
\label{lie_2}
[\lcx_\mu,\lcx_\nu]=0,\qquad  \lcx^2=0.
\end{align}

Dobrev {\it et al.}~\cite{Dobrev76a,Dobrev77,Karchev83,Craigie85} used 
the interior derivative in conformal OPE
in order to construct local composite tensor operators which 
transform under an elementary
representations of the conformal group $SO(2,2h)$. They formulated the 
condition for conformal invariance by means of the interior derivative 
on the light-cone.

For the first time, the application of the interior derivative in order to
satisfy the conditions of tracelessness on the light-cone, and for
antisymmetric tensors was given by Dobrev and Ganchev~\cite{Dobrev82}.

Additionally, the interior derivative has been used to formulate 
a conformal ``Lorentz condition'' in Conformal Quantum 
Electrodynamics~\cite{Todorov85,Furlan85}
and in manifestly $O(2,4)$-covariant formalism for the photon-Weyl graviton
system~\cite{Furlan87}.

Finally, I will show how one can use the interior derivatives for
the decomposition of the local light-cone operator 
(see also Refs.~\cite{Dobrev82,GL99c,GL01}),
\begin{align}
\label{O-2h}
O_{\alpha n}(y,\lcx)=\bar\psi(y)\gamma_{\alpha}(\lcx \Tensor{D})^n \psi(y),
\end{align}
into all its twist parts.
The operator~(\ref{O-2h}) has the canonical dimension $d_{O_{\alpha n}}=2h+n-1$.
We may uniquely decompose this light-cone operator according to
\begin{align}
O_{\alpha n}(y,\lcx)=O^{{\rm tw}(2h-2)}_{\alpha n}(y,\lcx)+
		     O^{{\rm tw}(2h-1)}_{\alpha n}(y,\lcx)+	
	             \lcx_\alpha\, O^{{\rm tw}(2h)}_{n-1}(y,\lcx).
\end{align}
Using the Eq.~(\ref{T^[n]_alpha(lcx)}) we get the
operator of maximal spin, $j=n+1$, and minimal twist $\tau=2h-2$
\begin{align}
\label{O-(2h-2)}
O^{{\rm tw}(2h-2)}_{\alpha n}(y,\lcx)&=\frac{1}{(n+1)(h+n-1)}\,\d_\alpha O_{n+1}(y,\lcx).
\end{align}
The operator of twist $\tau=2h-1$ and spin $j=n$ is obtained from
Eq.~(\ref{[n,1](n-1)lc}) as follows
\begin{align}
\label{O-(2h-1)}
O^{{\rm tw}(2h-1)}_{\alpha n}(y,\lcx)&=\frac{2}{(n+1)(h+n-2)}\,
\lcx^\beta\left\{\delta^\mu_{[\alpha}\d_{\beta]}+\frac{1}{2h+n-3}\, X_{[\alpha\beta]}\d^\mu\right\}
O_{\mu n}(y,\lcx),\nonumber\\
&=\frac{1}{(n+1)(h+n-2)}\,
\left\{\delta^\mu_{\alpha}\,(\lcx\d)-\lcx^\mu\d_\alpha-\frac{n-1}{2h+n-3}\, \lcx_\alpha\d^\mu\right\}
O_{\mu n}(y,\lcx).
\end{align}
From the trace terms of Eqs.~(\ref{O-(2h-2)}) and (\ref{O-(2h-1)}) we construct 
the operator with twist $\tau=2h$ and spin $j=n-1$ according to
\begin{align}
\label{O-(2h)}
O^{{\rm tw}(2h)}_{n-1}(y,\lcx)&=\frac{1}{(2h+n-3)(h+n-1)}\, \d^\mu O_{\mu n}(y,\lcx).
\end{align}
The light-cone operators of geometric twist satisfy the following conditions 
of tracelessness on the light-cone:
\begin{align}
\d^\alpha\, O^{{\rm tw}(2h-2)}_{\alpha n}(y,\lcx)&=0,\\
\d^\alpha\, O^{{\rm tw}(2h-1)}_{\alpha n}(y,\lcx)&=0.
\end{align}
The light-cone operator~(\ref{O-(2h)}) is traceless by construction.

\chapter{Target mass corrections for a scalar theory
in $2h$-dimensional spacetime}
\label{mass}
\renewcommand{\theequation}{\thechapter.\arabic{equation}}
\setcounter{equation}{0}
The purpose of this Appendix is the investigation of the off-forward 
scattering amplitude with kinematical target mass corrections from the minimal
twist contributions in $2h$-dimensional Euclidean spacetime ($2h\ge 3$).
For simplicity we consider a scalar field theory.
We will illustrate this with the simplest non-forward
scattering process: the (scalar) meson production
by two virtual photons
\begin{align}
\gamma^*(q_1)+\gamma^*(q_2)\rightarrow M(P).
\end{align}
We denote the momenta of the two incoming photons by
$q_1$ and $q_2$, and 
the momentum of the outgoing scalar meson by $P$.
The non-forward scattering of this process may be written as
\begin{align}
T(P,q) 
= \int \d^{2h}x \,\e^{\ii qx}\,
\langle 0|T (J(x) J(-x))|P\rangle,\qquad q\equiv\frac{1}{2}\big(q_2-q_1\big),
\end{align}
with the scalar current $J(x)=\phi(x)\phi(x)$.

Using the Wick theorem and Born approximation we obtain:
\begin{align}
T (J(x) J(-x))\approx D_h(x)
\big(N( x,- x)+N(- x, x)\big),
\end{align}
where $D_h(x)$ is the canonical Euclidean propagator for a scalar
field in $2h$-dimensional spacetime~\cite{Craigie85}
\begin{align}
D_h(x)=\frac{\Gamma(h-1)}{4\pi^h [x^2]^{h-1}}.
\end{align}
Now, we expand the nonlocal operator in an infinite tower of local 
ones\footnote{Alternatively, we could use nonlocal operators. 
Then the nonlocal leading twist operator is given in terms 
of ${\cal H}_{h-1}$.}
\begin{align}
N(x,-x)=\sum_{n=0}^\infty\frac{\ii^n}{n!}\phi(0)(\ii x\Tensor D)^n\phi(0)
\end{align} 
We perform the Fourier transformation by the help of the following 
formula\footnote{More strictly, we have: $[x^2]\rightarrow [x^2-\ii 0]$ and 
$(q^2)\rightarrow (q^2+\ii 0)$.}
\begin{align}
\int\d^{2h} x\,\e^{\ii qx} \frac{x_{\mu_1}\ldots x_{\mu_n}}{[x^2]^k} 
=(-\ii)^n\pi^h 2^{2h-2k+n}\frac{\Gamma(h-k+n)}{\Gamma(k)}\,
\frac{q_{\mu_1}\ldots q_{\mu_n}}{(q^2)^{h-k+n}}+O(g_{\mu_i\mu_j}),
\end{align}
in particular with 
\begin{align}
\int\d^{2h} x\, \frac{\e^{\ii qx}}{[x^2]^{h-1}} 
\langle 0|N_n(x)|P\rangle=(-\ii)^n\pi^h 2^{2+n}\frac{\Gamma(1+n)}{\Gamma(h-1)}
\frac{1}{(q^2)^{1+n}}
\langle 0|N_n(q)|P\rangle,
\end{align}
where we neglected all terms involving $\square_q$, since they vanish 
for leading twist.
The leading twist operator has $\tau=2(h-1)$. It is in the $q$-space
a harmonic polynomial of degree $n$ in $q$ which satisfies
$\square_q N_n^{\rm l.t.}(q)=0$. 
Eventually, we use the harmonic projector in the $q$-space and obtain for
the corresponding matrix element
\begin{align}
\label{N-ma}
\langle 0|N^{\rm l.t.}_n(q)|P\rangle
&=\sum_{k=0}^{[\frac{n}{2}]}\frac{(-1)^k(h+n-k-2)!}{4^k k! (h+n-2)!}\, q^{2k} \square_q^k
(qP)^n f_n\nonumber\\
&=\frac{n!}{(h+n-2)!}\sum_{k=0}^{[\frac{n}{2}]}\frac{(-1)^k(h+n-k-2)!}{4^k k! (n-2k)!}\, q^{2k}P^{2k}
(qP)^{n-2k} f_n\nonumber\\
&=\frac{n!}{(h-1)_n}\left(\frac{1}{2} \sqrt{q^2P^2}\right)^n 
C^{h-1}_n\left(\frac{qP}{\sqrt{q^2 P^2}}\right) f_n,
\end{align}
where $f_n$ is the reduced matrix element.
The rewriting from the first to the third line in Eq.~(\ref {N-ma}) is 
nothing else but another form of a harmonic polynomial which is related
to the irreducible representation of the group $SO(2h)$.  
Accordingly, 
the scattering amplitude in terms of local operators appears as an expansion 
with respect to Gegenbauer polynomials:
\begin{align}
\label{CA2-sc}
T(P,q)=
\sum_{n=0}^\infty
\frac{\Gamma(n+1)\Gamma(h-1)}{\Gamma(n+h-1)}\,\frac{1}{q^2}
\Bigg(\sqrt{\frac{P^2}{q^2}}\Bigg)^n 
C^{h-1}_n\left(\frac{qP}{\sqrt{q^2 P^2}}\right) 
(1+(-1)^n) f_n,
\end{align}
which is very close to Nachtmann's~\cite{Nachtmann73} target mass corrections 
in inclusive processes for a scalar field theory ($h=2$).
Using the integral representation of Euler's beta functions for the 
Gamma functions in Eq.~(\ref{CA2-sc})
\begin{align}
\frac{\Gamma(n+1)\Gamma(h-1)}{\Gamma(n+h-1)}&=
\frac{(n+h-1)!}{(n+h-2)!}\, B(n+1,h-1)\nonumber\\
&=(h-1+n)\int_0^1\d t\, t^n (1-t)^{h-2},
\end{align}
and replace $n t^n\rightarrow t\pd_t t^n$,
we get the following expression for the scattering amplitude
\begin{align}
\label{CA3-sc}
T(P,q)&=
\int_{-1}^1\d\xi\, f(\xi)
\int_0^1\d t\, (1-t)^{h-2} \big(h-1+t\pd_t\big)
\frac{1}{q^2}\nonumber\\
&\qquad\times
\sum_{n=0}^\infty
\Bigg(t\xi\sqrt{\frac{P^2}{q^2}}\Bigg)^n 
C^{h-1}_n\left(\frac{qP}{\sqrt{q^2 P^2}}\right) 
(1+(-1)^n).
\end{align}
The resummation in Eq.~(\ref{CA2-sc}) can be done by making use of the 
generating function for the Gegenbauer polynomials
\begin{align}
\sum_{n=0}^\infty a^n C^{h-1}_n(b)
&=(1-2ab +a^2)^{-(h-1)}\nonumber\\
&=\frac{(q^2)^{h-1}}{(q-t\xi P)^{2(h-1)}}.
\end{align}
Here $a=t\xi\sqrt{P^2/q^2}$ and $b=qP/\sqrt{q^2 P^2}$.
Finally, we obtain the
scattering amplitude with all mass-corrections on twist $\tau=2(h-1)$ level 
\begin{align}
\label{CA4-sc}
T(P,q)=
\int_{-1}^1\d\xi\, f(\xi)
\int_0^1\d t\, (1-t)^{h-2} (q^2)^{h-2}\big(h-1+t\pd_t\big)
\left(\frac{1}{(q+t\xi P)^{2(h-1)}}+\frac{1}{(q-t\xi P)^{2(h-1)}}\right).
\end{align}
If we carry out the derivative in Eq.~(\ref{CA4-sc}), we obtain
\begin{align}
\label{CA5-sc}
T(P,q)=
\int_{-1}^1\d\xi\, f(\xi)
\int_0^1\d t\, &(1-t)^{h-2} (q^2)^{h-2}
\bigg\{
\big(h-1+2t^2\big)
\left(\frac{1}{(q+t\xi P)^{2(h-1)}}+\frac{1}{(q-t\xi P)^{2(h-1)}}\right)\nonumber\\
&-2 t^2(h-1)
\left(\frac{t\xi\, qP+t^2\xi^2 P^2}{(q+t\xi P)^{2h}}
-\frac{t\xi\, qP-t^2\xi^2 P^2}{(q-t\xi P)^{2h}}\right)
\bigg\}.
\end{align}
Thus, we are able to express the scattering amplitude in terms of a
DA $f^{(\tau)}(\xi)$ with twist $\tau=2(h-1)$ and a mass-dependent
coefficient function.

For $h=2$ we perform the partial integration with respect to $t$ in 
Eq.~(\ref{CA4-sc}) and
obtain the simple expression for the scattering amplitude
\begin{align}
\label{CA6-sc}
T(P,q)=\int_{-1}^1\d\xi\, f(\xi)
\left(\frac{1}{(q+\xi P)^2}+\frac{1}{(q-\xi P)^{2}}\right).
\end{align}
From this point of view, one may introduce the following kinematical DA:
\begin{align}
{\cal F}(\xi,q,P)=
f(\xi)
\left(\frac{1}{(q+\xi P)^{2}}+\frac{1}{(q-\xi P)^{2}}\right),
\end{align}
which contains the power corrections of the DA $f(\xi)$ on twist
$\tau=2$ level.
If we expand the scattering amplitude (\ref{CA6-sc}) in powers of $P^2/q^2$, 
we finally get the simple formula:
\begin{align}
T(P,q)= \frac{2}{q^{2}}
\sum_{n=0}^\infty (n+1)\frac{P^n}{q^n}\, f_n,\qquad n-\text{even}.
\end{align}
\end{appendix}



\end{document}